\documentclass[a4paper,12pt,twoside,english]{myreport}
\usepackage{babel}
\usepackage{accents}
\usepackage{amsmath}
\usepackage{amssymb}
\usepackage[toc,page]{appendix}
\usepackage{dcolumn}
\newcommand{\rot}{\mathrm{rot}\,}
\newcommand{\laplace}{{\vartriangle}\,}
\newcommand{\eps}{\varepsilon}
\newcommand{\epso}{\varepsilon_{\mathrm{o}}}
\newcommand{\mb}{{\overline{m}}}
\newcommand{\frm}[1]{{\Theta^{(#1)}}}
\newcommand{\NP}[1]{{#1}_{\scriptscriptstyle\mathrm{NP}}}
\newcommand{\EM}{{\scriptscriptstyle\mathrm{EM}}}
\newcommand{\gyr}{{\scriptscriptstyle\mathrm{gyr}}}
\newcommand{\vrho}{\varrho_{\scriptscriptstyle\mathrm{EM}}}
\newcommand{\vs}{{\bf s}}
\newcommand{\trans}{{\scriptscriptstyle\top}}
\newcommand{\sreal}{\sigma}
\newcommand{\sslfdl}{s}
\newcommand{\ro}{{\mathrm{o}}}

\newcommand{\spcpnct}{\;}                              
\newcommand{\period}{{\mbox{\spcpnct.}\relax}}         
\newcommand{\commae}{{\mbox{\spcpnct,}\relax}}         
\newcommand{\comma}{{\mbox{\spcpnct,\quad}\relax}}     
\newcommand{\nquad}{{\mspace{-20mu}}}                  

\DeclareMathAccent{\frmarr}{\mathord}{letters}{"7E}

\newcommand{\cv}[1]{\partial_{#1}}
\newcommand{\lder}[1]{\mathcal{L}_{\displaystyle #1}}

\let\div\relax
\newcommand{\grad}[1][]{{}^\st\mspace{-2mu}\mathrm{d}_{#1}\mspace{-1mu}}
\DeclareMathOperator{\covd}{{}^\st\mspace{-1mu}\nabla\!}
\DeclareMathOperator{\div}{{}^\st\mspace{-2mu}div}

\newcommand{\Ric}{{}^\st\mathrm{Ric}}
\newcommand{\scR}{{}^\st\mspace{-2mu}\mathcal{R}}

\newcommand{\rsix}{{}^\sharp\!}
\newcommand{\lwix}{{}^\flat\!}

\newcommand{\rder}{\dot}
\newcommand{\rdder}{\ddot}
\newcommand{\uder}{\mathring}

\DeclareMathOperator{\Det}{Det}

\newcommand{\trpr}{{\scriptscriptstyle\textsf{T}}}

\newcommand{\trvol}{{\mathfrak{q}^{\frac12}}}
\newcommand{\trivol}{{\mathfrak{q}^{-\frac12}}}

\newcommand{\trgrad}[1][]{\mathrm{d}_{\mspace{1mu}#1}\mspace{-1mu}}
\DeclareMathOperator{\trcovd}{\nabla\!}
\DeclareMathOperator{\trdiv}{div}
\DeclareMathOperator{\trlapl}{\triangle}
\DeclareMathOperator{\trLB}{\nabla^2\!}
\newcommand{\trRiem}{R}
\newcommand{\trRic}{\mathrm{Ric}}
\newcommand{\trscR}{\mathcal{R}}

\newcommand{\kap}{\varkappa}

\newcommand{\folph}{{\displaystyle\centerdot}}

\newcommand{\mg}{{\scriptscriptstyle\mathrm{B}}}
\newcommand{\trspc}{{\scriptscriptstyle\mathrm{T}}}
\newcommand{\TF}{{\scriptscriptstyle\mathrm{tf}}}
\newcommand{\st}{{\scriptscriptstyle \mathit{D}}}
\newcommand{\tot}{{\scriptscriptstyle\mathrm{tot}}}

\usepackage{graphics}
\usepackage{epsfig}
\usepackage[cp1250]{inputenc}
\usepackage[text={5.7in,8.75in},twoside]{geometry}
\usepackage{fancyhdr}
\newcommand{\helv}{%
      \fontfamily{phv}\fontseries{b}\fontsize{9}{11}\selectfont}
\fancypagestyle{plain}{%
           \fancyhf{}%
           \fancyfoot[C]{\thepage}
           
           }

\def\titulpicEN{\thispagestyle{empty}
   \vspace*{10mm}
   \begin{center}
   {\large Charles University in Prague} \par \medskip
   {\large Faculty of Mathematics and Physics} \par \medskip
   \vspace*{10mm}
   \vbox to 30mm{\vspace*{\fill}
   {\huge\bfseries Doctoral thesis} \par \medskip
   \vspace*{\fill}}\par
   \end{center}
}

\def\autorpicEN#1#2#3#4#5{\thispagestyle{empty}
   \begin{center}
   \vspace*{10mm}
   {\large #1} \par \medskip
   {\large \it #2} \par \medskip
   {\large #3} \par \medskip
   {\large Supervisor: #4} \par \medskip
   {\large Branch of study: #5} \par \medskip
   \end{center}%
}

\def\abstrakten#1#2#3#4#5#6#7{\vspace*{10mm}\thispagestyle{empty}
    {\noindent Title: #5} \par\smallskip
    {\noindent Author: #1} \par\smallskip
    {\noindent Department: #2}\par\smallskip
    {\noindent Supervisor: #3}\par\smallskip
    {\noindent Supervisor's e-mail address: #4}\par\smallskip
    {\noindent Abstract: #6}\par\smallskip
    {\noindent Keywords: #7}\par\smallskip
    \vspace*{\fill}}
\def\abstrakt#1#2#3#4#5#6#7{\vspace*{10mm}\thispagestyle{empty}
    {\noindent N\'{a}zev pr\'{a}ce: #5} \par\smallskip
    {\noindent Autor: #1} \par\smallskip
    {\noindent \'{U}stav: #2}\par\smallskip
    {\noindent \v{S}kolitel: #3}\par\smallskip
    {\noindent \v{S}kolitelova e-mailov\'{a} adresa: #4}\par\smallskip
    {\noindent Abstrakt: #6}\par\smallskip
    {\noindent Kl\'{i}\v{c}ov\'{a} slova: #7}\par\smallskip
    \vspace*{\fill}}

\def\Contens{\cleardoublepage\thispagestyle{empty}\tableofcontents\cleardoublepage}
\begin{document}
\numberwithin{equation}{section}
 \pagestyle{empty}
 \cleardoublepage\pagestyle{empty}
 \titulpicEN{\parbox{\textwidth}{\centerline{\epsfig{file=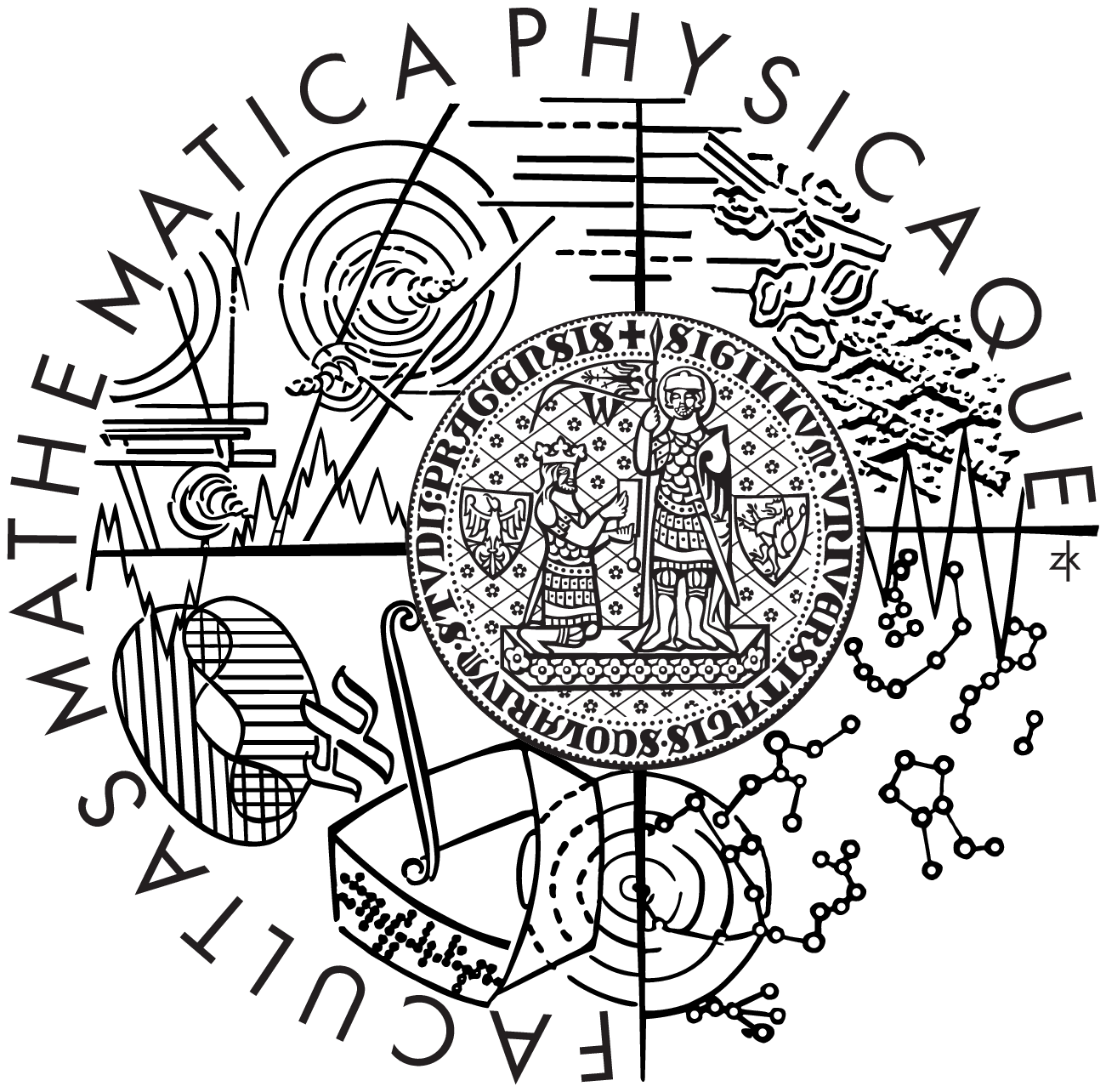, scale=0.45, angle=0}}}}
\par\medskip
\vspace*{\fill}\par
\autorpicEN{Hedvika Kadlecov\'{a}}{{\bf Gravitational field of gyratons on various background spacetimes}} { Institute of Theoretical Physics}{doc.~Pavel~Krtou\v{s}, Ph.D.}{F--1 Theoretical physics, astronomy and astrophysics}
\setcounter{page}{1}
\newpage
\begin{center}{\large\bf Acknowledgements}\end{center}\par
\medskip
First of all, I would like to thank to my supervisor, Pavel Krtou\v{s}, for his guidance during my Ph.D. studies.
I wish to thank to Andrei Zelnikov, Valeri Frolov and Don Page for their hospitality and for
discussions during my two research stays at the University of Alberta in Canada.
I am grateful to Ji\v{r}\'{i} Podolsk\'{y} and to Otakar Sv\'{i}tek for many helpful discussions.

Finally, (and most of all), I want to thank to my family, to my mother, grandmother, and to Tom\'{a}\v{s} Pech\'{a}\v{c}ek.
Without their love, support and motivation, these student's years would be much harder. \\

This work was supported by grants GA\v{C}R-202/09/H033, GAUK 12209 by the Czech Ministry of
Education, the project SVV 261301 of the Charles University in Prague and
the LC06014 project of the Center of Theoretical Astrophysics.
\newpage
A mathematician, like a painter or poet, is a maker\\
of patterns. If patterns are more permanent than theirs,\\
it is because they are made with ideas.\\

G. H. Hardy
\Contens
\cleardoublepage

\abstrakten{\it Hedvika Kadlecov\'{a}}{\it Institute of Theoretical Physics}{\it doc. Pavel Krtou\v{s}, Ph.D.}{\it Pavel.Krtous@mff.cuni.cz}{\it
Gravitational field of gyratons on various background spacetimes}{\it In this work we have found and analyzed several gyraton solutions on various non--trivial backgrounds in the large Kundt class of spacetimes. Namely, the gyraton solutions on direct product spacetimes, gyraton solutions on Melvin universe and its generalization which includes the cosmological constant. These solutions are of algebraic type II. Also we have investigated type III solutions within the Kundt class and we have found the gyratons on de Sitter spacetime. We have generalized the gyraton solutions on direct product spacetimes to higher dimensions.} {\it Kundt class of spacetimes, gravitational waves, Einstein--Maxwell equations, NP formalism}

\abstrakt{\it Hedvika Kadlecov\'{a}}{\it Institut teoretick\'{e} fyziky}{\it doc. Pavel Krtou\v{s}, Ph.D.}{\it Pavel.Krtous@mff.cuni.cz}{\it
Gravita\v{c}n\'{i} pole  gyraton\r{u} na pozad\'{i}ch r\r{u}zn\'{y}ch prostoro\v{c}as\r{u}}{\it V t\'{e}to pr\'{a}ci jsme nalezli a analyzovali n\v{e}kolik r\r{u}zn\'{y}ch gyratonov\'{y}ch \v{r}e\v{s}en\'{i} na r\r{u}zn\'{y}ch netrivi\'{a}ln\'{i}ch pozad\'{i}ch z \v{s}irok\'{e} t\v{r}\'{i}d\v{e} Kundtov\'{y}ch prostoro\v{c}as\r{u}: gyratonov\'{a} \v{r}e\v{s}en\'{i} na prostoro\v{c}asech tvo\v{r}en\'{y}ch p\v{r}\'{i}m\'{y}m sou\v{c}inem prostoro\v{c}as\r{u}, gyratonov\'{a} \v{r}e\v{s}en\'{i} na Melvinov\v{e} vesm\'{i}ru a
jeho zobecn\v{e}n\'{i} p\v{r}ipou\v{s}t\v{e}j\'{i}c\'{i} nenulovou kosmologickou konstantu. Tato  \v{r}e\v{s}en\'{i} jsou algebraick\'{e}ho typu II. Tak\'{e} jsme zkoumali \v{r}e\v{s}en\'{i} typu III v r\'{a}mci Kundtovy t\v{r}\'{i}dy prostoro\v{c}as\r{u} a na\v{s}li jsme gyratony na de Sitterov\v{e} prosotoro\v{c}ase. Zobecnili jsme gyratonov\'{a} \v{r}e\v{s}en\'{i} na prostoro\v{c}asech tvo\v{r}en\'{y}ch p\v{r}\'{i}m\'{y}m sou\v{c}inem prostoro\v{c}as\r{u} do vy\v{s}\v{s}\'{i}ch dimenz\'{i}.} {\it Kundtova t\v{r}\'{i}da prostoro\v{c}as\r{u}, gravita\v{c}n\'{i} vlny, Einstein--Maxwellovy rovnice, NP formalismus}
\cleardoublepage
\pagestyle{fancy}
\setcounter{page}{1}
\setcounter{chapter}{0}
\setcounter{section}{0}

\chapter{Introduction}

The class of exact solutions of Einstein--Maxwell equations which describes non--expanding gravitational waves is the Kundt class of spacetimes  \cite{Step:2003:Cam:,GrifPod:2009:Cam:}. It is characterized by the property that it admits a geodesics, shear--free, twist--free and non--expanding null congruence $k$. The class includes special subclasses  such as the well--known {\it pp}--waves, plane--wave spacetimes,
 generalized {\it pp}--waves, Kundt waves, VSI or CSI\footnote{These are abbreviations for Vanishing Scalar Invariants and Constant Scalar Invariants} spacetimes and gyratons.

The Kundt spacetimes are algebraically special, i.e. are of Petrov type II, D, III or N (or conformally flat) with $k$ being a repeated principal null direction of the Weyl tensor. Any Einstein--Maxwell and pure radiation fields are also aligned, i.e. $k$ is the common eigendirection of the Weyl and Ricci tensor.
The null congruence $k$ is not in general covariantly constant, therefore the rays of the corresponding non--expanding waves are not necessarily parallel, as in the case of {\it pp}--waves, and the wave fronts need not to be planar.

Recently, there has been a growing interest in the investigation of gyraton spacetimes because they have an interesting physical interpretation.
 The solutions represent the gravitational field of a localized source with an intrinsic rotation, moving at the speed of light. Such an idealized ultrarelativistic source, which can be modeled as a pulse of a spinning radiation beam, is accompanied by a sandwich or impulsive gravitational wave.

Historically, the gravitational fields generated by (nonrotating) light pulses and beams were already studied by Tolman \cite{Tol:1934:Oxf:} in 1934, who obtained the corresponding solution in the linear approximation of the Einstein theory. Exact solutions of the Einstein--Maxwell equations for such ``pencils of light'' were found and analyzed by Peres \cite{Peres:1960:PHYSR:} and Bonnor \cite{Bonnor:1969:COMMPH:,Bonnor:1969:INTHP:,Bonnor:1970a:INTHP:}. These solutions belong to a general family of {\it pp\,}-waves \cite{Step:2003:Cam:,GrifPod:2009:Cam:}.

The gyraton solution in the Minkowski spacetime reduces to the well-known Aichelburg--Sexl metric \cite{Aich-Sexl:1971:} in the impulsive limit (i.e. for an infinitely small cross-section of the beam, and for the delta-type distribution of the light-pulse in time) which describes the field of a point-like null particle. The Aichelburg--Sexl metric can be obtained by boosting the Schwarzschild metric to the speed of light, with the mass tending to zero so that the total energy is kept finite, i.e. in the Penrose limit. More general impulsive waves were subsequently obtained by boosting other black hole spacetimes with rotation, charge and a cosmological constant \cite{FerPen90,LouSan92,HotTan93,KBalNac95,KBalNac96,PodGri97,PodGri98prd} (for recent reviews, see \cite{Podolsky02,BarHog:2003:WorldSci:}).

Gyraton solutions are special sandwich or impulsive waves of the Kundt class (which generalize the {\it pp\,}-waves) such that the corresponding beam of radiation carries not only energy but also an additional angular momentum. Such spacetimes were first considered by Bonnor in \cite{Bonnor:1970b:INTHP:}, who studied the gravitational field created by a spinning null fluid. He called the corresponding particle made out of this continuum a ``spinning nullicon''. In some cases, this may be interpreted as a massless neutrino field \cite{Griffiths:1972a:INTHP:}.

These solutions are locally isometric to standard {\it pp\,}-waves in the exterior vacuum region outside the source. The interior region contains a nonexpanding null matter which has an intrinsic spin.
Gyratons are generally obtained by keeping these nondiagonal terms $g_{ui}$ in the Kundt form of the metric, where $u$ is the null coordinate and $x^i$ are coordinates on the transversal space. The corresponding energy-momentum tensor thus also contains an extra nondiagonal term $T_{ui}=j_{i}$. In four dimensional Minkowski spacetime, the terms $g_{ui}$ can be set to zero {\it locally}, using a suitable gauge transformation.

In  the section \eqref{sc:gyr} of this introduction we briefly review the known gyraton solutions, namely in subsection \eqref{ssc:gyrMin} the gyraton propagating in the Minkowski spacetime is presented, in \eqref{ssc:gyrSit} the gyraton in the asymptotically
anti--de Sitter spacetime is reviewed, in \eqref{scc:other} we discuss the gyratons which were subsequently derived in other theories.
Finally,  in the sections \eqref{sc:contr} and \eqref{sc:this_th} we describe the main goals and the overall layout of the theses.

\section{The gyraton solutions}\label{sc:gyr}

The gyraton solutions were until recently  known only on the Minkowski spacetime \cite{Fro-Fur:2005:PHYSR4:,Fro-Is-Zel:2005:PHYSR4:,Fro-Zel:2006:CLAQG:} and
in the asymptotically anti--de Sitter spacetime \cite{Fro-Zel:2005:PHYSR4:} in higher dimensions. The solutions have many similar properties:
The Einstein--Maxwell equations reduce to the set of linear equations on $(D-2)$-dimensional subspace, the scalar polynomial invariants are
vanishing (Minkowski case) or are constant (anti--de Sitter).
These solutions belong to the higher dimensional Kundt class of spacetimes which was recently presented in \cite{PodoZo:2009:CLAQG:}.

In the following we will review briefly the basic properties of these solutions.

\subsection{The gyratons on Minkowski background}\label{ssc:gyrMin}
The gyratons were investigated in a higher dimensional flat space in linear approximation in \cite{Fro-Fur:2005:PHYSR4:}. These gyratons represent a pulse of circularly polarized radiation or a modulated beam of ultrarelativistic particles with spin or other sources, which have finite energy~$E$ and finite total angular momentum~$J$. The gyraton itself is characterized by two arbitrary profile functions of $u$ which determine the energy density and angular momentum. The authors investigated the limit in which the source becomes infinitesimally small (with a negligible radius of the cross-section) and the profile functions are independent. They also studied the geodesic motion of test particles in the field of gyraton and demonstrated that, when the gyraton passes through the center of the ring of test particles, the particles start to rotate. In fact, the gyraton's angular momentum effectively creates a force which is similar to the usual centrifugal repulsive force, while the gyraton energy produces the attractive ``Newtonian'' force.

In \cite{Fro-Is-Zel:2005:PHYSR4:} they further investigated the exact gyraton solutions propagating in an asymptotically flat D-dimensional spacetime and proved that the Einstein's equations for gyratons reduce to a set of linear equations in the Euclidean ${(D-2)}$-dimensional space. They also showed that the gyraton metrics belong to a class of vanishing scalar curvature invariants (VSI) spacetimes for which all polynomial scalar invariants, constructed from the curvature and its covariant derivatives, vanish identically \cite{Prav-Prav:2002:CLAQG:}.  Subsequently, their charged version in arbitrary dimension was presented in \cite{Fro-Zel:2006:CLAQG:}.

Now, let us briefly review the basic properties of the gyratons in the Minkowski spacetime in D--dimensional spacetime explicitly.
The general ansatz for the gyraton metric in the D--dimensional Minkowski spacetime. The general ansatz is in fact
the null D-dimensional Brinkmann metric \cite{Brink:1925:MAAN:}
\begin{equation}
\trgrad s^2=-2\,\trgrad u\, \trgrad v + \trgrad {x}^2+\phi(u,{x})\,\trgrad u^2+2\,a_{i}\,\trgrad x^{i}\,\trgrad u,\label{gyrF}
\end{equation}
where we denote the flat transversal metric space $\trgrad {x}^2=\sum^{D}_{i=3} (\trgrad x^{i})^2$,
and the functions $\phi$ (gravitoelectric potential) and $a_{i}$ (gravitomagnetic potential) can be considered as a scalar and a vector field in the ($D-2$)-dimensional Euclidean space with Cartesian coordinates $x^{i}$. They depend also on an external parameter $u$ but not on $v$.
The metric \eqref{gyrF} reduces to Minkowski metric for ${a}_{i}={\phi}=0$ and it has vanishing scalar polynomial invariants  \cite{Prav-Prav:2002:CLAQG:}.

The null Killing vector is ${k}=k^{\mu}\partial_{\mu}=\partial_{v}$. In the null hypersurfaces $u=\text{const}$ we use as the coordinates the affine parameter $v$ and the  spatial coordinates $x^{i}$ ($i=3,\dots,D$). Instead of $u, v$ we also use coordinates $t,\;\xi$ given by $u=(t-\xi)/\sqrt{2}$, $v=(t+\xi)/\sqrt{2}$.
The metric \eqref{gyrF} then describes an object moving with the speed of light in the $\xi$ direction.
The coordinates $(x^{3},\dots, x^{D})$ are coordinates of an $D-2$-dimensional space which is transverse to the direction of motion.
 We denote covariant derivatives with respect to the flat spatial metric in the transverse space by semicolon, e.g. $()_{:a}$.

The null Killing vector ${k}$ is covariantly constant, i.e.
\begin{equation}\label{kov}
  k_{{\mu};{\nu}}=0.
\end{equation}

We introduce the antisymmetric tensor in the ($D-2$)-transversal plane,
\begin{equation}\label{F}
  f_{ij}=\partial_{i}a_{j}-\partial_{j}a_{i}.
\end{equation}
The Einstein equations reduce to two sets of equations in ($D-2$)-dimensional flat space
\begin{align}
  {f_{ij}}^{:j}&=j_{i},\label{jedna}\\
  \laplace \phi&=-j+\frac{1}{2}f_{ij}f^{ij}+2\,\partial_{u} \trdiv a,\label{druha}
\end{align}
where $j$ and $j_{i}$ are the gyraton sources.

The first set of equations \eqref{jedna} formally coincides with the Euclidean Maxwell equations in ($D-2$) dimensions
 where $j_{i}$ plays the role of the current. We need to find the static magnetic potential $a_{i}$ created by the gyraton source. The second equation \eqref{druha} is similar to the equation for the electric potential $\phi$ with the important difference that in addition to the charge distribution $j$ it contains an extra source proportional to $f_{ij}f^{ij}$.

The source terms $j$ and $j_{i}$ vanish outside the source of the gyraton.

In order to obtain a solution describing the total spacetime one needs to obtain a solution inside source of the gyraton. The specific gyraton models are discussed in \cite{Yosh:2007:PHYSR4:}. But if we obtain a general solution for the vacuum metric outside a gyraton source, it is guaranteed that for any gyraton model  there exists a corresponding solution.
Thus it is possible to find an exact solution of Einstein  equations for an arbitrary finite size source. Therefore it is reasonable to consider first point like source distributions in the transverse space. But the solutions may be only formal and may not have well-defined sense because $f_{ij}$ would have a singularity at ${x}=0$, see \cite{Fro-Is-Zel:2005:PHYSR4:}.

\subsection{The gyratons on anti--de Sitter background}\label{ssc:gyrSit}
In \cite{Fro-Zel:2005:PHYSR4:},  Frolov and Zelnikov took a cosmological constant into account and they explicitly found exact solution for gyratons in the asymptotically anti--de~Sitter spacetime. Namely, they obtained Siklos gyratons which generalize the Siklos family of nonexpanding waves \cite{Sik:1985:Cam:,Pod-rot:1998:CLAQG:}.

In this case, all polynomial scalar invariants are independent of the arbitrary metric functions which characterize the gyraton and have the same values as the corresponding invariants of pure anti--de~Sitter background. The AdS gyratons \cite{Fro-Zel:2005:PHYSR4:} thus belong to the class of spacetimes with constant scalar invariants (CSI) \cite{Coley-Her-Pel:2006:CLAQG,  Coley-Gib-Her-Pope:2008:CLAQG, Coley-Her-Pel:2008:CLAQG, Coley-Her-Pel:2009:CLAQG, Coley-Her-Pap-Pel:2009:}. In string theory it has been demonstrated that generalized {\it pp\,}-wave spacetimes do not get any quantum and $\alpha'$ corrections and hence are perturbatively exact. One may expect a similar property to be valid also for the gyratons, but more careful analy\-sis is required since even if all of the local counterterms in the effective action are trivial constants for CSI spacetimes, their metric variations can be nontrivial functions. Still, one can try to generalize the property of relatively simple quantum corrections to the case of semiuniversal metrics \cite{ Coley-Gib-Her-Pope:2008:CLAQG} when the Ricci tensor has a block-diagonal structure.

Let us consider a gyraton propagating in the $D-$dimensional asymptotically AdS background.

It is well-known that a pure AdS spacetime is conformal to the Minkowski spacetime, as
\begin{equation}\label{conformal trafo}
{\trgrad s^2_{\mathrm{AdS}}}=\Omega^2 \trgrad s^2_{\mathrm{Min}}
\end{equation}
where $\Omega$ is a specific conformal factor.
Using the conformal factor $\Omega=\frac{L^2}{z^2}$ with $z\equiv x^{3}$, the conformally related spacetime (AdS) remains homogeneous and isotropic, but instead of the Poincare group its isometry group is $SO(D-1,2)$.
The points $z=0$ correspond to the spatial infinity and $z=\infty$ is the horizon defined for the set of observers at the rest at constant $z$.

We multiply the metric \eqref{gyrF} in flat spacetime by conformal factor $\Omega=\frac{L^2}{z^2}$. The ansatz for the gyraton metric in AdS then has the following form
\begin{equation}
\trgrad s^2=\frac{L^2}{z^2}\left(-2\,\trgrad u \,\trgrad v + \trgrad {x}^2+\phi(u,{x})\,\trgrad u^2+2\,a_{i}\,\trgrad x^{i}\,\trgrad u \right). \label{gyrA}
\end{equation}
where $z\equiv x^{3}$ and the constant $L=\sqrt{-2\Lambda/(D-1)(D-2)}$ is the radius of the curvature of AdS, $\Lambda$ being negative cosmological constant.
The metric \eqref{gyrA} reduces to the pure AdS metric in absence of the gyraton $\phi={a}_{i}=0$. This property is preserved asymptotically
if we assume that both functions $\phi$ and ${a}_{i}$ vanish at the infinity of the transverse space.

An important property of the gyraton solutions in AdS is that all curvature invariants that can be built from the metric
are independent of the arbitrary functions in the metric ($\phi$ and $a_{i}$) that characterize the geometry, and acquire the same value as for the pure AdS spacetime. The gyratons in AdS are free of curvature singularities and are regular everywhere. So this implies these solutions do not get any $\alpha'$ corrections and are perturbatively exact in string theory.

By substituting the metric \eqref{gyrA} into Einstein equations with cosmological constant
we get two non-trivial equations
\begin{align}
  {f_{ij}}^{:j}-\frac{D-2}{z}f_{iz}&=j_{i},\label{vrr1}\\
  \laplace\phi-\frac{1}{2}f_{ij}f^{ij}-2\,\partial_{u}\trdiv a
  -\frac{D-2}{z}(\partial_{z}\phi&-2\,\partial_{u}a_{z})=-j.\label{vrr2}
\end{align}
The above solutions can be found explicitly using the Green functions for the AdS background spacetime.
First, we look for the solution of the equation \eqref{vrr1} and then we are able to solve the second
equation \eqref{vrr2} and find the function $\phi$. For explicit solutions, namely in four and five dimensions, see \cite{Fro-Zel:2005:PHYSR4:}.

\subsection{Other gyraton solutions -- applications}\label{scc:other}
The previously presented gyraton solutions have been applied to other theories.
 Supersymmetric gyraton solutions were obtained for a minimal five--dimensional gauged supergravity theory in \cite{Cald-Kle-Zor:2007:CLAQG:}, where they showed under which conditions the solution preserves part of the supersymmetry. The configuration represents a generalization of the Siklos waves with a nonzero angular momentum in anti--de~Sitter space and possess a Siklos--Virasoro reparametrization invariance.

The generalization of electrically charged gyratons to the theory of supergravity was found in \cite{Fro-Li:2006:PHYSR4:}.

The gravitational field generated by gyratons may be interesting for studies of production of mini black holes in colliders or in cosmic ray experiments. The problem of mini black hole formation in high energy particle collisions is an important issue of TeV gravity scenarios. The theory of such collisions, developed in \cite{Ear-Gid:2002:PHYSR4:,Yosh:2002:PHYSR4:, Yosh:2003:PHYSR4:, Yosh:2005:PHYSR4:, Yosh:2006:PHYSR4:}, was applied to gyraton models in \cite{Yosh:2007:PHYSR4:}.
The last paper studies head-on collisions of two gyratons and black hole formations in these processes. For simplification, several gyraton models were introduced, with special profiles of energy and spin density distribution. In the gyraton models the metric outside the source satisfies the vacuum Einstein equations and the gravitational field is distributed in the plane transverse to the direction of motion. It was demonstrated that it is sufficient to study the apparent horizon formation on the future edge of spacetime before interaction because the existence of an apparent horizon is a sufficient condition for the black hole formation. The apparent horizon forms only if the energy duration and the spin are smaller than some critical values.

\section{Our contribution}\label{sc:contr}
The main motivation for the study was to find new gyraton solutions on different backgrounds and give them proper interpretation because there are only few solutions of type II which are explicitly known in
the Kundt class of spacetimes.

In our first  paper \cite{Kadlecova:2009:PHYSR4:} we have found and analyzed new exact gyraton solutions of algebraic type II on backgrounds which are a direct-product of two 2-spaces of constant curvature. This family of (electro)vacuum
background spacetimes contains the Nariai, anti--Nariai and Pleba\'{n}ski--Hacyan universes of type~D,
or conformally flat Bertotti--Robinson and Minkowski spaces.

These gyraton solutions are given in a simple Kundt metric form and belong to the recently discussed class of spacetimes
with constant scalar invariants (CSI) of the curvature tensor. We have also shown that the Einstein equations reduce
to a set of linear equations on the transverse 2-space which can be explicitly solved using the Green functions.
In general, they have all basic characteristics as the previous gyraton solutions.

We were able to define the gyraton only on phenomenological level, i.e. by its stress-energy tensor which is assumed to be given. Then the aim was to determine the gyratonic influence on the metric and the electromagnetic field of the background solutions.
The gyratonic matter is again null, non--expanding with internal rotation, i.e. it has an
intrinsic spin.

In addition to the previous gyraton solutions we have investigated the gyratons in Newman--Penrose formalism and we have
demonstrated explicitly that the gyraton solutions on direct product spacetimes belong to the Kundt class of spacetimes and are of Petrov type II. Also this result is valid for the gyratons in the Minkowski spacetime since it is a special subcase of gyratons on direct product spacetimes.

Furthermore, we have found the gyraton solutions on Melvin universe \cite{Kadlecova:2010:PHYSR4:} of type II in four dimensions which is counterpart to our investigation
of gyraton on direct--product spacetimes. This work has been accepted for publication to Physical Review D. We also found gyraton solutions on generalized Melvin spacetime with non--vanishing cosmological constant \cite{KadlKrt:2010:CLAQG:}.

Next we investigate the higher dimensional generalization of the gyraton on direct--product spacetimes where
we discuss the possibility of general transversal spacetime.

The last goal of the thesis was to find gyraton also in de Sitter spacetime  and to find more gyraton solutions of type III in Kundt class,  following the discussion of the paper \cite{GrifDochPod:2004:CLAQG:}.

In our investigation we have found several new exact gyraton solutions of type II or III with similar and gradually more
complicated properties within the Kundt class of spacetimes (four and higher dimensional).
Our results would help in the further understanding of these interesting solutions. For example,
the fact that the invariants have the same values for the background spacetime itself and for the full gyratonic
metric is very interesting. This property will be very useful when gyratons would
be assumed in string theory. Other application is the better understanding of the particle collisions on different
backgrounds.

\section{In this thesis}\label{sc:this_th}

The thesis is organized as follows:

In section 2 we present the large class of gyraton solutions of type II on backgrounds which0
are formed by direct--product of two 2--spaces of constant curvature \cite{Kadlecova:2009:PHYSR4:}.
In fact, we have found all possible backgrounds of type D which are included in the considered
Kundt class of solutions. We include the paper which appeared in Physical Review D, let us note that
the extended version can be found in the arxiv:0905.2476.

In section 3 we present the gyraton solutions on Melvin background spacetime \cite{Kadlecova:2010:PHYSR4:}.
We include the extended version of the paper which can be found in the
arxiv:1006.1794v1, this work has been accepted for publication to Physical Review D.

In section 4 we present the gyraton solutions on generalized Melvin background spacetime which includes
also possible cosmological constant. In fact, this generalization contain the previous two solutions
as a subcases. This work is new and it will be soon prepared for publication \cite{KadlKrt:2010:CLAQG:}.

In section 5 we investigate the gyraton solutions of algebraical type III in
the Kundt class of spacetimes in four dimensions.
First, we review the basic theory about the Kundt solutions of type III, then
we derive various gyratons on conformally flat spacetimes ( Minkowski, anti--de
Sitter and de Sitter spacetimes) for different choices of $\tau$.
We also derive the Einstein equations for general $\tau$.
This section contains review of results from \cite{GrifDochPod:2004:CLAQG:} and some new results
specific for the gyratonic matter. It presents the work in progress.

In section 6 we analyze the higher dimensional generalization of the gyraton solutions
on direct product spacetimes. We introduce and discuss the higher dimensional form of
the metric from subclass of the Kundt family and we derive field equations.
It is the work in progress \cite{Krtous:2010:}.

\newpage\cleardoublepage
\chapter{The gyraton solutions on direct product spacetimes}
In this chapter we present the paper about the gyraton solutions on
backgrounds which are formed by direct product of 2--spaces of constant
curvature. In general, the solutions are of algebraic type II and the
backgrounds are type D.

We investigate the Einstein--Maxwell equations which reduce to 2--dimensional
transverse space and we discuss how to solve them. The metrics are also
studied in the Newman--Penrose formalism which enabled us to determine
the algebraic type of those solutions and we have found all possible
backgrounds of type D within the considered class of solutions.

\section{\label{sec:level1}Introduction}

Recently, there has been a growing interest in investigation of gyraton spacetimes. They represent the gravitational field of a localized source with an intrinsic rotation, moving at the speed of light. Such an idealized ultrarelativistic source, which can be modeled as a pulse of a spinning radiation beam, is accompanied by a sandwich or impulsive gravitational wave. 

In fact, gravitational fields generated by (nonrotating) light pulses and beams were already studied by Tolman \cite{Tol:1934:Oxf:} in 1934, who obtained the corresponding solution in the linear approximation of the Einstein theory. Exact solutions of the Einstein--Maxwell equations for such ``pencils of light'' were found and analyzed by Peres \cite{Peres:1960:PHYSR:} and Bonnor \cite{Bonnor:1969:COMMPH:,Bonnor:1969:INTHP:,Bonnor:1970a:INTHP:}. These solutions belong to a general family of {\it pp\,}-waves \cite{Step:2003:Cam:,GrifPod:2009:Cam:}.

In the impulsive limit (i.e. for an infinitely small cross-section of the beam, and for the delta-type distribution of the light-pulse in time), the simplest of these solutions reduces to the well-known Aichelburg--Sexl metric \cite{Aich-Sexl:1971:} which describes the field of a point-like null particle. It can be obtained by boosting the Schwarzschild metric to the speed of light, with the mass tending to zero so that the total energy is kept finite. More general impulsive waves were subsequently obtained by boosting other black hole spacetimes with rotation, charge and a cosmological constant \cite{FerPen90,LouSan92,HotTan93,KBalNac95,KBalNac96,PodGri97,PodGri98prd} (for recent reviews, see \cite{Podolsky02,BarHog:2003:WorldSci:}).

Gyraton solutions are special sandwich or impulsive waves of the Kundt class (which generalize the {\it pp\,}-waves) such that the corresponding beam of radiation carries not only energy but also an additional angular momentum. Such spacetimes were first considered by Bonnor in \cite{Bonnor:1970b:INTHP:}, who studied the gravitational field created by a spinning null fluid. He called the corresponding particle made out of this continuum a ``spinning nullicon''. In some cases, this may be interpreted as a massless neutrino field \cite{Griffiths:1972a:INTHP:}.

In the exterior vacuum region outside the source, these solutions are locally isometric to standard {\it pp\,}-waves. The interior region contains a nonexpanding null matter which has an intrinsic spin. In general, these solutions are obtained by keeping the nondiagonal terms $g_{ui}$ in the Brinkmann form \cite{Brink:1925:MAAN:} of the {\it pp\,}-wave solution, where $u$ is the null coordinate and $x^i$ are orthogonal spatial coordinates. The corresponding energy-momentum tensor thus also contains an extra nondiagonal term $T_{ui}=j_{i}$. In four dimensions, the terms $g_{ui}$ can be set to zero {\it locally}, using a suitable gauge transformation. However, they can not be {\it globally} removed because the gauge invariant contour integral $\oint g_{ui}(u,x^{j})\,\grad x^{i}$ around the position of the gyraton is proportional to the nonzero angular momentum density $j_{i}$, which is nonvanishing.

Similar gyratons in a higher dimensional flat space were investigated (in the linear approximation) by Frolov and Fursaev \cite{Fro-Fur:2005:PHYSR4:}. Such gyratons represent a pulse of circularly polarized radiation or a modulated beam of ultrarelativistic particles with spin or other sources, which have finite energy~$E$ and finite total angular momentum~$J$. The gyraton itself is characterized by two arbitrary profile functions of $u$ which determine the energy density and angular momentum. The authors investigated the limit in which the source becomes infinitesimally small (with a negligible radius of the cross-section) and the profile functions are independent. They also studied the geodesic motion of test particles in the field of gyraton and demonstrated that, when the gyraton passes through the center of the ring of test particles, the particles start to rotate. In fact, the gyraton's angular momentum effectively creates a force which is similar to the usual centrifugal repulsive force, while the gyraton energy produces the attractive ``Newtonian'' force.

Frolov, Israel, and Zelnikov \cite{Fro-Is-Zel:2005:PHYSR4:} further investigated the exact gyraton solutions propagating in an asymptotically flat D-dimensional spacetime and proved that the Einstein's equations for gyratons reduce to a set of linear equations in the Euclidean ${(D-2)}$-dimensional space. They also showed that the gyraton metrics belong to a class of vanishing scalar curvature invariants (VSI) spacetimes for which all polynomial scalar invariants, constructed from the curvature and its covariant derivatives, vanish identically \cite{Prav-Prav:2002:CLAQG:}. (For the discussion of spacetimes with nonvanishing but nonpolynomial scalar invariants of curvature, see \cite{Page:2009:}.) Subsequently, charged gyratons in Minkowski space in any dimension were presented in \cite{Fro-Zel:2006:CLAQG:}.

In \cite{Fro-Zel:2005:PHYSR4:},  Frolov and Zelnikov took a cosmological constant into account, and exact solution for gyratons in the asymptotically anti-de~Sitter spacetime were presented. Namely, they obtained Siklos gyratons which generalize the Siklos family of nonexpanding waves \cite{Sik:1985:Cam:} (investigated further in \cite{Pod-rot:1998:CLAQG:}). 

In this case, all polynomial scalar invariants are independent of the arbitrary metric functions which characterize the gyraton and have the same values as the corresponding invariants of pure anti-de~Sitter background. The AdS gyratons \cite{Fro-Zel:2005:PHYSR4:} thus belong to the class of spacetimes with constant scalar invariants (CSI) \cite{Coley-Her-Pel:2006:CLAQG,  Coley-Gib-Her-Pope:2008:CLAQG, Coley-Her-Pel:2008:CLAQG, Coley-Her-Pel:2009:CLAQG, Coley-Her-Pap-Pel:2009:}. In string theory it has been demonstrated that generalized {\it pp\,}-wave spacetimes do not get any quantum and $\alpha'$ corrections and hence are perturbatively exact. One may expect a similar property to be valid also for the gyratons, but more careful analy\-sis is required since even if all of the local counterterms in the effective action are trivial constants for CSI spacetimes, their metric variations can be nontrivial functions. Still, one can try to generalize the property of relatively simple quantum corrections to the case of {\it semiuniversal metrics} \cite{ Coley-Gib-Her-Pope:2008:CLAQG} when the Ricci tensor has a block-diagonal structure.

Let us also mention that string gyratons in supergravity were recently found in \cite{Fro-Li:2006:PHYSR4:}.  Supersymmetric gyraton solutions were also obtained for a minimal gauged theory in five dimensions in \cite{Cald-Kle-Zor:2007:CLAQG:}, where the configuration represents a generalization of the Siklos waves with a nonzero angular momentum in anti-de~Sitter space.

The gravitational field generated by gyratons may be interesting for studies of production of mini black holes in colliders (such as the LHC) or in cosmic ray experiments. The problem of mini black hole formation in high energy particle collisions is an important issue of TeV gravity. The theory of such collisions, developed in \cite{Ear-Gid:2002:PHYSR4:,Yosh:2002:PHYSR4:, Yosh:2003:PHYSR4:, Yosh:2005:PHYSR4:, Yosh:2006:PHYSR4:}, was applied to gyraton models in \cite{Yosh:2007:PHYSR4:}.


The purpose of our contribution is to further extend the family of gyratonic solutions, which are only known in Minkowski or anti-de~Sitter background spaces. In particular, we present a new large class of gyratons of algebraic type~II, propagating in less trivial universes which are a direct product of two 2-spaces of constant curvature. This family of vacuum and electrovacuum background spacetimes contains the Nariai \cite{Nariai:1951:}, anti-Nariai, and Pleba\'{n}ski--Hacyan universes  \cite{Pleb-Hacyan:1979:JMATHP:} of type~D, or conformally flat Bertotti--Robinson \cite{Bertotti:1959:,Robinson:1959:} and Minkowski spaces. These direct-product spacetimes with six isometries (see \cite{Step:2003:Cam:,GrifPod:2009:Cam:} for more details) recently attracted new interest because they can be  recovered as specific extreme limits of various black hole spacetimes in four or more dimensions \cite{GinspargPerry:1983:NUCLB:, BousoHawk:1996:PHYSR4:,Car-Dias-Lemos:2004:PHYSR4:,Dias-Lemos-1:2003:PHYSR4:}.

Impulsive gravitational and pure radiation waves in the (anti-)Nariai, Bertotti--Robinson, and Pleba\'{n}ski--Hacyan universes were presented and analyzed by Ortaggio and Podolsk\'{y} \cite{Ortaggio:2002:PHYSR4:, OrtagPodolsky:2002:CLAQG:}.  They showed, and subsequently analyzed in more detail in \cite{PodoOrtag:2003:CLAQG:}, that these solutions are straightforward impulsive limits of a more general class of Kundt spacetimes of type II with an arbitrary profile function, which can be interpreted as gravitational waves propagating on specific type~D or~O backgrounds, including those which are a direct product of two 2-spaces. In fact, the gyraton spacetimes investigated in this paper are generalizations of such Kundt waves when their ultrarelativistic source is made of a ``spinning matter''.

The paper is organized as follows. In Section \ref{sc:gyreq}, we  present the ansatz for the metric and fields. After a short review of the transverse space geometry we derive field equations and simplify them introducing the potentials. Next, we discuss the gauge freedom and suitable gauge fixings. The overview of the gyraton solutions is summarized in Section. \ref{sc:gyrsol}.

In Section \ref{sc:knownsol}, we give a survey of important special subclasses of our gyraton solution. They include direct-product spacetimes, all type~D vacuum backgrounds, and general Kundt waves on these backgrounds.
In Section \ref{sc:interpret5}, we concentrate on the interpretation and description of the gyratons. We discuss geometric properties of the principal null congruence, the Newman--Penrose (NP) quantities with respect to natural tetrads, and properties of the electromagnetic field.

The final Section \ref{sc:GreenFc} describes the Green functions required to solve the field equations.
The main results of the paper are summarized in concluding Section \ref{sc:conclusion5}. 

Some technical results needed to derive the field equations and NP quantities are left to Appendices \ref{apx:AppAa} and \ref{apx:NP5}. In Appendix \ref{apx:backgroundsTypeD}, we derive all electro-vacuum solutions of type D, and Appendix \ref{apx:Green} discusses further details concerning the Green functions.

\section{Gyratons on direct product spacetimes}\label{sc:gyreq}

\subsection{The ansatz for the metric and matter}\label{ssc:ansatz}

The aim of this paper is to derive and analyze the family of gyraton solutions describing a gyratonic matter which propagates, together with a related gravitational wave, through a direct-product spacetime filled with a ``uniform'' electromagnetic field.

We assume that such spacetimes belong to the Kundt class. It is characterized by a geometrical property that it admits a nonexpanding, nontwisting, and shear-free null congruence \cite{Step:2003:Cam:,GrifPod:2009:Cam:}. This congruence represents the null direction of propagation of the gyraton and of the accompanying gravitational wave.

In terms of canonical (real) coordinates $\{r,u,x,y\}$, such a metric reads
\begin{equation}\begin{split}\label{m1}
ds^2=&\frac{1}{P^2}\bigl(\trgrad x^2+ \trgrad y^2\bigr)-2\,\trgrad u\,\trgrad r-2H\,\trgrad u^2\\
&+2a_{x}\,\trgrad x\,\trgrad u+2a_{y}\,\trgrad y\,\trgrad u\;,
\end{split}\end{equation}
where ${H(r,u,x,y)}$ can depend on all coordinates, but the functions ${a_x(u,x,y)}$, ${a_y(u,x,y)}$, and ${P(u,x,y)}$ are \mbox{${r}$-independent}.
The restriction ${\partial_r P=0}$ follows from our assumption of vanishing expansion of the Kundt geometry, while the condition ${\partial_r a_i=0}$, where ${i=x,y}$, is necessary here to obtain a gyraton which propagates on a direct-product spacetime background. In fact, this condition is a consequence of the Maxwell equations in the case when the electromagnetic field is present. In the absence of the electromagnetic field, the vacuum Einstein equations admit that functions ${a_i}$ can be linear in~${r}$. However, geometrical properties of such solutions are substantially different from those of the direct-product spacetimes. Therefore, in the following we will always assume that
\begin{equation}\label{aiassumption}
\partial_r a_x=0\;,\quad \partial_r a_y=0\;.
\end{equation}
This assumption thus implies that such solutions belong to the special subclass of Kundt solutions (see section \ref{ssc:prop} for more details).

The metric should satisfy the Einstein equations with a stress-energy tensor generated by the electromagnetic field and the gyraton:
\begin{equation}\label{EinsteinEqq5}
G_{\mu\nu}+\Lambda g_{\mu\nu}=\varkappa \bigl( T^\EM_{\mu\nu}+T^{\gyr}_{\mu\nu}\bigr)\;.
\end{equation}
Here, $\Lambda$  and  $\varkappa=8\pi G$ are the cosmological and gravitational constants, respectively.

The spacetime can be filled with the background electromagnetic field, which is modified by a gravitational influence of the gyraton. We assume
\begin{equation}\begin{split}\label{EMFR}
F&=E\,\trgrad r\wedge\trgrad u + B\, \frac{1}{P^2}\trgrad x\wedge\trgrad y\\
 &\qquad\qquad+\sigma_{x}\,\trgrad u\wedge\trgrad x+\sigma_{y}\,\trgrad u\wedge\trgrad y\;,
\end{split}\end{equation}
where $E$ and $B$ are constants, so that the corresponding stress-energy tensor ${T^\EM_{\mu\nu}}$ has the form \eqref{EMTe}. This ansatz for the Maxwell tensor has been inspired by the electromagnetic field known in the Bertotti--Robinson \cite{Bertotti:1959:,Robinson:1959:} and Pleba\'{n}ski--Hacyan spacetimes \cite{Pleb-Hacyan:1979:JMATHP:}, to which we have added new terms proportional to functions ${\sigma_i(r,u,x,y)}$. In fact, terms with such a structure are generated if we demand a gauge symmetry of the electromagnetic field under gauge transformation discussed in \ref{ssc:gauges}.

Finally, we must characterize the gyratonic matter by specifying the structure of its stress-energy tensor. It is a generalization of a standard null fluid such that we additionally allow terms corresponding to `internal spatial rotation' of the gyraton source,\footnote{%
In all tensorial expressions for the metric and other symmetric tensors, we understand by, for example, ${\trgrad u \,\trgrad x}$ the \textit{symmetric} tensor product ${\frac12(\trgrad u \otimes\trgrad x + \trgrad x \otimes\trgrad u)}$.}
\begin{equation}\label{m77}
\varkappa\, T^{\gyr}=j_{u}\,\trgrad u^2+2j_x\,\trgrad u\,\trgrad x+2j_y\,\trgrad u\,\trgrad y\;.
\end{equation}
We admit a general coordinate dependence of the source functions ${j_u(r,u,x,y)}$ and ${j_i(r,u,x,y)}$. However, it will be shown below that the field equations enforce a rather trivial ${r}$-dependence of these functions. Let us note that previous papers on gyratons, namely \cite{Fro-Is-Zel:2005:PHYSR4:}, \cite{Fro-Zel:2005:PHYSR4:}, \cite{Fro-Zel:2006:CLAQG:}, assumed that the gyraton source is ${r}$-independent.

The gyraton source is thus described only on a phenomenological level, by its stress-energy tensor \eqref{m77}. We do not discuss a possible internal structure of the gyratonic matter, and we do not specify its own field equations. The gyraton stress-energy tensor is assumed to be given, and our aim here is to determine its influence on the metric and the electromagnetic field. However, we have to consider that the gyraton stress-energy tensor is locally conserved. It means that the functions ${j_u}$ and ${j_i}$ must satisfy the constraint given by
\begin{equation}\label{gyrenergycons5}
  T^{\gyr}_{\;\,\mu\nu}{}^{\>;\nu}=0\;.
\end{equation}
Of course, if we had considered a specific internal structure of the gyratonic matter, the local energy-momentum conservation would have been a consequence of field equations for the gyraton. Without that, we have to require \eqref{gyrenergycons5} explicitly.

To summarize, the fields are characterized by functions ${P}$, ${H}$, ${a_i}$, and ${\sigma_i}$ which must be determined by the field equations, provided the gyraton sources ${j_u}$ and ${j_i}$ and the constants ${E}$ and ${B}$ of the background electromagnetic field are prescribed.

As we will discuss in \ref{ssc:backgrounds}, pure background solutions are obtained when both  gyratons and the gravitational waves are absent, namely for ${T^{\gyr}_{\mu\nu}=0}$, ${a_i=0}$ and ${H \propto r^2}$. For the Minkowski and \mbox{(anti-)Nariai} backgrounds, $T^\EM_{\mu\nu}$ also vanishes, while it is nonzero for the Bertotti--Robinson and Pleba\'{n}ski--Hacyan spacetimes. 

Finally, for later convenience, we introduce a constant~${\rho}$, given by the parameters ${E}$ and ${B}$ of the electromagnetic field,
\begin{equation}\label{rhodef5}
\rho=\frac{\varkappa\epso}{2}(E^2+B^2)\;,
\end{equation}
(with ${\varkappa}$ and ${\epso}$ being gravitational and electromagnetic interaction constants,\footnote{%
There are two natural choices of geometrical units: the Gaussian with ${\varkappa=8\pi}$, and ${\epso=1/4\pi}$, and SI-like with ${\varkappa=\epso=1}$.} respectively)
and the constants ${\Lambda_+}$ and ${\Lambda_-}$, defined as
\begin{equation}\label{Lambdadef}
\Lambda_\pm=\Lambda\pm\rho\;.
\end{equation}

\subsection{Geometry of the transverse space}\label{ssc:transsp}

The geometrical structure of the Kundt metric \eqref{m1} identifies the null geodesic congruence generated by $\partial_r$ and parametrized by an affine time $r$, the family of null hypersurfaces ${u=\text{constant}}$, and two-dimensional \textit{transverse spaces} ${r,u=\text{constant}}$. It will be convenient to restrict various equations to these transverse spaces. For example, ${a_i}$ and ${\sigma_i}$ can be understood as components of ${u}$-dependent 1-forms on these two-dimensional spaces. Therefore, we now briefly review some formulas and definitions valid in such two-dimensional transverse geometry.

The transverse space is covered by two  spatial coordinates ${x^i}$, and we use the Latin indices ${i,j,\dots}$ to label the corresponding tensor components. The restriction of the metric \eqref{m1} to the transverse space is
\begin{equation}\label{trmetric}
ds^2_{\perp}=g^{ }_{\!{\perp} ij}\,\trgrad x^{i}\trgrad x^{j}=\frac{1}{P^2}\,(\trgrad x^2+\trgrad y^2)\;.
\end{equation}
Here we made a useful choice of coordinates ${x^i=\{x,y\}}$ in which ${ds^2_{\perp}}$ has a conformally flat form.\footnote{%
The conformally flat coordinates are not essential, but they simplify some expressions. In a two-dimensional space, a choice of such coordinates is always possible.}

The transverse curvature is fully characterized by the scalar curvature ${R_\perp}$, which in terms of conformally flat coordinates reads (cf.\ the definition \eqref{lapldef} below)
\begin{equation}\label{trsccurv}
\frac12 R_\perp\equiv\laplace\!\log P = P\bigl(P_{,xx}+P_{,yy}\bigr)-\bigl(P_{,x}^2+P_{,y}^2\bigr)\;.
\end{equation}
Inspecting the $ru$ component of the Einstein equations~\eqref{EinsteinEqq5}, we find that the transverse scalar curvature has to be constant,  
\begin{equation}\label{Peq}
  \frac12 R_\perp = \laplace\!\log P = \Lambda_+\;,
\end{equation}
cf.\ the first lines in equations \eqref{EMTe} and \eqref{EinsteinT5}, together with \eqref{Lambdadef}. The transverse spaces are thus the constant curvature 2-spaces, all with the \emph{same} curvature. Thanks to this property we can further simplify the choice of the transverse coordinates ${\{x,y\}}$ in such a way that the conformal factor ${P^{-2}}$ in \eqref{trmetric} is ${u}$-independent. Therefore, in the following we may assume
\begin{equation}\label{udepofP}
  \partial_r P= 0\;,\quad \partial_u P=0\;.
\end{equation}

Moreover, using a freedom in the choice of the transverse coordinates, we can also put the conformal factor ${P}$ to a canonical form. There are two standard choices solving \eqref{Peq}, namely,
\begin{equation}\label{P1}
  P=1+\frac14\Lambda_+(x^2+y^2)\;,
\end{equation}
and, for a negative ${\Lambda_+}$,
\begin{equation}\label{P2}
  P=\sqrt{-\Lambda_+}\; x\;.
\end{equation}
However, in the following, we do not need a particular form of ${P}$. It must just satisfy Eq.~\eqref{Peq}.

With the transverse metric \eqref{trmetric} we may associate the Levi-Civita tensor ${\epsilon_{ij}}$ (with ${\epsilon_{xy}=P^{-2}}$) and the covariant derivative denoted by a colon (e.g., ${a_{i:j}}$). We raise and lower the Latin indices using ${g^{ }_{\!{\perp} ij}}$, and we use a shorthand ${a^2\equiv a^i a_i=P^2(a_x^2+a_y^2)}$ for a square of the norm of a \\1-form ${a_i}$. In two dimensions, the Hodge duals of 0-,1- and 2-forms ${\varphi}$, ${a_i}$, and ${f_{ij}}$ read
\begin{equation}
(*\varphi)_{ij} = \varphi\, \epsilon_{ij}\;,\;\;
(*a)_i = a_j \epsilon^j{}_i\;,\;\;
*f = \frac12 f_{ij}\epsilon^{ij} = P^2 f_{xy}\;.
\end{equation}

For convenience, we also introduce an explicit notation for two-dimensional divergence and rotation of a transverse 1-form ${a_i}$,
\begin{align}
  \trdiv a &\equiv a_i{}^{:i} = P^2(a_{x,x}+a_{y,y})\;,\\
  \rot a &\equiv * \trgrad a = \epsilon^{ij} a_{j,i} = P^2(a_{y,x}-a_{x,y})\;,
\end{align}
and for the Laplace operator of a function ${\psi}$,
\begin{equation}\label{lapldef}
\laplace\psi =  \psi_{:i}{}^{:i} = P^2(\psi_{,xx}+\psi_{,yy})\;.
\end{equation}
Note that the divergence and rotation are related as ${\trdiv a = \rot {*}a}$.

Finally, we will generally assume that the transverse space is topologically simple in the sense that the space of harmonics is trivial. However, sometimes it will be physically relevant to consider also nontrivial solutions of the Laplace equation if we relax the boundary and asymptotical conditions in the noncompact case. For example, a solution of the Laplace equation around a localized source satisfies the homogeneous Laplace equation on the space with the source removed. Such a space is, however, noncompact and the solution is not vanishing on the boundary.

\subsection{The field equations}\label{scc:fequations}

After specifying the ansatz for our fields and reviewing the transverse geometry we can now derive the equations for the gyraton. We have to consider the Einstein equations \eqref{EinsteinEqq5} together with the Maxwell equations and the condition \eqref{gyrenergycons5} for the gyraton source.

We start with the cyclic Maxwell equation. Assuming \eqref{EMFR} and \eqref{udepofP} it reads
\begin{equation}\label{MXECE}
  0=\trgrad F = (\partial_r\sigma_i)\, \trgrad r\wedge\trgrad u\wedge\trgrad x^i -\rot\sigma\, \trgrad u \wedge \epsilon\;,
\end{equation}
where ${\epsilon=P^{-2}\trgrad x\wedge\trgrad y}$. We immediately infer that the 1-form ${\sigma_i}$ is ${r}$-independent, ${\partial_r \sigma_i=0}$, and rotation-free,
\begin{equation}\label{pot18}
  \rot\sigma=0\;.
\end{equation}
The second Maxwell equation ${F_{\mu\nu}{}^{;\nu}=0}$ has only\\ the ${u}$ component nonvanishing,\footnote{%
Here we used that ${a_i}$ and ${\sigma_i}$ are ${r}$-independent. If the condition \eqref{aiassumption} was not assumed before, it would follow from the transverse components of this Maxwell equation.}
which gives
\begin{equation}\label{pot28}
  \trdiv\sigma-E\,\trdiv a + B\,\rot a = 0\;.
\end{equation}
We call \eqref{pot18} and \eqref{pot28} the \emph{potential equations} since they guarantee the existence of potentials which will be discussed in detail in Section \ref{ssc:potss}. For this reason, it is useful to note that these equations imply the conditions 
\begin{equation}\label{potcond}
\begin{gathered}
 \trdiv\bigl[E\,(\sigma{-}E\,a)+B\,{*(\sigma{-}E\,a)}\bigr]=0\;,\\
 \rot\bigl[E\,(\sigma{-}B\,{*a})+B\,{*(\sigma{-}B\,{*a})}\bigr]=0\;,
\end{gathered}
\end{equation}
cf.\ equations~\eqref{dkappadlambda} below.

The Einstein equations can be derived from the Einstein tensor and the electromagnetic stress-energy tensor, which are given in Appendix \ref{apx:AppAx}. We have already discussed the $ru$ component which leads to the condition~\eqref{Peq}. The transverse diagonal components $xx$ and $yy$ give
\begin{equation}\label{Einsteinii}
  \partial^2_{r} H  =-\Lambda_-\;.
\end{equation}
We thus obtain the explicit ${r}$ dependence of the metric function ${H}$ as
\begin{equation}\label{Heq5}
  H = -\frac12 \Lambda_-\, r^2 + g\,r + h\;.
\end{equation}
where we have introduced ${r}$-independent functions ${g(u,x^j)}$ and ${h(u,x^j)}$.

Finally, the remaining nontrivial components of the Einstein equations are those involving the gyraton source \eqref{m77}. The $ui$ components give an equation related to ${j_i}$, which we call the \emph{first source equation},
\begin{equation}\label{jieq5}
\begin{split}
  j_i
  &= \frac12\, f_{ij}{}^{:j}  + g_{,i}-\Lambda_-\, a_i\\
  &\quad+\varkappa\epso\,\bigl[E\,(\sigma_i{-}E\,a_i)+B\,(\sigma_j{-}E\,a_j)\,\epsilon^j{}_i\bigr]\\
  &= \frac12\, f_{ij}{}^{:j}  + g_{,i}-\Lambda_+\, a_i\\
  &\quad+\varkappa\epso\,\bigl[E\,(\sigma_i{-}B\,{a_j}\epsilon^j{}_i)+B\,(\sigma_j{-}B\,a_k\epsilon^k{}_j)\epsilon^j{}_i\bigr]
  \;,
\end{split}\raisetag{44pt}
\end{equation}
where we have introduced the external derivative ${f_{ij}}$ of the 1-form ${a_i}$ as
\begin{equation}\label{fdef5}
  f_{ij} = a_{j,i}-a_{i,j} = (*\,\rot a)_{ij} \;.
\end{equation}
For convenience, we have written the equation \eqref{jieq5} in two equivalent forms. In the square brackets, they explicitly contain the terms which were already encountered in the equation \eqref{potcond}. We can thus easily split the first source equation into divergence and rotation parts:
\begin{align}
  -\trdiv j &= -\laplace g + \Lambda_-\,\trdiv a\;, \label{divjeq5}\\
  -\rot j &= \frac12\laplace b + \Lambda_+\, b \;, \label{rotjeq5}
\end{align}
where the function ${b(u,x^j)}$ is the Hodge dual of ${f_{ij}}$,
\begin{equation}\label{bdef5}
    b \equiv * f = \rot a\;.
\end{equation}
Equations \eqref{divjeq5} and \eqref{rotjeq5} carry essentially the same information
as the original source equation \eqref{jieq5}.\footnote{%
They are equivalent to \eqref{jieq5} if we ignore the possibility of harmonic 1-forms which can exist in topologically nontrivial spaces.}

Next, we examine the condition \eqref{gyrenergycons5} for the gyraton source. It gives
\begin{equation}\label{gyrenergycons23}
  -(\partial_r j_i)\,\trgrad x^i + \bigl(-\partial_r j_u+\trdiv j +a^i \partial_r j_i\bigr)\,\trgrad u =0\;,
\end{equation}
so that the source functions ${j_i}$ must be ${r}$-independent and ${j_u}$ has to have the structure
\begin{equation}\label{jdecomp}
  j_u = r\,\trdiv j + \iota\;.
\end{equation}
The gyraton source \eqref{m77} is thus fully determined by three \mbox{${r}$-independent} functions ${\iota(u,x^j)}$ and ${j_i(u,x^j)}$.

Finally, from the $uu$-component of the Einstein equation we obtain
\begin{equation}\label{jueq5}
\begin{split}
  j_u =\,&\bigl(\laplace g - \Lambda_-\trdiv a)\;r\\
      &+{\!}\laplace h + \frac12  b^2 - \Lambda_- a^2 + 2 a^i g_{,i}\\
      & +\partial_u(\trdiv a) + g\, \trdiv a\\
      &-\varkappa\epso \,(\sigma-Ea)^2\;.
\end{split}
\end{equation}
Comparing the coefficient in front of ${r}$ with \eqref{divjeq5}, we find that it consistently reproduces the structure \eqref{jdecomp}. The nontrivial ${r}$-independent part of \eqref{jueq5} gives the \emph{second source equation} which can be understood as the equation for the metric function ${h}$,
\begin{equation}\label{heq5}
\begin{split}
  \laplace h &=
      \iota \, - \frac12 b^2 + \Lambda_- a^2 - 2 a^i g_{,i} \\
      & +\varkappa\epso(\sigma-Ea)^2
      -\partial_u(\trdiv a)- g\, \trdiv a\;.
\end{split}
\end{equation}

\subsection{Potentials}\label{ssc:potss}

We have thus found that the Maxwell and Einstein equations reduce to two potential equations \eqref{pot18}, \eqref{pot28}, and two source equations \eqref{jieq5}, \eqref{jueq5}. These equations can further be considerably simplified by introducing potentials for the \mbox{1-forms} ${\sigma_i}$, ${a_i}$ and for the source ${j_i}$.

Indeed, the first potential equation \eqref{pot18} gives immediately that ${\sigma_i}$ has a potential ${\varphi(u,x^j)}$ such that
\begin{equation}\label{phipot5}
  \sigma_i = \varphi_{,i}\;.
\end{equation}
Using the Hodge decomposition we can express the 1-form ${a_i}$ using two scalar potentials ${\kappa(u,x^j)}$ and ${\lambda(u,x^j)}$:
\begin{equation}\label{klpotdef5}
    a_i = \kappa_{,i}+\epsilon_i{}^j\,\lambda_{,j}\;.
\end{equation}
These potentials control the divergence and the rotation of~${a_i}$ via
\begin{equation}\label{divrota5}
    \trdiv a = \laplace\kappa\;,\quad \rot a = -\laplace\lambda   \;.
\end{equation}

Equation \eqref{pot28} imposes a constraint among these three potentials ${\varphi}$, ${\kappa}$, and ${\lambda}$:
\begin{equation}\label{lapphi}
    \laplace\varphi \,=\, \laplace(E\kappa+B\lambda)\;.
\end{equation}
If the transverse space is compact (or if it is noncompact but sufficiently strong asymptotic conditions are imposed) the solution of the Laplace equation is trivial and we immediately obtain
\begin{equation}\label{phisol}
    \varphi = E\,\kappa + B\, \lambda\;.
\end{equation}

By using this constraint, it is possible to show that the potentials ${\kappa}$ and ${\lambda}$ solve the conditions \eqref{potcond}
\begin{equation}\label{dkappadlambda}
\begin{gathered}
  E\,(\sigma_i{-}E a_i)+B\,(\sigma_j{-}E a_j)\,\epsilon^j{}_i =
    (E^2+B^2)\, \lambda_{,j}\epsilon^j{}_i\;,\\
  E\,(\sigma_i{-}B{*a}_i)+B\,(\sigma_j{-}B{*a}_j)\,\epsilon^j{}_i =
    (E^2+B^2)\, \kappa_{,i}\;.
\end{gathered}
\end{equation}

In terms of the potentials, the first source equation \eqref{jieq5} can be written as
\begin{equation}\label{jieqpot5}
  j_i = \frac12\, f_{ij}{}^{:j} + \Lambda_+\,\lambda_{,j}\, \epsilon^j{}_i -\Lambda_-\,\kappa_{,i}  + g_{,i}\;.
\end{equation}
Its rotation part is the equation \eqref{rotjeq} for ${b}$, the solution of which can be used as a source for the equation for the potential ${\lambda}$,
\begin{equation}
  \laplace\lambda = -b\;.\label{lambdab5}
\end{equation}
The divergence part of \eqref{jieqpot5} can be written as a relation between the functions ${g}$, ${\kappa}$ and source ${\trdiv j}$,
\begin{equation}\label{kappageq}
  \laplace\!\bigl(g-\Lambda_-\kappa\bigr) = \trdiv j\;.
\end{equation}

The problem further simplifies if we introduce scalar potentials ${p(u,x^j)}$ and ${q(u,x^j)}$ for the gyraton source ${j_i}$,
\begin{equation}\label{sourcepot}
    j_i = p_{,i}+\epsilon_i{}^j\,q_{,j}\;,
\end{equation}
so that
\begin{equation}\label{sourcepotB}
    \rot j = -\laplace q\;,\quad\trdiv j = \laplace p\;.
\end{equation}

Substituting this to the field equation \eqref{jieqpot5} and splitting it into the gradient part and the rotation part (i.e., using the Hodge decomposition), we obtain
\begin{equation}\label{kappagp}
    g-\Lambda_-\kappa = p\;,
\end{equation}
and
\begin{equation}\label{lambdaeq5}
    \frac12\laplace\lambda+\Lambda_+\lambda=-q \;.
\end{equation}
Let us note that all of the potentials are defined up to an additive constant (which, however, can be ${u}$-dependent). In the derivation of \eqref{kappagp} and \eqref{lambdaeq5}, we have absorbed the integration constants into this nonuniqueness of potentials. In view of \eqref{lambdab5}, function ${b}$ is then given by
\begin{equation}\label{beq}
    b= 2(\Lambda_+\lambda+q)\;.
\end{equation}

We have thus reduced the field equations to simple algebraical relations \eqref{phisol}, \eqref{kappagp} between the potentials, to the Helmholtz--Poisson equation \eqref{lambdaeq5} for~${\lambda}$, and the Poisson equation \eqref{heq5} for~${h}$. The last one can be also rewritten using the potentials as
\begin{equation}\label{heqpot}
\laplace\hat h = \iota + q \,\laplace\lambda- p \,\laplace\kappa - 2 a^i p_{,i}\;,
\end{equation}
with ${\hat h}$ closely related to ${h}$:
\begin{equation}\label{hath}
\hat h = h +\partial_u\kappa+\frac12\Lambda_-\kappa^2-\frac12\Lambda_+\lambda^2\;.
\end{equation}

\subsection{Gauge transformation and the field equations in suitable gauges}\label{ssc:gauges}

\subsubsection*{Shift of the ${r}$ coordinate}

To find the gyraton solution explicitly, we need to determine the functions ${h}$, ${g}$, ${a_i}$, and ${\sigma_i}$, provided the gyraton sources ${j_i}$ and ${\iota}$ are prescribed. In terms of the potentials ${\kappa}$, ${\lambda}$, and ${\varphi}$, replacing the transverse 1-forms ${a_i}$ and ${\sigma_i}$, we have obtained equation \eqref{lambdaeq5} for ${\lambda}$, \eqref{phisol} for ${\varphi}$, and \eqref{heq5} for ${h}$. However, we have only one equation \eqref{kappageq} for ${\kappa}$ and ${g}$.

This deficiency of equations corresponds to the fact that our ansatz \eqref{m1}, \eqref{EMFR}, \eqref{m77} admits a gauge freedom.
Indeed, the coordinate transformation ${\tilde r\to r = \tilde r-\psi(u,x^j)}$, accompanied by the following redefinition of the metric functions and fields:
\begin{equation}\begin{gathered}\label{gauge2}
r=\tilde r-\psi\;,\\
g=\tilde g-\Lambda_-\psi\;,\quad
h=\tilde h-\frac{1}{2}\Lambda_- \psi^2 + \tilde g\,\psi + \partial_u\psi\;,\\
a_i=\tilde a_i-\psi_{,i}\;,\quad
\sigma_i=\tilde \sigma_i-E\,\psi_{,i}\;,\\
\kappa=\tilde \kappa-\psi\;,\quad
\lambda=\tilde\lambda\;,\quad
\varphi=\tilde\varphi-E\,\psi\;,\\
j_i=\tilde j_i\;,\quad
\iota=\tilde\iota+\psi\, \trdiv j\;,
\end{gathered}\end{equation}
leaves the metric, the Maxwell tensor, and the gyraton stress-energy tensor in the same form. Consequently, all of the field equations remain the same. Such a transformation is a pure gauge transformation and we can use it to simplify the solution of the equations.

This gauge transformation has a  geometrical meaning of shifting the origin of the affine parameter ${r}$ of the null congruence ${\partial_r}$.

Inspecting this gauge transformation, we find that the combination ${\,g-\Lambda_-\kappa\,}$ is gauge invariant. This combination enters the field equation \eqref{kappageq}, and only this combination is thus invariantly determined by the sources, namely, it is equal to ${p}$, cf.~\eqref{kappagp}. The particular splitting into ${g}$ and ${\kappa}$ parts is just a question of the gauge choice.

Indeed, it follows from \eqref{gauge2} that it is possible to modify one of the functions $g$, ${\kappa}$, or ${\varphi}$ to an arbitrary value or even to cancel it out from all the equations. Moreover, the freedom to choose one of these functions covers the gauge freedom fully. Therefore, we use them to control the gauge freedom: we may fix the gauge by setting ${g}$, ${\kappa}$, or ${\varphi}$ to be an arbitrarily chosen function. Any of these gauge conditions leads to the same family of solutions, only with a different parametrization of the gauge freedom.

\subsubsection*{Gauge fixing of ${g}$}\label{ssc:g}

Let us start with the gauge condition that the function ${g}$ is an arbitrary function. Then the equation \eqref{kappageq} should be understood as the Poisson equation for~${\kappa}$. In terms of the source potentials we have even the explicit solution given by \eqref{kappagp}. The potential ${\lambda}$ is determined by the equation \eqref{lambdaeq5}, and ${\varphi}$ by \eqref{phisol}. Finally, substituting these results into the second source equation \eqref{heq5} we obtain the Poisson equation for ${h}$.

In fact, the gauge fixing of ${g}$ can be used to eliminate the metric function ${g}$ completely. Setting
\begin{equation}\label{g0}
  g=0\;,
\end{equation}
the field equation for ${\kappa}$ reduces to
\begin{equation}\label{divjg0}
  \laplace\kappa=-\frac1{\Lambda_-}\,\trdiv j\;,
\end{equation}
with the solution
\begin{equation}\label{kappag0}
  \kappa=-\frac1{\Lambda_-}\,p\;.
\end{equation}
The equation for ${h}$ reads
\begin{equation}\label{heqg0}
\begin{split}
\laplace h &=
      \iota - 2\bigl(\Lambda_+\,\lambda+q\bigl)^2
      + \Lambda_+\, \lambda_{,i}\lambda_{,j}g^{ij}
      + 2\, \lambda_{,i}\, p_{,j}\,\epsilon^{ij} \\
      &\qquad\qquad+\frac1{\Lambda_-}\bigl(p_{,i} p_{,j} g^{ij}+\partial_u\laplace p\bigr)\;,
\end{split}
\end{equation}
and the metric function ${H}$ has only trivial quadratic dependence on ${r}$,
\begin{equation}\label{Hg0}
 H(r,u,x^j)=-\frac12 \Lambda_- r^2 + h(u,x^j)\;.
\end{equation}

\subsubsection*{Gauge fixing of ${\kappa}$}\label{ssc:kappa}

Alternatively, we can fix the potential ${\kappa}$, which is equivalent to the prescription of a value of ${\trdiv a}$. Equation \eqref{kappageq} is then the Poisson equation for ${g}$, otherwise the solution of the field equations proceed in the same way as above.

The special choice
\begin{equation}\label{kappa0}
  \kappa=0\;,\quad\text{i.e.}\quad\trdiv a = 0\;,
\end{equation}
implies simple relations for ${g}$:
\begin{gather}
  \laplace g =\trdiv j\;,\label{divjkappa0}\\
  g = p\;,\label{gkappa0}
\end{gather}
and between the 1-forms ${a_i}$ and ${\sigma_i}$ (or their potentials):
\begin{gather}
  \varphi = B\lambda\;,\label{phikappa0}\\
  \sigma_i = B\, {*a}_i\;.\label{sigmakappa0}
\end{gather}
The equation for ${h}$ now takes the form
\begin{equation}\label{hkappa0}
  \laplace h =
      \iota - 2\bigl(\Lambda_+\,\lambda+q\bigl)^2 + \Lambda_+\, \lambda_{,i}\lambda_{,j}g^{ij}
      + 2\, \lambda_{,i}\, p_{,j}\,\epsilon^{ij} \;.
\end{equation}

\subsubsection*{Gauge fixing of ${\varphi}$}\label{ssc:varphi}

The last natural gauge condition is a fixing of the potential ${\varphi}$. In this case, one first finds the potential ${\lambda}$ by solving the equation \eqref{lambdaeq5}. The relation \eqref{phisol} then gives the potential ${\kappa}$. Plugging this into \eqref{kappageq}, the equation for ${g}$ is obtained:
\begin{equation}
  \laplace g =\trdiv j + \frac{\Lambda_-}{E}\,\laplace\varphi-\frac{\Lambda_-B}{E}\,\laplace\lambda\;,\label{gphifixed}
\end{equation}
i.e.,
\begin{equation}
  g =p +\frac{\Lambda_-}{E}\,\varphi-\frac{\Lambda_-B}{E}\,\lambda\;.\label{gphifixed2}
\end{equation}
The 1-form ${a_i}$ can be written in terms of ${\sigma_i}$ and ${\lambda}$ as
\begin{equation}\label{aphifixed}
  E\, a_i =\sigma_i-B\,\lambda_{,i}-E\,\lambda_j\,\epsilon^j{}_i\;.
\end{equation}

Particularly, for
\begin{equation}\label{phi0}
  \varphi=0\;,\quad\text{i.e.}\quad\sigma_i = 0\;,
\end{equation}
we obtain
\begin{gather}
  \kappa=-\frac{B}{E}\, \lambda\;,\label{kappaphi0}\\
  g =p -\Lambda_-\frac{B}{E}\,\lambda\;,\label{gphi0}\\
  E\, a_i =-B\,\lambda_{,i}-E\,\lambda_j\,\epsilon^j{}_i\;.
\end{gather}
The choice ${\varphi=0}$  simplifies the Maxwell tensor \eqref{EMFR} to
\begin{equation}\label{EMFphi0}
F=E\,\trgrad r\wedge\trgrad u + B\, \frac{1}{P^2}\trgrad x\wedge\trgrad y\;.
\end{equation}

\subsubsection*{Reparametrization of the  ${u}$ coordinate}

After the above discussion of the gauge freedom corresponding to the transverse-dependent shift of the ${r}$ coordinate, we should also mention the remaining gauge freedom. The metric \eqref{m1}, the electromagnetic field \eqref{EMFR}, and the gyraton stress-energy tensor \eqref{m77} keep the same form under a general reparametrization of the ${u}$ coordinate ${\tilde u\to u=f(\tilde u)}$, accompanied by the rescaling ${\tilde r\to r=\tilde r/f'(\tilde u)}$ of  the ${r}$ coordinate. The metric functions and matter fields must be redefined as
\begin{equation}\begin{gathered}\label{urep}
u=f(\tilde u)\;,\quad
r=\frac{\tilde r}{f'(\tilde u)}\;,\\
g=\frac{\tilde g}{f'(\tilde u)}+\frac{f''(\tilde u)}{f'(\tilde u)^2}\;,\quad
h=\frac{\tilde h}{f'(\tilde u)^2}\;,\\
a_i=\frac{\tilde a_i}{f'(\tilde u)}\;,\quad
\sigma_i=\frac{\tilde \sigma_i}{f'(\tilde u)}\;,\\
\kappa=\frac{\tilde \kappa}{f'(\tilde u)}\;,\quad
\lambda=\frac{\tilde\lambda}{f'(\tilde u)}\;,\quad
\varphi=\frac{\tilde\varphi}{f'(\tilde u)}\;,\\
j_u=\frac{\tilde j_u}{f'(\tilde u)^2}\;,\quad
\iota=\frac{\tilde \iota}{f'(\tilde u)^2}\;,\quad
j_i=\frac{\tilde j_i}{f'(\tilde u)}\;.
\end{gathered}\end{equation}
It is worth to emphasize here that this reparametrization is independent of the transverse spatial coordinates. This gauge transformation is thus `global' from the point of view of the transverse space, and it does not influence the field equations (which we formulated as differential equations on the transverse space) in any significant way.

\subsection{Summary of the gyraton solutions}\label{sc:gyrsol}

Let us now summarize the main equations of the gyratons on direct-product backgrounds. These are spacetimes with the metric of the form
\begin{equation}\begin{split}
ds^2=\frac{1}{P^2}(\trgrad x^2{+}\trgrad y^2)-2\,\trgrad u\,\trgrad r-2H\,\trgrad u^2+2a_{i}\,\trgrad x^i \trgrad u\;,
\end{split}\end{equation}
filled with the electromagnetic field
\begin{equation}
F=E\,\trgrad r\wedge\trgrad u + B\, \frac{1}{P^2}\trgrad x\wedge\trgrad y+\trgrad u\wedge\sigma_{i}\,\trgrad x^i\;,
\end{equation}
and the gyratonic matter
\begin{equation}
\varkappa\, T^{\gyr}=j_{u}\,\trgrad u^2+2j_i\,\trgrad x^i\trgrad u\;.\end{equation}
The metric function ${H(r,u,x^i)}$ is quadratic in ${r}$,
\begin{equation}
  H = -\frac12 \Lambda_-\, r^2 + g\,r + h\;,
\end{equation}
the gyraton energy density ${j_u(r,u,x^i)}$ can be at most linear in ${r}$,
\begin{equation}
  j_u = r\,\trdiv j + \iota\;,
\end{equation}
and the functions ${g(u,x^i)}$, ${h(u,x^i)}$, ${a_j(u,x^i)}$, ${\sigma_j(u,x^i)}$,
${j_j(u,x^i)}$, and ${\iota(u,x^i)}$ are ${r}$-independent. The function ${P(x^i)}$ is ${r}$ and ${u}$-independent and it satisfies the equation
\begin{equation}
  \laplace\!\log P = \Lambda_+\;.
\end{equation}
It can be solved by ${P}$ of the form \eqref{P1} or, for ${\Lambda_+<0}$ by \eqref{P2}.

The transverse 1-forms ${a_i}$, ${\sigma_i}$, and ${j_i}$ can be written in terms of the scalar potentials ${\kappa}$, ${\lambda}$, ${\varphi}$, ${p}$, and ${q}$ as
\begin{gather}
  \sigma_i = \varphi_{,i}\;,\\
  a_i = \kappa_{,i}+\epsilon_i{}^j\,\lambda_{,j}\;,\\
  j_i = p_{,i}+\epsilon_i{}^j\,q_{,j}\;.
\end{gather}
These potentials are unique up to (unphysical) constants on the transverse space, which can always be gauged away.

Finally, the functions ${h}$, ${g}$, ${\kappa}$, ${\lambda}$, and ${\varphi}$ must satisfy the linear field equations
\begin{gather}
  \varphi = E\,\kappa + B\, \lambda\;,\\
  g-\Lambda_-\kappa = p\;,\label{sumgkpeq}\\
  \frac12\laplace\lambda+\Lambda_+\lambda=-q \;,\label{sumlqeq}\\
  \laplace\hat h = \iota + q \,\laplace\lambda- p \,\laplace\kappa - 2 a^i p_{,i}\;,\label{sumheq}
\end{gather}
where
\begin{equation}\label{sumhhatdef}
  \hat h = h +\partial_u\kappa+\frac12\Lambda_-\kappa^2-\frac12\Lambda_+\lambda^2\;.
\end{equation}

\section{Important special subclasses}\label{sc:knownsol}

The large family of solutions of Einstein--Maxwell equations discussed above belongs to the Kundt class \eqref{m1} of nonexpanding, shear-free and twist-free spacetimes \cite{Kundt:1961:ZEPH:, Kundt:1962:PRS:, Step:2003:Cam:}, namely to its subclass characterized by the condition \eqref{aiassumption}. As we have seen in Section \ref{ssc:ansatz}, and will be discuss more in Section \ref{sc:interpret5}, the gyratonic matter \eqref{m77} is the ``rotating'' generalization of a null fluid. As special cases, this family of solutions contains some previously known spacetimes from the Kundt family which correspond to electro-vacuum or pure (null) radiation. In this section we will  shortly discuss such important subcases.

\subsection{Direct-product background spacetimes}\label{ssc:backgrounds}

It is natural to start with the simplest case of highly symmetric spacetimes. Considering the metric function ${H}$ of the form \eqref{Heq5}, setting ${\,a_{i}=0}$, and ${\,g=0=h}$,
and choosing the expression \eqref{P1} for ${P}$, the metric \eqref{m1} reduces to
\begin{equation}\label{n3}
d s^2=\frac{\trgrad x^2+ \trgrad y^2}{[1+\frac{1}{4}\Lambda_{+}\,(x^2+y^2)]^2}-2\trgrad u\,\trgrad r+\Lambda_{-}\,r^2\trgrad u^2\;.
\end{equation}
It describes backgrounds on which the gyratons propagate. By performing the transformation ${r=v(1-\frac12\Lambda_{-}\, u v)^{-1}}$ with ${u=(t-z)/\sqrt 2}$, and ${v=(t+z)/\sqrt 2}$, the metric becomes
\begin{equation}\label{n3tr}
d s^2=\frac{\trgrad x^2+ \trgrad y^2}{[1+\frac{1}{4}\Lambda_{+}\,(x^2+y^2)]^2}
+\frac{\trgrad z^2-\trgrad t^2}{[1+\frac{1}{4}\Lambda_{-}\,(z^2-t^2)]^2}\;.
\end{equation}
The background spacetimes thus have geometry of a direct product of two 2-spaces of constant curvature $\Lambda_{+}$ and $\Lambda_{-}$, respectively. The first is the space spanned by two spatial coordinates so that it is flat Euclidean space $E^{2}$, {2-sphere} $S^{2}$, or 2-hyperboloid $H^{2}$, according to the sign of the constant $\Lambda_{+}$. The second is the (1+1)-dimensional spacetime spanned by a timelike coordinate and one spatial coordinate. According to the sign of the constant~$\Lambda_{-}$, it is Minkowski 2-space $M_{2}$, de Sitter space $dS_{2}$, or anti-de~Sitter space $AdS_{2}$.

Therefore, there are nine theoretically possible distinct subclasses given by the choice of $\Lambda_{+}$ and $\Lambda_{-}$, but only six of them are physically relevant because the energy density $\rho$ must be non-negative, which eliminates three cases. The most important of such background spacetimes are summarized in Table~\ref{table15}. In addition, there are more general Bertotti--Robinson direct product space-times for which the constants $\Lambda_{+}$ and $\Lambda_{-}$ are independent and nontrivial; i.e., the cosmological constant $\Lambda$ and the energy density ${\rho>0}$ of the electromagnetic field can be chosen arbitrarily.

\begin{table}
\caption{\label{table15}Some of possible background spacetimes which are the direct product of two 2-spaces of constant curvature. Here $\Lambda$ is a cosmological constant, and $\rho$ is a constant energy density of the electromagnetic field.}
\begin{ruledtabular}
\begin{tabular}{cccccccc}
 $\Lambda_{+}$ &  $\Lambda_{-}$ & \text{geometry} & \text{spacetime}  & $\Lambda$ & $\rho$\\
\hline
 0 & 0                  & ${E^{2}\times M_{2}}$  & Minkowski        & $=0$ & $=0$ \\
 $\Lambda$ & $\Lambda$  & ${S^{2}\times dS_{2}}$  & Nariai          & $>0$ & $=0$ \\
 $\Lambda$ & $\Lambda$  & ${H^{2}\times AdS_{2}}$  & anti-Nariai   & $<0$ & $=0$ \\
 $\rho$ & $-\rho$ & ${S^{2}\times AdS_{2}}$  & Bertotti--Robinson   & $=0$ & $>0$ \\
 $2\Lambda$ & 0   & ${S^{2}\times M_{2}}$  & Pleba\'{n}ski--Hacyan   & $>0$ & $=\Lambda$ \\
 0 &  $2\Lambda$  & ${E^{2}\times AdS_{2}}$  & Pleba\'{n}ski--Hacyan & $<0$ & $=|\Lambda|$
\end{tabular}
\end{ruledtabular}
\end{table}

In a natural null tetrad, the only nonvanishing NP Weyl and curvature scalars are (see Section~\ref{ssc:NP})
\begin{equation}\begin{aligned}\label{f18back}
\Psi_{2}&=-\frac{1}{3}\Lambda\,, \qquad R=4\Lambda\,, \qquad \Phi_{11}=\frac{1}{2} \rho\,,\\
\end{aligned}\end{equation}
where ${\Lambda=\frac12(\Lambda_{+}+\Lambda_{-})}$ and ${\rho=\frac12(\Lambda_{+}-\Lambda_{-})}$, together with ${\Phi_{1}=\frac{1}{2} (E+iB)}$. These electro-vacuum solutions are thus of algebraic type~D, unless ${\Lambda=0}$ which applies to a conformally flat Bertotti--Robinson universe and flat Minkowski space. Vacuum direct product spacetimes (with ${\rho=0}$) are Minkowski and (anti-)Nariai spaces. For the two Pleba\'{n}ski--Hacyan spacetimes, one and only one of the 2-spaces is flat. Therefore, ${\Phi_{11}=\tfrac{1}{2}|\Lambda|}$, so that the condition $2\Phi_{11}\pm 3\Psi_{2}=0$ is satisfied.

More details about some of these background spacetimes can be found in the original works \cite{Nariai:1951:,Bertotti:1959:,Robinson:1959:,Pleb-Hacyan:1979:JMATHP:}, reviews \cite{Step:2003:Cam:, GrifPod:2009:Cam:} or, e.g., in \cite{Ortaggio:2002:PHYSR4:,OrtagPodolsky:2002:CLAQG:,Dias-Lemos:2003:PHYSR4:}.

\subsection{Type D background spacetimes}\label{ssc:backgroundsTypeD}

As will be seen in Section \ref{ssc:NP}, the gyraton spacetimes are in general of algebraic type~II. However, they contain a wider subclass of electro-vacuum solutions of type~D, which can also be naturally regarded as possible background geometries.

Type~D electro-vacuum solutions of Einstein's equations are known \cite{Step:2003:Cam:, GrifPod:2009:Cam:, Kinnersley:, Carter:, Plebanski:1979:,Pleb-Hacyan:1979:JMATHP:}. However, their forms are usually different from the  parametrization of the geometry used here. For this reason we will write those type D spacetimes, which belong to our subclass of the Kundt family, explicitly. All such spacetimes are derived in Appendix \ref{apx:backgroundsTypeD}; here we only summarize the results.

Although these spacetimes have the same curvature scalars as in \eqref{f18back}, they are not, in general, direct-product spaces. In particular, they have a lower symmetry than the highly symmetric backgrounds discussed above.

\subsubsection*{The $\Lambda_+=0$ case\\(exceptional Pleba\'{n}ski--Hacyan spacetime)}

As shown in Appendix \ref{apx:backgroundsTypeD}, all type D solutions naturally split into two cases. For ${\Lambda_+=0}$ (i.e., ${\Lambda=-\rho<0}$, ${\Lambda_-=2\Lambda}$) we find a generalization of the exceptional Pleba\'{n}ski--Hacyan type~D electro-vacuum spacetime \cite{Pleb-Hacyan:1979:JMATHP:,OrtagPodolsky:2002:CLAQG:,PodoOrtag:2003:CLAQG:}. The metric reads
\begin{equation}\label{genPlHac}
\begin{split}
d s^2&=
   \trgrad x^2+ \trgrad y^2
   +2\bigl(\Lambda\, r^2 - L_{x}\,x - L_{y}\,y\bigr) \trgrad u^2\\
   &\qquad\qquad\quad
   -2\trgrad u\trgrad r
   +2\bigl(a_x \trgrad x  + a_y\trgrad y \bigr)\trgrad u
   \;,
\end{split}
\end{equation}
where ${L_{i}(u)}$ and ${a_i(u)}$, ${i=x,y}$, are arbitrary functions of the coordinate $u$ only (i.e., constants on each transverse space). This corresponds to the metric \eqref{m1} with ${P=1}$, ${g=0}$, ${h}$ linear in ${x,y}$, and ${a_i}$ independent of ${x,y}$.

For ${a_i=0}$ it reduces to the exceptional Pleba\'{n}ski--Hacyan spacetime. It further reduces to the direct product spacetime~\eqref{n3} when also both ${L_{i}}$ vanish. Although the functions ${L_{i}\not=0}$ do not enter the curvature scalars \eqref{f18back}, the geometry of this spacetime is different from that of the direct-product spacetimes (for example, it contains another shear-free but non-geodesic null direction).

Nontrivial coefficients ${a_i}$ can be gauged away using the transformation \eqref{gauge2}. However, such a transformation generates a nonvanishing metric function ${g}$ and a quadratic dependence of ${h}$ on ${x,y}$. It thus seems that the case ${a_i\neq0}$ is indeed a nontrivial generalization of the exceptional Pleba\'{n}ski--Hacyan spacetime.

\subsubsection*{The $\Lambda_+\neq0$ case}

In the case when the transverse space has a nonvanishing curvature ${\Lambda_+}$, the metric of  type D electro-vacuum solutions is given by the metric functions
\begin{gather}
  \lambda=\frac{Q}{P}\;,\quad\kappa=0\;,\label{gentypeDkl}\\
  h=\frac12\Lambda_+\lambda^2\;,\quad g=0\;.\label{gentypeDh}
\end{gather}
Here the functions ${P}$ and $Q$ can be written as
\begin{equation}\label{gentypeDPQ}
\begin{gathered}
  P=1+\frac14\Lambda_+(x^2+y^2)\;,\\
  Q=q_0\bigl(1-\frac14\Lambda_+(x^2+y^2)\bigr)+q_x x +q_y y\;,
\end{gathered}
\end{equation}
respectively, where ${q_0(u),\,q_x(u)}$ and $q_y(u)$ are constant on the transverse space.

In the case ${\Lambda_+>0}$, when the transverse space is a sphere, the solution for ${\lambda}$ can also be rewritten as 
\begin{equation}\label{reglambdasph}
\lambda=C\,\bigl[\cos\theta\cos\theta'+\sin\theta\sin\theta'\cos(\phi-\phi')\bigr]\;,
\end{equation}
where ${\theta}$ and ${\phi}$ are standard spherical coordinates, cf.~\eqref{s2x}, and ${C}$, ${\theta'}$, and ${\phi'}$ are (possibly ${u}$-dependent) transverse constants equivalent to ${q_0,\,q_x}$  and ${\,q_y}$.

A slightly more general parametrization of these spacetimes can be found in Appendix \ref{apx:backgroundsTypeD}.

\subsection{Kundt waves without gyratons (${j_i=0}$)}\label{ssc:KundtWaves}

\subsubsection*{Kundt waves on direct product spacetimes}

Now we briefly describe more general Kundt spacetimes of the form \eqref{m1} which, however, still do \emph{not} contain a gyratonic matter. In such a case, the source functions $j_i$ vanish, i.e., ${p, q=0}$, and the field equation~\eqref{lambdaeq5} reduces to
\begin{equation}\label{vaclambdaeq}
  \laplace \lambda + 2\Lambda_+\, \lambda =0\;.
\end{equation}
Let us first consider the trivial solution ${\lambda=0}$; the general case is discussed below.

Since ${p=0}$, it is possible to use the gauge transformation \eqref{gauge2} to eliminate both ${\kappa}$ and ${g}$, cf.~\eqref{kappagp}. Consequently, we obtain ${a_i=0}$ everywhere, and the metric simplifies to
\begin{equation}\label{n3waves}
d s^2=d s^2_{\mathrm{bg}} -2\,h(u,x, y)\,\trgrad u^2\,,
\end{equation}
where $d s^2_{\mathrm{bg}}$ is the metric of direct product spacetimes (some of which are listed in Table~\ref{table15}) given by \eqref{n3}. For nontrivial profile functions $h$, this class of solutions can be interpreted as specific exact Kundt gravitational waves which propagate in flat, (anti-)Nariai, Bertotti--Robinson, or  Pleba\'{n}ski--Hacyan universes (see \cite{PodoOrtag:2003:CLAQG:} and, for the limit of impulsive waves, \cite{Ortaggio:2002:PHYSR4:,OrtagPodolsky:2002:CLAQG:}).

Indeed, from the corresponding NP scalars \eqref{RicciPhipot}, \eqref{WeylPsipot} (cf. Section~\ref{ssc:NP}), by using \eqref{Delta} it follows that \eqref{f18back} remains unchanged, and, in addition, there is
\begin{equation}\label{f17alter}
\Psi_{4}=(P^2h_{,\zeta})_{,\zeta}\,,\quad\Phi_{22}=P^2h_{,\zeta\bar{\zeta}}\,,
\end{equation}
where ${\zeta=(x+iy)/\sqrt{2}}$. When ${\Psi_{4}\not=0}$, such spacetimes are thus of types~II or~N, and in general, contain a null radiation field characterized by $\Phi_{22}$. In particular, \emph{pure vacuum} gravitational waves of this type (which propagate on a vacuum or electrovacuum background space) are given by the condition ${\Phi_{22}=0}$, so that their profile functions $h$ must be of the form
\begin{equation}\label{vacuumwaves}
h=\mathcal{F}(u,\zeta)+\bar{\mathcal{F}}(u,\bar\zeta)\,,
\end{equation}
where $\mathcal{F}(u,\zeta)$ is any function, holomorphic in~$\zeta$.

\subsubsection*{Kundt waves on type~D backgrounds}

Similarly, we can also describe gravitational waves propagating on general type D backgrounds discussed above. Indeed, the field equations for ${\lambda}$, ${\kappa}$, and ${g}$ are linear, and the equation for ${h}$ is linear in ${h}$ (with nonlinear terms with ${\lambda}$ and ${\kappa}$ as a `source'). We can thus easily superpose a pure gravitational-wave contribution of the form \eqref{vacuumwaves} on top of \emph{any} background metric function ${h}$, keeping the values of ${\lambda, \kappa}$, and ${g}$ unchanged. In particular, considering the exceptional Pleba\'{n}ski--Hacyan type~D background \eqref{genPlHac}, the family of gravitational waves described by \cite{ GarciaAlvarez:1984:} is obtained.

\subsubsection*{More general Kundt waves}\label{sc:genKundt}

The equation \eqref{vaclambdaeq} is the special Helmholtz equation on the transverse space such that the coefficient of the `mass' term is exactly given by the curvature of the transverse space. Its general solution can thus be parametrized by a single function ${\mathcal{L}(u,\zeta)}$, holomorphic in ${\zeta}$, as
\begin{equation}\label{lambdasol}
  \lambda = \mathcal{L}_{,\zeta} + \bar{\mathcal{L}}_{,\bar\zeta} 
    - 2\mathcal{L}\,\bigl(\log P\bigr)_{\!,\zeta} - 2\bar{\mathcal{L}}\,\bigl(\log P\bigr)_{\!,\bar\zeta}\;.
\end{equation}
Again, the functions ${\kappa}$ and ${g}$ can be gauged away, ${\kappa=g=0}$, and the electro-vacuum condition ${\iota=p=q=0}$ implies  ${\laplace\hat h=0}$, see~\eqref{heqpot}. However, now we have an additional contribution to ${h}$ thanks to a nontrivial ${\lambda}$, cf.~\eqref{hath}:
\begin{equation}\label{hsolgenwave}
  h= \mathcal{F}(u,\zeta)+\bar{\mathcal{F}}(u,\bar\zeta) + \frac12\Lambda_+ \lambda^2\;.
\end{equation}

We have thus obtained an explicit form of a general Kundt electro-vacuum spacetime \eqref{m1}. Apart from nontrivial $H$, these most general gravitational waves within our class also have nontrivial metric functions $a_i$, 

given as ${a_\zeta= -i\Lambda_+P^{-2}(\mathcal{L}+\bar{\mathcal{L}})}$. We are not aware of a discussion of such waves in the literature. 

It should, however, be mentioned that some of these solutions have unphysical behavior of the metric functions---typical solutions of \eqref{lambdasol} and \eqref{hsolgenwave} have singularities or diverge in transverse directions. They thus cannot be interpreted as globally well-behaved gravitational waves. Nevertheless, some of them can be interpreted as external vacuum solutions around a localized matter source, e.g., around a beam of null radiation or gyratonic matter. Such solutions will be discussed in the next section. Here we only note that they can be constructed from given matter sources using the Green functions. They are regular and satisfy vacuum equations outside the sources.

Since the `mass' term in \eqref{vaclambdaeq} has a special value, this equation also admits \emph{globally regular} solutions. Regular solutions for ${\lambda}$ are exactly those discussed for the type~D backgrounds, namely, given by \eqref{gentypeDkl}, \eqref{gentypeDPQ} (or \eqref{reglambdasph}). The solution for ${h}$ which leads to the regular geometry is given by \eqref{hsolgenwave} with sufficiently smooth ${\mathcal{F}}$, e.g., when it is quadratic in ${\zeta}$.

\subsection{Gyratons on the flat background}\label{ssc:flatgyr}

Our class of solutions also contains, as a subcase, the original gyraton on a flat background \cite{Bonnor:1970b:INTHP:}. Indeed, for a vanishing cosmological constant and  electromagnetic field absent, the background is Minkowski space. If we admit only an ${r}$ independent gyraton source (i.e., if we assume ${\trdiv j=0}$, ${j_u=\iota}$) and if we employ the gauge ${g=0}$, we immediately obtain the solution discussed in \cite{Fro-Is-Zel:2005:PHYSR4:, Fro-Fur:2005:PHYSR4:}.

\section{Properties of the gyraton solutions}\label{sc:interpret5}

\subsection{Character of the gyratons}\label{ssc:gyraton}

Now we concentrate on nontrivial gyratons contained in the above class. A characteristic feature of the gyratonic matter is a nonvanishing source ${j_i}$ in \eqref{m77} or, equivalently, its two potentials ${p}$ and ${q}$, cf.~\eqref{sourcepot}. The gyratonic matter moves with the speed of light, as can be identified by inspecting the dependence of the metric function on the coordinate ${u}$. From the form of the metric \eqref{m1} we infer that ${u}$ is a null coordinate, with null generators given by the principal null congruence ${\partial_r}$. All of the metric functions can depend on this coordinate, and this dependence is not restricted by the field equations. It means that the profile of the gyraton in the ${u}$ direction can be prescribed arbitrarily. Thanks to a trivial ${r}$ dependence of the fields, such a profile remains essentially unaltered (except for the ``cooling effect'' discussed below). This can be understood as a motion of the gyraton in the direction of the null congruence ${\partial_r}$.

The characteristic spatial components of the gyraton stress-energy tensor represent a possibility of an internal energy flow of otherwise null radiation. It can be naturally split into two components.

The divergence-free component, controlled by the source potential ${q}$, corresponds to a ``rotational'' part of the energy flow. However, since the gyratonic matter is null, the nature of the ``rotation'' must be internal---it describes a spin of the null fluid. This kind of the source was discussed in the context of the gyratons in flat spacetime \cite{Fro-Is-Zel:2005:PHYSR4:, Fro-Fur:2005:PHYSR4:} and in anti-de~Sitter space \cite{Fro-Zel:2005:PHYSR4:}. From the field equation \eqref{lambdaeq5} we observe that this ``rotational'' part of the source gives rise to the component of the metric function ${a_i}$ determined by the potential ${\lambda}$ via \eqref{klpotdef5}. This component is independent of the gauge, so the presence of the ``rotational'' part of the gyraton source necessary leads to the nondiagonal component ${a_i}$ in the metric ${\eqref{m1}}$.

The rotation-free component of the gyraton source, controlled by the potential ${p}$, has a different character. As can be read out from the conservation law \eqref{jdecomp}, the source with a nontrivial divergence ${\trdiv j}$ describes an internal flow of the energy in the gyraton beam which changes its internal energy ${j_u}$ with ${r}$. We could thus understand the ${p}$ component of the source as some kind of ``cooling'' which steadily decreases the energy density of the gyraton beam. Such a kind of the energy transfer is not very plausible physically, mainly because the cooling should occur in matter moving with the speed of light. It inevitably leads to an unnatural causal behavior of the source.

Indeed, it is easy to check that the gyraton stress-energy tensor \eqref{m77} (composed by either a ${p}$ or a \mbox{${q}$ component}) does not satisfy neither a null, weak, strong nor dominant energy condition. However, for a spinning matter it is not so surprising---bad causal behavior is typical for spinning relativistic objects when they are idealized excessively. 

From the equation \eqref{kappagp} we also observe that the ${p}$ part of the source controls the combination ${g-\Lambda_-\kappa}$ of the metric functions. Splitting its influence between ${g}$ and ${\kappa}$ is just a matter of a gauge choice. We have already discussed that it is possible to eliminate either of them but not simultaneously. A gyraton source composed just from the rotation-free component (${p\neq0}$, ${q=0}$) thus does not necessary lead to a nondiagonal component ${a_i}$ in the metric---its influence can be gauged away entirely into the metric function ${g}$, and vice versa.

\subsection{Geometrical properties of the principal null
congruence}\label{ssc:prop}

Let us now briefly discuss geometrical properties of the gyraton solutions.
The additional property \eqref{aiassumption}, ${\partial_r a_i=0}$,
characteristic for the subclass of spacetimes discussed here, has a
consequence that the null vector ${k}$ is \emph{recurrent},
\cite{Prav-Prav:2002:CLAQG:, Step:2003:Cam:}, namely,
\begin{equation}\label{recurrentk5}
k_{\alpha;\beta}=(-\partial_{r}H) k_{\alpha}k_{\beta}.
\end{equation}
The null character of $k$ and the condition \eqref{recurrentk5}
also imply that the null congruence with tangent vector $k$ is
geodesic, expansion-free, sheer-free, and twist-free and thus belongs 
to the Kundt class. 

The condition \eqref{aiassumption} and the condition \eqref{Heq5} 
(the function ${H}$ is at most quadratic in $r$, with
a constant coefficient in front of $r^2$) guarantee that these Kundt
metrics are of the CSI type \cite{Coley-Her-Pel:2006:CLAQG}. For these
metrics it was also shown that there exists a ($u$-dependent) 
diffeomorfhism $\tilde{x}^i=\tilde{x}^{i}(u,x^{k})$ such
that the transverse metric \eqref{trmetric} can be made $u$-independent. We have already used
this property at the very beginning when we applied the conditions
\eqref{udepofP}. Moreover, the transverse space is locally homogeneous. 

For the complete \mbox{four-dimensional} gyraton spacetime, 
it was demonstrated in \cite{Coley-Gib-Her-Pope:2008:CLAQG} that there 
always exists a related locally homogeneous spacetime 
which has invariants that are identical to those of the Kundt CSI metric. This
``background'' metric can be obtained by setting ${a_{i}=g=h=0}$, 
\begin{equation}\label{hom}
d s_{\mathrm{bg}}^2= d s_{\perp}^2 -2\,\trgrad u\,\trgrad r + \Lambda_- r^2 \trgrad u^2\;,
\end{equation}
which is exactly the metric for direct-product background spacetimes
\eqref{n3}.

The condition \eqref{aiassumption} is also equivalent to the fact that the 2-spaces orthogonal to the transverse spaces are surface-forming.

It is a general property of the Kundt family that the vector ${k}$ is the principal null direction of the spacetime. By determining its degeneracy we can thus identify the algebraic type. To proceed, it will be convenient to introduce an aligned complex null tetrad ${\{k,\,l,\,m,\,\mb\}}$. There exists a standard choice of such a tetrad in the context of the Kundt family of spacetimes \cite{Step:2003:Cam:}, namely
\begin{equation}\label{b-vectors5}
\begin{aligned}
  k&=\partial_{r}\;,\\
  l&=\partial_{u}
     -P^2\bigl(a_{x}\partial_{x}+a_{y}\partial_{y}\bigr)
     -\bigl(H+{\textstyle\frac{1}{2}}a^2\bigr)\,\partial_{r}\;,\\
  m&=\frac{P}{\sqrt{2}}(\partial_{x}+i\partial_{y})\;,\\
  \mb&=\frac{P}{\sqrt{2}}(\partial_{x}-i\partial_{y})\;.
\end{aligned}
\end{equation}
The spacelike complex vectors ${m^a}$ and ${\mb^a}$ are tangent to the transverse space.
Clearly, ${[k,l]=-(\partial_rH)\,k}$, so the space spanned on ${k}$ and ${l}$ is indeed surface-forming.
The dual frame in the space of 1-forms reads
\begin{equation}\label{b-forms5}
\begin{aligned}
\frm{k}&=\trgrad r + (H+{\textstyle\frac{1}{2}}a^2)\, \trgrad u\;,\\
\frm{l}&=\trgrad u\;,\\
\frm{m}&=\frac{1}{\sqrt{2}P}(\trgrad x{-}i\trgrad y)+\frac{P}{\sqrt{2}}(a_{x}{-}ia_{y})\,\trgrad u\;,\\
\frm{\mb}&=\frac{1}{\sqrt{2}P}(\trgrad x{+}i\trgrad y)+\frac{P}{\sqrt{2}}(a_{x}{+}ia_{y})\,\trgrad u\;.
\end{aligned}\end{equation}

Calculating the Newman--Penrose spin coefficients with respect to this tetrad (see the following section for the nontrivial ones), we recover again the general properties that the congruence is nonexpanding and nontwisting (${\NP\rho=0}$), sheer-free (${\NP\sigma=0}$), geodesic and affinely parametrized (${\NP\kappa=\NP\eps=0}$). In addition, from ${\NP\kappa=\NP\pi=\NP\eps=0}$, it follows that the tetrad \eqref{b-vectors5} is parallelly transported along the null congruence. Moreover, the condition \eqref{aiassumption} is directly related to the vanishing coefficient ${\NP\tau=0}$.

\subsection{NP formalism in complex coordinates}\label{ssc:NP}

It turns out to be more convenient (and common in the literature on Kundt spacetimes) to introduce complex coordinates in the transverse space. Instead of conformally flat real coordinates ${x}$ and ${y}$ we will now use the complex coordinates ${\zeta}$ and ${\bar{\zeta}}$ such that
\begin{equation}\label{f1}
 \zeta=\frac{1}{\sqrt{2}}(x+iy)\;.
\end{equation}
The coordinate 1-forms and vector fields transform as
\begin{equation}\label{f2}
\begin{aligned}
  \trgrad \zeta&=\frac{1}{\sqrt{2}}(\trgrad x+i\trgrad y)\;,&
  \partial_{\zeta} &=\frac{1}{\sqrt{2}}(\partial_{x}-i\partial_{y})\;,\\
  \trgrad \bar{\zeta}&=\frac{1}{\sqrt{2}}(\trgrad x-i\trgrad y)\;,&
  \partial_{\bar\zeta}&=\frac{1}{\sqrt{2}}(\partial_{x}+i\partial_{y})\;,
\end{aligned}
\end{equation}
and the transverse Laplace operator \eqref{lapldef} on any scalar ${\psi}$ becomes
\begin{equation}\label{Delta}
  \laplace\psi = 2P^2\psi_{,\zeta\bar\zeta}\;.
\end{equation}

Instead of the real 1-form components ${a_i}$ it is customary to introduce a complex function ${W(u,\zeta,\bar\zeta)}$ by
\begin{equation}\label{f4}
W=-a_\zeta=-{\textstyle\frac{1}{\sqrt{2}}}(a_{x}{-}ia_{y})\;,\quad
\overline{W}=-a_{\bar\zeta}
\;.
\end{equation}
Substituting for ${a_i}$ the potentials via \eqref{klpotdef5} and using
${\epsilon^\zeta{}_{\zeta}=-\epsilon^{\bar\zeta}{}_{\bar\zeta}=-i}$,
${\epsilon^\zeta{}_{\bar\zeta}=\epsilon^{\bar\zeta}{}_{\zeta}=0}$,
we find that
\begin{equation}\label{Wkl}
W=-(\kappa+i\lambda)_{,\zeta}\;,\quad
\overline{W}=-(\kappa-i\lambda)_{,\bar\zeta}\;.
\end{equation}
We also obtain
\begin{equation}\label{Wrels}
\begin{gathered}
a^2=2P^2W\overline{W}\;,\\
\laplace\kappa=\trdiv a = - P^2 \bigl(\overline{W}_{\!,\zeta}+W_{\!,\bar{\zeta}}\bigr)\;,\\
\laplace\lambda=-\rot a = -b 
   = i P^2 f_{\zeta\bar\zeta} 
   = i P^2 \bigl({W}_{\!,\bar\zeta}-\overline{W}_{\!,\zeta}\bigr)\;.
\end{gathered}
\end{equation}

The metric \eqref{m1} in complex coordinates then reads
\begin{equation}\label{metriccx}
d s^2=
   \frac{2}{P^2}\trgrad \zeta \trgrad \bar{\zeta}-2\trgrad u\trgrad r-2H\trgrad u^2   -2\bigl(W\trgrad \zeta {+} \overline{W}\trgrad \bar{\zeta}\bigr) \trgrad u\;,
\end{equation}
the canonical form \eqref{P1} of ${P}$ is
\begin{equation}\label{f9}
P=1+\frac12{\Lambda_{+}}\zeta \bar{\zeta}\;,
\end{equation}
and the Maxwell tensor \eqref{EMFR} takes form
\begin{equation}\begin{split}\label{EMFcx}
F&=E\,\trgrad r\wedge\trgrad u + B\, \frac{i}{P^2}\trgrad \zeta\wedge\trgrad \bar\zeta\\
 &\qquad\qquad+\sigma_{\zeta}\,\trgrad u\wedge\trgrad \zeta+\sigma_{\bar\zeta}\,\trgrad u\wedge\trgrad \bar\zeta\;,
\end{split}\end{equation}
where ${\sigma_\zeta=(\sigma_x-i\sigma_y)/\sqrt{2}}$.

The tetrads \eqref{b-vectors5} and \eqref{b-forms5} are closely related to the introduced complex coordinates:
\begin{equation}\label{vectors15}
\begin{aligned}
  k&=\partial_{r}\;,\\
  l&=\partial_{u}+P^2\bigl(W\partial_{\bar{\zeta}}+\overline{W}\partial_{\zeta}\bigr)
     -(H+P^2W\overline{W})\partial_{r}\;,\\
  m&=P\partial_{\bar{\zeta}},\\
  \mb&=P\partial_{\zeta},
\end{aligned}
\end{equation}
and
\begin{equation}\label{forms1}
\begin{aligned}
\frm{k}&=\trgrad r+(H+P^2W\overline{W}) \trgrad u\;,\\
\frm{l}&=\trgrad u\;,\\
\frm{m}&=\frac{1}{P}\,\trgrad \bar{\zeta}-PW\trgrad u\;,\\
\frm{\mb}&=\frac{1}{P}\,\trgrad \zeta-P\overline{W}\trgrad u\;.
\end{aligned}
\end{equation}
The list of nontrivial NP coefficients is then
\begin{equation}\label{f10}
\begin{aligned}
\NP\lambda&=(P^2W)_{,\zeta}\;,\\
\NP\mu&=\frac{1}{2}P^2\bigl(W_{,\bar{\zeta}}+\overline{W}_{,\zeta}\bigr)\;,\\
\NP\nu&=P(H+P^2W\overline{W})_{,\zeta}\;,\\
\NP\gamma&={\textstyle\frac{1}{2}}\Bigl[\partial_{r}H
   +{\textstyle\frac{1}{2}}\bigl((P^2\overline{W})_{,\zeta}-(P^2W)_{,\bar{\zeta}}\bigr)\Bigr]\;,\\
\NP\alpha&={\textstyle\frac{1}{2}}P_{,\zeta}\;,\\
\NP\beta&=-{\textstyle\frac{1}{2}}P_{,\bar{\zeta}}\;.
\end{aligned}
\end{equation}

The source equations can be recovered in the Newman--Penrose formalism by comparing the components of the Ricci tensor with the corresponding components of the electromagnetic and gyraton stress-energy tensor. The general form of nonvanishing Ricci scalars for the metric \eqref{metriccx} is listed in Appendix \ref{apx:NP5} in equations \eqref{RicciPhi}. In terms of potentials these have a form
\begin{equation}\label{RicciPhipot}
\begin{aligned}
\Phi_{11}&=\frac{1}{2}\rho\;,\\
\Phi_{12}&=\frac{P}{2}\Bigl[
   -2\rho\kappa
   +\bigl(g-\Lambda_{-}\kappa\bigr)
   +i\bigl(\frac{1}{2}\laplace\lambda+\Lambda_+\lambda\bigr)
   \Bigr]_{,\bar\zeta}\;,\\
\Phi_{22}&=\frac{1}{2}\Bigl[
   r\,\laplace\! (g-\Lambda_- \kappa)
   +iP^2\bigl(a_{\zeta} b_{,\bar{\zeta}}{-}a_{\bar{\zeta}} b_{,\zeta}\bigr)\\
   &\quad
   +\laplace h
   +\frac{1}{2}\,b^2
   +\Lambda_{+}a^2
   +g\laplace\!\kappa +\partial_u\laplace\!\kappa
   \Bigr]\;,\\
R&=24\NP\Lambda =4\Lambda\;.
\end{aligned}
\end{equation}
The constants ${\Lambda}$ and ${\rho}$ have entered these expressions via combinations ${\Lambda=\frac12(\Lambda_++\Lambda_-)}$ and ${\rho=\frac12(\Lambda_+-\Lambda_-)}$ of the constants ${\Lambda_\pm}$ which parametrize the metric \eqref{metriccx} through \eqref{Heq5} and \eqref{Peq}. Their relation to the cosmological constant ${\Lambda}$ and the electromagnetic energy density ${\rho}$ is established by comparing these components to the cosmological term and to the corresponding components of the stress-energy tensors. For the electromagnetic field, the nonvanishing components are
\begin{equation}\label{EMTPhi}
\begin{aligned}
\Phi^\EM_{11}&=\frac{1}{2}\rho\;,\\
\Phi^\EM_{12}&=-P\rho\,\kappa_{,\bar\zeta}\;,\\
\Phi^\EM_{22}&=2P^2\rho\, \kappa_{,\zeta}\,\kappa_{,\bar\zeta}\;,
\end{aligned}
\end{equation}
with ${\rho}$ given by \eqref{rhodef5}. Similarly, for the gyratonic matter we obtain
\begin{equation}\label{gyrTPhi}
\begin{aligned}
\Phi^\gyr_{11}&=0\;,\\
\Phi^\gyr_{12}&=\frac{P}{2}j_{\bar\zeta}\;,\\
\Phi^\gyr_{22}&=\frac{1}{2}j_u-P^2\bigl(a_{\bar\zeta}j_\zeta + a_{\zeta}j_{\bar\zeta}\bigr)\;.
\end{aligned}
\end{equation}
The first and second source equations \eqref{kappageq} and \eqref{heqpot} are obtained from the above components ${\Phi_{12}}$ and ${\Phi_{22}}$, respectively, by realizing that
\begin{equation}\label{sourcecx}
j_\zeta = (p+iq)_{,\zeta}\;,\quad
j_{\bar\zeta} = (p-iq)_{,\bar\zeta}\;.
\end{equation}

Finally, in Appendix \ref{apx:NP5} we also present the Weyl scalars \eqref{WeylPsi} in the form which follows directly from the metric \eqref{metriccx} without using the field equations. If we introduce the potentials and employ the field equations, the nontrivial scalars reduce to
\begin{align}
 \Psi_{2}&=-\frac{1}{3}\Lambda\;,\notag\\
 \Psi_{3}&=P\Lambda\,\kappa_{,\zeta}+\frac{P}{2}(p-iq)_{,\zeta}\;,\label{WeylPsipot}\\
 \Psi_{4}&=
         r\bigl(P^2 (p-i\Lambda_- \lambda)_{,\zeta}\bigr)_{,\zeta}
         +\bigr(P^2\hat h_{,\zeta}\bigr)_{,\zeta}\notag\\
     &\quad
         +\bigl[\Lambda_+\lambda+i\Lambda_-\kappa+i p+i\partial_u\bigr]
         \bigl(P^2\lambda_{,\zeta}\bigr)_{,\zeta}\notag\\
     &\quad
         +2i P^2 q_{,\zeta} (\kappa+i\lambda)_{,\zeta}
         +p \bigl(P^2\kappa_{,\zeta}\bigr)_{,\zeta}
         -2\Lambda P^2 (\kappa_{,\zeta})^2\;,\notag
\end{align}
where ${\hat h}$ is given by \eqref{hath}.

Since ${\Psi_0=\Psi_1=0}$ and, for a nonvanishing cosmological constant, ${\Psi_{2}\neq0}$, we conclude that the vector ${k}$ is the double degenerate principal null direction and the gyraton spacetime is of the algebraical type II. The conditions for further algebraic degeneracy to the type D are, in the vacuum case, discussed in Section \ref{ssc:backgroundsTypeD} and in Appendix~ \ref{apx:backgroundsTypeD}. In the nonvacuum case these conditions are rather strong: for example, there are no nontrivial type~D gyratons with ${p=0}$.

For the vanishing cosmological constant ${\Lambda=0}$, the presence of a nontrivial rotational gyratonic matter (given by the potential ${q}$; terms with ${\kappa}$ and ${p}$ are not significant, as they can be cancelled by a suitable gauge) guarantees that the spacetime is of type III. The spacetime reduces to type N only for ${\Lambda=0}$ and ${q=0}$.

Comparing \eqref{RicciPhipot} with \eqref{WeylPsipot}, or directly \eqref{RicciPhi} with \eqref{WeylPsi}, we find that ${\Phi_{12}}$ and ${\Psi_3}$ are closely related, namely
\begin{equation}\label{f16}
\Phi_{12}+{\overline{\Psi}}_{3}=Pg_{,\bar{\zeta}}\,.
\end{equation}
The radiative characteristic of the gravitational field ${\Psi_3}$ is thus determined by the matter component ${\Phi_{12}}$, up to the term which can be controlled by the gauge. In the gauge ${g=0}$, we have directly ${\Psi_3=-\overline{\Phi}_{12}}$.

\subsection{NP components in the gauge invariant tetrad}\label{ssc:NPginv}

The tetrad \eqref{b-vectors5} introduced in Section \ref{ssc:prop} is parallelly transported and transverse-surface forming, however it is not gauge invariant under the transformation \eqref{gauge2}. Consequently, the spin coefficients and curvature scalars can have rather nontrivial dependence on the gauge transformation. It is possible to choose another tetrad which is gauge independent. Actually, it is a tetrad which is also well known in the context of the Kundt family of solutions \cite{Step:2003:Cam:}. This is related to the tetrad \eqref{b-vectors5} by a null rotation with the vector ${k}$ fixed, i.e., it is also aligned with the principal null congruence. Explicitly, this null rotation is 
\begin{equation}\begin{aligned}\label{t1}
k'&=k\;,\\
m'&=m + K k\;,\\
l'&=l + K \mb + \overline{K} m + K\overline{K}  k\;,
\end{aligned}\end{equation}
with the complex parameter ${K=-P\overline{W}}$.
Thus, the gauge invariant tetrad of null vectors is
\begin{equation}\label{vectors11}
\begin{gathered}
k'=\partial_{r}\;,\quad
l'=\partial_{u}-H\partial_{r}\;,\\
m'=P(\partial_{\bar{\zeta}}-\overline{W}\partial_{r})\;,\quad
\mb'=P(\partial_{\zeta}-W\partial_{r})\;,
\end{gathered}
\end{equation}
and for the dual frame of 1-forms we obtain

\begin{gather}
\frm{k}{}'=\trgrad r+W\trgrad\zeta +\overline{W}\trgrad\bar{\zeta}+H\trgrad u\;,\quad
\frm{l}{}'=\trgrad u\;,\notag\\
\frm{m}{}'=\frac{1}{P}\trgrad\bar\zeta\;,\quad
\frm{\mb}{}'=\frac{1}{P}\trgrad\zeta\;.\label{forms11}
\end{gather}

The corresponding nontrivial spin coefficients are
\begin{equation}\begin{aligned}\label{sc105}
&\NP\lambda'=0\;,\\
&\NP\mu'=-\frac{i}{2}b\;,\\
&\NP\nu'=P\bigl(H_{,\zeta}-W\partial_{r}H -\partial_{u}W\bigr)\;,\\
&\NP\gamma'=\frac{1}{2}\bigl(\partial_{r}H-\frac12ib\bigr)\;,\\
&\NP\alpha'=\frac{1}{2}P_{,\zeta}\;,\\
&\NP\beta'=-\frac{1}{2}P_{,\bar{\zeta}}\;,
\end{aligned}\end{equation}
and the nonvanishing  Ricci scalars are
\begin{equation}\label{sc35}
\begin{aligned}
\Phi_{11}&=\frac{1}{2}\rho\;,\\
\Phi_{12}
&=\frac{P}{2}\Bigl[
   -2i\rho\lambda
   +\bigl(g-\Lambda_{-}\kappa\bigr)
   +i\bigl(\frac{1}{2}\laplace\lambda+\Lambda_+\lambda\bigr)
   \Bigr]_{,\bar\zeta}\;,\\
\Phi_{22}
&=\frac{1}{2}\Bigl[
   r\,\laplace\!(g-\Lambda_-\kappa)
   +2P^2\bigl(a_{\bar{\zeta}}g_{,\zeta}{+}a_{\zeta} g_{,\bar{\zeta}}\bigr)\\
   &\quad+\frac{1}{2}b^2
   +\laplace h
   -\Lambda_{-}a^2
   +g\,\laplace\kappa +\partial_u\laplace\kappa
   \Bigr]\;.\\
\end{aligned}
\end{equation}
For the electromagnetic field, the nonvanishing components are
\begin{equation}\label{EMTPhi2}
\begin{aligned}
\Phi^\EM_{11}&=\frac{1}{2}\rho\;,\\
\Phi^\EM_{12}&=-iP\rho\,\lambda_{,\bar\zeta}\;,\\
\Phi^\EM_{22}&=2P^2\rho\, \lambda_{,\zeta}\,\lambda_{,\bar\zeta}\;.
\end{aligned}
\end{equation}
Similarly, for the gyratonic matter we obtain
\begin{equation}\label{gyrTPhi2}
\begin{aligned}
\Phi^\gyr_{11}&=0\;,\\
\Phi^\gyr_{12}&=\frac{P}{2}j_{\bar\zeta}\;,\\
\Phi^\gyr_{22}&=\frac{1}{2}j_u\;.
\end{aligned}
\end{equation}
The nonvanishing Weyl scalars read
\begin{align}
 \Psi_{2}&=-\frac{1}{3}\Lambda\;,\notag\\
 \Psi_{3}&=-iP\Lambda\lambda_{,\zeta}+\frac{P}{2}(p-iq)_{,\zeta}\;,\notag\\
 \Psi_{4}&=
         r\bigl(P^2 (p-i\Lambda_- \lambda)_{,\zeta}\bigr)_{,\zeta}
         +\bigr(P^2\hat h_{,\zeta}\bigr)_{,\zeta}\label{sc3a}\\
     &\quad
         +\bigl[\Lambda_-\lambda+i\Lambda_-\kappa+i p+i\partial_u\bigr]
         \bigl(P^2\lambda_{,\zeta}\bigr)_{,\zeta}\notag\\
     &\quad
         +2P^2 p_{,\zeta} (\kappa+i\lambda)_{,\zeta}
         +p \bigl(P^2\kappa_{,\zeta}\bigr)_{,\zeta}
         +2\Lambda_{-}P^2 (\lambda_{,\zeta})^2\;.\notag
\end{align}

\subsection{Electromagnetic field}\label{ssc:elmag5}

In our ansatz made in Section \ref{ssc:ansatz} we allowed the spacetime to be filled with the electromagnetic field \eqref{EMFR}. This field does not have its own dynamical degrees of freedom---it is specified just by two constants ${E}$ and ${B}$. In the presence of a gyraton, this electromagnetic field is modified through the ${\trgrad u\wedge \sigma_i\trgrad x^i}$ terms. However, the transverse 1-form ${\sigma_i}$ is uniquely determined by the gyraton, see equations \eqref{phipot5}, \eqref{phisol}.

The Maxwell tensor \eqref{EMFR} can be split into two parts
\begin{equation}\label{EMFsplit}
 F = E \bigl(\trgrad r\wedge\trgrad u+\trgrad u \wedge \trgrad\kappa\bigr)
    +B \bigl(P^{-2}\trgrad x\wedge\trgrad y+\trgrad u \wedge \trgrad\lambda\bigr)\;.
\end{equation}
It is interesting to observe that the 2-form proportional to the constant ${B}$ is the four-dimensional Hodge dual of the 2-form proportional to ${E}$. Thus, the Maxwell tensor has a familiar structure of a linear combination of dual `electric' and `magnetic' parts. Moreover, after substituting \eqref{phisol} for ${\varphi}$, the field equations depend only on the `weights' ${E}$ and ${B}$ of the electric and magnetic parts through the constant ${\rho=\frac{\varkappa\epso}{2}(E^2+B^2)}$ (via the constants ${\Lambda_\pm=\Lambda\pm\rho}$). The geometry of the spacetimes thus does not depend on a particular splitting of the electromagnetic field.

To inspect the algebraic structure of the electromagnetic field, we need the tetrad components ${\Phi_A}$ of the Maxwell tensor. With respect to the parallelly transported tetrad \eqref{vectors15}, we obtain
\begin{equation}\begin{aligned}\label{EMPhi}
&\Phi_{0}=0\;,\\
&\Phi_{1}=\frac{1}{2}(E+iB)\;,\\
&\Phi_{2}=-P(E+iB)\,\kappa_{,\zeta}\;,
\end{aligned}\end{equation}
while with respect to the gauge invariant tetrad \eqref{vectors11} we get\begin{equation}\begin{aligned}\label{EMPhiginv5}
&\Phi_{0}=0\;,\\
&\Phi_{1}=\frac{1}{2}(E+iB)\;,\\
&\Phi_{2}=iP(E+iB)\,\lambda_{,\zeta}\;.
\end{aligned}\end{equation}
It follows that the electromagnetic field is aligned with the principal null direction ${k}$ of the gravitation field, but this vector is not a double degenerate vector of the field.

The corresponding tetrad components of the electromagnetic stress-energy tensor ${\Phi^\EM_{AB}=\varkappa\epso\Phi_A{\bar\Phi}_B}$ have been listed in \eqref{EMTPhi} and \eqref{EMTPhi2}. Notice that the ${\Phi^\EM_{12}}$ and ${\Phi^\EM_{22}}$ components of \eqref{EMTPhi} can be simultaneously canceled by the gauge choice ${\kappa=0}$. This choice also cancels the component ${\Phi_2}$ in \eqref{EMPhi}.

\section{Green functions}\label{sc:GreenFc}

In Section \ref{sc:gyreq} we demonstrated that for our ansatz the Einstein--Maxwell equations effectively reduce to the Poisson equations 
\begin{equation}\label{poisson}
  \laplace\psi = -s\;,
\end{equation}
(e.g., equations \eqref{lambdab5}, \eqref{heqpot} for ${\lambda}$ and ${\hat h}$), and to the Helmholtz--Poisson equations
\begin{equation}\label{helmholtz}
  \laplace\psi + R_\perp\psi = -s\;,
\end{equation} 
(equations \eqref{lambdaeq5} and \eqref{rotjeq5} for ${\lambda}$ or ${b}$). These equations on the two-dimensional transverse space can be solved using the Green functions ${G_{(0)}}$ and ${G_{(1)}}$, respectively. Such functions satisfy\footnote{%
In the case when there exist normalizable zero modes, one has to subtract a projector to the space of these modes from the delta function on the right-hand side. See a discussion in Appendix~\ref{apx:Green}.}  
\begin{gather}
  \laplace G_{(0)}(x,x') = -\delta(x,x')\;,\\
  [ \laplace + R_\perp ] G_{(1)}(x,x') = -\delta(x,x')\;, 
\end{gather} 
where ${x}$ and ${x'}$ are points in the transverse space.
The solutions are then given by the integral over the corresponding sources
\begin{equation}
\psi(x)=\int G_{(\nu)}(x,x')\,s(x')\,\sqrt{g'_{\perp}}\,d ^2{x'}\;.
\end{equation}
In particular,
\begin{equation}\begin{aligned}\label{grfc0}
\lambda(u,x)&=\int G_{(0)}(x,x')\,b(u,x')\,\sqrt{g'_{\perp}}\,d ^2{x'}\;,\\
\hat h(u,x)&=\int G_{(0)}(x,x')\,\chi(u,x')\,\sqrt{g'_{\perp}}\,d ^2{x'}\;,
\end{aligned}\end{equation}
with the source 
\begin{equation}\label{m41}
\chi = -\iota - q \,\laplace\lambda+ p \,\laplace\kappa +2 a^i p_{,i}\;,
\end{equation}
and
\begin{equation}\begin{aligned}\label{grfc1}
\lambda(u,x)&= 2\int G_{(1)}(x,x')\,q(u,x')\,\sqrt{g'_{\perp}}\,d ^2{x'}\;,\\
b(u,x)&=       2\int G_{(1)}(x,x')\,\rot j(u,x')\,\sqrt{g'_{\perp}}\,d ^2{x'}\;.
\end{aligned}\end{equation}

It follows from the Einstein equations that the two-dimensional transverse space is a maximally symmetric space of constant curvature ${R_\perp = 2\Lambda_+}$, i.e., a plane ${E^2}$ for ${\Lambda_+=0}$, a sphere ${S^2}$ for ${\Lambda_+>0}$, and a hyperboloid ${H^2}$ (Lobachevsky plane) for ${\Lambda_+<0}$. The corresponding Green functions are known explicitly (see, e.g., \cite{Zelnikov:2008:JHEP:}) and they are discussed in more detail in  Appendix \ref{apx:Green}. Here we present only those results which are important for solution of our problem.

\subsection{Green functions for 2-plane $E^2$}\label{sc:plane}

For ${\Lambda_+\!\!=0}$ the transverse space is the flat plane. Both the Green functions coincide and they have the form
\begin{equation}\label{FrFcE2}
  G_{(\nu)}(x,x') = -\frac{1}{2\pi}\,\,\log\ell(x,x')\;.
\end{equation}
Here ${\ell(x,x')}$ is the distance between the points ${x}$ and~${x'}$.
\subsection{Green functions for 2-hyperboloid $H^2$}\label{sc:hyper}

For ${\Lambda_{+}=-1/L^2 <0}$ the transverse space \eqref{trmetric} is hyperboloid of a constant negative curvature, $L$ being the curvature radius. It can be parametrized by different useful coordinate systems. Here, we list some of them which are frequently used in the literature, namely, hyperspherical, Poincar\'e, Lobachevsky, and projective coordinates, respectively:
\begin{equation}\begin{aligned}\label{h1}
\trgrad s_{\perp}^2
&=L^2\left(\,\trgrad \rho^2+\sinh^2\rho~\trgrad \phi^2\right)\\
&={\frac{L^2}{z^2}}(\,\trgrad t^2+\trgrad z^2)\\
&={L^2}\left(\,\trgrad \mu^2+\cosh^2\mu~\trgrad \tau^2\right)\\
&={\frac{1}{\left(1{+}\frac14{\Lambda_{+}}(x^2{+}y^2)\right)^2}}\,\bigl(\,\trgrad x^2+\trgrad y^2\bigr)\;.
\end{aligned}\end{equation}
Relations of the coordinates to the projective ones are
\begin{equation}\label{lobcoors}
\begin{aligned}
  x&=2L \tanh\frac\rho2\cos\phi\;,&
  y&=2L \tanh\frac\rho2\sin\phi\;,\\
  x&=L\frac{4t}{t^2+(1+z)^2}\;,&
  y&=L\frac{t^2+z^2-1}{t^2+(1+z)^2}\;,\\
  x&=\frac{2L\cosh\mu\sinh\tau}{\cosh\mu\cosh\tau+1}\;,&
  y&=\frac{2L\sinh\mu}{\cosh\mu\cosh\tau+1}\;.
\end{aligned}
\end{equation}
Clearly, the conformally flat coordinates used in the text correspond to the projective coordinates with the choice \eqref{P1} and to the Poincar\'e coordinates with the choice \eqref{P2}.

Because of the maximal symmetry of the space, the Green functions can be expressed only in terms of the geodesic distance between the points ${\ell(x,x')}$ or, more conveniently, of its function
\begin{equation}\begin{aligned}\label{h2}
\eta(x,x')=\cosh{\left(\sqrt{-\Lambda_{+}}\,\ell(x,x')\right)}\;.
\end{aligned}\end{equation}
The function ${\eta}$ in an explicit form reads
\begin{equation}\begin{aligned}\label{h3}
\eta
&=\cosh\rho\cosh\rho'-\sinh\rho\sinh\rho'\cos(\phi-\phi')\\
&=1+{(t-t')^2+\frac{(z-z')^2}{2zz'}}\\
&=\cosh\mu\cosh\mu'\cosh(\tau-\tau')-\sinh\mu\sinh\mu'\\
&=\frac{\left(1{-}\frac{\Lambda_{+}(x^2{+}y^2)}4\right)\left(1{-}\frac{\Lambda_{+}(x'^2{+}y'^2)}4\right)+\Lambda_{+}(xx'{+}yy')}
  {\left(1{+}\frac{\Lambda_{+}(x^2{+}y^2)}4\right)\left(1{+}\frac{\Lambda_{+}(x'^2{+}y'^2)}4\right)}\;.
\end{aligned}\end{equation}
Clearly, $\eta\in [1,\infty)$, with $\eta=1$ corresponding to coincident points, and ${\eta\to\infty}$ to an infinite distance. 
Using these quantities, the Green functions in question are
\begin{equation}\begin{aligned}\label{h4}
G_{(0)}({x,x'})&=-{\frac1{4\pi}}\log\left(\eta-\frac1\eta+1\right)\;,\\G_{(1)}({x,x'})&=-{\frac1{4\pi}}\left(\eta\log\left(\eta-\frac1\eta+1\right)+2\right)\;.
\end{aligned}\end{equation}
When $\eta\rightarrow\infty$, the Green functions tend to zero.

\subsection{Green functions for 2-sphere $S^2$}\label{sc:sphere}

When ${\Lambda_{+}=1/L^2>0}$ the metric \eqref{trmetric} describes a sphere
\begin{equation}\label{s1x}
ds_{\perp}^2
={L^2}\bigl(\trgrad \theta^2\!+\sin^2\theta\trgrad \phi^2\bigr)
=\frac{\trgrad x^2\!+\trgrad y^2}{\left(1{+}\frac14{\Lambda_{+}(x^2\!+y^2)}\right)^2}\;.
\end{equation}
Spherical coordinates $(\theta,\phi)$  are related to the projective coordinates $(x,y)$, used in \eqref{m1} with ${P}$ given by \eqref{P1}, via the coordinate transformation
\begin{equation}\label{s2x}
x=2L\tan\frac\theta2\,\cos\phi\;,\quad
y=2L\tan\frac\theta2\,\sin\phi\;.
\end{equation}
Then the Green functions are functions of
\begin{equation}\label{s3x}
\eta({x,x'})=\cos{\bigl(\sqrt{\Lambda_+}\,\ell({x,x'})\bigr)}\;.
\end{equation}
Here $\eta$ varies in the interval $[-1,1]$ and has the form
\begin{equation}\begin{aligned}\label{s4}
\eta&=\cos\theta \cos\theta'+\sin\theta\sin\theta'\cos(\phi-\phi')\\
&=\frac{\left(1{-}\frac{\Lambda_{+}(x^2{+}y^2)}4\right)\left(1{-}\frac{\Lambda_{+}(x'^2{+}y'^2)}4\right)+\Lambda_{+}(xx'{+}yy')}
  {\left(1{+}\frac{\Lambda_{+}(x^2{+}y^2)}4\right)\left(1{+}\frac{\Lambda_{+}(x'^2{+}y'^2)}4\right)}\;.
\end{aligned}\end{equation}
The generic solution for the Green functions on a sphere is a linear combination of the Legendre functions ${\rm Q}_{\nu}(\eta)$ and ${\rm P}_{\nu}(\eta)$. However, the requirement of regularity at the antipodal point $\eta=-1$  singles out their particular combination. Also, one has to be cautious since both equations \eqref{poisson} and \eqref{helmholtz} on the compact sphere have normalizable zero modes---see Appendix \ref{apx:Green} for more details. Eventually, the Green functions read
\begin{equation}\begin{aligned}\label{s55}
G_{(0)}({ x,x'})&=-{\frac1{4\pi}}\log\left(1-\eta\right)\;,\\
G_{(1)}({ x,x'})&=-{\frac1{4\pi}}\Bigl(\eta\log\left({1-\eta}\right)+1\Bigr)\;.
\end{aligned}\end{equation}

Moreover, the left-hand side of \eqref{heqpot} is the Laplacian defined on a compact sphere. The integral of the Laplacian over the sphere has to be zero. This property imposes an integral condition on physically acceptable distributions of the stress-energy tensor, namely
\begin{equation}\begin{aligned}\label{s6}
\int_{S^2} \chi\,\sqrt{g_{\perp}}\,{d^2x}=0\;.
\end{aligned}\end{equation}
Similarly, for the equation \eqref{rotjeq5} we also get an integral constraint
\begin{equation}\begin{aligned}\label{s7}
\int_{S^2} \cos\theta\,\,\rot j\,\sqrt{g_{\perp}}\,{d^2x}=0\;.
\end{aligned}\end{equation}
Because of this property the zero modes do not contribu te to the components $a_i$ of the metric.
The constraints \eqref{s6} and \eqref{s7} appear only because $S^2$ is compact and are analogous to the property that closed worlds must have zero total energy, charge, or angular momentum \cite{Mar-Fro:1970:TMF:, BicakKrtous:2001:}.

\section{Conclusion}\label{sc:conclusion5}

We presented a new class of gyraton solutions on electro-vacuum background spacetimes which are formed by a direct product of two constant-curvature 2-spaces. These involve the (anti-)Nariai, Bertotti--Robinson, and Pleba\'{n}ski--Hacyan spacetimes in four dimensions. The background geometries are solutions of the Einstein--Maxwell equations corresponding to the uniform background electric and magnetic fields. The gyraton solutions are of Petrov type II and belong to the Kundt family of shear-free and twist-free nonexpanding spacetimes. 

Gyratons describe the gravitational field created by a stress-energy tensor of a spinning (circularly polarized) high-frequency beam of electromagnetic radiation, neutrino, or any other massless fields. They also provide a good approximation for the gravitational field of a beam of ultrarelativistic particles with a spin. The gyratons generalize standard {\it pp\,}-waves or Kundt waves by admitting a nonzero angular momentum of the source. This leads to other nontrivial components of the Einstein equations, namely ${G_{ui}+\Lambda g_{ui}=\varkappa T_{ui}}$, in addition to the pure radiation $uu$-component which appears for {\it pp\,}-waves or Kundt waves.

We have shown that all of the Einstein--Maxwell equations can be solved exactly  for any distribution of the matter sources (see Section \ref{sc:gyrsol} for a summary), and the problem has been reduced to finding the scalar Green functions on a two-dimensional sphere, plane or hyperboloid. These Green functions have been presented in detail in Section \ref{sc:GreenFc}.
Special cases of these gyraton solutions and their properties are discussed in Sections \ref{sc:knownsol} and \ref{sc:interpret5}.

We have also studied the gyraton solutions using the Newman--Penrose formalism. The characteristic  term $a_{i}$, describing the rotational part of the gyraton, generates the nontrivial Ricci $\Phi_{12}$ and Weyl $\Psi_{3}$ scalars, in addition to the case of pure {\it pp\,} and Kundt waves. Curiously, there exists a very simple relation \eqref{f16} between them.

To complete our investigation, we have also studied gyratons on more general type-D backgrounds (including the exceptional Pleba\'{n}ski--Hacyan spacetime) which are not direct-product spaces. In addition, in Section \ref{sc:genKundt} we have identified a special subclass of the gyraton solutions---general vacuum Kundt waves which also contain cases previously not discussed in the literature.

A natural next step would be the study of gyratons in a full family of Kundt spacetimes, especially on conformally flat backgrounds, including the (anti-)de~Sitter universe. Another generalization could be their extension to higher dimensions, where, however, one has to deal with a richer possible structure of the transverse geometries.

\begin{center}{\large\bf Acknowledgements}\end{center}\par
H.~K. was supported by Grant No.~GA\v{C}R-205/09/H033 and by the Czech Ministry of Education under Project No.~LC06014. A.~Z. was financially supported by the Killam Trust and partly by the Natural Sciences and Engineering Research Council of Canada. P.~K. was supported by Grant No.~GA\v{C}R-202/09/0772, and J.~P. by Grant No.~GA\v{C}R-202/08/0187.
H.~K., P.~K., and J.P. thank the University of Alberta (Edmonton, Canada) for the hospitality
during their stays where this work has started and has been finished. A.~Z. is grateful to the Charles University (Prague, Czech Republic) for hospitality during his work on this paper. The authors are also grateful to Valeri Frolov and Dmitri Pogosyan for stimulating discussions.

\addcontentsline{toc} {section}{Appendix}
\begin{subappendices}
\section{The Einstein equations}\label{apx:AppAa}

Here we present geometric quantities which appear in the Einstein field equations, namely the Einstein tensor of the general metric \eqref{m1} and the electromagnetic stress-energy tensor corresponding to the field \eqref{EMFR}.

The inverse to the metric \eqref{m1} is
\begin{equation}
\begin{split}\label{a01}
g^{\mu\nu}&\partial_{\mu}\partial_{\nu}=
  P^2\bigl(\partial_{x}\partial_{x}+\partial_{y}\partial_{y}\bigr)
  -2\partial_{u}\partial_{r}\\
  &+2P^2\bigl(a_{x}\partial_{x}+a_{y}P^2\partial_{y}\bigr)\partial_{r}
  +2(H+\frac{1}{2} a^2)\,\partial_{r}\partial_{r}.
\end{split}\raisetag{8ex}
\end{equation}

The stress-energy tensor of the electromagnetic field $T^\EM$,
according to the definition
\begin{equation}\label{defem5}
T^\EM_{\mu\nu}=\epso\Bigl(
   F_{\mu\kappa}F_{\nu\lambda}\,g^{\kappa\lambda}
   -\frac{1}{4}g_{\mu\nu}\,F_{\kappa\lambda}F^{\kappa\lambda}
   \Bigr)\;,
\end{equation}
has nonzero components
\begin{align}
\varkappa T^\EM_{ur}&=\rho\;,\notag\\
\varkappa T^\EM_{uu}&=2H\rho+
   \varkappa\epso(\sigma-E a)^2\;,\notag\\
\varkappa T^\EM_{ux}&=\varkappa\epso\Bigl(\frac12(E^2{-}B^2)\,a_{x}-EB\, a_{y}
   -E\sigma_{x}+B \sigma_{y}\Bigr)\;,\notag\\
\varkappa T^\EM_{uy}&=\varkappa\epso\Bigl(\frac12(E^2{-}B^2)\,a_{y}+EB a_{x}
   -E\sigma_{y}-B \sigma_{x}\Bigr)\;,\notag\\
\varkappa T^\EM_{xx}&=\frac{\rho}{P^2}\;,\label{EMTe}\\
\varkappa T^\EM_{yy}&=\frac{\rho}{P^2}\;,\notag
\end{align}
where the density ${\rho=\frac{\varkappa\epso}{2}(E^2{+}B^2)}$ was defined in~\eqref{rhodef5}.

The Einstein tensor for the metric \eqref{m1} reads
\begin{equation}\label{EinsteinT5}
\begin{aligned}
G_{ur}&= \laplace\!\log P\;,\\
G_{uu}&=\frac{1}{2}b^2+\laplace H +(\partial^2_{r}H)a^2+2a^i\partial_{r}H_{,i}\\
&\quad +(\partial_r H)\,\trdiv a+\partial_{u}\trdiv a
+2H\laplace\!\log P\;,\\
G_{ux}&=\frac{1}{2}b_{,y}-a_{x}\bigl(\laplace\!\log P-\partial^2_{r}H\bigr)+\partial_{r}H_{,x}\;,\\
G_{uy}&=-\frac{1}{2}b_{,x}-a_{y}\bigl(\laplace\!\log P-\partial^2_{r}H\bigr)+\partial_{r}H_{,y}\;,\\
G_{xx}&=\frac{1}{P^2}\,\partial^2_{r}H\;,\\
G_{yy}&=\frac{1}{P^2}\,\partial^2_{r}H\;,
\end{aligned}
\end{equation}
where
\begin{equation}\label{lapllogpcx5}
  \laplace\!\log P=P\bigl(P_{,xx}+P_{,yy}\bigr)-\bigl(P_{,x}^2+P_{,y}^2\bigr)\;,
\end{equation}
cf.~\eqref{trsccurv}.
Here we have used only the metric \eqref{m1}, without assuming any other information about the metric functions. In particular, we have not used the field equations. To be more precise, in the components ${G_{ui}}$ and ${G_{ij}}$ we employed the fact that ${P}$ is ${u}$-independent; cf.\ relation \eqref{udepofP}. However, as we already mentioned in section \ref{ssc:transsp}, such a choice is always possible provided that the transverse spaces have the same homogeneous geometry---which can be derived just from the component ${G_{ur}}$.

\section{Newman--Penrose quantities}\label{apx:NP5}

In this appendix we present the Ricci and Weyl curvature scalars corresponding to our subclass of the Kundt family given by the metric \eqref{metriccx}, 
\begin{equation}
d s^2=
   \frac{2}{P^2}\trgrad \zeta \trgrad \bar{\zeta}-2\trgrad u\trgrad r-2H\trgrad u^2   -2\bigl(W\trgrad \zeta {+} \overline{W}\trgrad \bar{\zeta}\bigr) \trgrad u\;,
\end{equation}
with functions ${W(u,\zeta,\bar\zeta)}$ and ${P(\zeta,\bar\zeta)}$ independent of ${r}$, or ${r}$ and ${u}$, respectively. The expressions below have not employed any field equations.

\subsection*{The standard tetrad}
With respect to the tetrad \eqref{vectors15} the only nonvanishing  Ricci scalars are
\begin{equation}\label{RicciPhi}
\begin{aligned}
\Phi_{11}&=\frac{1}{4}(\partial^2_{r}H+\laplace\!\log P)\;,\\
\Phi_{12}
&=\frac{P}{2}\biggl[\partial_{r}H_{,\bar{\zeta}}
  +(\laplace\!\log P)\overline{W}
  -\frac{1}{2}i b_{,\bar{\zeta}}\biggr]\;,\\
\Phi_{22}
&=\frac{P^2}{2}\biggl[\frac{b^2}{2P^2}
  +2H_{,\zeta\bar{\zeta}}
  -i(W b_{,\bar{\zeta}}{-}\overline{W} b_{,\zeta})\\
  &\quad+2(\laplace\!\log P) W\overline{W}
  -(\partial_{r}H\!+\partial_{u})(\overline{W}_{,\zeta}{+}{W}_{,\bar{\zeta}})\biggr]\;,\\
R&=24{\Lambda}_{\rm NP}=2(-\partial^2_{r}H+\laplace\!\log P)\;,
\end{aligned}
\end{equation}
and nonvanishing Weyl scalars read
\begin{equation}
\begin{aligned}\label{WeylPsi}
\Psi_{2}&=\frac{1}{6}(\partial^2_{r}H-\laplace\!\log P)\;,\\
\Psi_{3}
 &=\frac{P}{2}\left[\partial_{r}H_{,\zeta}-(\laplace\!\log P)W-\frac{1}{2}\,i\,b_{,\zeta}\right]\;,\\
\Psi_{4}
&=-iP^2W b_{,\zeta}-(\laplace\!\log P)P^2W^2-(P^2\partial_{u}W)_{,\zeta}\\
  &\quad+(P^2H_{,\zeta})_{,\zeta}-(\partial_{r}H)(P^2W)_{,\zeta}\;.\\
\end{aligned}
\end{equation}
Here, ${\laplace\!\log P}$ is given in \eqref{lapllogpcx5}, and ${b=iP^2(\overline{W}_{,\zeta}-W_{,\bar\zeta})}$ as in \eqref{Wrels}.
The components of the electromagnetic field are 
\begin{equation}\begin{aligned}\label{f10a}
&\Phi_{0}=0,\\
&\Phi_{1}=\frac{1}{2}(E+iB),\\
&\Phi_{2}=P\left(-\sigma_{\zeta}+iBW\right)\;,
\end{aligned}\end{equation}
where ${\sigma_{\zeta}=\frac{1}{\sqrt{2}}(\sigma_{x}-i\sigma_{y})}$. 

\subsection*{The gauge invariant tetrad}

With respect to the gauge invariant tetrad \eqref{vectors11} we obtain that the only nonvanishing  Ricci scalars are
\begin{equation}\begin{aligned}\label{ff1}
\Phi_{11}&=\frac{1}{4}(\partial^2_{r}H+\laplace\!\log P)\;,\\
\Phi_{12}
  &=\frac{P}{2}\left[-\overline{W}(\partial^2_{r}H)
    +\partial_{r}H_{,\bar{\zeta}}
    -\frac{1}{2}i b_{,\bar{\zeta}}\right]\;,\\
\Phi_{22}
  &=\frac{P^2}{2}\biggl[\frac{b^2}{2P^2}+2H_{,\zeta\bar{\zeta}}
    -2(\overline{W}\partial_{r}H_{,\zeta}{+}W\partial_{r}H_{,\bar{\zeta}})\\
    &\quad+2(\partial^2_{r}H)W\overline{W}
    -(\partial_{r}H\!+\partial_{u})(\overline{W}_{\!,\zeta}{+}W_{\!,\bar{\zeta}})\biggr]\;,\\
\end{aligned}\end{equation}
the nonvanishing Weyl scalars read
\begin{equation}\begin{aligned}\label{ff2}
\Psi_{2}&=\frac{1}{6}(\partial^2_{r}H-\laplace\!\log P)\;,\\
\Psi_{3}
  &=\frac{P}{2}\left[-W(\partial^2_{r}H)+\partial_{r}H_{,\zeta}
    -\frac{1}{2}ib_{,\zeta}\right]\;,\\
\Psi_{4}
  &=P^2W^2(\partial^2_{r}H)-(P^2\partial_{u}W)_{,\zeta}+(P^2H_{,\zeta})_{,\zeta}\\
    &\quad-P^2 W (\partial_{r}H)_{,\zeta}
    -(P^2W\partial_{r}H)_{,\zeta}\;,\\
\end{aligned}\end{equation}
and, finally, the scalars for electromagnetic field are
\begin{equation}\begin{aligned}\label{sc2}
&\Phi_{0}=0\;,\\
&\Phi_{1}=\frac{1}{2}(E+iB)\;,\\
&\Phi_{2}=-P\left(EW+\sigma_{\zeta}\right)\;.
\end{aligned}\end{equation}

\section{All spacetimes of type D}\label{apx:backgroundsTypeD}

The purpose of this appendix is to derive all electro-vacuum solutions of the algebraic type~D within the class considered. Inspecting the field equations \mbox{\eqref{sumgkpeq}--\eqref{sumheq}} without the gyratonic matter, ${\iota=p=q=0}$, we find
\begin{gather}
g= \Lambda_-\kappa\;,\\
P^2\lambda_{,\zeta\bar\zeta}+\Lambda_+\lambda=0\;,\label{vaclambda}\\
P^2\hat h_{,\zeta\bar\zeta}=0\;,\label{vach}
\end{gather}
with ${\hat h}$ given by \eqref{sumhhatdef}. Using the gauge transformation \eqref{gauge2}, we could eliminate both ${\kappa}$ and ${g}$, but this is not necessary in the following.

From \eqref{WeylPsipot} we infer that the vector ${k=\partial_r}$ is a double degenerate principal null direction. When ${\Psi_2\neq0}$, i.e., for a nonvanishing cosmological constant, the condition that there exists another degenerate null direction is
\begin{equation}
3\Psi_4\Psi_2=2\Psi_3^2\;,
\end{equation}
cf.~\cite{Step:2003:Cam:}. This reduces to the relation
\begin{equation}
\begin{split}
0&=i r \Lambda_- \bigl(P^2\lambda_{,\zeta}\bigr)_{,\zeta}\\
 &\quad +\bigl(P^2 \hat h_{,\zeta}\bigr)_{,\zeta}
  +\bigl[\Lambda_+\lambda+\Lambda_-\kappa+i\partial_u\bigr] \bigl(P^2\lambda_{,\zeta}\bigr)_{,\zeta}
  \;.
\end{split}
\end{equation}
Taking into account the ${r}$ dependence, we obtain the following two conditions\footnote{The case ${\Lambda_-=0}$ can be easily discussed separately.}:
\begin{gather}
\bigl(P^2\lambda_{,\zeta}\bigr)_{,\zeta}=0\;,\label{typeDlambda}\\
\bigl(P^2\hat h_{,\zeta}\bigr)_{,\zeta}=0\;.\label{typeDh}
\end{gather}
These equations must be accompanied by the condition \eqref{Peq} for the metric function ${P}$, namely,
\begin{equation}
P^2(\log P^2)_{,\zeta\bar\zeta}=\Lambda_+\;.\label{Peqapx}
\end{equation}

\subsection*{The case ${\Lambda_+\neq0}$}

Integrating \eqref{vaclambda}, we obtain
\begin{equation}\label{lambda,zeta}
\lambda_{,\zeta}=\frac{\bar{\mathcal{L}}}{P^2}\;,
\end{equation}
where the arbitrary function ${\bar{\mathcal{L}}(u,\bar\zeta)}$ can depend only on ${\bar\zeta}$ and ${u}$. First assuming ${\Lambda_+\neq0}$, we can substitute the ${\bar\zeta}$ derivative of \eqref{lambda,zeta} into \eqref{vaclambda}, which leads to
\begin{equation}\label{typeDlambdasol}
\lambda=\frac{\bar{\mathcal{L}}\,P^2{}_{,\bar\zeta}-\bar{\mathcal{L}}_{,\bar\zeta}\,P^2}{\Lambda_+ P^2}\;.
\end{equation}
Now we have to check the consistency of this general solution for ${\lambda}$ with \eqref{lambda,zeta} (since we have used the equation \eqref{vaclambda} to obtain ${\lambda}$). It turns out that the solution is consistent only thanks to the fact that the constants ${\Lambda_+}$ which appears in \eqref{vaclambda} and \eqref{Peqapx} are the same.

However, the function ${\bar{\mathcal{L}}}$ in \eqref{typeDlambdasol} is not arbitrary. The last condition which must be satisfied is that ${\lambda}$ is real. It is not straightforward to find the consequences for ${\bar{\mathcal{L}}}$ explicitly in a general case. But we can use the freedom in transverse diffeomorfhism to transform the solution of \eqref{Peqapx} into a particular form. It will be useful to use such transverse coordinates for which ${P}$ is linear  both in ${\zeta}$ and~${\bar\zeta}$: 
\begin{equation}\label{Plincond}
P_{,\zeta\zeta}=0\;.
\end{equation}
Explicitly,
\begin{equation}
P=p_0+p_1\zeta+\bar p_1\bar\zeta+p_2\zeta\bar\zeta\;,
\end{equation}
with ${p_0}$ and ${p_2}$ real constants, and ${p_1\in\mathbb{C}}$, satisfying (as a consequence of \eqref{Peqapx}) the relation
\begin{equation}\label{Pcoefcond}
p_0p_2-p_1\bar p_1=\frac12\Lambda_+\;.
\end{equation}
The solutions \eqref{P1} and \eqref{P2} are particular examples of such a choice.

Assuming \eqref{Plincond}, we can now easily find the reality condition for ${\lambda}$ given by \eqref{typeDlambdasol}. Since ${P}$ is real, it requires that ${Q\equiv P \lambda=(2\bar{\mathcal{L}}\,P_{,\bar\zeta}-\bar{\mathcal{L}}_{,\bar\zeta}\,P)/\Lambda_+}$ is also real. Taking the derivative ${Q_{,\zeta\zeta}}$, we find that it vanishes, and ${Q}$ is thus linear in both ${\zeta}$ and ${\bar\zeta}$:
\begin{equation}\label{Qlincond}
Q=q_0+q_1\zeta+\bar q_1\bar\zeta+q_2\zeta\bar\zeta\;.
\end{equation}
Here ${q_0}$ and ${q_2}$ must be real and ${q_1}$ complex transverse constants, but they can by ${u}$-dependent. The field equation \eqref{vaclambda} now gives the restriction
\begin{equation}\label{Qcoefcond}
p_0q_2+p_2q_0=p_1\bar q_1+\bar p_1 q_1\;.
\end{equation}
Substituting in \eqref{lambda,zeta}, we find
\begin{equation}
\begin{split}
\bar{\mathcal{L}}&=
  (p_0q_1-q_0p_1)+(\bar p_1 q_2-p_2\bar q_1)\bar\zeta^2\\
  &\quad+(p_0q_2-p_2q_0-p_1\bar q_1+\bar p_1 q_1)\bar\zeta\;.
\end{split}
\end{equation}

In particular, for ${P}$ given by \eqref{P1} 
we get
\begin{align}
P&=1+\frac12\Lambda_+\zeta\bar\zeta\;,\\
Q&=q_0(1-\frac12\Lambda_+\zeta\bar\zeta)+q_1\zeta+\bar q_1\bar\zeta\;,
\end{align}
which is equivalent to \eqref{gentypeDPQ} in real coordinates ${x}$ and ${y}$.
For ${P=\sqrt{-\Lambda_+}\,x}$, which is a generic example of the choice \eqref{P2}, we have ${Q=c\sqrt{-\Lambda_+}\,y}$ and
\begin{equation}
\lambda=-i c\,\frac{\zeta-\bar\zeta}{\zeta+\bar\zeta}=c\,\frac{y}{x}\;.
\end{equation}

Finally, we should solve the equations \eqref{vach} and \eqref{typeDh} for ${\hat h}$. As for ${\lambda}$, we find that ${\hat h_{,\zeta} = P^{-2}\bar{\mathcal{H}}}$, with ${\bar{\mathcal{H}}_{,\zeta}=0}$. By substituting into \eqref{vach}, this leads to
\begin{equation}
0=P^2\hat h_{,\zeta\bar\zeta}
 = \bar{\mathcal{H}}_{,\bar\zeta}-\bar{\mathcal{H}}\bigl(\log P^2\bigr)_{,\bar\zeta}\;.
\end{equation}
Taking the derivative with respect to ${\zeta}$ and using \eqref{Peqapx}, we find
\begin{equation}
0=\bar{\mathcal{H}}(\log P^2)_{,\zeta\bar\zeta}=\frac{\Lambda_+\bar{\mathcal{H}}}{P^2}\;.
\end{equation}
For ${\Lambda_+\neq0}$ we thus obtain that the conditions \eqref{vach} and \eqref{typeDh} admit only the trivial solution ${\bar{\mathcal{H}}=0}$. 
Thanks to \eqref{hath}, for ${h}$ we obtain
\begin{equation}
h=\frac12\Lambda_+\lambda^2-\frac12\Lambda_-\kappa^2-\partial_u\kappa\;,
\end{equation}
in which ${\kappa}$ can be set to zero by a proper gauge.

\subsection*{The case ${\Lambda_+=0}$}

If the transverse space is flat, i.e. ${\Lambda_+=0}$, we naturally chose Cartesian transverse coordinates for which ${P=1}$. We thus immediately get the conditions
\begin{equation}
  \lambda_{,\zeta\bar\zeta}=0\;,\quad   \lambda_{,\zeta\zeta}=0\;,
\end{equation}
which imply ${\lambda_{,\zeta}=q_1}$ and
\begin{equation}\label{lambdalin}
  \lambda=q_0+q_1\zeta+\bar q_1\bar\zeta=q_0+q_x x + q_y y\;,
\end{equation}
where ${q_0(u), q_x(u)}$ and  ${q_y(u)}$ are real and ${q_1(u)}$ complex functions of ${u}$ only. The metric 1-form ${a_i}$ is then
\begin{equation}
  a_i = (\kappa+q_y x - q_x y)_{,i }\;.
\end{equation}
Thanks to such a very special form \eqref{lambdalin}, it is a gradient, and therefore it could be transformed away by a suitable choice of ${\kappa}$ using the gauge transformation. However, it would generate a nontrivial function ${g}$ and contributions to the equation for ${\hat h}$. Therefore, such a gauge fixing may not be the best choice.

Since in this case ${\hat h}$ satisfies the same equation as ${\lambda}$, we can write
\begin{equation}\label{hathlin}
  \hat h=L_0+L_1\zeta+\bar L_1\bar\zeta=L_0+L_x x + L_y y\;,
\end{equation}
with ${L_0(u)}$, ${L_x(u)}$ and ${L_y(u)}$ real and ${L_1(u)}$ complex transverse constants. With the gauge ${\kappa=g=0}$, the relation \eqref{hath} gives
\begin{equation}\label{hlin}
  h=L_x x+L_y y\;,
\end{equation}
where the constant ${L_0}$ has been eliminated by rescaling the coordinate ${u}$, i.e., incorporating the gauge transformation \eqref{urep}. We have thus obtained the metric \eqref{genPlHac} which is the ${a_i\neq0}$ generalization of the exceptional Pleba\'nski--Hacyan spacetime.

\section{Scalar Green functions on constant curvature spaces}\label{apx:Green}

The Green functions for the Poisson and Helmholtz-Poisson equations \eqref{poisson} and \eqref{helmholtz} discussed in Section~ \ref{sc:GreenFc} are special cases of the Green function satisfying 
\begin{equation}\label{G1}
\left[\laplace-\xi R_\perp\right]\,G({x,x'})=-\delta({x,x'})\;,
\end{equation}
with ${\xi=0}$ and ${\xi=-1}$, respectively. In our case, the scalar curvature ${R_{\perp}=2\Lambda_+}$ is constant, and the manifold is either a 2-plane, a 2-sphere or a 2-hyperboloid, depending on the sign of $\Lambda_+$. Because of the symmetry of these spaces the Green function $G({x,x'})$ is only the function of the geodesic distance $\ell({x,x'})$ between the points ${x}$ and ${x'}$. In this appendix we present the Green functions for an arbitrary parameter~$\xi$. 

Since the case of flat plane is trivial, we will discuss only the cases of a hyperboloid and a sphere. Let us note that on a sphere, for particular values ${\xi=-\frac12 l(l+1)}$ with integer $l\ge 0$ the equation \eqref{G1} has to be modified in order to take into account normalizable zero modes. Namely, it is neccessary to subtract from the $\delta$-function a projector to the space of zero modes.

\subsection*{2-hyperboloid $H^2$}

For ${\Lambda_+<0}$, the Green functions can be expressed in terms of the Legendre functions $Q_{\nu}$ (see, e.g., \cite{Zelnikov:2008:JHEP:})
\begin{equation}\label{G2}
G_{(\nu)}({x,x'})={\frac1{2\pi}}\,Q_{\nu}(\eta)\;,
\end{equation}
where
\begin{gather}
\nu=-{\frac12}+\sqrt{{\frac14}-2\xi}\;,\label{G3a}
\end{gather}
and ${\eta(x,x')}$ is given by the equations \eqref{h2}, \eqref{h3}.
When the separation between the two points increases, $\eta\rightarrow\infty$ and the Green function \eqref{G2} tends to zero. This is a proper boundary condition for Green functions on hyperbolic spaces.
In our paper only the ${\nu=0,1}$ cases are important, and the corresponding Green functions can be expressed in terms of elementary functions:
\begin{equation}\label{G5}
G_{(0)}({x,x'})={\frac1{2\pi}}Q_{0}(\eta)=-{\frac1{4\pi}}\log\Bigr(\frac{\eta-1}{\eta+1}\Bigr)\;,
\end{equation}
\begin{equation}\label{G6}
G_{(1)}({x,x'})={\frac1{2\pi}}Q_{1}(\eta)=-{\frac1{4\pi}}\left(\eta\log\Bigr(\frac{\eta-1}{\eta+1}\Bigr)+2\right)\;.
\end{equation}

\subsection*{2-sphere $S^2$}

When ${\Lambda_+>0}$, the Green function \eqref{G1} for
arbitrary $\xi$ can be expressed in terms of the Legendre function ${\rm P}_{\nu}$:
\begin{equation}\begin{aligned}\label{G7}
G_{(\nu)}({ x,x'})&=-{\frac1{4\sin{\pi\nu}}}{\rm P}_{\nu}(-\eta)\\
&={\frac1{2\pi}}{\rm Q}_{\nu}(\eta)-{\frac14}\cot({\pi\nu})\,{\rm P}_{\nu}(\eta)\;,
\end{aligned}\end{equation}
see \cite{Zelnikov:2008:JHEP:}, with ${\nu}$ given by \eqref{G3a} and ${\eta(x,x')}$ by the equations \eqref{s3x} and \eqref{s4}.
The Legendre functions ${\rm P}_\nu,\,{\rm Q}_\nu$ are defined on the interval $[-1,1]$ and are related to the Legendre functions ${P_\nu,\,Q_\nu}$ on the complex plane \\
as ${{\rm P}_\nu(\eta)=\frac12[P_\nu(\eta{+}i0)+P_\nu(\eta{-}i0)]}$, and similarly for ${Q}$s \cite{Bateman:book:}. When ${0<\eta<1}$ we can observe from the second line of \eqref{G7} that there is the required singularity at ${\eta=1}$, when both points coincide. For ${-1<\eta<0}$, the first line gives the regularity of the Green function at ${\eta=-1}$ corresponding to antipodal points.


If $\nu$ is an integer number, one has to be cautious because of a contribution of normalizable zero modes. The zero modes must be removed from the spectrum. This case thus must be treated separately.

When $\nu=0$, the corresponding Green function then satisfies the equation
\begin{equation}\begin{aligned}\label{G10}
\laplace G_{(0)}({x,x'})=-\delta({x,x'})+{\frac{\Lambda_+}{4\pi}}\;,
\end{aligned}\end{equation}
where ${\frac{4\pi}{\Lambda_+}}=\int_{S^2}\sqrt{g_{\perp}}{d^2{x}}$ is the total area of the compact transverse space \eqref{trmetric}. 
The term subtracted from the delta function is exactly the projector on the zero mode, which for ${\nu=0}$ is constant.

The solution of this problem, which is regular at antipodal points ($\eta=-1$) reads
\begin{equation}\begin{aligned}\label{G11}
G_{(0)}({x,x'})&=-{\frac1{4\pi}}\log\left(1-\eta\right)\;,
\end{aligned}\end{equation}
with $\eta$ given again by \eqref{s3x}, \eqref{s4}. One can shift this Green function by a constant to satisfy different boundary conditions at the antipodal points, but this constant is unimportant since it will drop out of the equation \eqref{heq} anyway due to the constraint \eqref{s6}.

In the case of the equation \eqref{rotjeq} we have $\nu=1$. 
In this case the projector on zero modes is proportional to ${\rm P}_1(\eta({ x,x'}))=\eta({x,x'})$ and the equation for the Green function reads
\begin{equation}\begin{aligned}\label{G12}
\left[\laplace +2\Lambda_+\right]\,G_{(1)}({x,x'})=-\delta({x,x'})+\frac{3\Lambda_+}{4\pi}\,\eta({x,x'})\;.
\end{aligned}\end{equation}
The solution can be found explicitly as
\begin{equation}\begin{aligned}\label{G13}
G_{(1)}({x,x'})&=-{\frac1{4\pi}}\Bigl(\eta\log\left({1-\eta}\right)+1\Bigr)\;.
\end{aligned}\end{equation}
Similarly to the ${\nu=0}$ case, one can add to the Green function the projector to the zero modes with an arbitrary coefficient to match different boundary conditions at the antipodal point. But the component $T^{\rm gyr}_{ui}=j_{i}=\epsilon_{i}{}^k q_{,k}$ of the stress-energy tensor of the gyraton has to satisfy the integral constraint similar to \eqref{s6}, namely
\begin{equation}\begin{aligned}\label{G14}
\int_{S^2} \cos\theta\;q\,\sqrt{g_{\perp}}\, {d^2{x}}=0\;.
\end{aligned}\end{equation}
Because of this property, zero modes do not contribute to the components $a_i$ of the metric.
Constraints \eqref{s6}, \eqref{s7} appear only because $S^2$ is a compact manifold.

Let us finally note that in the case of a general integer ${\nu=l}$ the modified Green function equation is
\begin{equation}\label{G15}
\begin{split}
&\left[\laplace\!\!+l(l{+}1)\Lambda_+\right]\,G_{(l)}({ x,x'})\\
&\qquad=-\delta({ x,x'})+\Lambda_+\frac{2l{+}1}{4\pi}\,{\rm P}_l(\eta(x,x'))\;.
\end{split}
\end{equation}
Its solution can be written as
\begin{equation}\label{G16}
G_{(l)}({x,x'})=-{\frac1{4\pi}}\Bigl({\rm P}_l(\eta)\log\left({1-\eta}\right)+U_l(\eta)\Bigr)\;,
\end{equation}
with $U_l(\eta)$ being a specific polynomial in $\eta$ of the \mbox{${(l{-}1)}$-th} order. 
In particular, 
${U_0=0}$,
${U_1=1}$,
${U_2=\frac{3}{2}\eta+\frac{1}{3}}$,
${U_3=\frac{5}{2}\eta^2+\frac{3}{5}\eta-\frac{2}{3}}$, etc.

\end{subappendices}

\newpage\cleardoublepage
\chapter{The gyratons on the Melvin spacetime}
\section{The gyratons on the Melvin spacetime}\label{sc:gyreqq}
\subsection{\label{sec:level11}Introduction}

Gyraton solutions represent the gravitational field of
a localized matter source with an intrinsic rotation
which is moving at the speed of light.
Such an idealized ultrarelativistic source can describe
a pulse of a spinning radiation beam and it is accompanied
by a sandwich or impulsive gravitational wave.

The gravitational fields generated by (nonrotating) light
pulses and beams were already studied by Tolman \cite{Tol:1934:Oxf:}
in 1934, who obtained the corresponding solution in the
linear approximation of the Einstein theory.
Exact solutions of the Einstein--Maxwell equations for such `pencils of light'
were found and analyzed by Peres \cite{Peres:1960:PHYSR:} and Bonnor
\cite{Bonnor:1969:COMMPH:,Bonnor:1969:INTHP:,Bonnor:1970a:INTHP:}.
These solutions belong to a general family of {\it pp\,}-waves
\cite{Step:2003:Cam:,GrifPod:2009:Cam:}.

In the impulsive limit (i.e., for an infinitely thin beam,
and for the delta-type distribution of the light-pulse in time),
the simplest of these solutions represents the well-known
Aichelburg--Sexl metric \cite{Aich-Sexl:1971:} which describes
the field of a pointlike null particle.  Subsequently,
more general impulsive waves were found
\cite{FerPen90,LouSan92,HotTan93,KBalNac95,KBalNac96,PodGri97,PodGri98prd}
(for recent reviews see \cite{Podolsky02,BarHog:2003:WorldSci:}).

The gyraton solutions are generalization of
{\it pp\,}-waves which belong to the Kundt class
for which the source---the beam of radiation---carries not only energy,
but also an additional angular momentum. Such spacetimes were
first considered by Bonnor in \cite{Bonnor:1970b:INTHP:},
who studied the gravitational field created by a spinning null fluid.
In some cases, this may be interpreted as a
massless neutrino field~\cite{Griffiths:1972a:INTHP:}.

Gyratons on Minkowski background are locally isometric to standard
{\it pp\,}-waves in the exterior vacuum region, outside the source.
The interior region contains a nonexpanding null matter which possesses
an intrinsic spin. In general, these solutions are obtained by keeping
nondiagonal terms $g_{ui}$ in the Brinkmann form \cite{Brink:1925:MAAN:}
of the {\it pp\,}-wave solution, where $u$ is the null coordinate
and $x^i$ are orthogonal spatial coordinates. The corresponding
energy-momentum tensor thus also contains an extra nondiagonal term
$T_{ui}=j_{i}$. In four dimensions, the terms $g_{ui}$ can be set
to zero {\it locally}, using a suitable gauge transformation.
However, they can not be {\it globally} removed because the gauge
invariant contour integral $\oint g_{ui}(u,x^{j})\,\trgrad x^{i}$
around the position of the gyraton is proportional to the nonzero
angular momentum density $j_{i}$, which is nonvanishing.

These gyratons were investigated (in the linear approximation)
in \cite{Fro-Fur:2005:PHYSR4:} in higher dimensional flat space,
the exact gyraton solutions propagating in an asymptotically
flat $D$-dimensional spacetime were further investigated
in \cite{Fro-Is-Zel:2005:PHYSR4:}. They proved that the Einstein's
equations for gyratons reduce to a set of linear equations
in the Euclidean ${(D-2)}$-dimensional space and showed that
the gyraton metrics belong to a class of so called \emph{VSI spacetimes}
for which all polynomial scalar invariants, constructed from the curvature
and its covariant derivatives, vanish identically
\cite{Prav-Prav:2002:CLAQG:}. (For the discussion of spacetimes
with nonvanishing but nonpolynomial scalar invariants of curvature,
see \cite{Page:2009:}.) Subsequently, charged gyratons in Minkowski
space in any dimension were presented in \cite{Fro-Zel:2006:CLAQG:}.

In \cite{Fro-Zel:2005:PHYSR4:}, the exact gyraton solutions
in the asymptotically anti-de~Sitter spacetime were found.
Namely, they obtained Siklos gyratons which generalize the
Siklos family of nonexpanding waves \cite{Sik:1985:Cam:}
(investigated further in \cite{Pod-rot:1998:CLAQG:})
which belong to the class of spacetimes with constant scalar invariants
(\emph{CSI spacetimes}) \cite{Coley-Her-Pel:2006:CLAQG, Coley-Gib-Her-Pope:2008:CLAQG, Coley-Her-Pel:2008:CLAQG, Coley-Her-Pel:2009:CLAQG}.

Recently, the large class of gyratons on the direct-product
spacetimes was found in \cite{Kadlecova:2009:PHYSR4:}, where we showed that
this class of gyratons has similar properties as the previous
gyratonic solutions: the Einstein's equations reduce to a set of
linear equations in transversal 2-space and these spacetimes
belong to the CSI class of spacetimes.

Let us also mention that string gyratons in supergravity
were recently found in \cite{Fro-Li:2006:PHYSR4:}.
Supersymmetric gyraton solutions were also obtained in minimal gauged theory
in five dimensions in
\cite{Cald-Kle-Zor:2007:CLAQG:}, where the configuration represents
a generalization of the Siklos waves with a nonzero
angular momentum in anti-de~Sitter space.

The gyratons are important in studies of production of mini
black holes or in cosmic ray experiments.
The theory of high energy particle collisions was developed in
\cite{Yosh:2005:PHYSR4:,Yosh:2006:PHYSR4:}
and was applied to gyraton models in \cite{Yosh:2007:PHYSR4:}.

The main purpose of this paper is to further extend the family of
gyratonic solutions. In particular, we present new gyraton solutions
of algebraic type~II, propagating in the Melvin universe.

The Melvin universe \cite{Bonnor:1954:PRS:,Melvin:1965:PHYSR:} is a nonsingular
electro-vacuum solution with physical properties which are interesting
both from a classical and a quantum point of view. The spacetime represents
a parallel bundle of magnetic (or electric) flux held together by its
own gravitational attraction. The transverse space orthogonal to the direction of the flux
has a nontrivial spatial geometry. It was represented in \cite{Thorne:1965:PHYSR:}
by a suitable embedding diagram which resembles a tall narrow-necked vase.
Also in \cite{Thorne:1965:PHYSR:,MelvinWallingford:1966:JMATHP:} it was shown
that no motion can get too far from the axis of symmetry.
This aspect is analogous to the attractive effect of a negative cosmological
constant in the anti-de Sitter universe.

The Melvin universe was considered as
an important model in astrophysical processes related to gravitational
collapse because of its stability. It was shown in \cite{Melvin:1965:PHYSR:,Thorne:1965:PHYSR:}
that the spacetime is surprisingly stable against small radial perturbations
and also against large perturbations which are concentrated in
a finite region about the axis of symmetry. The asymmetries are radiated
away in gravitational and electromagnetic waves
\cite{GarfinkleMelvin:1992:PHYSR4:,Ortaggio:2004:PHYSR4:}.

The Melvin universe also appears as a limit in more
complicated solutions, in \cite{HavrdovaKrtous:2007:GENRG2:} it is obtained
as a specific limit of a charged C-metric.
The Melvin universe has been generalized to Kaluza-Klein and dilaton
theories \cite{GibbonsMaeda:1988:NUCLB:}, to nonlinear electrodynamics
\cite{GibbonsHerdeiro:2001:CLAQG:}, and has important applications
in the study of quantum black hole pair creation in a background electromagnetic field
\cite{Gib:1986:Sin:,GarfinkleStrominger:1991:PHRELEB:,GarfinkleGidStro:1994:PHYSR4:,%
DowkerGauntKaTra:1994:PHYSR4:,DowkerGauntGiddHorowitz:1994:PHYSR4:,%
HawkingRoss:1995:PHYSR4:,Emparan:1995:PHRELE:}.

The Melvin universe recently attracted a new interest because it is possible
to find gravitational waves in the Melvin universe
\cite{GarfinkleMelvin:1992:PHYSR4:} by an ultrarelativistic boost of the
Schwarzschild--Melvin black hole metric \cite{Ortaggio:2004:PHYSR4:}.
It was shown that these wave solutions are straightforward impulsive
limits of a more general class of Kundt spacetimes of type II with
an arbitrary profile function, which can be interpreted as gravitational
waves propagating on the Melvin spacetime. The gyraton spacetimes
investigated in this paper are generalizations of such Kundt waves
when their ultrarelativistic source is made of a `spinning matter'.


The paper is organized as follows. In Section \ref{sc:gyreqq} we review
basic information about the Melvin universe which will be useful in the paper.
We derive the ansatz for the gyraton metric by a direct transformation
from the Kundt form of the metric to Melvin's coordinates.
We also review the transverse space geometry of the wave front.

In Section \ref{sc:fieldeq}, we derive the field equations
and we simplify them introducing potentials. We discuss the structure of
the equations and the gauge freedom of the solutions.

Next, in Section \ref{sc:examples}, we solve the Einstein--Maxwell equations
in the special case of the $\phi$-independent spacetimes,
especially with a thin matter source localized on the axis of symmetry.

In Section \ref{sc:interpret} we concentrate on the interpretation of the
gyraton solutions. We discuss the properties of the scalar polynomial
invariants and the geometric properties of the principal null congruence.
We evaluate the curvature tensor in an appropriate tetrad, discuss the
Petrov type, the matter content of the spacetime, and properties of the electromagnetic field.

The main results of the paper are summarized in concluding Section \ref{sc:conclusionX}.
Some technical results needed to derive the field equations,
spin coefficients and invariants are left to
Appendices \ref{apx:AppAx}, \ref{apx:NPx}, and \ref{apx:Invars}.
\subsection{The Melvin universe}\label{scc:melvin}
In this section we briefly review basic properties of the Melvin spacetime
\cite{Bonnor:1954:PRS:,Melvin:1965:PHYSR:,Ortaggio:2004:PHYSR4:} which
will be useful throughout the paper.
The Melvin universe describes an axial electromagnetic field concentrating under
influence of its self-gravity. The strength of the electromagnetic
field is determined by the parameters $E$ and $B$.
In cylindrical coordinates $(t,z,\rho,\phi)$, the metric and the Maxwell tensor read
\begin{gather}
d s^2=\Sigma^2(-\trgrad t^2+\trgrad z^2+\trgrad \rho^2)+\Sigma^{-2}\rho^2\trgrad \phi^2\;,\label{o1}\\
{F}=E\,\trgrad z \wedge \trgrad t + B\Sigma^{-2}\rho\,\trgrad \rho \wedge \trgrad \phi\;,\quad\label{Fintz}
\end{gather}
where
\begin{equation}\label{sigma}
\Sigma=1+\frac{1}{4}\vrho\rho^2\;.
\end{equation}
The constant $\vrho$ is given by the parameters $E$, $B$ as\footnote{%
$\varkappa=8\pi G$ and $\epso$ are gravitational and electromagnetic constants.
There are two standard choices of geometrical units: the Gaussian with $\varkappa=8\pi$ and
$\varepsilon_{\rm o}=1/4\pi$, and SI like with $\varkappa=\varepsilon_{\rm o}=1$.}
\begin{equation}\label{rhodef}
\vrho=\frac{\varkappa\epso}{2}(E^2+B^2)
\;.
\end{equation}

Introducing double null coordinates,
\begin{equation}\label{null}
u=\frac{1}{\sqrt{2}}(t-z)\;,\quad v=\frac{1}{\sqrt{2}}(t+z)\;,
\end{equation}
we obtain an alternative expression for metric \eqref{o1}
and the Maxwell tensor \eqref{Fintz}:
\begin{gather}
d s^2=\Sigma^2(-2\trgrad u\,\trgrad v+\trgrad \rho^2)+\Sigma^{-2}\rho^2\trgrad \phi^2\;,\label{o2}\\
{F}=E\,\trgrad v \wedge \trgrad u + B\Sigma^{-2}\rho\,\trgrad \rho \wedge \trgrad \phi\;.\quad\label{physF}
\end{gather}
The electromagnetic field can be rewritten also in the complex self-dual form,\footnote{%
We follow the notation of \cite{Step:2003:Cam:}, namely, $\mathcal{F}\equiv {F}+i\,{{\star}F}$
is a complex self-dual Maxwell tensor, where the 4-dimensional Hodge dual is
${\star}F_{\mu\nu}=\frac{1}{2}\varepsilon_{\mu\nu\rho\sigma}F^{\rho\sigma}$.
The self-dual condition reads ${\star}\mathcal{F}=-i\mathcal{F}$.
The orientation of the 4-dimensional Levi-Civita tensor
is fixed by the sign of the component $\varepsilon_{vu\rho\phi}=\rho\Sigma^2$.
The energy-momentum tensor of the electromagnetic field is given by
$T_{\mu\nu}=\frac{\varepsilon_{\rm o}}{2}\mathcal{F}_{\mu}{}^{\rho}\overline{\mathcal{F}}_{\nu\rho}$.}
\begin{equation}
\mathcal{F}=\mathcal{B}\,\bigl(\trgrad v \wedge \trgrad u - i\Sigma^{-2}\rho\,\trgrad \rho \wedge \trgrad \phi\bigr)\label{mF}\;.
\end{equation}
with the complex constant $\mathcal{B}$ defined as
\begin{equation}\label{B}
\mathcal{B}=E+iB\;.
\end{equation}

The metric \eqref{o1} resembles a vacuum solution of the Levi-Civita family \cite{Step:2003:Cam:} for a large
value of $\rho$. For ${\vrho=0}$ (${E,B=0}$) spacetime reduces to the Minkowski spacetime in cylindrical coordinates.
For ${E\neq0}$, ${B=0}$ the Maxwell tensor describes
an electric field pointing along the $z$-direction,
whereas for ${E=0}$, ${B\neq0}$ we get a purely magnetic field oriented along the $z$-direction.

The metric admits the four Killing vectors
\begin{equation}
\partial_{t}\;,\quad \partial_{z}\;,\quad \partial_{\phi}\;,\quad z\partial_{t}+t\partial_{z}=v\partial_{v}-u\partial_{u}\;,
\end{equation}
which correspond to staticity, cylindrical symmetry, and invariance under a boost transformation.

Using the adapted null tetrad ${{k}=\partial_{v}}$, ${{l}=\Sigma^{-2}\partial_{u}}$, and
${{m}=\frac{1}{\sqrt{2}}(\Sigma^{-1}\partial_{\rho}-i\Sigma\rho^{-1}\partial_{\phi})}$,
the only non--vanishing components of Weyl and Ricci tensors are
\begin{equation}\begin{aligned}\label{Melpsi}
\Psi_{2}&=\frac{1}{2}\vrho\,(-1+\frac{1}{4}\vrho\rho^2)\Sigma^{-4}\;,\\
\Phi_{11}&=\frac{1}{2}\vrho\,\Sigma^{-4}\;.
\end{aligned}\end{equation}
This demonstrates that the Melvin universe is a non--vacuum solution
of the Petrov type D, except at points satisfying $\rho=2/\sqrt{\vrho}$, where the Weyl tensor vanishes.
It is interesting to note that the scalar curvature vanishes, $R=0$.

To conclude, the Melvin spacetime belongs to the family of non-expanding, non-twisting type D electrovacuum
solutions investigated by Pleba\'{n}ski \cite{Plebanski:1979:JMATHP:}. As a consequence, it also belongs
to the general Kundt class \cite{Step:2003:Cam:,GrifPod:2009:Cam:} which will be important in the following text.

\subsection{The ansatz for the gyratons on Melvin universe}\label{ssc:def}

Gyratons are generalized gravitational waves corresponding to null sources with intrinsic rotation.
In general, the gyraton solutions are obtained by adding non-diagonal terms to the metric of the standard
gravitational wave solutions, or in other words, by keeping the non-diagonal terms $g_{ui}$ in
the standard Kundt metric \cite{Step:2003:Cam:}.
Therefore, we derive the ansatz for the gyraton on Melvin spacetime by adding such new
terms to the Kundt form of the Melvin metric.
It can be explicitly obtained from \eqref{o2} by transformation
\begin{equation}\label{r-v}
v=\Sigma^{-2}r\;,
\end{equation}
which leads to
\begin{equation}\label{MK}
d s^2=-2\trgrad u\,\trgrad r+d s^2_{\trans}+2 r W_i\,\trgrad u\trgrad x^i\;.
\end{equation}
Here we introduced a 2-dimensional metric
\begin{equation}\label{transmM}
d s^2_{\trans}=\Sigma^2\trgrad \rho^2+\Sigma^{-2}\rho^2\trgrad \phi^2
\end{equation}
and ${r}$-independent 1-form $W=W_i\trgrad x^i$,
\begin{equation}
W=2\frac{\Sigma_{,\rho}}{\Sigma}\trgrad \rho\;.
\end{equation}
These tensors can be understood as tensors on space spanned by two coordinates ${\rho,\,\phi}$. This space
can be covered by other suitable spatial coordinates ${x^i}$, and we will use the Latin indices
${i,j,\dots}$ to label the corresponding tensor components.

By an appropriate transformation of coordinates
\cite{Ortaggio:2004:PHYSR4:,GrifPod:2009:Cam:}, the 2-dimensional metric $ds^2_{\trans}$
can be transformed into a conformally flat form,
which in the standard complex null coordinates $\zeta,\,\bar{\zeta}$ reads
\begin{equation}\label{trmetric3}
d s^2_{\trans}= \frac{2}{P^2}\trgrad \zeta \trgrad \bar{\zeta}\;.
\end{equation}
Such a transformation brings the metric \eqref{MK} into the Kundt form.

The gyraton generalization of \eqref{MK} then reads
\begin{equation}\begin{split}\label{WR}
d s^2&=-2\Sigma^2{H}\,\trgrad u^2 - 2 \trgrad u \trgrad r+d s^2_{\perp}\\
&\quad+ 2(r W_i {-} \Sigma^2 a_i)\; \trgrad u \trgrad x^i\;.
\end{split}\end{equation}
We have added the term ${-2\Sigma^2{H}\,\trgrad u^2}$ which represents a gravitational wave on the
Melvin universe \cite{GarfinkleMelvin:1992:PHYSR4:,Ortaggio:2004:PHYSR4:} with an arbitrary profile function ${H}$,
and the non-diagonal terms ${-2\Sigma^{2}a_i\trgrad u\trgrad x^i}$ characteristic for gyratons.
It will be shown in the following that these terms can be generated by specific gyratonic matter.

Transforming back to the Melvin coordinate $v$ and cylindrical coordinates ${\rho,\,\phi}$,
we obtain the ansatz for the metric describing the gyraton on Melvin spacetime,\footnote{%
Here we use notation different from \cite{GarfinkleMelvin:1992:PHYSR4:,Ortaggio:2004:PHYSR4:},
we use $-2H\trgrad u^2$ instead of $-H\trgrad u^2$ to match our notation in \cite{Kadlecova:2009:PHYSR4:}.}
\begin{equation}\begin{split}\label{s5}
d s^2&=-2\Sigma^2 H \trgrad u^2-2\Sigma^2\trgrad u\,\trgrad v  + \bigl(\Sigma^2\trgrad \rho^2 + \Sigma^{-2}\rho^2\trgrad \phi^2\bigr) \\
&\quad+2\,\Sigma^2 \bigl(a_{\rho}\,\trgrad u\,\trgrad \rho+ a_{\phi}\,\trgrad u\,\trgrad \phi\bigr)\;.
\end{split}\raisetag{13pt}\end{equation}

The function $H(u,v,\rho,\phi)$ can depend on all coordinates, but we assume that the
functions ${a_{i}(u,\rho,\phi)}$ are ${\mbox{${v}$ independent}}$
(it actually follows from the Maxwell equations as will be shown below).
Let us note that the previously cited works assumed also the function $H$  \mbox{$v$ independent}.

\subsection{The ansatz for the matter}\label{ssc:ansatz1}

The metric should satisfy the Einstein equations with a stress-energy tensor generated by
the electromagnetic field and the gyratonic source,
\begin{equation}\label{EinsteinEq}
G_{\mu\nu}=\varkappa \bigl( T^\EM_{\mu\nu}+T^{\gyr}_{\mu\nu}\bigr)\;.
\end{equation}
We assume that the electromagnetic field \eqref{physF} modified by the gyraton is given by
\begin{equation}\label{realF}
{F}=E\,\trgrad v \wedge\trgrad u+B\Sigma^{-2}{\rho}\,\trgrad \rho\wedge\trgrad \phi+\sreal_{j}\trgrad u\wedge\trgrad x^{j}\;.
\end{equation}
Similar to \cite{Kadlecova:2009:PHYSR4:}, we have added the term  $\sreal_{j}\,\trgrad u\wedge\trgrad x^{j}$.

To evaluate Maxwell equations, it is useful to write down the self-dual form of the Maxwell tensor $\mathcal{F}$.
The Hodge dual of \eqref{realF} reads
\begin{equation}\begin{split}\label{realF1}
{\star}{F}&=B\,\trgrad v\wedge\trgrad u-E\Sigma^{-2}{\rho}\,\trgrad \rho\wedge\trgrad \phi\\
&\qquad\qquad+({*\sreal}+Ba-E\,{*a})_{j}\,\trgrad u\wedge\trgrad x^{j}\;,
\end{split}\end{equation}
where the star $*$ means the 2-dimensional Hodge dual defined on the transversal space,
see the next section \eqref{ssc:transsp1}. For the self-dual Maxwell tensor ${\mathcal{F}}$ we thus obtain
\begin{equation}\begin{split}\label{EMFq}
\mathcal{F}&=\mathcal{B}\,\Bigl(\trgrad v\wedge\trgrad u - i\Sigma^{-2}\rho\,\trgrad \rho\wedge\trgrad \phi\\
 &\qquad\qquad\qquad+(\mathcal{S}-i\,{*a})_{j}\,\trgrad u\wedge\trgrad x^{j}\,\Bigr)\;.
\end{split}\end{equation}
where we introduced a complex transverse \mbox{1-form}~$\mathcal{S}_{j}$,
\begin{equation}
\mathcal{B}\mathcal{S}_{j}=(\sreal+i{*\sreal})_{j}+iB(a+i{*a})_{j}\;.
\end{equation}
This form is self-dual with respect to the Hodge duality ${*}$ on the transversal space,
\begin{equation}
*\mathcal{S}_{j}=-i\,{\mathcal{S}}_{j}\;,
\end{equation}
and therefore it can be written using a real \mbox{1-form}~$\sslfdl_{j}$:
\begin{equation}
\mathcal{S}_{j}=(\sslfdl+i\,{*\sslfdl})_{j}\;.
\end{equation}
The original 1-form $\sreal_{i}$ can be expressed in terms of $\sslfdl_{j}$ as
\begin{equation}\label{si}
\sreal_{j}=E\,\sslfdl_{j}-B\,{*(\sslfdl-a)_{j}}\;.
\end{equation}
In the following we use $\sslfdl_{j}(v,u,\rho,\phi)$ as a basic variable for the
electromagnetic field.

The stress-energy tensor ${T^\EM_{\mu\nu}}$ corresponding to the field \eqref{EMFq} is given in \eqref{EMTe}.

Finally, we must define the gyratonic matter by specifying the structure of its stress-energy tensor.
It is obtained from the standard stress-energy tensor of a null fluid by adding terms corresponding
to `internal spatial rotation' of the fluid:
\begin{equation}\label{m7}
\varkappa\, T^{\gyr}=j_{u}\,\trgrad u^2+2j_{\rho}\,\trgrad u\,\trgrad \rho+2j_{\phi}\,\trgrad u\,\trgrad \phi\;.
\end{equation}
We admit a general coordinate dependence of the source functions ${j_u(v,u,\rho,\phi)}$ and ${j_j(v,u,\rho,\phi)}$.
However, it will be shown below that the field equations enforce a trivial ${v}$ dependence.

The gyraton source is described only on a phenomenological level, by its
stress-energy tensor \eqref{m7}, which is assumed to be given, and our aim is to
determine its influence on the metric and the electromagnetic field.
However, we have to consider that the gyraton stress-energy tensor is locally conserved.
It means that the functions ${j_u}$ and ${j_i}$ must satisfy the constraint given by
\begin{equation}\label{gyrenergycons}
  T^{\gyr}_{\;\,\mu\nu}{}^{\>;\nu}=0\;.
\end{equation}
Of course, if we had considered a specific internal structure of the gyratonic matter,
the local energy-momentum conservation would have been a consequence of field equations
for the gyraton. Without that, we have to require \eqref{gyrenergycons} explicitly.

To conclude, the fields are characterized by functions ${\Sigma}$, ${H}$, ${a_j}$, and ${\sslfdl_j}$,
which must be determined by the field equations provided the gyraton sources ${j_u}$ and ${j_j}$
and the constants ${E}$ and ${B}$ are given.

\subsection{The geometry of the transverse space}\label{ssc:transsp1}

The geometry \eqref{s5} identifies the null geodesic congruence
generated by $\partial_v$ which is parametrized by an affine parameter $v$,
the family of null hypersurfaces ${u=\text{constant}}$, and 2-dimensional
\textit{transverse spaces} ${u,v=\text{constant}}$.

The gravitational wave moves along the null direction ${\partial_v}$,
i.e., it propagates with the speed of light along the $z$-direction,
which is the direction of the electromagnetic field.
The hypersurface ${u=\text{constant}}$ corresponds to the surface of the constant `phase', and
the transverse spaces ${u,v=\text{constant}}$ are spatial wave fronts of the wave.

Physical quantities do not depend on the affine parameter ${v}$, or this dependence is trivial
and it will be explicitly found. Specifically, the geometry of the transverse space is \mbox{${v}$ independent}.

It turns out to be convenient to restrict various quantities to the transverse space.
For example, we can interpret ${a_i}$ and ${\sslfdl_i}$ as components of
$u$-dependent ${\mbox{${1}$-forms}}$ on the transverse space.
Our goal is to formulate all equations for physical quantities
on the transverse spaces. For that we need to review some properties of the
transverse geometry. It was studied in detail in \cite{Ortaggio:2004:PHYSR4:},
and on a general level in \cite{Kadlecova:2009:PHYSR4:}, nevertheless
it will be useful to mention some of the properties explicitly.

The transverse metric is obtained by restriction of the full metric \eqref{s5}
to the transverse space and it is given by the expression \eqref{transmM},
\begin{equation}\label{transm}
d s^2_{\trans}=g_\trans{}_{ij}\trgrad x^i \trgrad x^j = \Sigma^2\trgrad \rho^2+\Sigma^{-2}\rho^2\trgrad \phi^2
\end{equation}

The associated Gauss curvature is given by the scalar curvature $R_{\trans}$,
\begin{equation}\label{trsccurv1}
K=\frac12 R_\trans=\frac{\vrho}{\Sigma^{4}}\Bigl(2-{\textstyle\frac{1}{4}}\vrho\rho^2\Bigr)\;.
\end{equation}
It is obvious that only the electromagnetic field is responsible for the
non-flatness of the transversal space---it is insensitive to the presence of the gyraton.
For $\vrho=0$ the curvature vanishes and we get the flat plane.

In general, the curvature is not constant and it is finite everywhere.
The Gauss curvature $K$ has maximum on the axis $\rho=0$ where it is equal to $2\vrho$;
it is positive for $0\leq\rho<2\sqrt{2}/\sqrt{\vrho}$, and vanishes on the circle at $\rho=2\sqrt{2}/\sqrt{\vrho}$.
For $\rho>2\sqrt{2}/\sqrt{\vrho}$, it goes to negative values and it has its minimum $K=-\frac{\vrho}{256}$ at $\rho=2\sqrt{3}/\sqrt{\vrho}$.
Then it grows again, and as $\rho\rightarrow +\infty$ the curvature vanishes, $K\rightarrow 0^{-}$.

The circumference of a circle of constant radius $\rho$ is vanishing when $\rho\rightarrow+\infty$.
Therefore, we measure a much shorter circumference for larger $\rho$---as if we would move
``along the stem of the wine-glass toward the narrowing end,''
\cite{Melvin:1965:PHYSR:,Thorne:1965:PHYSR:,Ortaggio:2004:PHYSR4:}.

The 2-dimensional Levi-Civita tensor ${\epsilon_{ij}}$ associated with the metric \eqref{transm}
is ${\epsilon=\rho\,\trgrad\rho\wedge\trgrad\phi}$. The covariant derivative will be denoted by
a colon, e.g., ${a_{i:j}}$. We raise and lower the Latin indices using ${g_\trans{}_{ij}}$,
which differs from lowering indices using $g_{\alpha\beta}$ thanks to non-vanishing terms $g_{ui}$.
We use a shorthand for a transverse square of the norm of a 1-form ${a_i}$ as
\begin{equation}
a^2\equiv a^i a_i=\Sigma^{-2}a_\rho^2+\rho^{-2}\Sigma^2 a_\phi^2\;.
\end{equation}

In two dimensions, the Hodge duals of 0, 1 and 2-forms ${\varphi}$, ${a_i}$, and ${f_{ij}}$, respectively, read
\begin{equation}
(*\varphi)_{ij} = \varphi\, \epsilon_{ij}\;,\;\;
(*a)_i = a_j \epsilon^j{}_i\;,\;\;
*f = \frac12 f_{ij}\epsilon^{ij} =\frac{1}{\rho}f_{\rho\phi} \;.
\end{equation}

For convenience, we also introduce an explicit notation for 2-dimensional divergence and rotation of a transverse 1-form ${a_i}$,
\begin{align}
  &\trdiv a \equiv a_i{}^{:i}= \frac{1}{\Sigma^2}a_{\rho,\rho}+\frac{\Sigma^2}{\rho^2}a_{\phi,\phi}+\frac{1}{\rho\Sigma^2}a_{\rho}-\frac{\Sigma^2_{,\rho}}{\Sigma^4}a_{\rho}\;,\notag\\
&\rot a \equiv \,\epsilon^{ij} a_{j,i} = \frac{1}{\rho}(a_{\phi,\rho}-a_{\rho,\phi})\;.
\end{align}
For 2-form $f_{ij}$ we get
\begin{equation}
  \trdiv f \equiv f_{ij}{}^{:j} = \frac{1}{\Sigma^2}(f_{\phi\rho,\rho}-\frac{1}{\rho}f_{\phi\rho})\trgrad \phi+\frac{\Sigma^2}{\rho^2}f_{\rho\phi,\phi}\trgrad \rho\;,
\end{equation}
and ${\rot f=0}$. We can generalize the action of divergence and rotation also on a scalar function $f$ as
${\trdiv f = 0}$ and ${\rot f = -{*}\trgrad f}$.
Note that the divergence and rotation are related as ${\trdiv a = \rot {*}a}$, and
the relation to the transverse exterior derivative is ${\trgrad a=*\rot a}$.
Clearly, ${\trdiv\div a=0}$, ${\trdiv\rot a=0}$, and ${\rot \trgrad a=0}$.

The Laplace operator of a function ${\psi}$ reads
\begin{equation}\label{lapldefo}
\laplace\psi =  \psi_{:i}{}^{:i} = \frac{1}{\Sigma^2}\psi_{,\rho\rho}+\frac{\Sigma^2}{\rho^2}\psi_{,\phi\phi}+\frac{1}{\rho\Sigma^2}\psi_{,\rho}-\frac{\Sigma^2_{,\rho}}{\Sigma^4}\psi_{,\rho}\;,
\end{equation}
and for a transverse 1-form~${\eta}$ it is defined as
${\laplace\eta \equiv \trgrad\,\trdiv\eta-\rot\rot\eta}$.

Finally, the transverse space is topologically trivial since it has topology of a plane. We can thus assume that the
Poincare lemma (${\trgrad\omega=0} \Rightarrow {\omega=\trgrad\sigma}$) holds, which in terms of rotation and divergence means that ${\trdiv\omega=0}$ implies ${\omega = \rot\sigma}$. However, since the transverse space is non-compact and we do not know a priori boundary conditions
for various quantities at infinity $\rho\to\infty$, we have to admit non-trivial harmonics. Therefore, we cannot assume
a uniqueness of the Hodge decomposition. Moreover, in some cases it can be physically relevant to consider also
topologically nontrivial harmonics which are singular, e.g., at the origin ${\rho=0}$. Such solutions would correspond
to fields around singular sources localized on the axis. However, we will ignore these cases in a general discussion.

\section{The field equations}\label{sc:fieldeq}

\subsection{The field equations for matter}\label{scc:fequationse}
Now, we will investigate the equations for matter, i.e., the Maxwell equations for electromagnetic field and the condition \eqref{gyrenergycons}
for the gyraton source.

Both Maxwell equations for real Maxwell tensor are equivalent to the cyclic Maxwell equation for the self-dual Maxwell tensor \eqref{EMFq},
\begin{equation}\begin{aligned}\label{MXECX}
  \trgrad {\mathcal{F}}
   =\mathcal{B}&\Bigl[\,\partial_v\bigl(\sslfdl+i\,{*(\sslfdl-a)}\bigr)_{j}\,\trgrad v\wedge\trgrad u\wedge\trgrad x^j \\
   &-\bigl(\rot\sslfdl+i\,\trdiv(\sslfdl-a)\bigr)\, \trgrad u \wedge \epsilon\,\Bigr]=0\;.
\end{aligned}\end{equation}
From the real part we immediately get that the 1-form~${\sslfdl_i}$ is ${v}$ independent, $\partial_v \sslfdl_i=0$, and rotation-free,
\begin{equation}\label{pot1}
  \rot\sslfdl=0\;.
\end{equation}
From the imaginary part  it follows that
the 1-form $a_{i}$ is also ${v}$ independent (as we have already mentioned above) and it satisfies
\begin{equation}\label{pot2}
  \trdiv(\sslfdl- a) = 0\;.
\end{equation}

Equations \eqref{pot1} and \eqref{pot2} guarantee the existence and determine the structure of potentials
which will be discussed in detail in Section \ref{ssc:potss}.

Next, we analyze the condition \eqref{gyrenergycons} for the gyraton source. When translated to the transverse space, it gives
\begin{equation}\label{gyrenergycons2}
  -\partial_v j_i\,\trgrad x^i + \left(-\partial_v j_u+\trdiv(\Sigma^2 j)+\Sigma^2 a^i \partial_v j_i\right)\trgrad u =0\;.
\end{equation}
The source functions ${j_i}$ must be thus ${v}$ independent and ${j_u}$ has to have the structure
\begin{equation}\label{jdecomp5}
  j_u = v\,\trdiv(\Sigma^2 j) + \iota\;,
\end{equation}
where $\iota(u,x^{i})$ is a $v$ independent function.
The gyraton source \eqref{m7} is therefore fully determined by
three \mbox{${v}$-independent} functions ${\iota(u,x^j)}$ and ${j_i(u,x^j)}$.

Equation \eqref{jdecomp5} gives us also an insight into interpretation of the gyratonic terms
${j_i}$. They are composed from two contributions: one is related to a kind of `heat flow' which
changes energy ${j_u}$ of the fluid, and the other which is related to intrinsic rotation of the
fluid. The source representing `heat flow' has thus non-vanishing divergence ${\trdiv(\Sigma^2 j)}$
and we require a vanishing rotational part ${\rot(\Sigma^2 j)}$. In opposite,
the source representing intrinsic rotation has vanishing heat flow, i.e., it satisfies
\begin{equation}\label{noheat}
  \trdiv(\Sigma^2 j) =0\;.
\end{equation}
Such a source can be written in terms of a rotational potential ${\nu}$
as ${j=-\Sigma^{-2}\rot\nu}$. In components it means
\begin{equation}\label{noheatj}
   j_\rho = -\frac1\rho\nu_{,\phi}\;,\quad
   j_\phi = \frac\rho{\Sigma^4}\nu_{,\rho}\;.
\end{equation}

Physically more relevant is the rotational part of the source, since it can describe the spin of the
null fluid, or, in a specific limit, of the polarized beam of light. Terms related to heating flow
have bad causal behavior and therefore typically do not satisfy
various energy conditions and they are thus rather unphysical.
Interpretation of the gyraton source was discussed previously also in
\cite{Fro-Is-Zel:2005:PHYSR4:, Fro-Fur:2005:PHYSR4:,Fro-Zel:2005:PHYSR4:,Kadlecova:2009:PHYSR4:}.

\subsection{The Einstein equations}\label{scc:EinsteinEqs}

The Einstein gravitational law \eqref{EinsteinEq} needs the Einstein tensor and the electromagnetic and gyraton
stress-energy tensors. These quantities can be found in the Appendix \ref{apx:AppAa}.
We can combine them and inspect various components of the Einstein equations.

The $vu$-component determines the function $\Sigma$, namely it gives the condition
\begin{equation}\label{Sig}
-\rho(\Sigma_{,\rho})^2+2\Sigma\Sigma_{,\rho}=\vrho\,\rho\;.
\end{equation}
It is straightforward to check that it is satisfied again by $\Sigma$ in the Melvin form \eqref{sigma}.

The transverse diagonal components $\rho\rho$ and $\phi\phi$ require
\begin{equation}\label{Einsteinii5}
  \partial^2_{v} H  =0\;,
\end{equation}
thus we obtain the explicit ${v}$ dependence of the metric function ${H}$ as
\begin{equation}\label{Heq5x}
  H = g\,v + h\;,
\end{equation}
where we have introduced ${v}$-independent functions ${g(u,x^j)}$ and ${h(u,x^j)}$.

The remaining nontrivial components of the Einstein equations are those involving the gyraton source \eqref{m7}.
The $ui$-components give the equation related to ${j_i}$,
\begin{equation}\label{jieq}
  j_i= \frac12\, \Sigma^2 f_{ij}{}^{:j}  -(\Sigma^2)_{,k}\,g^{kj}f_{ji}+ g_{,i}+\frac{2\vrho}{\Sigma^2}(\sslfdl_{i}-a_{i})\;,
\end{equation}
where we introduce the exterior derivative ${f_{ij}}$ of the transverse \mbox{1-form} ${a_i}$ and its
Hodge dual ${b(u,x^j)}$,
\begin{gather}\label{fdef}
f_{ij} \equiv a_{j,i}-a_{i,j} = (*\,b)_{ij}\;,\\
b \equiv * f = \rot a\;.\label{bdef}
\end{gather}
In terms of ${b}$, equation \eqref{jieq} can be rewritten as
\begin{equation}\label{jieqpot}
  \Sigma^2\,j = \frac12\,\rot(\Sigma^4\, b) + \Sigma^2{\trgrad}g+2\vrho(\sslfdl-a)\;.
\end{equation}
Here and in the following $\trgrad g$ represents just the transverse gradient ${\trgrad g=g_{,i}\,\trgrad x^{i}}$,
and for simplicity we skipped the transverse indices.

It is useful to split the equation into divergence and rotation parts by applying ${\trdiv}$ and ${\rot}$:
\begin{align}
  \trdiv{(\Sigma^2\, j)}&=\trdiv {(\Sigma^2{\trgrad}g)},\label{divjeq}\\
  \rot{(\Sigma^2 j)}&= - \frac12 \laplace(\Sigma^4 b) + \rot(\Sigma^2{\trgrad}g)-2\vrho\,b\;. \label{rotjeq}
\end{align}
Here we have used the relations \eqref{pot1} and \eqref{bdef}.
The formula \eqref{divjeq} is the equation for $g$, \eqref{rotjeq} is the equation
for $b$ and together with \eqref{bdef} it determines $a_{i}$. In the next section we will return
to these equations introducing suitable potentials which allow us to escape the necessity
of taking an additional derivative of \eqref{jieqpot} when deriving the equation for ${a_i}$.

Finally, from the $uu$-component of the Einstein equation we obtain
\begin{equation}
\begin{split}
  j_u =\,&\trdiv {(\Sigma^2{\trgrad}g)}\;v
      +\Sigma^2(\laplace h - {(\Sigma^{-2})_{,\rho}}h_{,\rho})\\
      &+\frac12\Sigma^4 b^2+ 2\Sigma^2 a^i g_{,i}+\partial_u\trdiv(\Sigma^2 a) + g\, \trdiv(\Sigma^2 a)\\
      &-2\vrho\,(\sslfdl-a)^2\;.\label{jueq}
\end{split}\raisetag{14pt}
\end{equation}
When we compare the coefficient in front of ${v}$ with \eqref{divjeq} we find that it has structure consistent with \eqref{jdecomp5}.
The nontrivial ${v}$-independent part of \eqref{jueq} gives the equation for the metric function~${h}$,
\begin{equation}\label{heq}
\begin{split}
\Sigma^2&\bigl(\laplace h - (\Sigma^{-2})_{,\rho}\,h_{,\rho}\bigr)=
      \iota \, -\frac12\Sigma^4 b^2- 2\Sigma^2 a^i g_{,i}\\
      & -\partial_u\trdiv(\Sigma^2 a) - g\, \trdiv(\Sigma^2 a)+2\vrho\,(\sslfdl-a)^2\;.
\end{split}\raisetag{40pt}
\end{equation}

\subsection{Introducing potentials}\label{ssc:potsT}

In the previous section we have found that the Maxwell and Einstein equations
reduce to two potential equations \eqref{pot1}, \eqref{pot2},
and two source equations \eqref{jieq}, \eqref{jueq}.

According to the two dimensional Hodge decomposition we can express the 1-form ${a_i}$
using two scalar potentials ${\kappa(u,x^j)}$ and ${\lambda(u,x^j)}$,
\begin{equation}\label{klpotdef}
    a = \trgrad\kappa+\rot\lambda\;.
\end{equation}
These potentials control the divergence and the rotation of~${a_i}$ as
\begin{equation}\label{divrota}
    \trdiv a = \laplace\kappa\;,\quad \rot a = -\laplace\lambda\;.
\end{equation}
Comparing with \eqref{bdef} we thus obtain the equation for $\lambda$ in terms of $b$,
\begin{equation}
  \laplace\lambda = -b\;.\label{lambdab}
\end{equation}

The first potential equation \eqref{pot1} gives immediately that ${\sslfdl_i}$ has a potential ${\varphi(u,x^j)}$,
\begin{equation}\label{phipot}
  \sslfdl = {\trgrad}\varphi\;.
\end{equation}
Equation \eqref{pot2} implies that there exists a potential  ${\psi(u,x^j)}$ satisfying
\begin{equation}\label{potpsi}
   \sslfdl-a=-\rot\psi\;.
\end{equation}
In terms of these potentials the 1-form \eqref{si} from the real Maxwell tensor \eqref{physF} reads
\begin{equation}\label{sreal}
   \sreal=E\trgrad\varphi + B\trgrad\psi\;.
\end{equation}

The potential ${\varphi}$ and ${\psi}$ are not, however, independent.
Substituting \eqref{klpotdef} and \eqref{phipot} into \eqref{potpsi}
we obtain the key relation among the potentials:
\begin{equation}\label{potrel}
   {\trgrad}(\varphi -\kappa)+\rot(\psi-\lambda)=0\;.
\end{equation}
If the Hodge decomposition was unique, the gradient and rotational parts would be vanishing separately,
i.e., we would get ${\varphi = \kappa}$ and ${\psi = \lambda}$ (up to unimportant constants).
The non-uniqueness of the Hodge decomposition is linked to
the possible existence of a non-trivial harmonic 1-form $\eta_i(u,x^{j})$,
\begin{equation}\label{eta}
   \laplace\eta=0\;,
\end{equation}
in terms of which the gradient and rotation parts of \eqref{potrel} can be expressed as
\begin{align}
   \varphi-\kappa&=\trdiv\eta\;,\label{phikappa}\\
   \psi-\lambda&=-\rot\eta\;.\label{psilambda}
\end{align}

These are equations for electromagnetic potentials ${\varphi}$ (or ${\psi}$, respectively)
in terms of the metric potentials ${\kappa}$ and ${\lambda}$. The 1-form ${\eta}$ encodes an extra freedom,
which allows a nontrivial electromagnetic field not uniquely determined by the metric. (Such contributions
would allow one to take into account, for example, an additional electromagnetic charge localized at the origin
of the transverse space,
cf.\ the discussion of particular cases in section \ref{sc:examples}.) However, sufficiently restrictive
conditions at the infinity and the smoothness on the whole transverse space for the potentials would
eliminate this freedom, so the case ${\eta=0}$ is a rather representative choice.

After eliminating the electromagnetic potentials, we need
to formulate the equations for ${\kappa}$ and ${\lambda}$. 
We start with the divergence part of \eqref{jieqpot} which can be rewritten using
the modified Laplace operator acting on the metric function $g$:
\begin{equation}\label{geq}
\Sigma^2\bigl(\laplace g - (\Sigma^{-2})_{,\rho}\, g_{,\rho}\bigr)=\trdiv\bigl(\Sigma^2 j\bigr)\;.
\end{equation}
However, for ${g}$ solving this equation, \eqref{jieqpot} is also
the integrability condition for the quantity ${\Sigma^2(\trgrad g-j)}$.
It can thus be written in terms of a
potential ${\omega}$,
\begin{equation}\label{eq2}
\Sigma^2({\trgrad}g-j)=\rot{\omega}\;,
\end{equation}
or, more explicitly,
\begin{equation}\label{eq3}
{\trgrad}\omega=-\Sigma^2(\rot g+{*}j)\;.
\end{equation}

For the source \eqref{noheatj} without intrinsic `heating', the right-hand side of \eqref{geq}
is zero and the function ${\omega}$ has an additive contribution from the rotational potential
${\nu}$, namely it has to satisfy
\begin{equation}\label{omeganoheat}
{\trgrad}(\omega-\nu)=-\Sigma^2\rot g\;.
\end{equation}

The function $\omega$ contains information from the source $j$ and from the metric function $g$
relevant for the rotational part of equation \eqref{jieqpot}.
Indeed, substituting the potentials and ${\rot\omega}$
into \eqref{jieqpot} we obtain
\begin{equation}\label{rotjeqq1}
\rot\Bigl(\frac12 \Sigma^4 b-2\vrho\psi+\omega\Bigr)=0\;.
\end{equation}
Substituting \eqref{lambdab}, \eqref{psilambda}, and integrating
(absorbing an integration constant to ${\omega}$),
we derive the equation for the potential ${\lambda}$
\begin{equation}\label{lambdaeq}
\frac12 \Sigma^4\laplace\lambda+2\vrho\lambda=\omega + 2\vrho\,\rot\eta\;.
\end{equation}
Taking into account relations \eqref{psilambda} and \eqref{eta}, we obtain the alternative
equation for ${\psi}$
\begin{equation}\label{psieq}
\frac12 \Sigma^4\laplace\psi+2\vrho\psi=\omega \;.
\end{equation}

The other metric potential ${\kappa}$ remains unrestricted. This non-uniqueness is related to
the gauge freedom discussed in detail in section \ref{ssc:gauger}.
This coordinate freedom allows us to set the potential ${\kappa}$ to an arbitrary convenient form,
e.g., to eliminate it completely.

\subsection{Discussion of the field equations}\label{ssc:fieldeqssum}

We have thus formulated all field equations as equations on the transverse space. They are written
in a separated form, i.e., they can be solved one after the other:
First, one has to find harmonic 1-form ${\eta}$ satisfying equation~\eqref{eta}
and metric function ${g}$ satisfying \eqref{divjeq}. It allows one to integrate the function ${\omega}$
which together with ${\rot\eta}$ appears as a source in equation \eqref{lambdaeq} for the potential
${\lambda}$. Using the gauge freedom one can choose the other potential ${\kappa}$. Equations
\eqref{phikappa} and \eqref{psilambda} determine electromagnetic potentials and through equation
\eqref{si} the electromagnetic field. Finally, the remaining metric function ${h}$ is determined by
equation \eqref{heq}, in which the previously computed quantities contribute to the source on the right-hand side.

The most complicated field equations---\eqref{geq}, \eqref{heq}, and \eqref{lambdaeq}---are partial
differential equations on the \mbox{2-dimensional} space, which are solvable, at least in principle. They all contain
a modified Laplace operator, equations \eqref{geq} and \eqref{heq} for ${g}$ and ${h}$ the same one, namely
\begin{equation}\label{modlaplop}
  \Sigma^2\bigl(\laplace f - (\Sigma^{-2})_{,\rho}\, f_{,\rho}\bigr)\equiv\trdiv\bigl(\Sigma^2 \trgrad f\bigr)\;.
\end{equation}
Solutions for particular cases (assuming rotational symmetry) will be discussed in section~\ref{sc:examples}.

It is important to observe that, except equation \eqref{heq} for ${h}$,
the field equations are linear. We can thus superpose two solutions
simply by adding the fields together. Only in the last step, when computing the source
for equation \eqref{heq}, one has to include total superposition of the fields
${g}$, ${a_i}$, and ${\sreal_i}$ since the expression for the source is non-linear.

Finally, we have not paid much attention to the
\mbox{$u$ dependence} of the studied quantities.
All metric functions, matter fields and sources can depend on the coordinate $u$
and this dependence does not enter the field equations except in one term on the right-hand side of equation \eqref{heq}.
The profile of the gyraton in the ${u}$ direction can thus be specified arbitrarily.
It corresponds to the fact that both matter and gravitational field move with the speed of light
and information on different hypersurfaces $u=\text{constant}$ evolves rather
independently.

Also the dependence of the metric and fields on the coordinate $v$ is very simple
and it was found for all quantities explicitly.

\subsection{The gauge transformation}\label{ssc:gauger}

The coordinate transformation ${\tilde v\to v = \tilde v-\chi(u,x^j)}$ accompanied
by the following redefinition of the metric functions and matter fields:
\begin{equation}\begin{gathered}\label{gauge}
v=\tilde v-\chi\;,\\
g=\tilde g\;,\quad
h=\tilde h+\tilde g\,\chi + \partial_u\chi\;,\quad
a_i=\tilde a_i-\chi_{,i}\;,\\
\sslfdl_{i}=\tilde{\sslfdl}_{i}-\chi_{,i}\;,\quad
\sreal_i=\tilde{\sreal}_i-E\,\chi_{,i}\;,\\
\kappa=\tilde \kappa-\chi\;,\quad \lambda=\tilde\lambda\;,\quad
\varphi=\tilde\varphi-\chi\;,\quad \psi=\tilde{\psi}\;,\\
\omega=\tilde{\omega}\;, \quad \eta=\tilde{\eta}\;,\\
j_i=\tilde j_i\;,\quad \iota=\tilde\iota+\chi\, \trdiv j\;,
\end{gathered}\end{equation}
leaves the metric, the Maxwell tensor, and the gyraton stress-energy
tensor in the same form. Therefore, all the field equations remain the same.
This transformation is thus a pure gauge transformation and we can use it
to simplify the solution of the equations.

There are two natural choices of gauge: we can eliminate either
the metric potential ${\kappa}$ or the electromagnetic potential ${\varphi}$.
In the first case ${a=\rot\lambda}$ and ${\varphi=\trdiv\eta}$.
In the latter case ${\sslfdl=0}$, ${\sreal=B\trgrad\psi}$, ${\kappa=-\trdiv\eta}$, and
${a=\rot\psi}$.

Let us mention that in  \cite{Kadlecova:2009:PHYSR4:} an analogous gauge transformation
allowed us to choose also the metric function ${g}$. For the gyraton on the Melvin universe the metric
function ${g}$ decouples from ${\kappa}$ and it is gauge independent.


The discussed gauge transformation has a clear geometrical meaning: it corresponds to a shift of the
origin of the affine parameter ${v}$ of the null congruence ${\partial_v}$. Note that such a
change redefines transverse spaces.


\section{Special cases}\label{sc:examples}

In this section we will study the special solutions of the field equations. Namely, we restrict
to the axially symmetric situation, i.e., to the case when the geometry and the fields are invariant
under action generated by the rotational vector ${\partial_\phi}$. Further, we concentrate on
the gyraton generated by a thin beam of matter concentrated at the origin of the transverse
space which means on the axis of symmetry.

Thank to linearity mentioned at the end of section \ref{ssc:fieldeqssum},
we can discuss various special cases separately. However,
the geometry of the spacetime with gyraton is not merely a superposition of the individual contribution
since the nonlinear coupling in the metric function ${h}$.

We do not discuss in detail the ${u}$ dependence of the fields.
It does not enter the field equations, except in the source term
of equation \eqref{heq} for ${h}$. The \mbox{${u}$ dependence} of the gyraton
sources and corresponding dependence of other fields can thus be chosen arbitrarily.


The symmetry assumption enforces that quantities
${a_i}$, ${g}$, ${h}$, ${\sreal_i}$, ${j_i}$, and ${\iota}$
are ${\phi}$-independent. It can induce a slightly weaker
condition on the potentials: typically, they have
a linear dependence on ${\phi}$.

The thin beam approximation requires that gyratonic matter
is concentrated at the origin ${o}$ of the transverse space given by ${\rho=0}$, i.e.,
${j_i}$ and ${\iota}$ should be distributions with the support at the origin.
However, we relax this condition slightly in the case  of
gyratonic `heat flow' discussed in section \ref{ssc:heating}.

In all discussed cases we use a natural gauge
\begin{equation}\label{gaugekappa0}
    \kappa = 0\;,\quad\text{i.e.,}\quad
    a = \rot \lambda\;.
\end{equation}
Together with the symmetry assumptions it implies ${\lambda_{,\phi}=\text{constant}}$.

\subsection{Pure gravitational gyraton}
\label{ssc:gravgyr}

We start with the simplest vacuum case: we set ${j_i=0}$ and ${\iota=0}$
and we assume no pure electromagnetic contribution, i.e., ${\eta=0}$.
The equation for ${g}$ has only a trivial solution ${g=g_\ro=\text{constant}}$,
the function ${\omega}$ is also a trivial constant and we obtain equation
\eqref{lambdaeq} with vanishing right-hand side. Taking into account that
${\lambda_{,\phi}=\text{constant}}$ we obtain that ${\lambda}$ must be
${\phi}$ independent (a ${\phi}$-linear term in ${\lambda}$ would require
an analogous term in the source) and we obtain an ordinary differential equation
\begin{equation}\label{lambdaeqrho}
    \frac1\rho\Bigl(\frac\rho{\Sigma^2}\lambda_{,\rho}\Bigr)_{,\rho}+\frac{4\vrho}{\Sigma^4}\lambda = 0\;.
\end{equation}
Substituting \eqref{sigma} for ${\Sigma}$, it is possible to
obtain two independent solutions, one regular at the origin,
\begin{equation}\label{lambdagravgyr}
    \lambda=-\gamma\Sigma^{-1}\Bigl(1-{\textstyle\frac34}\vrho\rho^2\Bigr)\;,
\end{equation}
and the other behaving as ${\log\rho}$ near the origin, which corresponds to
a delta source at the origin and it will be discussed in section \ref{ssc:spin}.

The metric 1-form ${a=\rot\lambda}$ has thus components
\begin{equation}\label{agravgyr}
    a_\rho = 0\;,\quad a_\phi= -\frac\rho{\Sigma^2}\lambda_{,\rho}=-\gamma\frac{2\vrho}{\Sigma^4}\rho^2\;,
\end{equation}
and its `strength' ${b=\rot a}$ is then
\begin{equation}\label{bgravgyr}
    b = -\gamma \frac{4\vrho}{\Sigma^5}\Bigl(1-{\textstyle\frac34}\vrho\rho^2\Bigr)\;.
\end{equation}
These can be plugged into \eqref{heq} which turns out to be
\begin{equation}\label{hsourcegravgyr}
\frac1\rho\bigl(\rho h_{,\rho}\bigr){}_{,\rho}=-\gamma^2\frac{8\vrho^2}{\Sigma^6}
   \Bigl(1-{\textstyle\frac14}\vrho\rho^2\Bigr)\Bigl(1-{\textstyle\frac94}\vrho\rho^2\Bigr)\;.
\end{equation}
It can be integrated explicitly; however, the result is rather long, so we skip it.

\subsection{Non-spinning beam with `heat' flow}
\label{ssc:heating}

As we discussed in section \ref{scc:fequationse}, the gyraton source can have two contributions:
one, which changes gyratonic energy density ${j_u}$ and the other corresponding to
intrinsic rotation. Let us investigate the case when the gyratonic energy is concentrated
at the origin (a thin beam) but there is axially symmetric energy flow in the ${\partial_\rho}$
direction which accumulates energy at the beam (a `heating' process).
Namely, we assume that ${\partial_v j_u = \trdiv(\Sigma^2 j)}$ is nonzero
only at the origin, so elsewhere the transverse flow ${j_i}$ must satisfy
 equation \eqref{noheat}. Such ${j_i}$ has the form
\begin{equation}\label{jheating}
    j_\rho = \frac{\alpha}{2\pi}\frac1\rho\;,\quad j_\phi = 0\;,
\end{equation}
and the increasing energy of the gyraton is given by
\begin{equation}\label{juheating}
    j_u = \alpha\, v\, \delta_o\;.
\end{equation}
Here, ${\delta_o}$ stands for the transverse delta function localized
at the origin ${o}$, normalized to the standard metric volume element
on the transverse space.

Since the `heating' is localized only at the origin,
the gyraton source \eqref{jheating} can be locally written using
the source potential ${\nu}$, cf. \eqref{noheatj}, as
\begin{equation}\label{nuheating}
    \nu = -\frac\alpha{2\pi}\phi\;.
\end{equation}
Note however, that the potential cannot be defined globally and it is not
well behaved at the origin.

We again assume no pure electromagnetic contribution, i.e., ${\eta=0}$.
The requirement of the axial symmetry enforces that the difference ${\Sigma^{-2}\rot\omega}$
between ${\trgrad g}$ and ${j}$, cf.~\eqref{eq2}, must be zero, i.e., ${\omega=0}$.
The ${\phi}$-independent metric function ${g}$ must thus satisfy ${g_{,\rho} = j_\rho}$ which gives
\begin{equation}\label{gheating}
    g = \frac\alpha{2\pi}\log\rho+g_\ro\;.
\end{equation}

The equation for ${\lambda}$ has again the form \eqref{lambdaeqrho} and in this case
we choose the trivial solution ${\lambda=0}$. With the gauge ${\kappa=0}$ we thus obtain
the only nontrivial field to be the metric function ${g}$, otherwise ${a_i=0}$ and ${\sigma_i=0}$.
The source for equation \eqref{heq} is also trivial, given only by ${\iota}$,
and we will study it in the next case.

\subsection{Non-spinning light beam}
\label{ssc:nospin}

A particular example of the gyraton source is standard null fluid. The thin non-spinning beam
localized at the origin is described by the source
\begin{equation}\label{jjunospin}
    \iota = \varepsilon \delta_o\;,\quad j_i =0\;.
\end{equation}
We can set all the fields except ${h}$ to be zero: ${g=0}$, ${a_i=0}$, and ${\sigma_i=0}$.
The equation for ${h}$ outside the origin is ${\rho^{-1}(\rho h_{,\rho}){}_{,\rho}= 0}$
which (with proper fixing of the source constant) gives
\begin{equation}\label{hnospin}
    h = \frac\varepsilon{2\pi}\log\rho\;.
\end{equation}

\subsection{Thin gyraton---spinning light beam}
\label{ssc:spin}

Finally we proceed with the most characteristic representant of the gyratonic matter.
It is a simple null beam of energy localized at the origin with no heating, which, however,
contains an intrinsic energy rotation. Since we have a point-like source at the transverse space,
we can speak about inner spin instead of a global rotational energy flow.

The gyraton source has a form\footnote{%
The exact structure of the singular source ${j_i}$ at the origin
can be read out from the singular solution \eqref{lambdaspin} of equation \eqref{lambdaeq} for
${\lambda}$ below.}
\begin{gather}
    \iota = \varepsilon \delta_o\;,\label{iotaspin}\\
    j= -\Sigma^{-2}\rot\nu\;,\quad\text{with}\quad \nu = \beta\delta_o\;.\label{jspin}
\end{gather}
The symmetry assumptions together with the no-heating requirement implies  ${g=\text{constant}}$, which
we choose to be zero in this case. Equation \eqref{eq2} also implies that ${\omega=\nu}$.
Ignoring again the electromagnetic contribution, ${\eta=0}$, we obtain the equation for ${\lambda}$
\begin{equation}\label{lambdaeqrhonu}
    \frac1\rho\Bigl(\frac\rho{\Sigma^2}\lambda_{,\rho}\Bigr)_{,\rho}+\frac{4\vrho}{\Sigma^4}\lambda = \Sigma^{-4}\nu\;.
\end{equation}
The solution of the homogeneous equation with a singular behavior ${\sim\log\rho}$
corresponding to the delta function \eqref{jspin} at the origin reads
\begin{equation}\begin{split}\label{lambdaspin}
    \lambda=-\frac{\beta}{2\pi}\Sigma^{-1}&\Bigl(1+{\textstyle\frac12}\vrho\rho^2-{\textstyle\frac3{64}}\vrho^2\rho^4-{\textstyle\frac1{768}}\vrho^3\rho^4\\
        &+{\textstyle\frac12}\bigl(1{-}{\textstyle\frac34}\vrho\rho^2\bigr)\log\bigl({\textstyle\frac14}\vrho\rho^2\bigr)\Bigr)\;.
\end{split}\raisetag{18pt}
\end{equation}
The multiplicative constant in the argument of the logarithm can be chosen arbitrary since it generates only
an additional homogeneous contribution of the form \eqref{lambdagravgyr}.

The metric 1-form ${a_i}$ and its rotation ${b}$ are
\begin{equation}\label{aspin}
\begin{split}
    a_\rho &=0 \;,\\
    a_\phi &= -\frac\beta{2\pi}\Sigma^{-4}\Bigl(
        1-{\textstyle\frac38}\vrho^2\rho^4-{\textstyle\frac1{32}}\vrho^3\rho^6-{\textstyle\frac1{768}}\vrho^4\rho^8\\
        &\qquad\qquad\quad-\vrho\rho^2\log\bigl({\textstyle\frac14}\vrho\rho^2\bigr)\Bigr)\;,
\end{split}\raisetag{18pt}
\end{equation}
and
\begin{equation}\label{bspin}
\begin{split}
    b &= 2\frac{\beta}{2\pi}\vrho\Sigma^{-5}\Bigl(
         \bigl(1{-}{\textstyle\frac34}\vrho\rho^2\bigr)\log\bigl({\textstyle\frac14}\vrho\rho^2\bigr)\\
         &\qquad+2+\vrho\rho^2-{\textstyle\frac3{32}}\vrho^2\rho^4-{\textstyle\frac1{384}}\vrho^3\rho^6\Bigr)\;.
\end{split}
\end{equation}
The source for equation \eqref{heq} becomes cumbersome and lengthy, but treatable, in principle.

\subsection{Electromagnetic wave}
\label{ssc:EMw}

In the previous examples we have ignored the possibility of a nontrivial electromagnetic field.
Namely, we have assumed that the electromagnetic field is given by the metric potentials
via relations ${\varphi=\kappa}$ and ${\psi=\lambda}$. However, we have already observed that
equation \eqref{potrel} admits also other solutions which we have parametrized using
a harmonic 1-form ${\eta}$. To include such solutions we should classify all 1-form harmonics
on the transverse space. But if we restrict to the axially symmetric fields we can solve
the potential equation \eqref{potrel} directly, without referring to ${\eta}$ explicitly.

Let us study a pure electromagnetic contribution to the matter, i.e., we assume ${\iota=0}$, ${j_i=0}$ here.
We can thus take a trivial vanishing solution for ${g}$ which implies ${\omega=0}$.

The symmetry assumptions tell us that 1-forms ${\sslfdl}$, ${\sreal}$, and ${a}$ are ${\phi}$ independent,
which implies that all potentials including ${\varphi-\kappa}$ and ${\psi-\lambda}$ can
be at most linear in ${\phi}$:
\begin{equation}\label{tildephipsi}
    \varphi-\kappa = \hat\varphi(\rho)+\varphi_\phi\,\phi\;,\quad
    \psi-\lambda = \hat\psi(\rho)+\psi_\phi\,\phi\;,
\end{equation}
${\varphi_\phi}$, ${\psi_\phi}$ being constants and ${\hat\varphi}$ and ${\hat\psi}$ functions of $\rho$ only.
Substituting into \eqref{potrel} we get
\begin{equation}\label{potrelsym}
    \Bigl(\hat\varphi_{,\rho}-\frac{\Sigma^2}{\rho}\psi_\phi\Bigr)\trgrad\rho+
    \Bigl(\varphi_\phi+\frac\rho{\Sigma^2}\hat\psi_{,\rho}\Bigr)\trgrad\phi=0\;,
\end{equation}
which implies
\begin{equation}\label{potrelsymcom}
    \hat\varphi_{,\rho}=\frac{\Sigma^2}{\rho}\psi_\phi\;,\quad
    \hat\psi_{,\rho}=-\frac{\Sigma^2}{\rho}\varphi_\phi\;.
\end{equation}
Taking into account \eqref{sigma}, it can be easily integrated and substituting
back to \eqref{tildephipsi} we obtain
\begin{equation}\label{tildephipsii}
\begin{aligned}
    \varphi &= \kappa+\psi_\phi\bigl(\log\rho+{\textstyle\frac18}\vrho\rho^2\bigr)+\varphi_\phi\,\phi\;,\\
    \psi &= \lambda-\varphi_\phi\bigl(\log\rho+{\textstyle\frac18}\vrho\rho^2\bigr)+\psi_\phi\,\phi\;.
\end{aligned}
\end{equation}

At this moment it is easier to solve equation \eqref{psieq} for ${\psi}$ with vanishing
right-hand side instead of equation \eqref{lambdaeq} for ${\lambda}$. It has solutions in the form
\eqref{lambdagravgyr} and \eqref{lambdaspin}. The metric potential ${\lambda}$ is then given
by the second of the equations \eqref{tildephipsi}. The metric potential ${\kappa}$ is vanishing thanks
to our gauge.

After choosing the solution for ${\psi}$, we can thus compute
all quantities ${a}$, ${b}$, ${\sslfdl}$, and ${\sreal}$
and substitute them to equation \eqref{heq} for ${h}$.

Let us mention that solution \eqref{tildephipsi} is singular at the origin.
A careful distributional calculation would show that equations for the potentials
\eqref{phipot} and \eqref{potpsi} may not be satisfied at the origin.
Tracing this singular term back to Maxwell equations, it could lead to non-vanishing
electric charges localized at the origin. However, since we have not
written down the Maxwell equations with sources, we do not discuss these
terms in more detail.

\section{Properties of the gyraton spacetimes}\label{sc:interpret}

\subsection{Gravitational field}

In this section we discuss some of the geometrical properties of the gyratonic solutions.

One of the important characteristics of spacetimes are the scalar polynomial invariants
which are constructed only from the curvature and its covariant derivatives.
It was shown that gyratons in the Minkowski spacetime \cite{Fro-Is-Zel:2005:PHYSR4:}
have all the scalar polynomial invariants vanishing (VSI spacetimes) \cite{Prav-Prav:2002:CLAQG:},
the gyratons in the anti-de Sitter \cite{Fro-Zel:2005:PHYSR4:} and direct
product spacetimes \cite{Kadlecova:2009:PHYSR4:} have all invariants constant
(CSI spacetimes) \cite{Coley-Her-Pel:2006:CLAQG}.

In these cases, the invariants are independent of all metric functions
which characterize the gyraton, and have the same
values as the corresponding invariants of the background spacetime.
We observe that similar property is valid also for the gyraton on
Melvin spacetime, however, in this case the invariants are generally \emph{non-constant},
namely, they depend on the coordinate $\rho$. This property is a consequence of the general theorem
holding for the relevant subclass of the Kundt solution, see Theorem II.7 in \cite{ColeyEtal:2010}.

Values of some of the scalar curvature invariants can be found in Appendix \ref{apx:Invars}.

The metric \eqref{s5} admits the null vector ${k}=\partial_{v}$.
It is a Killing vector for ${g=0}$ and ${\trdiv(\Sigma^2j)=0}$, i.e., for the no `heating' part in the gyratonic source.
The covariant derivative of ${k}$ is given by
\begin{equation}\label{recurrentk}
k_{\alpha;\beta}=-\Sigma^{-2}(\partial_{v}H) k_{\alpha}k_{\beta}+\Sigma^{-2}{k_{[\alpha}\nabla_{\beta]}\Sigma^2}\;.
\end{equation}
We observe that the congruence is not even recurrent \cite{Step:2003:Cam:}
as in the case of a gyraton on direct product spacetimes \cite{Kadlecova:2009:PHYSR4:}.
For $\partial_{v}H=0$ we recover the formula in Garfinkle and Melvin \cite{GarfinkleMelvin:1992:PHYSR4:}.
In general, the non-recurrency of the congruence is related to
the non-vanishing spin coefficient $\NP\tau=-\frac{1}{\sqrt{2}}\frac{1}{\Sigma^2}\Sigma_{,\rho}$,
cf.~\cite{Prav-Prav:2002:CLAQG:}, calculated in Appendix \ref{apx:NP5}.

The null character of $k$ and the condition \eqref{recurrentk}
imply that the null congruence with tangent vector $k$ is
geodesic, expansion-free, sheer-free and twist-free, and the spacetime thus belongs
to the Kundt class.

Next we calculate components of the curvature tensors with respect to the
following adapted null tetrad ${\{k,\,l,\,m,\,\mb\}}$ \cite{Step:2003:Cam:}
\begin{equation}\label{b-vectors}
\begin{aligned}
  {k}&=\partial_{v}\;,\\
  {l}&=\frac{1}{\Sigma^2}(\partial_{u}-H\partial_{v})\;,\\
  {m}&=\frac{1}{\sqrt{2}\Sigma}\Bigl(a_{m}\,\partial_{v}
       +\partial_{\rho}-i\,\frac{\Sigma^2}{\rho}\partial_{\phi}\Bigr)\;.
\end{aligned}
\end{equation}
Here, we have introduced the projection of a transverse 1-form $a$ on the vector ${m}$
\begin{equation}
a_{m}= m^i a_i = a_\rho - i\,\frac{\Sigma^2}{\rho} a_\phi = (a +i\,{*a})_{\rho}\;,
\end{equation}
and we will use an analogous notation also for components of
the transverse gradient of a real function ${f}$
\begin{equation}
f_{,m}=m^i f_{,i} = f_{,\rho}-i\,\frac{\Sigma^2}{\rho}f_{,\phi}\;,\quad
f_{,\mb}=\overline{f_{,m}}\;.
\end{equation}
The dual tetrad of 1-forms ${\{\frm{k},\,\frm{l},\,\frm{m},\,\frm{\mb}\}}$ has a simple form
\begin{gather}\label{b-forms}
\frm{k}=\trgrad v + H\trgrad u-a\;,\quad
\frm{l}=\Sigma^2\trgrad u\;,\\
\frm{m}=\frac{\Sigma}{\sqrt{2}}\bigl(\trgrad \rho-i\,{*\trgrad \rho}\bigr)\;,\quad
\frm{\mb}=\frac{\Sigma}{\sqrt{2}}\bigl(\trgrad \rho+i\,{*\trgrad \rho}\bigr)\;.\notag
\end{gather}

With respect to this tetrad, we have found that the non-vanishing curvature components
are given by four new components and by those which are the same for the Melvin universe \eqref{o2}.
The non-vanishing  Ricci scalars are
\begin{equation}\label{sc3}
\begin{aligned}
\Phi_{12}
&=\frac{1}{4\sqrt{2}}\frac{1}{\Sigma^3}\Bigl(\Sigma^2 i b_{,m}+2ib(\Sigma^2)_{,\rho}+2g_{,m}\Bigr),\\
\Phi_{22}
&=\frac{1}{2}\frac{1}{\Sigma^2}\Bigl(\laplace H - (\Sigma^{-2})_{,\rho} H_{,\rho}+{\textstyle\frac12}\Sigma^2 b^2\\
&\qquad\qquad+2a^{i}g_{,i}+\Sigma^{-2}(g+\partial_{u})\trdiv(\Sigma^2 a)  \Bigr)\;.\\
\end{aligned}
\end{equation}
and the non-vanishing Weyl scalars read
\begin{equation}\begin{split}\label{Weyl}
 \Psi_{3}&=\frac{1}{4\sqrt{2}}\frac{1}{\Sigma^3}\Bigl(\Sigma^2 i b_{,\mb}+ib(\Sigma^2)_{,\rho}+2g_{,\mb}\Bigr)\;,\\
 \Psi_{4}&=\frac{1}{2}\frac{1}{\Sigma^5}\biggl[2\Sigma\, a_{\mb}\,g_{,\mb}+\Sigma\biggl(H_{,\rho\rho}{-}\frac{\Sigma^4}{\rho^2}H_{,\phi\phi}{+}2i\frac{\Sigma^2}{\rho}H_{,\phi\rho}\biggr)\\
 &+\Sigma(g+\partial_{u})\left(a_{\rho,\rho}-\frac{\Sigma^4}{\rho^2}a_{\phi,\phi}+i\frac{\Sigma^2}{\rho}(a_{\rho,\phi}{+}a_{\phi,\rho})\right)\\
 &+\left(2\Sigma_{,\rho}-\frac{\Sigma}{\rho}\right)\biggl(H_{,\rho}+\partial_{u}a_{\rho}+g a_{\rho}\\
 &\qquad\qquad\qquad\qquad\quad +2i\frac{\Sigma^2}{\rho}(H_{,\phi}{+}\partial_{u}a_{\phi}{+}g a_{\phi})\biggr)\biggr]\;.\\
\end{split}\raisetag{48pt}
\end{equation}
In particular, there exists a relation between $\Psi_{3}$ and $\Phi_{12}$:
\begin{equation}\label{relpsiphi}
\overline{\Phi}_{12}+\Psi_{3}=\frac{1}{4\sqrt{2}}\frac{1}{\Sigma^3}\left[-i(\Sigma^2)_{,\rho}b+4g_{,\mb}\right].
\end{equation}

Therefore, the metric \eqref{s5} describes the transversal gravitational wave ($\Psi_{4}$ term) in the ${k}$ direction with
a longitudal wave component ($\Psi_{3}$ term). The gravitational wave is accompanied by an aligned pure
radiation field ($\Phi_{22}$ term) with non-null component ($\Phi_{12}$ term) propagating in the Melvin universe.
In fact, the scalars $\Psi_{3}$ and $\Phi_{12}$ are generated by the gyratonic functions $a_{i}$ and the function $g$.
In general the spacetime \eqref{s5} is of Petrov type $II$.

Let us now investigate the subcases of our solutions.
When we set the gyratonic functions $a_{i}=0$, the Ricci scalars become
\begin{equation}\begin{aligned} \label{ScRic}
\Phi_{12}&=\frac{1}{2\sqrt{2}}\frac{1}{\Sigma^3}g_{,m}\;,\\
 \Phi_{22}&=\frac{1}{2}\frac{1}{\Sigma^4\rho^2}\left[\rho(\rho H_{,\rho})_{,\rho}+\Sigma^4 H_{,\phi\phi}\right]\;\\
\end{aligned}\end{equation}
and the Weyl scalars then read
\begin{equation}\begin{aligned} \label{ScWey}
 \Psi_{3}&=\frac{1}{2\sqrt{2}}\frac{1}{\Sigma^3}g_{,\mb}\;,\\
 \Psi_{4}&=\frac{1}{2}\frac{1}{\Sigma^5\rho}\left[\Sigma\rho H_{,\rho\rho}+(2\rho\Sigma_{,\rho}-\Sigma)H_{,\rho}\right]\\
 &-\frac{1}{2}\frac{1}{\rho^2}H_{,\phi\phi}+i\frac{1}{2}\frac{1}{\Sigma^3\rho^2}\left[2\partial_{\rho}(\Sigma g H)-3\Sigma H\right]_{,\phi}.
\end{aligned}\end{equation}
We again obtain the spacetime with similar characteristics as for the full gyratonic  metric \eqref{s5}.
However, the scalars $\Psi_{3}$ and $\Phi_{12}$, which now depend only on function $g$, are now related by an even simpler relation:
\begin{equation}
\overline{\Phi}_{12}={\Psi}_{3}.
\end{equation}

If we assume additionally $H$ to be $v$-independent (i.e., $g=0$) we obtain the only
non-vanishing scalars $\Phi_{22}$ and $\Psi_{4}$ in the same form as in \eqref{ScRic} and \eqref{ScWey}.
This case and its subcases  were thoroughly discussed in \cite{Ortaggio:2004:PHYSR4:}.\footnote{%
The terms $\Phi_{22}$ and $\Psi_{4}$ have a little different form which is caused
by the slightly different choice of null tetrad in our paper.}

Finally, let us mention that for $\vrho=0$ the background reduces to Minkowski spacetime
and we thus recover the gyraton moving on the Minkowski background as an important subcase of our solutions.

\subsection{Electromagnetic field}\label{ssc:elmag4}

The gyraton propagates in a non-null electromagnetic field \eqref{EMFq}, the influence of
which on the geometry is characterized by its density $\vrho$ \eqref{rhodef}.
The electromagnetic field is modified by the gyraton through the ${\sreal_i\trgrad u\wedge\trgrad x^i}$
terms. The electromagnetic field can be rewritten in terms of potentials using \eqref{sreal} as
\begin{equation}\begin{split}\label{realFS}
{F}&=E\bigl(\trgrad v \wedge\trgrad u+\trgrad u\wedge\trgrad \varphi\bigr)\\
&\quad+B\bigl(\Sigma^{-2}\rho\,\trgrad \rho\wedge\trgrad \phi+\trgrad u\wedge\trgrad \psi\bigr)\;.
\end{split}\end{equation}
It describes a superposition of electric and magnetic fields, both pointing
along the $z$ direction, which are modified by the gravitation field of the gyraton.
The additional term does not have a simple structure of electric or magnetic field,
however both are of the form ${\trgrad u\wedge\trgrad f}$ with ${f}$ being the proper potential.

The electromagnetic field  projected on the null tetrad \eqref{b-vectors}
is characterized by three scalars $\Phi_{i}$,
\begin{equation}\label{EMPhiginvX}
\Phi_{0}=0\;,\quad
\Phi_{1}=\frac{\overline{\mathcal{B}}}{2\Sigma^2}\;,\quad
\Phi_{2}=\frac{\overline{\mathcal{B}}}{\sqrt{2}\Sigma^3}\left[ a_{\mb}-\overline{\mathcal{S}}_{\rho}\right]\;.\\
\end{equation}
It follows that the non-null electromagnetic field is aligned with the principal null direction ${k}$ of the gravitation field, but this vector is not a double degenerate vector of the field.

\section{Conclusion}\label{sc:conclusionX}

We have derived and analyzed new gyraton solutions moving with the speed of light on
electro-vacuum Melvin background spacetime in four dimensions. This solution extends
the gyraton solutions previously known on the Nariai, anti-Nariai, and Pleba\'{n}ski--Hacyan
universes of type~D, and on conformally flat Bertotti--Robinson and Minkowski space.

The gyraton solutions describe a gravitational field created by a stress-energy tensor of
a spinning (circularly polarized) high-frequency beam of electromagnetic radiation,
neutrino, or any other massless fields. The gyratons generalize standard gravitational
{\it pp\,}-waves or Kundt waves by admitting a non-zero angular momentum of the source.
The interpretation is that the null matter in the interior of the gyratonic source
possesses an intrinsic spin (or non-zero angular momentum). This leads to other nontrivial
components of the Einstein equations, namely, ${G_{ui}}$ in addition to the pure
radiation $uu$-component which appears for {\it pp\,}-waves or Kundt waves.

We have shown that it is possible to define the gyraton by adding the gyratonic terms $a_{i}$
to the gravitational wave on the Melvin spacetime in cylindrical coordinates in a similar way
as we have defined them in the general Kundt class. We were able to find an ansatz for the
gyraton metric on the Melvin spacetime by direct transformation from the Kundt class of
metrics \eqref{ssc:def}.

We have further demonstrated that the Einstein--Maxwell equations reduce to the set of linear
equations on the 2-dimensional transverse spacetime which has a non-trivial geometry given
by the transverse metric. These equations can be solved exactly for any distribution
of the matter sources. In general, the problem has been thus reduced to a construction of
scalar Green functions for certain differential operators on the transverse space.

We have solved and analyzed the field equations for particular examples with the axial symmetry.
In these cases the equations reduce to ordinary differential equations.

We have analyzed geometric properties of the principal null congruence and we have found
that it is not recurrent contrary to the case of gyratons on direct product spacetimes.
We have explicitly calculated the curvature tensor and  determined that the gyratons on
Melvin spacetime are of Petrov type II and belong to the Kundt family of
shear-free and twist-free nonexpanding spacetimes.
The gyratonic term $a_{i}$ generates the non-trivial Ricci $\Phi_{12}$
and Weyl $\Psi_{3}$ scalars, in addition to the gravitational waves investigated in
\cite{Ortaggio:2004:PHYSR4:}. We found also a very simple
relation \eqref{relpsiphi} between these components. By studying particular subclasses
we have shown that our solutions are generalizations of those from \cite{Ortaggio:2004:PHYSR4:}.

The scalar polynomial invariants of the metric \eqref{s5} are in
general non-constant (although, some of them are zero)---they depend on the coordinate $\rho$.
The invariants are not affected by the presence of the gyratons,
they are the same as for the Melvin background. The same property was proved for
gyratons on backgrounds belonging to VSI or CSI families of spacetimes.

It would be interesting to investigate a generalization of our ansatz for more complicated
spacetimes which could allow, e.g., an inclusion of a cosmological constant.

\begin{center}{\large\bf Acknowledgements}\end{center}\par
We wish to thank to Tom\'{a}\v{s} Pech\'{a}\v{c}ek and Otakar Sv\'{i}tek for helpful discussions
and to Marcello Ortaggio for his paper about gravitational waves in the Melvin universe which
motivated our work.
H.~K.  was supported by Grants No. GA\v{C}R-202/09/H033, No. GAUK~12209,
and  Project No. SVV~261301 of the Charles University in Prague.
P.~K. was supported by Grant No. GA\v{C}R~202/09/0772, and both authors thank  the
Project No. LC06014 of the Center of Theoretical Astrophysics.

\newpage
\addcontentsline{toc} {section}{Appendix}
\begin{subappendices}
\section{The Einstein equations}\label{apx:AppAx}

Here we present quantities needed for evaluation of the Einstein equations.

The inverse to the metric \eqref{s5} is
\begin{equation}
\begin{split}\label{a03}
g^{\mu\nu}&\partial_{\mu}\partial_{\nu}=
  \frac{1}{\Sigma^2}\partial_{\rho}\partial_{\rho}+\frac{\Sigma^2}{\rho^2}\partial_{\phi}\partial_{\phi}
  -\frac{2}{\Sigma^2}\partial_{u}\partial_{v}\\
  &+2\bigl(a_{\rho}\frac{1}{\Sigma^2}\partial_{\rho}+a_{\phi}\frac{\Sigma^2}{\rho^2}\partial_{\phi}\bigr)\partial_{v}
  +(2\frac{H}{\Sigma^2}+a^2)\,\partial_{v}\partial_{v}.
\end{split}\raisetag{8ex}
\end{equation}

The stress-energy tensor $T^\EM$ of the electromagnetic field \eqref{EMFq} can be defined as
\begin{equation}\label{defemx}
T^\EM_{\mu\nu}=\frac{\epso}{2}{\mathcal{F}_{\mu}}^{\rho}\overline{\mathcal{F}}_{\nu\rho}
\end{equation}
where $\mathcal{F}\equiv {F}+i{{\star}F}$ is the complex self-dual Maxwell tensor.
The 4-dimensional Hodge dual is defined by
${\star}F_{\mu\nu}=\frac{1}{2}\varepsilon_{\mu\nu\rho\sigma}F^{\rho\sigma}$
and the Maxwell tensor ${{\mathcal{F}_{\mu\nu}}}$ satisfies the self-duality
condition  ${\star}\mathcal{F}=-i\mathcal{F}$.

The non-vanishing components of the stress-energy tensor \eqref{defemx} are
\begin{align}
\varkappa T^\EM_{uv}&=\frac{\vrho}{\Sigma^2},\notag\\
\varkappa T^\EM_{uu}&=2\vrho\left(\frac{H}{\Sigma^2}+(\sslfdl-a)^2\right)\;,\notag\\
\varkappa T^\EM_{u\rho}&=\frac{\vrho}{\Sigma^2}(a_{\rho}-2\sslfdl_{\rho}),\notag\\
\varkappa T^\EM_{u\phi}&=\frac{\vrho}{\Sigma^2}(a_{\phi}-2\sslfdl_{\phi}),\notag\\
\varkappa T^\EM_{\rho\rho}&=\frac{\vrho}{\Sigma^2}=\frac{\vrho}{\Sigma^4}g_{\rho\rho}\;,\label{EMTW}\\
\varkappa T^\EM_{\phi\phi}&=\frac{\vrho\rho^2}{\Sigma^6}=\frac{\vrho}{\Sigma^4}g_{\phi\phi}\;,\notag
\end{align}
where the density ${\vrho}$ was defined in~\eqref{rhodef}.

The non-vanishing components of the stress-energy tensor \eqref{m7} of the gyratonic matter are
\begin{equation}
\begin{gathered}
\varkappa T^\gyr_{uu}=j_u=v\,\trdiv(\Sigma^2 j)+\iota\;,\\
\varkappa T^\gyr_{ui}=j_i\;.
\end{gathered}
\end{equation}

The Einstein tensor for the metric \eqref{s5} reads
\begin{align}
G_{uv}&= \frac{1}{\Sigma^2\rho}\bigl(-\rho(\Sigma_{,\rho})^2+2\Sigma(\Sigma_{,\rho})\bigr)\;,\notag\\
G_{uu}&=\frac{1}{2}\Sigma^4 b^2+\Sigma^2(\laplace H+\frac{(\Sigma^2)_{,\rho}}{\Sigma^4}H_{,\rho})
  +\Sigma^2(\partial^2_{v}H)a^2\notag\\
  &\quad +2\Sigma^2a^i\partial_{v}H_{,i}+(\partial_v H+\partial_{u})\,\trdiv(\Sigma^2 a)\notag\\
  &\quad + 2HG_{uv}\;,\notag\\
G_{u\rho}&=\frac{1}{2}\frac{\Sigma^4}{\rho}b_{,\phi}-a_{\rho}\bigl(G_{uv}-\partial^2_v H\bigr)
    +\partial_{v}H_{,\rho}\;,\label{EinsteinTx}\\
G_{u\phi}&=-\frac{1}{2}\rho(b_{,\rho}+\frac{4\Sigma_{,\rho}}{\Sigma}b)
    -a_{\phi}\bigl(G_{uv}-\partial^2_v H\bigr)+\partial_{v}H_{,\phi}\;,\notag\\
G_{\rho\rho}&=G_{uv}+\partial^2_v H\;,\notag\\
G_{\phi\phi}&=\frac{\rho^2}{\Sigma^4}(G_{uv}+\partial^2_v H)\;\notag.
\end{align}
Here we have used only the metric \eqref{s5}, without any usage of the field equations.


\section{The NP formalism}\label{apx:NPx}

Calculating the Newman--Penrose spin coefficients with respect to the tetrad \eqref{b-vectors},
we recover again that the congruence ${k}$ is nonexpanding and
nontwisting (${\NP\rho=0}$), sheer-free (${\NP\sigma=0}$), geodesic and affine parameterized
(${\NP\kappa=\NP\eps=0}$). In addition, the tetrad is gauge invariant and it is not parallelly
transported along the null congruence because it does not satisfy ${\NP\kappa=\NP\pi=\NP\eps=0}$.

The remaining spin coefficients are
\begin{equation}\begin{gathered}\label{sc10x}
\NP\lambda=0\;,\quad
\NP\mu=\frac{i}{2}b\;,\\
\NP\gamma=\frac{1}{4}\frac{1}{\Sigma^2}\bigl(2g+i\Sigma^2b)\;,\\
\NP\nu=\frac{1}{\sqrt{2}}\frac{1}{\Sigma^3}\left\{(g+\partial_{u})a_{\mb}+g_{,\mb}\right\}\;,\\
\NP\tau=-\frac{1}{\sqrt{2}}\frac{1}{\Sigma^2}\Sigma_{,\rho}\;,\quad
\NP\pi=+\frac{1}{\sqrt{2}}\frac{1}{\Sigma^2}\Sigma_{,\rho}\;,\\
\NP\alpha=\frac{1}{2\sqrt{2}}\frac{1}{\Sigma^2\rho}(2\rho\Sigma_{,\rho}-\Sigma)\;,\quad
\NP\beta=\frac{1}{2\sqrt{2}}\frac{1}{\Sigma\rho}\;.
\end{gathered}\end{equation}


\section{Scalar polynomial curvature invariants}\label{apx:Invars}


As we have already mentioned, the scalar curvature invariants are independent of all metric functions
which characterize the gyraton, and have the same values  as the corresponding invariants of the Melvin universe (cf.~\cite{ColeyEtal:2010}).
Let us stress here, however, that the invariants are generally \emph{non-constant},
namely, they depend on the coordinate $\rho$. In this appendix we list some of the curvature invariants.

The scalar curvature for the whole gyraton metric \eqref{s5} is zero, $R=0$.
Next, we define the following scalar polynomial invariants constructed
from the Riemann tensor:
\begin{equation}\begin{aligned}
R^{(2)}&=R^{ab}{}_{cd}R^{cd}{}_{ab}\;,\\
R^{(3)}&=R^{ab}{}_{cd}R^{cd}{}_{ef}R^{ef}{}_{ab}\;,\\
R^{(4)}&=R^{ab}{}_{cd}R^{cd}{}_{ef}R^{ef}{}_{pq}R^{pq}{}_{ab}\;,\\
R^{(5)}&=R^{ab}{}_{cd}R^{cd}{}_{ef}R^{ef}{}_{pq}R^{pq}{}_{rs}R^{rs}{}_{ab}\;.
\end{aligned}\end{equation}
Using the {\it GRtensor} package in {\it Maple}, we get the explicit expressions:
\begin{align}
R^{(2)}&=\frac{2\vrho^2}{\Sigma^8}\Bigl({\textstyle\frac38}\vrho^2\rho^4-3\vrho\rho^2+10\Bigr)\;,\notag\\
R^{(3)}&=-\frac{3\vrho^3}{\Sigma^{12}}\Bigl({\textstyle\frac1{16}}\vrho^3\rho^6
       -{\textstyle\frac34}\vrho^2\rho^4+7\vrho\rho^2-20\Bigr)\;,\notag\\
\begin{split}
R^{(4)}&=\frac{4\vrho^4}{\Sigma^{16}}\Bigl({\textstyle\frac9{256}}\vrho^4\rho^8
       -{\textstyle\frac{9}{16}}\vrho^3\rho^6\\
     &\qquad\qquad+{\textstyle\frac{51}8}\vrho^2\rho^4-33\vrho\rho^2+65\Bigr)\;,
\end{split}\label{invariants}\\
\begin{split}
R^{(5)}&=-\frac{5\vrho^5}{\Sigma^{20}}\Bigl({\textstyle\frac3{256}}\vrho^5\rho^{10}
       -{\textstyle\frac{15}{64}}\vrho^4\rho^8+{\textstyle\frac{31}{8}}\vrho^3\rho^6\\
     &\qquad\qquad-{\textstyle\frac{63}{2}}\vrho^2\rho^4+64\cdot127\vrho\rho^2-204\Bigr)\;.
\end{split}\notag
\end{align}
We explicitly observe that these invariants do not depend on any of the metric
functions $a_{i}$, ${g}$, and ${h}$ which characterize the gyraton.

The invariants \eqref{invariants} mimic the behavior of the Gauss curvature of the
transverse space discussed in detail in \eqref{ssc:transsp1}.
They have their maximum on the axis $\rho=0$ and they are vanishing as
``the neck of the vase closes off asymptotically'' as $\rho$ tends to infinity.
For $\vrho=0$ we get the identically vanishing invariants,
i.e., the invariants for the gyratons on Minkowski background (VSI).

In {\it Maple} tensor package {\it GRtensor} there is defined a set of curvature invariants \texttt{CMinvars}.
For completeness, we present the explicit expressions for them:
\begin{equation}\begin{gathered}
R=R_{2}={W1I}={W2I}=0\;,\\
R_{1}=\frac{\vrho^2}{\Sigma^8}\;,\quad{W1R}=\frac{3\vrho^2}{2^5\Sigma^{8}}(\vrho\rho^2-4)^2\;,\\
R_{3}=\frac{\vrho^4}{4\Sigma^{16}}\;,\quad{W2R}=\frac{3\vrho^3}{2^8\Sigma^{12}}(\vrho\rho^2-4)^3\;,\\[1ex]
{M1I}={M2I}={M4}={M5I}=0\;,\\
{M1R}=\frac{\vrho^3}{2^2\Sigma^{12}}(\vrho\rho^2-4)\;,\\
{M2R}={M3}=\frac{\vrho^4}{2^4\Sigma^{16}}(\vrho\rho^2-4)^2\;,\\
{M5R}=\frac{\vrho^5}{2^{6}\Sigma^{20}}(\vrho\rho^2-4)^3\;.
\end{gathered}\end{equation}\\[-12pt]

\end{subappendices}

\newpage
\chapter{The gyraton solutions on generalized Melvin universe with cosmological constant}
\section{Introduction}
In this chapter we present and analyze new exact gyraton solutions of algebraic type II on generalized
 Melvin universe of type D which admit non--vanishing cosmological constant $\Lambda$. We show that it
 generalizes both, gyraton solutions on Melvin and on direct product spacetimes.
 When we set $\Lambda=0$ we get solutions on Melvin spacetime
 and for $\Sigma=1$ we obtain solutions on direct product spacetimes.
 We demonstrate that the solutions are member of the Kundt family of spacetimes as its subcases.
 We show that the Einstein equations reduce to a set of equations on the transverse 2-space.
 We also discuss the polynomial scalar invariants which are non--constant in general but constant
 for sub--solutions on direct product spacetimes.
 The contents of this chapter is work in progress \cite{KadlKrt:2010:CLAQG:}.

The chapter is organized as follows.
In Section \ref{sc:def} we present our ansatz for the gyraton metric on generalized Melvin universe and the generalized electromagnetic tensor. In Section \ref{scc:fequationsss} we briefly review the derivation of the Einstein--Maxwell equations. The source--free Einstein equations determine the functions $\Sigma$ and $S$, in particular, there exists a relation between them. Next we derive the non--trivial source equations. The Einstein--Maxwell equations do decouple for the gyraton metric on generalized Melvin universe as
 for its subcase solutions on Melvin and on direct product spacetimes.
In Section \ref{sc:interprett} we concentrate on the interpretation of our solutions. Especially, we discuss the geometry of the transverse metric of the generalized Melvin universe in detail for different values of the cosmological constant. We show explicitly that the Melvin universe and direct product spacetimes are special cases of our solutions.
In Section \ref{sc:Invariants} we discuss the properties of the scalar polynomial invariants which are functions of $\rho$ but for subcase solutions on direct product spacetimes $(\Sigma=1)$ the invariants are constant.
The notation on the transversal spacetime is reviewed in Appendix \ref{AppO}, the necessities to derive the field equations and NP formalism are left to Appendices \ref{apx:AppAA} and \ref{apx:prop}, these sections should be understood as an appendix to this chapter.
\section{The ansatz for the gyratons on generalized Melvin universe}\label{sc:def}

The ansatz for the gyraton metric on the generalized Melvin spacetime is the following,
\begin{equation}\label{s1w}
{\bf g}=-2\Sigma^2 H \trgrad u^2-\Sigma^2\trgrad u\vee\trgrad v + {\bf q} +\Sigma^2 \trgrad u\vee {\bf a},
\end{equation}
where we have introduced the 2--dimensional transversal metric ${\bf q}$  on transverse spaces $u,\,v=$constant as
\begin{equation}\label{trmetric4}
{\bf q}=\Sigma^2\trgrad\rho^2+\frac{S(\rho)^2}{\Sigma^2}\trgrad \phi^2.
\end{equation}
We have assumed that the metric \eqref{s1w} belongs to the Kundt class of spacetimes and
that the transversal metric ${\bf q}$ has one Killing vector
\begin{equation}\label{qqq}
\mathcal{L}_{\frac{\partial}{\partial \phi}}{\bf q}=0.
\end{equation}
The metric \eqref{s1w} represents gyraton propagating on the background which is formed by generalized Melvin
spacetime. The metric \eqref{s1w} generalizes only the transversal metric therefore the algebraical type is $II$ as
for the gyraton on the Melvin spacetime \cite{Kadlecova:2010:PHYSR4:}, the NP quantities are listed in Appendix \ref{apx:prop}.

We have generalized the transversal metric for the Melvin universe by assuming
general function $S=S(\rho)$ instead of the simple coordinate $\rho$ in front of the term $\trgrad\phi^2$, see \cite{Kadlecova:2010:PHYSR4:}.
We will show that these general functions $\Sigma(\rho)$ and $S(\rho)$ are determined by the Einstein--Maxwell equations and
have proper interpretation. The presence of cosmological constant $\Lambda$ is not allowed for the solution on pure Melvin background \cite{Kadlecova:2010:PHYSR4:}.

The transverse space is covered by two  spatial coordinates ${x^i}$ $(i=\rho,\,\phi)$ and it is convenient to introduce
suitable notation on it. Because this chapter is generalization of two previous papers \cite{Kadlecova:2009:PHYSR4:, Kadlecova:2010:PHYSR4:} we will skip many technical details and therefore we have added the notation to the Appendix \ref{AppO}.
The function $H(u,v,{\bf x})$ in the metric \eqref{s1w} can depend on all coordinates, but the functions ${a(u,{\bf x})}$ are ${\mbox{${v}$-independent}}$.

The derivation of the Einstein--Maxwell equations is almost identical with the previous paper \cite{Kadlecova:2010:PHYSR4:} and the changes regarding the new function $S$ influence only the notation on the transversal spacetime in Appendix \ref{AppO}. Therefore we will describe the derivation of Einstein--Maxwell equations very briefly.

The metric should satisfy the Einstein equations with cosmological constant $\Lambda$ and with a  stress-energy tensor generated by the electromagnetic field of the background Melvin spacetime  ${\bf T}^\EM$ and the gyratonic source ${\bf T}^{\gyr}$ as\footnote{$\varkappa=8\pi G$ and $\epso$ are gravitational and electromagnetic constants. There are two general choices of geometrical units: the gaussian with $\varkappa=8\pi$ and $\varepsilon_{\rm o}=1/4\pi$, and SI-like with $\varkappa=\varepsilon_{\rm o}=1$.}

\begin{equation}\label{EinsteinEqqw}
{\bf G}+\Lambda\,{\bf g}=\varkappa \bigl( {\bf T}^\EM+{\bf T}^{\gyr}\bigr)\;.
\end{equation}
We assume the electromagnetic field is given by
\begin{equation}\label{realFF}
{\bf F}=E\trgrad v \wedge\trgrad u+\frac{B}{\Sigma^2}{\boldsymbol{\epsilon}}+\trgrad u\wedge (E\,\vs-B{*(\vs-{\bf a})})\;,
\end{equation}
where  $E$ and $B$ are parameters of electromagnetic field.
The self--dual complex form of the Maxwell\footnote{We will follow the notation of \cite{Step:2003:Cam:}. Namely, $\boldsymbol{\mathcal{F}}\equiv {\bf F}+i{{\star}\bf{F}}$ is complex self--dual Maxwell tensor, where the 4--dimesional Hodge dual is ${\star}{F}_{\mu\nu}=\frac{1}{2}\varepsilon_{\mu\nu\rho\sigma}{F}^{\rho\sigma}$. The self--dual condition reads ${{\star}\boldsymbol{\mathcal{F}}}=-i\boldsymbol{\mathcal{F}}$. The orientation of the 4--dimensional Levi--Civita tensor
is fixed by the sign of the component $\varepsilon_{vu\rho\phi}=S\Sigma^2$.
The energy--momentum tensor of the electromagnetic field is given by $T_{\mu\nu}=\frac{\varepsilon_{\rm o}}{2}\mathcal{F}_{\mu}^{\rho}\overline{\mathcal{F}}_{\nu\rho}$.}
 tensor is
\begin{equation}\begin{split}\label{EMFa}
\boldsymbol{\mathcal{F}}&=\mathcal{B}(\trgrad v\wedge\trgrad u - \frac{i}{\Sigma^2}{\boldsymbol \epsilon}+\trgrad u\wedge[\vs+i{*(\vs}-{\boldsymbol{a}})]),
\end{split}\end{equation}
for details see \cite{Kadlecova:2010:PHYSR4:}.

We have denoted the complex constant
\begin{equation}
\mathcal{B}=E+iB,
\end{equation}
and we have introduced a constant $\vrho$,
\begin{equation}\label{rhodeff}
\vrho=\frac{\varkappa\epso}{2}(E^2+B^2).
\end{equation}

We define the gyratonic matter only on a phenomenological level as
\begin{equation}\label{mm7}
\varkappa\, {\bf T}^{\gyr}=j_{u}\,\trgrad u^2+\trgrad u\vee {\bf j}\;,
\end{equation}
where the source functions ${j_u(v,u,{\bf x})}$ and ${j(v,u,{\bf x})}$.
We assume that the gyraton stress-energy tensor is locally conserved,
\begin{equation}\label{gyrcon}
  \nabla {\cdot} {\bf T}^{\gyr}=0\;.
\end{equation}

To conclude, the fields are characterized by functions ${\Sigma}$, $S$, ${H}$, ${\bf a}$, and ${\vs}$ which must be determined by the field equations and the gyraton sources ${j_u}$ and ${\bf j}$ and the constants ${E}$ and ${B}$ of the background electromagnetic field are prescribed.
\section{The Einstein--Maxwell field equations}\label{scc:fequationsss}
First, we will start to solve the Maxwell equations, it is sufficient to calculate the cyclic Maxwell equation for the self--dual Maxwell tensor \eqref{EMFa}
\begin{equation}\begin{aligned}\label{MXECC}
  0=\trgrad {\mathcal{F}} =\mathcal{B}&\left\{\partial_v(\vs+i{*(\vs-{\bf a})})\, \trgrad v\wedge\trgrad u\wedge\trgrad {\bf x}\right.\\
   &\left.-[\rot\vs+i\,\trdiv(\vs-{\bf a})]\, \trgrad u \wedge \boldsymbol{\epsilon}\right\}\;.
\end{aligned}\end{equation}

From the real part we immediately get that the 1-forms ${\vs}$ is ${v}$-independent, and rotation free
\begin{equation}\label{pot11}
  \rot\vs=0\;.
\end{equation}
From imaginary part it follows that the 1--form ${\bf a}$ is also independent and it satisfies
\begin{equation}\label{pot22}
\trdiv(\vs-{\bf a}) = 0\;.
\end{equation}

\subsection{The trivial  Einstein--Maxwell equations--determining the function $\Sigma$ and $S$}
Next we will derive the Einstein--Maxwell equations from the Einstein tensor and the electromagnetic stress-energy tensor, which are listed in
Appendix \ref{apx:AppAA}.

First we will  solve the equations which are source free and we will be able to determine
the analytic formula for the functions $\Sigma$ and $S$.

The first equation we obtain from the $vu$-component,
\begin{equation}\label{r1}
-\frac{(\Sigma_{,\rho})^2}{\Sigma^2}+2\frac{\Sigma_{,\rho}}{\Sigma}\frac{S_{,\rho}}{S}-\frac{S_{,\rho\rho}}{S}=\Lambda\Sigma^2+\frac{\vrho}{\Sigma^2},
\end{equation}
the next two equations we get from the transverse diagonal components $\rho\rho$ and $\phi\phi$ ,
\begin{align}
-\frac{(\Sigma_{,\rho})^2}{\Sigma^2}+2\frac{\Sigma_{,\rho}}{\Sigma}\frac{S_{,\rho}}{S}+\partial^2_{v}H&=-\Lambda\Sigma^2+\frac{\vrho}{\Sigma^2}\;,\label{r2}\\
-\frac{(\Sigma_{,\rho})^2}{\Sigma^2}+2\frac{\Sigma_{,\rho\rho}}{\Sigma}+\partial^2_{v}H&=-\Lambda\Sigma^2+\frac{\vrho}{\Sigma^2}\;.\label{r3}
\end{align}
When we compare the equation \eqref{r2} and \eqref{r3} we immediately get
the relation between the functions $\Sigma$ and $S$,
\begin{equation}\label{s2q}
\Sigma_{,\rho}\frac{S_{,\rho}}{S}=\Sigma_{,\rho\rho},
\end{equation}
and thus we are able to determine their explicit relation ($\Sigma_{,\rho}\neq0$) as
\begin{equation}\label{s3q}
\Sigma_{,\rho}=\gamma S,
\end{equation}
where $\gamma$ is an integration constant.

After substituting the relation \eqref{s3q} into equation \eqref{r1} then we get equation
\begin{equation}
-\frac{(\Sigma_{,\rho})^2}{\Sigma^2}+2\frac{\Sigma_{,\rho\rho}}{\Sigma}+\frac{\Sigma_{,\rho\rho\rho}}{\Sigma_{,\rho}}=\Lambda\Sigma^2+\frac{\vrho}{\Sigma^2}\;,\label{r4}
\end{equation}
which will be useful later.

To determine the function $H$ it is useful to substitute \eqref{s3q} into the equation \eqref{r3} and then multiply it by $\frac12\frac{\Sigma}{\Sigma_{,\rho}}$,
we get
\begin{equation}\label{r5}
\frac12(\partial^2_{v}H)_{,\rho}\frac{\Sigma}{\Sigma_{,\rho}}-2\frac{\Sigma_{,\rho\rho}}{\Sigma}+\frac{(\Sigma_{,\rho})^2}{\Sigma^2}+\frac{\Sigma_{,\rho\rho\rho}}{\Sigma_{,\rho}}=-\Lambda\Sigma^2-\frac{\vrho}{\Sigma^2}\;.
\end{equation}
Now, we add the equation \eqref{r1} to \eqref{r5} and  obtain,
\begin{equation}\label{H}
\frac12(\partial^2_{v}H)_{,\rho}\frac{\Sigma}{\Sigma_{,\rho}}=0,
\end{equation}
then for $\Sigma_{,\rho}\neq 0$ we can write that
\begin{equation}\label{alphaH}
  \partial^2_{v} H  =-\alpha\;,
\end{equation}
where $\alpha$ is a constant.

Thus the metric function ${H}$ has a structure
\begin{equation}\label{Heq}
  H = -\frac{1}{2}\alpha v^2 +g\,v + h\;,
\end{equation}
where we have introduced ${v}$-independent functions ${g(u,{\bf x})}$ and ${h(u,{\bf x})}$.

In the following we want to determine an analytical expression for $\Sigma$, in order to do that
 we substitute the result \eqref{alphaH} into \eqref{r3},
\begin{equation}\label{r6}
2\frac{\Sigma_{,\rho\rho}}{\Sigma}-\frac{(\Sigma_{,\rho})^2}{\Sigma^2}=-\Lambda\Sigma^2+\frac{\vrho}{\Sigma^2}+\alpha\;.
\end{equation}
When we add  the expression \eqref{r4} to \eqref{r6}, we obtain that
\begin{equation}\label{r7}
\Sigma_{,\rho\rho\rho}=-2\Lambda\Sigma^2{\Sigma_{,\rho}}+\alpha{\Sigma_{,\rho}}\;.
\end{equation}
We can rewrite the previous equation as $
\Sigma_{,\rho\rho\rho}=-\frac{2}{3}\Lambda(\Sigma^3)_{,\rho}+\alpha{\Sigma_{,\rho}}$ to be able to integrate it again as
\begin{equation}\label{r8}
\Sigma_{,\rho\rho}=-\frac{2}{3}\Lambda\Sigma^3+\alpha\Sigma+\frac12\beta\;,
\end{equation}
which we can rewrite as
\begin{equation}\label{r9}
\frac12[(\Sigma_{,\rho})^2]_{,\rho}=-\frac{1}{6}\Lambda(\Sigma^4)_{,\rho}+\alpha(\Sigma^2)_{,\rho}+\frac12\beta\Sigma_{,\rho}\;.
\end{equation}
After another integration we get the final formula for the derivative of the function $\Sigma$,
\begin{equation}\label{r10}
(\Sigma_{,\rho})^2=-\frac{1}{3}\Lambda\Sigma^4+\alpha\Sigma^2+\beta\Sigma+c\;,
\end{equation}
and it can be rewritten using \eqref{s3q}  as
\begin{equation}\label{r11}
\gamma S=\left[-\frac{1}{3}\Lambda\Sigma^4+\alpha\Sigma^2+\beta\Sigma+c\right]^{1/2}\;,
\end{equation}
where $\alpha$, $\beta$ and $c$ are integration constants which should be determined.

Furthermore, we are able to determine the constant $c$ explicitly.
When we substitute the result \eqref{r10} and \eqref{r8} into \eqref{r6} we
immediately obtain that
\begin{equation}\label{c}
c=-\vrho.
\end{equation}
The constants $\alpha$ and $\beta$ will be determined in the section \ref{sc:interpret}.

\subsection{The Einstein--Maxwell equations for the sources}\label{sources}
The remaining nontrivial components of the Einstein equations are those involving the gyraton source \eqref{mm7}.
To write the source equation we have to evaluate the component $G_{uv}$ using the expressions for derivatives of $\Sigma$.
Then the component $G_{uv}$ has the explicit form
\begin{equation}
G_{uv}=\Lambda\Sigma^2+\frac{\vrho}{\Sigma^2}.
\end{equation}

The $ui$-components give equations related to ${\bf j}$,
\begin{equation}\label{jieqpott}
  \Sigma^2\,{\bf j} = \frac12\rot(\Sigma^4\, b) + \Sigma^2\trgrad g-\alpha\Sigma^2{\bf a}+2\vrho(\vs-{\bf a})\;,
\end{equation}
where
\begin{equation}
b = \rot{\bf a}\;.\label{bdeff}
\end{equation}

It is useful to split the source equation into divergence and rotation parts:
\begin{align}
  \trdiv{(\Sigma^2\, {\bf j})}&=\trdiv {\Sigma^2(\trgrad g-\alpha\,{\bf a})},\label{divjeqq}\\
  \rot(\Sigma^2 {\bf j})&= - \frac12 \laplace(\Sigma^4 b) + \rot(\Sigma^2\trgrad g)\notag\\
  &-\alpha\rot(\Sigma^2{\bf a})-2\vrho\,b\;.\label{rotjeqq}
\end{align}
These are coupled equations for ${g}$ and ${{\bf a}}$. We will return to them below.

The condition \eqref{gyrcon} for the gyraton source gives,
that the sources ${\bf j}$ must be ${v}$-independent and ${j_u}$ has the structure
\begin{equation}\label{jdecomp1}
  j_u = v\,\trdiv(\Sigma^2 {\bf j}) + \iota\;,
\end{equation}
where $\iota(u,{\bf x})$ is $v$--independent function, see \cite{Kadlecova:2010:PHYSR4:} Eq. 2.51.
The gyraton source \eqref{mm7} is therefore  determined by three \mbox{${v}$-independent} functions ${\iota(u,{\bf x})}$ and ${j(u,{\bf x})}$.

The $uu$-component of the Einstein equation gives
\begin{equation}\label{jueq1}
\begin{split}
  j_u =\,&v\left[\trdiv {(\Sigma^2\trgrad g)}-\alpha\trdiv(\Sigma^2 {\bf a})\right]
      +\Sigma^2(\laplace h - (\Sigma^{-2})_{,\rho}h_{,\rho})\\
      &+\frac12\Sigma^4 b^2+ 2\Sigma^2 {\bf a}{\cdot}\trgrad g+(\partial_u+g)\trdiv(\Sigma^2 {\bf a}) \\
      &-\alpha\Sigma^2{\bf a}^2-2\vrho\,(\vs-{\bf a})^2\;.
\end{split}
\end{equation}
Then we can compare the coefficient in front of ${v}$ with \eqref{divjeqq} and we get consistent strurotjeqcture  with \eqref{jdecomp1}. The nontrivial ${v}$-independent part of \eqref{jueq1} gives the equation for the metric function ${h}$,
\begin{equation}\label{heq1}
\begin{split}
 \Sigma^2&(\laplace h -(\Sigma^{-2})_{,\rho}h_{,\rho})=
      \iota \, -\frac12\Sigma^4 b^2- 2\Sigma^2 {\bf a}{\cdot }\trgrad g\\
      & -(\partial_u+g)\trdiv(\Sigma^2 {\bf a})+\alpha\Sigma^2{\bf a}^2+2\vrho\,(\vs-{\bf a})^2\;.
\end{split}
\end{equation}

Now, let us return to solution of equations \eqref{divjeqq} and \eqref{rotjeqq}. The first equation simplifies if
we use gauge condition
\begin{equation}\label{gaugefix}
\trdiv\bigl(\Sigma^2{\bf a}\bigr)=0\;.
\end{equation}
It can be satisfied due to gauge freedom ${v\to v-\chi}$, ${{\bf a}\to{\bf a}-\trgrad\chi}$, cf.\ the discussion in \cite{Kadlecova:2010:PHYSR4:}.
Such a condition implies the existence of a potential ${\tilde\lambda}$,
\begin{equation}\label{lambdadef}
    \Sigma^2{\bf a} = \rot\tilde\lambda\;.
\end{equation}

The equation \eqref{divjeqq} now reduces to
\begin{equation}\label{divjeqgauged}
    \trdiv(\Sigma^2\trgrad g-\Sigma^2 {\bf j})=0\;.
\end{equation}
It guarantees the existence of a scalar ${\omega}$ such that
\begin{equation}\label{omegadef}
    \trgrad g =  {\bf j} + \Sigma^{-2}\rot\omega\;.
\end{equation}
However, we have to enforce the integrability conditions
\begin{equation}\label{integrability}
    \rot\trgrad g=0\;,
\end{equation}
which turns out to be the equation for ${\omega}$:
\begin{equation}\label{omegaeq}
    \trdiv\bigl(\Sigma^{-2}\trgrad\omega\bigr) = \rot{\bf j}\;.
\end{equation}
We thus obtained the decoupled equations \eqref{omegadef} and \eqref{omegaeq} which determine the metric function ${g}$.

Substituting \eqref{lambdadef} and \eqref{omegadef} to \eqref{rotjeqq}, and using identity
\begin{equation}\label{btlambda}
    b = \rot\bigl(\Sigma^{-2}\rot\tilde\lambda\bigr)\;,
\end{equation}
we get the decoupled equation for ${\tilde\lambda}$:
\begin{equation}\label{lambdaeqq}
\begin{split}
    &\frac12\laplace\Bigl(\Sigma^4\rot\bigl(\Sigma^{-2}\rot\tilde\lambda\bigr)\Bigr)\\
    &\qquad+2\vrho\rot\bigl(\Sigma^{-2}\rot\tilde\lambda\bigr)
       -\alpha\laplace\tilde\lambda=-\laplace\omega\;.
\end{split}
\end{equation}
It is a complicated equation of the forth order. It can be simplified to an ordinary differential equation if we assume
the additional symmetry properties of the fields, e.g., the rotational symmetry around the axis.
The potential ${\tilde\lambda}$ then determines the metric 1-form ${{\bf a}}$ through \eqref{lambdadef}.

After finding ${\bf a}$ one can solve the field equations for ${{\bf s}}$.
The potential equations \eqref{pot11} give immediately that
\begin{equation}\label{phipott}
  \vs = {\rm d}\varphi\;.
\end{equation}
Substituting to the condition \eqref{pot22} we get the Poisson equation for ${\varphi}$:
\begin{equation}\label{phieq}
    \laplace\varphi = \trdiv{\bf a}\;.
\end{equation}

Finally, the remaining metric function $h$ is determined by the equation \eqref{heq1}.

\section{The interpretation of the solutions}\label{sc:interprett}
\subsection{The geometries of the transversal spacetime}
In this section we will investigate the geometry of the transversal metric ${\bf q}$ (the wave fronts) \eqref{trmetric4} and
we will determine the constants $\alpha,\,\beta$ in the final equation \eqref{r10}. Subsequently, we
will discuss the various geometries of ${\bf q}$ in proper parametrization and we will determine
the meaning of the parameter $\gamma$.

We impose conditions to the derivatives of $\Sigma$ (i.e., $S$) \eqref{r10}, \eqref{r8} and \eqref{r7} while
using the relation \eqref{s3q} between $\Sigma_{,\rho}$ and $S$ to determine $\alpha$ and $\beta$.

First, we impose conditions at the axis $\rho=0$.
We assume that $S$ and $\Sigma_{,\rho}$ vanish at the axis $\rho=0$,
\begin{equation}\label{SS}
S=0,\qquad \Sigma_{,\rho}=0,
\end{equation}
second, we can always rescale the metric \eqref{trmetric4} to get
\begin{equation}\label{SS1}
\Sigma|_{\rho=0}=1,
\end{equation}
third, we want no conical  singularities there, therefore we assume
\begin{equation}\label{SS2}
\Sigma_{,\rho\rho}|_{\rho=0}=\gamma,
\end{equation}
which we can be justified by computation of the ratio of the circumference $o$ divided by $2\pi$ times radius
in limit $\rho\rightarrow 0$,
\begin{equation}\label{SS3}
\frac{o}{2\pi r}=\frac{2\pi\frac{S}{\Sigma}}{2\pi\int\Sigma\trgrad \rho}=\frac{1}{\Sigma}\left(\frac{S}{\Sigma}\right)_{,\rho}=\frac{1}{\gamma}\frac{\Sigma_{,\rho\rho}\Sigma-(\Sigma_{,\rho})^2}{\Sigma^3}=1.
\end{equation}
Applying the conditions \eqref{SS}, \eqref{SS1} and \eqref{SS2}, we obtain
\begin{equation}\begin{aligned}
-&\frac{1}{3}\Lambda+\alpha+\beta-\vrho=0,\\
-&\frac{2}{3}\Lambda+\alpha+\frac12\beta=\gamma.
\end{aligned}\end{equation}

We can then determine the constants $\alpha$ and $\beta$ explicitly
in terms of the cosmological constant $\Lambda$, the density of electromagnetic
field $\vrho$ and the parameter $\gamma$,
\begin{equation}\begin{aligned}\label{glab}
\alpha&=\Lambda-\vrho+2\gamma,\\
\beta&=-\frac{2}{3}\Lambda+2\vrho-2\gamma.
\end{aligned}\end{equation}
We can conveniently rewrite \eqref{s3q},
\begin{equation}\begin{aligned}\label{s33}
&(\gamma S)^2=(\Sigma_{,\rho})^2=\\
&=\left[-\frac{1}{3}\frac{\Lambda}{\gamma^2}(\Sigma^2-2)\Sigma-\frac{\vrho}{\gamma^2}(\Sigma-1)+\frac{2}{\gamma}\Sigma\right](\Sigma-1).
\end{aligned}\end{equation}

Now we know explicitly the constants in the derivative of $\Sigma$ and we can
investigate the interpretation of the generalized Melvin spacetime.
It is convenient to introduce  new coordinate $x$ as
\begin{equation}\label{x}
\Sigma=1+\gamma x,
\end{equation}
then we can write that
\begin{equation}\label{sx}
S=x_{,\rho},\quad\Sigma_{,\rho}=\gamma x_{,\rho}\;.
\end{equation}

The transversal metric ${\bf q}$ \eqref{trmetric4} then
can be rewritten as
\begin{equation}\label{q1}
{\bf q}=\left(\frac{\Sigma}{S}\right)^2\trgrad x^2+\left(\frac{S}{\Sigma}\right)^2\trgrad \phi^2=\frac{1}{G}\trgrad x^2+G\trgrad \phi^2,
\end{equation}
where we can express the new function $G$ as
\begin{equation}\label{q2}
G=\left(\frac{S}{\Sigma}\right)^2=-\frac{1}{3}\frac{\Lambda}{\gamma^2}\Sigma^2+\frac{\alpha}{\gamma^2}+\frac{\beta}{\gamma^2}\frac{1}{\Sigma}-\frac{\vrho}{\gamma^2}\frac{1}{\Sigma^2},
\end{equation}
and
\begin{equation}\label{q3}
S^2=\mp \ell^2\gamma^2 x^4 \mp \ell^2\gamma x^3 +(\mp 3\ell^2-\vrho+2\gamma)x^2+2x,
\end{equation}
where we denoted
\begin{equation}\label{q4}
\mp \ell^2=\frac{\Lambda}{3},\quad \pm=\text{sign}\,\Lambda.
\end{equation}

Before we will discuss the possible geometries given by the transversal metric ${\bf q}$
\eqref{trmetric4} and interpret them accordingly
we introduce important characteristics for the generalized Melvin spacetime.

The radial radius is then defined as
\begin{equation}\label{q5}
r=\int_{0}^{x}\frac{1}{\sqrt{G}}\trgrad x,
\end{equation}
the circumference radius is simply given
by the function $G$,
\begin{equation}\label{q6}
R=\sqrt{G}.
\end{equation}
Interestingly, the ratio of the radia is then determined
by the derivative of $G$,
\begin{equation}\label{q7}
\frac{\trgrad R}{\trgrad r}=\sqrt{G}\frac{\trgrad \sqrt{G}}{\trgrad x}=\frac12 G_{,x}.
\end{equation}

The scalar curvature of ${\bf q}$ which can be calculated also from \eqref{R1},
is given by
\begin{equation}\begin{aligned}\label{RR}
\mathcal{R}=-G_{,xx}&=-\frac{2}{\Sigma^4}\left[3\Sigma_{,\rho}+\frac{2}{3}\Lambda\Sigma^4-3\alpha\Sigma^2-2\beta\Sigma\right].
\end{aligned}\end{equation}

The geometries of the transversal spacetime ${\bf q}$ can be illustrated
by investigating the function $G$ and its roots when we consider different
values of $\Lambda$,\, $\vrho$ and of the parameter $\gamma$.

\begin{figure}[htp]
\begin{center}
\includegraphics{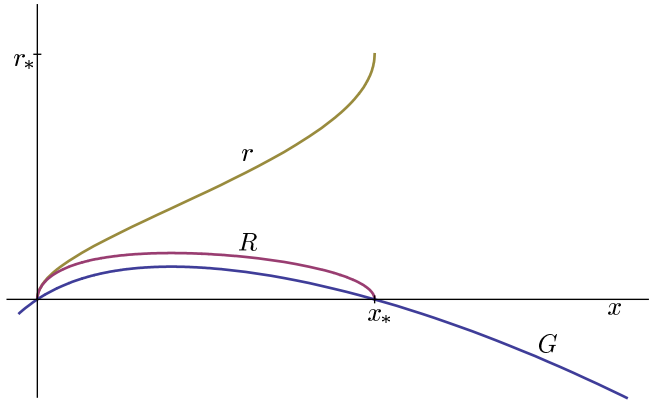}
\end{center}
\caption{\label{fig:graf1}%
The case when $\Lambda>0$ which represents closed spacetime. The function $G$ is visualized for
any value of $\vrho$ and $\gamma$. The coordinate $\rho$ ranges $\rho\in(0,\rho_{*})$ where the $\rho_{*}$ is the first root of $G$ where the spacetime closes.}
\end{figure}

First, we consider positive cosmological constant $\Lambda>0$ for any $\vrho$ and $\gamma$ we obtain {\it closed space} where $\rho\in(0,\rho_{*})$ and $\rho_{*}$ represents the first positive root of $G$ where in fact the spacetime closes itself. The other characteristics are:
 the radial radius tends to a finite value $r\rightarrow r_{*}$ at the $\rho_{*}$  and the circumference radius vanishes $R\rightarrow 0$ when $\rho\rightarrow \rho_{*}$ .
This special case is visualized in the graph \ref{fig:graf1}.

For the vanishing cosmological constant $\Lambda=0$ we obtain three possible
spacetimes according to the values of $\vrho$ and $\gamma$.

\begin{figure}[htp]
\begin{center}
\includegraphics{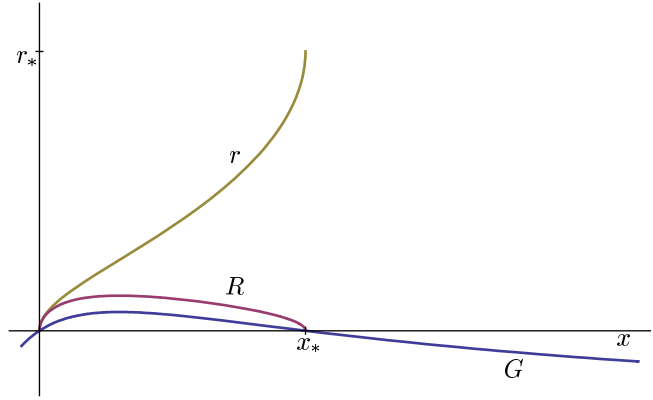}
\end{center}
\caption{\label{fig:graf2}%
The case when $\Lambda=0$ and $\vrho>2\gamma$ represents the closed spacetime. The function $G$ is visualized for $\vrho>2\gamma$ and the coordinate $\rho$ ranges $\rho\in(0,\rho_{*})$ where the $\rho_{*}$ is the root of $G$ where the spacetime closes.}
\end{figure}

When $\vrho>2\gamma$ then we get {\it closed space} where the range of the coordinate $\rho$ goes again
as $\rho\in(0,\rho_{*})$ and $\rho_{*}$ is then the root of $G$ and it is the closing point of the universe.
The radia are then $r\rightarrow r_{*}$ and $R\rightarrow 0$ when $\rho\rightarrow \rho_{*}$, see the graph \ref{fig:graf2}.

\begin{figure}[htp]
\begin{center}
\includegraphics{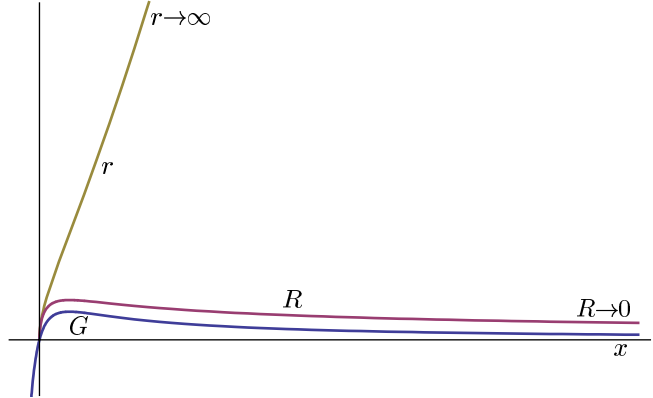}
\end{center}
\caption{\label{fig:graf3}%
The case when $\Lambda=0$ and $\vrho=2\gamma$ then  represents the closed spacetime with an infinite peak. The function $G$ is visualized for $\vrho=2\gamma$ and the coordinate $\rho$ ranges $\rho\rightarrow\infty$.}
\end{figure}

When $\vrho=2\gamma$ then we obtain {\it closed space with and infinite peak} for $\rho\rightarrow \infty$.
Therefore, when $\rho\rightarrow \infty$ the radial radius tends to infinity $r\rightarrow \infty$ and the circumference radius goes to zero $R\rightarrow 0$, see the graph \ref{fig:graf3}.
This case represents the pure Melvin spacetime \cite{Bonnor:1954:PRS:,Melvin:1965:PHYSR:} which we discussed in \cite{Kadlecova:2010:PHYSR4:}.

\begin{figure}[htp]
\begin{center}
\includegraphics{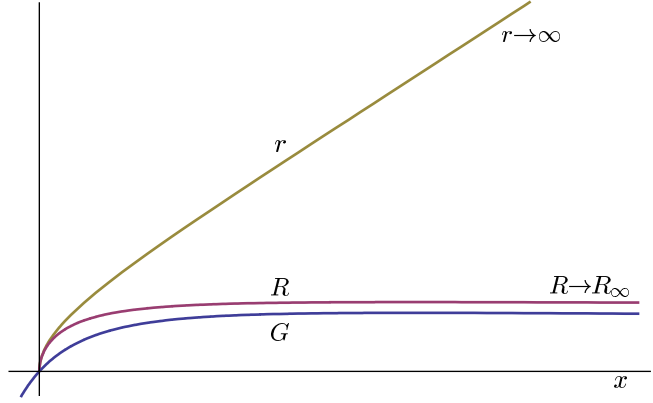}
\end{center}
\caption{\label{fig:graf4}%
The case when $\Lambda=0$ and $\vrho<2\gamma$ then  represents the open spacetime. The function $G$ is visualized for $\vrho<2\gamma$ and the coordinate $\rho$ ranges $\rho\rightarrow\infty$.}
\end{figure}

When $\vrho<2\gamma$ then we obtain {\it an open space} for $\rho\in(0,\infty)$.
When $\rho\rightarrow \infty$, the radial radius tends to infinity $r\rightarrow \infty$; however, the circumference radius goes to a finite value, $R\rightarrow R_{\infty}$, see the graph \ref{fig:graf4}.

\begin{figure}[htp]
\begin{center}
\includegraphics{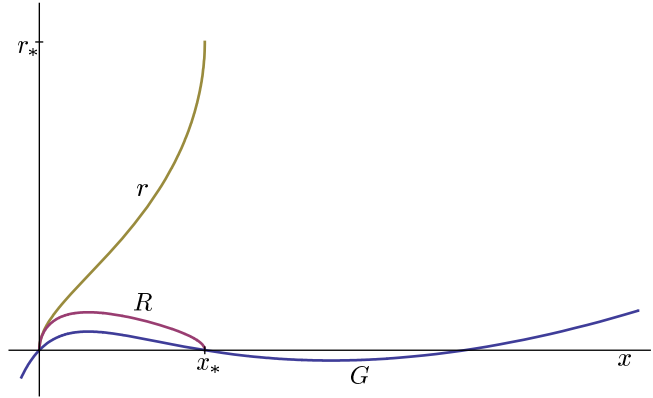}
\end{center}
\caption{\label{fig:graf5}%
The case when $\Lambda<0$ and $\gamma<\gamma_{\mathrm cr}$ represents the closed spacetime. The coordinate $\rho$ ranges $\rho\in(0,\rho_{*})$ where the $\rho_{*}$ is the root of $G$ where the spacetime closes.}
\end{figure}

When we consider the negative cosmological constant $\Lambda<0$ we obtain three possible
spacetimes according to the values of $\gamma$.
For ${\gamma}$ smaller than certain critical value ${\gamma_{\mathrm cr}}$ (which depends on ${\Lambda}$ and ${\vrho}$), we get {\it closed space} where the range of the coordinate $\rho$ goes again as $\rho\in(0,\rho_{*})$ and $\rho_{*}$ is then the root of $G$ and the closing point of the universe.
The radia are then $r\rightarrow r_{*}$ and $R\rightarrow 0$ when $\rho\rightarrow \rho_{*}$, see the graph \ref{fig:graf5}.

\begin{figure}[htp]
\begin{center}
\includegraphics{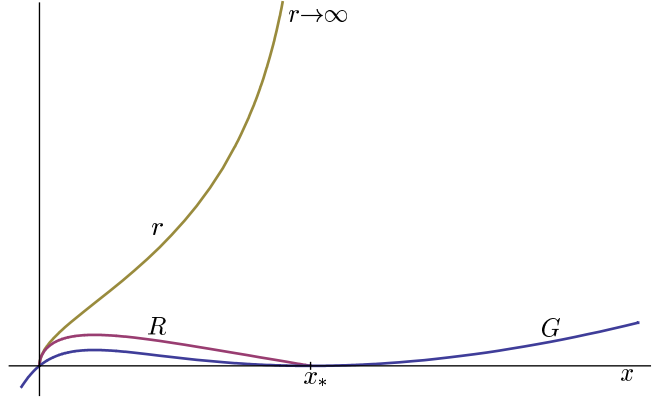}
\end{center}
\caption{\label{fig:graf6}%
The case when $\Lambda<0$ and $\gamma=\gamma_{\mathrm cr}$ represents the asymptotically closed spacetime. The coordinate $\rho$ ranges $\rho\in(0,\rho_{*})$ where the $\rho_{*}$ is the root of $G$. The radial distance tends to infinity and the circumference shrinks to zero.}
\end{figure}

When $\gamma=\gamma_{\mathrm cr}$, we obtain {\it closed space with and infinite peak} where the range of the coordinate $\rho$ goes
as $\rho\in(0,\rho_{*})$ and $\rho_{*}$ is the root of $G$.
The radia are then $r\rightarrow\infty$ and $R\rightarrow 0$ when $\rho\rightarrow \rho_{*}$, see the graph \ref{fig:graf6}.

\begin{figure}[htp]
\begin{center}
\includegraphics{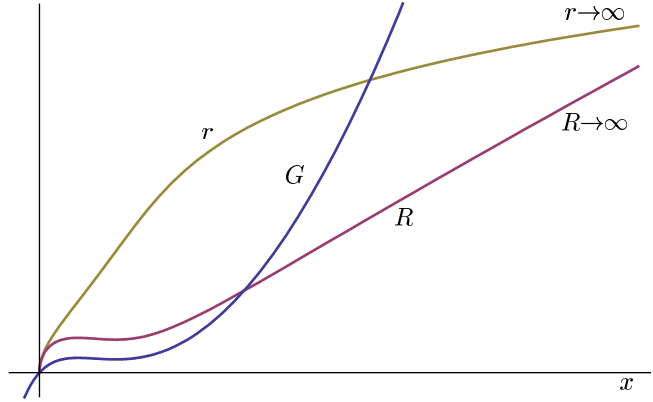}
\end{center}
\caption{\label{fig:graf7}%
The case when $\Lambda<0$ and $\gamma>\gamma_{\mathrm cr}$ represents the open spacetime. The coordinate ${\rho}$ takes positive real values.
For $\rho\rightarrow \infty$,\,$r\rightarrow \infty$, and $R\rightarrow R_{\infty}$, see the graph \ref{fig:graf7}.}
\end{figure}

When $\gamma>\gamma_{\mathrm cr}$, we obtain {\it open space} for $\rho\in(0,\infty)$.
For $\rho\rightarrow \infty$,\,$r\rightarrow \infty$, and $R\rightarrow R_{\infty}$, see the graph \ref{fig:graf7}.

\begin{table}
\caption{\label{table16} Possible geometries of the transversal spacetime ${\bf q}$. Here $\Lambda$ is a cosmological constant, $\vrho$ is energy density of the electromagnetic field and $\gamma$ is the parameter of `Melviniztion' of the spacetime. Critical value ${\gamma_{\mathrm cr}(\Lambda,\vrho)}$} is determined by the condition that the function ${G}$ has degenerated root at ${\rho_*}$.
\begin{ruledtabular}
\begin{tabular}{cccccc}
 $\Lambda$ &  $\vrho$,$\gamma$ & \text{transversal spacetime}  & $\rho$ & $r|_{\rho\rightarrow\rho_{*}}$ & $R|_{\rho\rightarrow\rho_{*}}$ \\
\hline
 $\Lambda>0$ & \text{any} & closed space       & $(0,\rho_{*})$ & $r_{*}$& $0$ \\
 \hline
 & $\gamma<\vrho/2$   & closed space         & $(0,\rho_{*})$ & $r_{*}$& $0$ \\
 $\Lambda=0$ & $\gamma=\vrho/2$    & Melvin universe  & $\mathbb{R}^{+}$ & ${\infty}$ & $0$ \\
 & $\gamma>\vrho/2$   & open space  & $\mathbb{R}^{+}$ & $\infty$ & $R_{\infty}$\\
 \hline
 & $\gamma<\gamma_{\mathrm cr}$   & closed space  & $(0,\rho_{*})$ & $r_{*}$ & $0$  \\
$\Lambda<0$ &  $\gamma=\gamma_{\mathrm cr}$    & closed with $\infty$ peak & $(0,\rho_{*})$ & $\infty$ & $0$ \\
  &  $\gamma>\gamma_{\mathrm cr}$    & open space & $\mathbb{R}^{+}$ & $\infty$ & $\infty$\\
\end{tabular}
\end{ruledtabular}
\end{table}

We have summarized our resulting geometries arising from the generalized Melvin universe in a Table \ref{table16}.

To conclude this section, we have investigated the transversal spacetime of the generalized Melvin universe.
We have identified the constants $\alpha$ and $\beta$, interpreted them in terms of the cosmological constant $\Lambda$, $\vrho$ and $\gamma$.
After suitable parametrization of the transversal spacetime we have discussed all possible cases of universes which are contained
in the generalized Melvin universe. The Melvin universe occurs as a special case. We have visualized these cases in graphs and summarized them in the Table \ref{table16}.

The parameter $\gamma$ changes the character of the  influence of the electromagnetic field on the geometry. With larger ${\gamma}$ the influence is stronger and for $\Lambda \leq 0$ it can even change the global structure of the spacetime, what exactly happens for the critical value ${\gamma_{\mathrm cr}}$ (for $\Lambda=0$ $\gamma_{\mathrm cr}=\vrho/2$).

\subsection{The backgrounds for our solutions}
The background spacetimes are defined as a limit when $h=g=0$ and ${\bf a}=0$, then
the metric \eqref{s1w} reduces to
\begin{equation}\label{ss1}
{\bf g}={\bf q}-\Sigma^2\trgrad u\vee\trgrad v +\alpha v^2\Sigma^2\trgrad u^2.
\end{equation}

The metric \eqref{ss1} admits one killing vector
\begin{equation}
\partial_{\phi}
\end{equation}
which correspond to cylindrical symmetry.

Using the adapted null tetrad ${\bf k}=\partial_{v},\,{\bf l}=\Sigma^{-2}(\partial_{u}+\frac12\alpha v^2\partial_{v}),\,{\bf m}=\frac{1}{\sqrt{2}}(\Sigma^{-1}\partial_{\rho}-i\Sigma S^{-1}\partial_{\phi})$, the only non--vanishing components of Weyl and Ricci tensors are,
\begin{equation}\begin{aligned}\label{Melpsii}
\Psi_{2}&=\frac{1}{2\Sigma^4}\left(\beta\Sigma-2\vrho\right),\\
\Phi_{11}&=\frac{1}{2\Sigma^4}\vrho.
\end{aligned}\end{equation}
This demonstrates that the generalized Melvin universe is a non--vacuum solution
of type D, except the points where $\Psi_{2}=0$.

The background metric \eqref{ss1} contains several sub--solutions.
For $\Lambda=0$ and $\vrho=2\gamma$ we obtain the Melvin universe which
serves as a background in \cite{Kadlecova:2010:PHYSR4:} and the
the only non--vanishing Weyl and curvature scalars are
\begin{equation}\begin{aligned}\label{Mel1}
\Phi_{2}&=-\frac{\vrho}{2\Sigma^4}(2-\Sigma)=\frac{1}{2}\frac{\vrho}{\Sigma^4}(-1+\frac{1}{4}\vrho\rho^2),\\
\Psi_{11}&=\frac{1}{2\Sigma^4}\vrho,
\end{aligned}\end{equation}
where we have used the $\Sigma=1+\frac{1}{4}\vrho\rho^2$ which specifies the Melvin
spacetime.
The scalar curvature of the transversal spacetime ${\bf q}$ \eqref{RR} then
becomes
\begin{equation}
\mathcal{R}=0,
\end{equation}
which agrees with \cite{Kadlecova:2010:PHYSR4:}.

For $\Sigma=1$ we get the direct product background
spacetimes, the metric \eqref{ss1} reduces to
\begin{equation}\label{ss2}
{\bf g}={\bf q}-\trgrad u\vee\trgrad v +\alpha v^2\trgrad u^2,
\end{equation}
the only non--vanishing Weyl and curvature scalars then are
\begin{equation}\begin{aligned}\label{Mel2}
\Psi_{2}=\frac{1}{2}\left(\beta-2\vrho\right)=-\frac{\Lambda}{3},\,\Phi_{11}=\frac{1}{2}\vrho.
\end{aligned}\end{equation}
The scalar curvature of the transversal spacetime ${\bf q}$ \eqref{RR} then
becomes
\begin{equation}
\mathcal{R}=2(\Lambda+\vrho),
\end{equation}
which agrees with \cite{Kadlecova:2009:PHYSR4:}.

\begin{table}
\caption{\label{table2}Some of possible background spacetimes in the case $\gamma=0$ which  represents the direct product of two 2-spaces of constant curvature. The parameter
$\Lambda_{+}=\Lambda+\vrho$ gives the geometry of the wave front and $\Lambda_{-}=\Lambda-\vrho$  determines the conformal structure of the background.}
\begin{ruledtabular}
\begin{tabular}{cccccccc}
 $\Lambda_{+}$ &  $\Lambda_{-}$ & \text{geometry} & \text{background}  & $\Lambda$ & $\vrho$\\
\hline
 0 & 0                  & ${E^{2}\times M_{2}}$  & Minkowski        & $=0$ & $=0$ \\
 $\Lambda$ & $\Lambda$  & ${S^{2}\times dS_{2}}$  & Nariai          & $>0$ & $=0$ \\
 $\Lambda$ & $\Lambda$  & ${H^{2}\times AdS_{2}}$  & anti-Nariai   & $<0$ & $=0$ \\
 $\vrho$ & $-\vrho$ & ${S^{2}\times AdS_{2}}$  & Bertotti--Robinson   & $=0$ & $>0$ \\
 $2\Lambda$ & 0   & ${S^{2}\times M_{2}}$  & Pleba\'{n}ski--Hacyan   & $>0$ & $=\Lambda$ \\
 0 &  $2\Lambda$  & ${E^{2}\times AdS_{2}}$  & Pleba\'{n}ski--Hacyan & $<0$ & $=|\Lambda|$
\end{tabular}
\end{ruledtabular}
\end{table}

We can visualize the gyraton solutions on the most important
direct product background spacetimes which we have summarize them in the table \ref{table2} \cite{Kadlecova:2009:PHYSR4:}. The parameter $\Lambda_{+}=\Lambda+\vrho$
defines the geometry of the wave front and  $\Lambda_{-}=\Lambda-\vrho$  determines the conformal structure of the background.

The gyratons can be visualized in three dimensions as propagating curved wave fronts with matter which contains an intrinsic spin and with the direction of motion. The conformal background can not be visualized therefore we have only mentioned them for completeness.
The shape of the wave front does not change in time propagation and its boundary stays the same because the solutions are non--expanding.
The gyratons can be summarized according to the geometry of the wave front as can be observed from the table \ref{table2}.

\begin{figure}[htp]
\begin{center}
\includegraphics[scale=0.5]{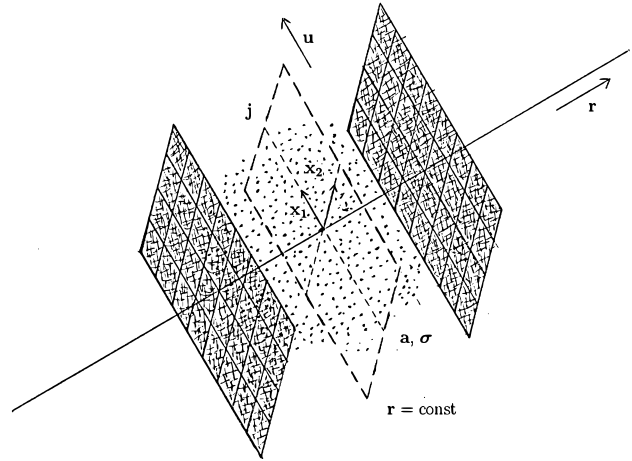}
\end{center}
\caption{\label{fig:gyr1}%
The gyratons with flat $\mathbb{E}^2$ geometry of the wave front. The gyraton on Minkowski spacetime propagates also in the Minkowski background geometry and gyraton on Plebanski--Hacyan $(\Lambda<0)$ spacetime propagates in conformally anti--de Sitter $AdS_{2}$ background.
The 1--forms ${\bf a},\,{\vs}$ and the sources ${\bf j}$ are defined on the dashed transversal spacetime.}
\end{figure}

\begin{figure}
\begin{center}
\includegraphics[scale=0.5]{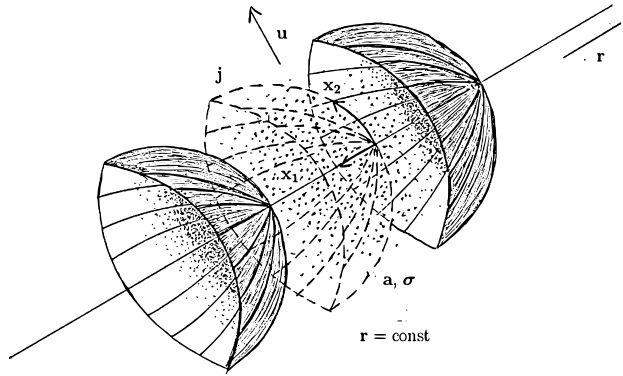}
\end{center}
\caption{\label{fig:gyr2}%
The gyratons with 2--sphere $\mathbb{S}^2$ geometry of the wave front. The gyraton on Narai spacetime propagates in conformally de--Sitter background geometry, gyraton on Bertotti--Robinson spacetime propagates in conformally anti--de Sitter $AdS_{2}$ background and the gyraton
on Plebanski--Hacyan $(\Lambda>0)$ propagates in the Minkowski flat background.
The 1--forms ${\bf a},\,{\vs}$ and the sources ${\bf j}$ are defined on the dashed transversal spacetime.}
\end{figure}

The gyratons with flat $\mathbb{E}^2$ geometry of the wave front are visualized in graph \ref{fig:gyr1}.
The gyratons with 2--sphere $\mathbb{S}^2$ geometry of the wave front are visualized in graph \ref{fig:gyr2}.
The gyraton with 2--hyperboloid $\mathbb{H}^2$ geometry of the wave front occurs only in the case of  the gyraton on anti--Narai spacetime which propagates  in conformally de--Sitter background geometry. The hyperbolic geometry is not easy to visualize therefore we do not present it here.

To summarize the background metric \eqref{ss1} generalizes the metric for the pure  Melvin universe and
the direct product spacetimes into one background metric and combines their
properties. We have visualized some important gyratons on direct product spacetimes.

\section{The scalar polynomial invariants}\label{sc:Invariants}
In this section we will investigate the scalar polynomial invariants which are constructed only from curvature and its
covariant derivatives. The scalar invariants are important characteristics of gyraton spacetimes.
The gyratons in the Minkowski spacetime \cite{Fro-Is-Zel:2005:PHYSR4:} have vanishing invariants (VSI) \cite{Prav-Prav:2002:CLAQG:}, the gyratons in the AdS \cite{Fro-Zel:2005:PHYSR4:} and direct
product spacetimes \cite{Kadlecova:2009:PHYSR4:} have all invariants constant (CSI) \cite{Coley-Her-Pel:2006:CLAQG}.
The invariants are independent of all metric functions $a_{i}$ which characterize the gyraton, and have the same
values as the corresponding invariants of the background spacetime.
We have shown that similar property is valid also for the gyraton on Melvin spacetime \cite{Kadlecova:2010:PHYSR4:}, but the invariants
are functions of the coordinate $\rho$ and depend on the constant density $\vrho$.

In these cases, the invariants are independent of all metric functions
which characterize the gyraton, and have the same
values as the corresponding invariants of the background spacetime.
We observed that similar property is valid also for the gyraton on
Melvin spacetime and it is valid also for its generalization with $\Lambda$, however, in this case the invariants are generally \emph{non-constant},
namely, they depend on the coordinate $\rho$. This property is a consequence of general theorem
holding for the relevant subclass of Kundt solution, see Theorem II.7 in \cite{ColeyEtal:2010}.

Nevertheless, we will present here the $0$th order invariants explicitly, according to \cite{ColeyEtal:2010} the
property about invariants is valid also for invariants of higher orders.

In next text we investigate the invariants for the gyratons on generalized Melvin spacetime, they generalize the invariants for the gyratons on Melvin spacetime and the direct product spacetime.

We present the set of invariants {\tt CMinvars} in package {\it GRtensor} in {\it Maple}, the other invariants defined and presented in \cite{Kadlecova:2010:PHYSR4:} are too complicated therefore we don't present them here.
The scalar curvatures for the background spacetimes \eqref{ss1} and the full gyraton metric \eqref{s1w} are equal and constant,
\begin{equation}\label{RbgRc}
\mathcal{R}_{bg}=\mathcal{R}_{c}=4\Lambda,
\end{equation}
where the curvatures are identical to those in direct product case \cite{Kadlecova:2009:PHYSR4:}.

The following relations occur in the invariants and they reduce to simple expressions after we have used \eqref{r10}, \eqref{r8} and \eqref{r7},
\begin{align}\label{zj}
2\alpha\Sigma^2&-8\Sigma_{,\rho\rho}\Sigma+4(\Sigma_{,\rho})^2+2\Sigma^2\frac{\Sigma_{,\rho\rho\rho}}{\Sigma_{,\rho}}=-4\vrho,\notag\\
-2\alpha\Sigma^2&+12(\Sigma_{,\rho})^2-12\Sigma_{,\rho\rho}\Sigma+2\Sigma^2\frac{\Sigma_{,\rho\rho\rho}}{\Sigma_{,\rho}}=6(\beta\Sigma-2\vrho),
\end{align}
thanks to them we have simplified the expressions for invariants to
\begin{equation}\begin{aligned}
R&=R_{c};\,R_{2}={W1I}={W2I}=0\;,\\
R_{1}&=\frac{1}{\Sigma^8}\vrho^2,\,{W1R}=\frac{3}{2\Sigma^{8}}(\beta\Sigma-2\vrho)^2\;,\\
R_{3}&=\frac{1}{2^{2}\Sigma^{16}}\vrho^4,\,{W2R}=\frac{3}{2^2\Sigma^{12}}(\beta\Sigma-2\vrho)^3\;.
\end{aligned}\end{equation}
The other invariants are
\begin{equation}\begin{aligned}
{M1I}&={M2I}={M4}={M5I}=0\,,\\
{M1R}&=\frac{1}{\Sigma^{12}}(\beta\Sigma-2\vrho)\vrho^2\\
{M2R}&={M3}=\frac{1}{\Sigma^{16}}(\beta\Sigma-2\vrho)^2\vrho^2\;,\\
{M5R}&=\frac{1}{\Sigma^{20}}(\beta\Sigma-2\vrho)^3 \vrho^2\;,
\end{aligned}\end{equation}
We observe that the invariants for the full gyraton metric and the background metric are again identical.

For $\Lambda=0$  and $\vrho=2\gamma$ we will get the invariants for the gyratons on Melvin universe \cite{Kadlecova:2010:PHYSR4:}
when we assume the particular form of $\Sigma=1+\frac{1}{4}\vrho\rho^2$ which identifies the Melvin solution and
the following relation reduces to
\begin{equation}\label{MNMelvin}
\beta\Sigma-2\vrho=\vrho(\Sigma-2)=\frac{1}{4}\vrho(\vrho\rho^2-4).
\end{equation}

For $\Sigma=1\,(\gamma=0)$ we get the constant invariants for the gyratons on direct product spacetime, some
of them are zero and the non--trivial depend only on the cosmological constant $\Lambda$ and the density $\vrho$.

The relation  \eqref{MNMelvin} then becomes
\begin{equation}\label{MNDirPr}
\beta\Sigma-2\vrho=-\frac{2}{3}\Lambda,
\end{equation}
and the invariants are then
\begin{equation}\begin{aligned}\label{i1}
R&=R_{c};\,R_{2}={W1I}={W2I}=0\;,\\
R_{1}&=\vrho^2,\,{W1R}=\frac{2}{3}\Lambda^2\;,\\
R_{3}&=\frac{1}{2^{2}}\vrho^4,\,{W2R}=-\frac{2}{3^2}\Lambda^3\;,
\end{aligned}\end{equation}
the other invariants are
\begin{equation}\begin{aligned}\label{i2}
{M1I}&={M2I}={M4}={M5I}=0\,,\\
{M1R}&=-\frac{2}{3}\Lambda \vrho^2\\
{M2R}&={M3}=\frac{2^2}{3^2}\Lambda^2\vrho^2\;,\\
{M5R}&=-\frac{2^3}{3^3}\Lambda^3\vrho^2\;.
\end{aligned}\end{equation}
We observe that the invariants for the gyratons on direct spacetimes \cite{Kadlecova:2009:PHYSR4:} are truly constant.
We can rewrite the invariants in terms of constants $\Lambda_{+}$ and $\Lambda_{-}$ which we have used in \cite{Kadlecova:2009:PHYSR4:},
see  Appendix \ref{apx:inv}.

To conclude this section we have shown that the invariants for the background spacetimes and the
full gyraton metric are the same which is the common property of known gyraton solutions.
In general, the scalar polynomial curvature invariants are functions of the coordinate $\rho$ and depend
on the constants $\Lambda$ and the electromagnetic density $\vrho$ (direct product density) and $\gamma$ (Melvin parameter) .
\section{Conclusion}\label{sc:conclusionE}
Our work generalizes the studies of the gyraton on the Melvin universe \cite{Kadlecova:2010:PHYSR4:}.
Namely we have generalized the transversal background metric for the pure Melvin universe where instead of
the coordinate $\rho$ we have assumed general function $S$ dependent only on the coordinate $\rho$. This
change enabled us to find new solutions with possible non--zero cosmological constant. This is not
allowed for the pure Melvin background spacetime.
We were able to derive relation between metric functions $\Sigma$ and $S$ from the source free  Einstein--Maxwell equations.
The derivative of the function $\Sigma_{,\rho}$ is then polynomial in the function $\Sigma$ itself and contains
four parameters. We have showed that these parameters can be expressed using constants $\Lambda$, $\vrho$ and $\gamma$.

The Einstein--Maxwell equations reduce again to the set of linear equations on the 2--dimensional transverse spacetime which
has non--trivial geometry given by the generalized Melvin spacetime \eqref{trmetric4}.
Fortunately, these equations do decouple and they can be solved least in principle for any distribution of the matter sources.

In detail, we have studied the transversal geometries of  generalized Melvin spacetime \eqref{trmetric4}.
We have discussed the various possible values of constants $\Lambda$, $\vrho$ and $\gamma$. It occurs that for $\Lambda>0$ the transversal geometry represents only one type of space, the case $\Lambda=0$ includes three different spaces, one of them corresponds to the Melvin spacetime as a special case. The case $\Lambda<0$ also describes three types of spaces. We have visualized them in several graphs in Section \ref{sc:interprett}
and summarized them in the Table \ref{table16}.
Thanks to this discussion we were able to interpret the parameter $\gamma$ as the parameter which makes the electromagnetic field of the direct product spacetimes stronger.

We have investigated the polynomial scalar invariants. In this generalized case, the invariants are again
not constant and they are functions of the metric function $\Sigma$ and the full gyratonic
metric has the same invariants as the background metric.

It would be interesting to investigate even more general ansatz for the transversal metric.
\addcontentsline{toc} {section}{Appendix}
\begin{subappendices}
\section{The notation on transversal space ${\bf q}$}\label{AppO}

The inverse metric to \eqref{s1w} is the following
\begin{equation}\label{a00}
{\bf g}^{-1}=(2\frac{H}{\Sigma^2}+a^2)\,\partial_{v}\partial_{v}-\frac{1}{\Sigma^2}\partial_{u}\vee\partial_{v}+{\bf q}^{-1}+\partial_{v}\vee \overline{\bf a},
\end{equation}
where the inverse transversal space ${\bf q}^{-1}$ reads,
\begin{equation}
{\bf q}^{-1}=\frac{1}{\Sigma^2}\partial_{\rho}\partial_{\rho}+\frac{\Sigma^2}{S^2}\partial_{\phi}\partial_{\phi}
\end{equation}
and
\begin{equation}
\overline{\bf a}={\bf q}^{-1}{\cdot}{\bf a}.
\end{equation}

With the transverse metric \eqref{trmetric4} we may associate the Levi-Civita tensor ${\boldsymbol{\epsilon}=S\,\trgrad\rho\wedge\trgrad\phi}$.  We will denote  raising the indices as "$\sharp$" and lowering as "$\flat$" using ${\bf q}$, which differs from lowering indices using ${\bf g}$ thanks to non--vanishing term $g_{u{\bf x}}$, and we use a shorthand for a square of the norm of a 1-form ${\bf a}$ as
\begin{equation}
{\bf a}^2\equiv {\bf a}{\cdot}{\bf q}^{-1}{\cdot}{\bf a}=a_\rho^2\frac{1}{\Sigma^2}+a_\phi^2\frac{\Sigma^2}{S^2}.
\end{equation}

In two dimensions, the Hodge duals of 0,1 and 2-forms ${\varphi}$, ${\bf a}$, and ${\bf f}$ read
\begin{equation}
(*\varphi) = \varphi\, {\boldsymbol{\epsilon}}\;,\;\;
{*{\bf a}} = \overline{\bf a}{\cdot}\boldsymbol{\epsilon},\;\;
{*{\bf f}} = \frac12 {\bf f}{\cdot}{^{\sharp}\boldsymbol{\epsilon}} =\frac{1}{S}f_{\rho\phi} \;.
\end{equation}

For convenience, we also introduce an explicit notation for 2-dimensional divergence and rotation of a transverse 1-form ${\bf a}$,
\begin{equation}\begin{aligned}
  &\trdiv{\bf a} \equiv \nabla{\cdot}{\bf a}= \frac{1}{\Sigma^2}a_{\rho,\rho}+\frac{\Sigma^2}{S^2}a_{\phi,\phi}+\frac{S_{,\rho}}{S\Sigma^2}a_{\rho}-\frac{\Sigma^2_{,\rho}}{\Sigma^4}a_{\rho}\;,\\
&\rot{\bf a} \equiv \trgrad {\bf a}{\cdot}{^{\sharp}\epsilon}  = \frac{1}{S}(a_{\phi,\rho}-a_{\rho,\phi})\;.
\end{aligned}\end{equation}
For 2-form ${\bf f}$, we get
\begin{equation}
  \trdiv{\bf f} \equiv \nabla{\cdot}{\bf f} = \frac{1}{\Sigma^2}(f_{\phi\rho,\rho}-\frac{S_{,\rho}}{S}f_{\phi\rho})\trgrad \phi+\frac{\Sigma^2}{S^2}f_{\rho\phi,\phi}\trgrad \rho\;.
\end{equation}
and $\rot f=0$. We can generalize action of divergence and rotation also on a scalar function $f$ as $\trdiv f=0$ and $\rot f=-*\trgrad f$.
Note that the divergence and rotation are related as ${\trdiv{\bf a} = \rot {*}{\bf a}}$ and
the relation to the transverse exterior derivative is $\trgrad a=*\rot{\bf a}$. Obviously, $\trdiv\trdiv{\bf a}=\trdiv\rot{\bf a}=0$,
and $\rot\grad{\bf a}=0$.

The Laplace operator of the function ${\psi}$ reads,
\begin{equation}\label{lapldeff}
\laplace\psi  = \frac{1}{\Sigma^2}\psi_{,\rho\rho}+\frac{\Sigma^2}{S^2}\psi_{,\phi\phi}+\frac{S_{,\rho}}{S\Sigma^2}\psi_{,\rho}-\frac{\Sigma^2_{,\rho}}{\Sigma^4}\psi_{,\rho}\;,
\end{equation}
and for a transverse 1--form ${\boldsymbol{\eta}}$ it is defined as $\laplace\boldsymbol{\eta}\equiv\trgrad\trdiv{\boldsymbol{\eta}}-\rot\rot{\boldsymbol{\eta}}$.

Another useful expressions are
\begin{align}
\trdiv (\Sigma^2 {\bf a})&=\Sigma^2\,\trdiv{\bf a} + \frac{\Sigma^2_{,\rho}}{\Sigma^2}\,a_{\rho},\label{div}\\
\rot (\Sigma^2 {\bf a})&=\Sigma^2\,\rot{\bf a} + \frac{\Sigma^2_{,\rho}}{S}\,a_{\phi},\label{rot}\\
\trdiv (\Sigma^4 {f})&=\Sigma^4\,\trdiv f + 2(\Sigma^2)_{,\rho}f_{\phi\rho}, \label{div4}
\end{align}
which are used in simplifying the Einstein equations.

\section{The Einstein equations}\label{apx:AppAA}

Here we present  the Einstein tensor of the metric \eqref{s1w} and the electromagnetic stress-energy tensor corresponding to the field \eqref{EMFa}.

The non--vanishing components of the stress--energy tensor ${\bf T}^{EM}$ are
\begin{align}
\varkappa T^\EM_{uv}&=\frac{\vrho}{\Sigma^2},\notag\\
\varkappa T^\EM_{uu}&=2\vrho\left(\frac{H}{\Sigma^2}+(\vs-{\bf a})^2\right)\;,\notag\\
\varkappa T^\EM_{u{\bf x}}&=\frac{\vrho}{\Sigma^2}({\bf a}-2\vs),\notag\\
\varkappa T^\EM_{\rho\rho}&=\frac{\vrho}{\Sigma^2}=\frac{\vrho}{\Sigma^4}g_{\rho\rho}\;,\label{EMTT}\\
\varkappa T^\EM_{\phi\phi}&=\frac{\vrho S^2}{\Sigma^6}=\frac{\vrho}{\Sigma^4}g_{\phi\phi}\;,\notag
\end{align}
where the density ${\vrho}$ was defined in~\eqref{rhodeff}.

The non--vanishing components of the stress--energy tensor \eqref{EMFa} of the gyratonic matter are
\begin{equation}\begin{aligned}\label{uu}
\varkappa T_{uu}^{\gyr}&=j_{u}=v\,\trdiv(\Sigma^2{\bf j})+\iota,\\
&\varkappa T_{u{\bf x}}^{\gyr}={\bf j}.
\end{aligned}\end{equation}

The Einstein tensor for the metric \eqref{s1w} reads
\begin{equation}\label{EinsteinTr}
\begin{aligned}
G_{uv}&= \frac{1}{\Sigma^2 S}\bigl[-S(\Sigma_{,\rho})^2+2(\Sigma_{,\rho})(S_{,\rho})\Sigma-\Sigma^2(\partial^2_{,\rho}S)\bigr],\\
G_{uu}&=\frac{1}{2}\Sigma^4 b^2+\Sigma^2(\laplace H-(\Sigma^{-2})_{,\rho}H_{,\rho}) +\Sigma^2(\partial^2_{v}H)a^2\\
&\quad +2\Sigma^2a^i\partial_{v}H_{,i}+(\partial_v H+\partial_{u})\,\trdiv(\Sigma^2 a)\\
&\quad + 2HG_{uv},\\
G_{u\rho}&=\frac{1}{2}\frac{\Sigma^4}{S}b_{,\phi}-a_{\rho}\bigl(G_{uv}-\partial^2_v H\bigr)+\partial_{v}H_{,\rho},\\
G_{u\phi}&=-\frac{1}{2}S(b_{,\rho}+\frac{4\Sigma_{,\rho}}{\Sigma}b)-a_{\phi}\bigl(G_{uv}-\partial^2_v H\bigr)+\partial_{v}H_{,\phi}\;,\\
G_{\rho\rho}&=\frac{1}{S}[S G_{uv}+\partial^2_{\rho}S+S(\partial^2_v H)],\\
G_{\phi\phi}&=\frac{S}{\Sigma^6}\left[\Sigma^2(S G_{uv}+\partial^2_{\rho}S)+\Sigma^2 S(\partial^2_v H)\right.\\
&\left.\quad+2\Sigma[S(\partial^2_{\rho}\Sigma)-\Sigma_{,\rho}S_{,\rho}]\right].
\end{aligned}
\end{equation}

The scalar curvature $\mathcal{R}$ for the transversal spacetime ${\bf q}$ is given by
\begin{equation}\label{R1}
\mathcal{R}=-\frac{2}{S\Sigma^4}[3S(\Sigma_{,\rho})^2-3\Sigma\Sigma_{,\rho}S_{,\rho}-\Sigma S(\partial^2_{\rho}\Sigma)+\Sigma^2(\partial^2_{\rho}S)],
\end{equation}
and the scalar curvature for the background metric \eqref{ss1} is then
\begin{equation}\label{R4}
\mathcal{R}_{bg}=-\frac{2}{S\Sigma^3}[S\Sigma(\partial^2_{v}H)+S(\partial^2_{\rho}\Sigma)-\Sigma_{,\rho}S_{,\rho}+\Sigma(\partial^2_{\rho}S)].
\end{equation}
and the scalar curvature for the whole gyratonic metric \eqref{s1w},
\begin{equation}\label{R3}
\mathcal{R}_{c}=-\frac{2}{S\Sigma^3}[S\Sigma(\partial^2_{v}H)+S(\partial^2_{\rho}\Sigma)-\Sigma_{,\rho}S_{,\rho}+\Sigma(\partial^2_{\rho}S)].
\end{equation}

Here we have used only the metric \eqref{s1w}, without any usage of the field equations.

\section{The gyraton solutions on generalized Melvin background as type II solutions of the Kundt class}\label{apx:prop}
In this section we will review geometrical properties of the gyraton solutions on generalized Melvin spacetime.

The covariant derivative of ${k}$ is non--recurrent and given by
\begin{equation}\label{recurrentkk}
k_{\alpha;\beta}=-\Sigma^{-2}(\partial_{v}H) k_{\alpha}k_{\beta}+\Sigma^{-2}k_{[\alpha}\nabla_{\beta]}\Sigma^2,
\end{equation}
which is identical result as for the gyraton on the Melvin universe.

In the following we calculate the curvature tensor associated to \eqref{s1w} while using the adapted null tetrad
\begin{equation}\label{b-vectorss}
\begin{aligned}
  {\bf k}&=\partial_{v}\;,\\
  {\bf l}&=\frac{1}{\Sigma^2}(\partial_{u}-H\partial_{v})\;,\\
  {\bf m}&=\frac{1}{\sqrt{2}\Sigma}\left\{a_{m}\,\partial_{v}+\partial_{\rho}-i\frac{\Sigma^2}{S}\partial_{\phi}\right\}\;.
\end{aligned}
\end{equation}
Here we have introduced the projection of a transverse  1--form $a$ on vector ${m}$ as
\begin{equation}
a_{m}=m^{i}a_{i}=a_{\rho}-i\frac{\Sigma^2}{S}a_{\phi}=(a+i{*a})_{\rho},
\end{equation}
and we used also analogous notation for components of the transverse gradient of a real function f
\begin{equation}
f_{,m}=m^{i}f_{,i}=f_{,\rho}-i\frac{\Sigma^2}{S}f_{,\phi},\,f_{,\overline{m}}=\overline{f_{,m}}.
\end{equation}

The non--vanishing spin coefficients are
\begin{equation}\begin{aligned}\label{sc100}
&\NP\mu=\frac{i}{2}b,\,\NP\gamma=\frac{1}{4}\frac{1}{\Sigma^2}\bigl(2H_{,v}+i\Sigma^2 b)\;,\\
&\NP\nu=\frac{1}{\sqrt{2}}\frac{1}{\Sigma^3}\left\{(H_{,v}+\partial_{u})a_{\overline{m}}+H_{,\overline{m}}\right\}\;,\\
&\NP\tau=-\frac{1}{\sqrt{2}}\frac{1}{\Sigma^2}\Sigma_{,\rho}\;,\,\NP\pi=+\frac{1}{\sqrt{2}}\frac{1}{\Sigma^2}\Sigma_{,\rho},\\
&\NP\alpha=\frac{1}{2\sqrt{2}}\frac{1}{\Sigma^2S}(2 S \Sigma_{,\rho}-S_{,\rho}\Sigma),\,\NP\beta=\frac{1}{2\sqrt{2}}\frac{1}{\Sigma S}S_{,\rho}\;.
\end{aligned}\end{equation}

We found that the only non--vanishing components are the same as for the background universe \eqref{ss1}, plus four other terms.
The nonvanishing  Ricci scalars are,
\begin{equation}\label{sc33}
\begin{aligned}
\Phi_{12}
&=\frac{1}{4\sqrt{2}}\frac{1}{\Sigma^3}\left[\Sigma^2 i b_{,m}+2ib(\Sigma^2)_{,\rho}+2g_{,m}-2\alpha a_{m}\right],\\
\Phi_{22}
&=\frac{1}{2}\frac{1}{\Sigma^2}\left(\laplace H - (\Sigma^{-2})_{,\rho}H_{,\rho}+\frac12\Sigma^2 b^2
   +2a^{i}g_{,i}\right.\\
&\left.-\alpha a^2+\frac{1}{\Sigma^2}(g+\partial_{u})\trdiv(\Sigma^2 a)  \right)\;.\\
\end{aligned}
\end{equation}
and the nonvanishing Weyl scalars read,
\begin{equation}\begin{aligned}\label{Weyl1}
 \Psi_{3}&=\frac{1}{4\sqrt{2}}\frac{1}{\Sigma^3}\left[\Sigma^2 i b_{,\overline{m}}+ib(\Sigma^2)_{,\rho}+2g_{,\overline{m}}-2\alpha a_{\overline{m}}\right]\;,
 \end{aligned}\end{equation}
 and
\begin{equation}\begin{aligned}\label{Weyl2}
 \Psi_{4}&=\frac{1}{2}\frac{1}{\Sigma^5}\left[2\Sigma\, a_{\overline{m}}\,g_{,\overline{m}}-\alpha\Sigma a^2_{\overline{m}}\right.\\
&\left.+\Sigma(H_{,\rho\rho}-\frac{\Sigma^4}{S^2}H_{,\phi\phi}+2i\frac{\Sigma^2}{S}H_{,\phi\rho})\right.\\
 &\left.+\Sigma(\partial_{v}H+\partial_{u})\left(a_{\rho,\rho}-\frac{\Sigma^4}{S^2}{a}_{\phi,\phi}+i\frac{\Sigma^2}{S}(a_{\rho,\phi}+a_{\phi,\rho})\right)\right.\\
 &\left.+\left(2\Sigma_{,\rho}-\Sigma\frac{S_{,\rho}}{S}\right)\times\right.\\
 &\left.\times\left(H_{,\rho}+\partial_{u}a_{\rho}+(\partial_{v}H)a_{\rho}+2i\frac{\Sigma^2}{S}(H_{,\phi}+\partial_{u}a_{\phi}+(\partial_{v}H) a_{\phi})\right)\right]\;.
\end{aligned}\end{equation}
In particular, there exists a relation between $\Psi_{3}$ and $\Phi_{12}$ as 
\begin{equation}\label{relpsiphii}
\overline{\Phi}_{12}+\Psi_{3}=\frac{1}{4\sqrt{2}}\frac{1}{\Sigma^3}\left[-i(\Sigma^2)_{,\rho}b+4g_{,\overline{m}}-4\alpha a_{\overline{m}}\right].
\end{equation}
This solution is  of Petrov type II and then describes the gravitational wave $\Psi_{4}$ with component $\Psi_{3}$ with  an aligned pure radiation $\Phi_{22}$ and component $\Phi_{12}$ which propagates on the generalized Melvin spacetime of Petrov type $D$.


\section{The invariants for gyraton on direct product spacetimes}\label{apx:inv}
Here, we present the rewritten form of invariants \eqref{i1} and \eqref{i2} where we
used the constants $\Lambda_{+}=\Lambda+\vrho$ and $\Lambda_{-}=\Lambda-\vrho$ to
match our previous work \cite{Kadlecova:2009:PHYSR4:}.

The invariants have the form
\begin{equation}\begin{aligned}
R&=2(\Lambda_{+}+\Lambda_{-});\,R_{2}={W1I}={W2I}=0\;,\\
R_{1}&=\frac{1}{2^2}(\Lambda_{-}-\Lambda_{+})^2,\,{W1R}=\frac{1}{6}(\Lambda_{-}+\Lambda_{+})^2\;,\\
R_{3}&=\frac{1}{2^{6}}(\Lambda_{-}-\Lambda_{+})^4,\,{W2R}=-\frac{1}{6^2}(\Lambda_{-}+\Lambda_{+})^3\;,
\end{aligned}\end{equation}
the other invariants are
\begin{equation}\begin{aligned}
{M1I}&={M2I}={M4}={M5I}=0\,,\\
{M1R}&=-\frac{1}{12}(\Lambda_{-}-\Lambda_{+})^2(\Lambda_{-}+\Lambda_{+})\\
{M2R}&={M3}=\frac{1}{6^2}(\Lambda_{-}-\Lambda_{+})^2(\Lambda_{-}+\Lambda_{+})^2\;,\\
{M5R}&=-\frac{1}{108}(\Lambda_{-}-\Lambda_{+})^2(\Lambda_{-}+\Lambda_{+})^3\;.
\end{aligned}\end{equation}
We obtain again the invariants \eqref{i1} and \eqref{i2} when we use the relations
\begin{equation}
\Lambda_{+}+\Lambda_{-}=2\Lambda,\,\Lambda_{-}-\Lambda_{+}=-2\vrho,
\end{equation}
as a substitution.
\end{subappendices}

\cleardoublepage
\chapter{The gyraton solutions of algebraic type III in the Kundt class of spacetimes}
In this chapter we mainly review the general theory of the Kundt solutions of type III and N
with possible non--vanishing cosmological constant acccording to \cite{GrifDochPod:2004:CLAQG:}.
We present results of the paper \cite{GrifDochPod:2004:CLAQG:} in real notation which we want to use in our work
(the real notation is useful either due to possible generalization in higher dimensions).
We want to look for new gyratonic solutions of type III in the Kundt class which is the work in progress.
We present here only the source equations in Section \ref{equations} for illustration.

Originally, the main aim of the investigation was to find the gyratons on de Sitter spacetime.

\section{The Kundt solutions of type III}\label{KIII}
The general line element of Kundt class of spacetimes \cite{Step:2003:Cam:, GrifPod:2009:Cam:} which admits a geodesics, shear--free, twist--free and non--expanding null congruence can be expressed in the form in real coordinates $(r,u,x,y)$,
\begin{equation}\label{m0}
\trgrad s^2=\frac{1}{P^2}(\trgrad x^2+\trgrad y^2) -2\trgrad u(\trgrad r + H\trgrad u-a_{x}\trgrad x-a_{y}\trgrad y),
\end{equation}
where $P(u,x,y)$, $H(r,u,x,y)$ and $a_{i}(r,u,x,y)$ are real functions,
which are to be determined by the field equations.
The coordinate $u$ labels the null surfaces, $r$ is an affine parameter along the repeated principal null congruence
$k=\partial_{r}$ and $x$ and $y$ are coordinates which span the transverse spatial 2--space.
For any line element of the form \eqref{m0}, the wave surfaces given by constant $u$ (at any $r$) are spacelike with the metric
$\trgrad s^2=\frac{1}{P^2}(\trgrad x^2+\trgrad y^2)$. The Gaussian curvature $K(u,x,y)$ of such wave surfaces
is given by $K=\laplace\log P$. It is independent of $r$ which demonstrates explicitly the non--expanding
character of the Kundt spacetimes.

A complete class of  type III, type N, and conformally flat Kundt spacetimes with possibly non--vanishing
cosmological constant $\Lambda$ and pure radiation
is characterized by the vanishing Weyl scalars $\Psi_{1}=\Psi_{2}=0$ which
are calculated in the natural null tetrad $k=\partial_{r}$,\,$l=\partial_{u}-H\partial_{r}$,\,$m=\frac{P}{\sqrt{2}}(\partial_{x}+i\partial_{y})+\frac{P}{\sqrt{2}}(a_{x}+ia_{y})\partial_{r}$ adapted to the repeated null congruence.

Namely, the components of field equations $ri$ \; imply
that the functions $a_{i}$ should be at most linear in r, i.e.
the $a_{i}$ have the structure \cite{GrifDochPod:2004:CLAQG:, GrifPod:2009:Cam:},
\begin{equation}\label{m00}
a_{i}=-\frac{2\tau_{i}}{P}r+a^{0}_{i},
\end{equation}
where $a^{0}_{i}$ and $\tau_{i}$ are independent of $r$. It turns out that $\tau_{i}$ is one of the spin coefficients.
The field equations imply that the function $P$
should satisfy the equation
\begin{equation}\label{m001}
R_{\perp}=\frac{1}{2}\laplace\log P=\frac{\Lambda}{6},
\end{equation}
which says that the Gaussian curvature of the wave surfaces is constant $K=2R_{\perp}=\frac{\Lambda}{3}$, the $R_{\perp}$ is the curvature of the transverse space. Therefore we can use coordinate freedom to put $P$ into the canonical form,
\begin{equation}\label{m01}
P=1+\frac{\Lambda}{12}(x^2+y^2).
\end{equation}
When $\Lambda=0$ the wave surfaces are planes, for $\Lambda>0$ they are 2--spheres while for $\Lambda<0$ the surfaces have constant
negative curvature, i.e. hyperboloidal Lobachevski planes.
The field equations also determine the structure of function $H$,
\begin{equation}
H=-\bigl(\frac{1}{2}(\tau^2_{x}+\tau^2_{y})+\frac{\Lambda}{6}\bigr)r^2+2g r+h,
\end{equation}
where the functions  $g$ and $h$ are independent of $r$.

With the choice \eqref{m01} the field equations lead to the following general form of the spin coefficient $\tau$
\begin{equation}\label{m02}
\tau_{i}=P(\log \frac{P}{Q} )_{,i}
\end{equation}
where the function $Q$ has a form,
\begin{equation}\label{m021}
Q=a(1-\frac{\Lambda}{12}(x^2+y^2))+b_{x}x+b_{y}y.
\end{equation}
Here
$a(u)$ and $b_{i}(u)$ are arbitrary real functions of $u$.
The components of $\tau_{i}$ have the explicit form
\begin{equation}\begin{aligned}\label{m022}
\tau_{x}=\frac{-b_{x}+\frac{\Lambda}{3} a x+\frac{\Lambda}{12}(b_{x}(x^2-y^2)+2xy b_{y})}{a+b_{x}x+b_{y}y-a\frac{\Lambda}{12}(x^2+y^2)},\\
\tau_{y}=\frac{-b_{y}+\frac{\Lambda}{3} a y+\frac{\Lambda}{12}(b_{y}(y^2-x^2)+2xy b_{x})}{a+b_{x}x+b_{y}y-a\frac{\Lambda}{12}(x^2+y^2)}.
\end{aligned}\end{equation}

The spin coefficient $\tau$ can be interpreted as a measure of the rotation of the principal null congruence about a spacelike
direction, \cite{GrifDochPod:2004:CLAQG:}.

There also exists another solution of the equation for $P$,
\begin{equation}\label{m011}
P=\sqrt{-\tfrac{\Lambda}{6}}\,x,
\end{equation}
which is valid only for a negative cosmological constant. The above form of $P$
is connected with the $P$ \eqref{m01} by  a simple transformation. Solutions with this
expression for $P$ \eqref{m011} lead to Siklos solutions, \cite{Sik:1985:Cam:}. It means
that the solutions have conformal factor which multiplies the standard {\it pp}--wave solution.
The Siklos gyratons were found in \cite{Fro-Zel:2005:PHYSR4:}.

It is convenient to introduce the expression
\begin{equation}\label{m03}
k\equiv\frac{\Lambda}{6} a^2+\frac{1}{2}(b^2_{x}+b^2_{y}),
\end{equation}
which has been identified as a quantity whose sign is invariant \cite{OzsvathRobRozga:1985:JMATHP:}. Therefore it was used
in classification of Kundt's family of solutions. For each subfamily which is defined by the sign of $k$, the function $\tau$ can be expressed
in an canonical form, \cite{GrifPod:2009:Cam:,GrifDochPod:2004:CLAQG:}.
The two choices of $\tau$ are the following:

For $b=0$ we obtain $\tau$ as a {\it case 1},

\begin{equation}\label{jajaja}
\tau_{x}= \frac{\frac{\Lambda}{3} x}{1-\frac{\Lambda}{12}(x^2+y^2)},\,\tau_{y}= \frac{\frac{\Lambda}{3} y}{1-\frac{\Lambda}{12}(x^2+y^2)}.
\end{equation}
Similarly, if $a=0$  then
we get {\it case 2},

\begin{equation}
\tau_{x}=-\frac{1-\frac{\Lambda}{12}(x^2-y^2)}{x},\,\tau_{y}= \tfrac{\Lambda}{6} y.
\end{equation}
These expressions reduce to the two standard cases which are well known when $\Lambda=0$.

When $\Lambda=0$ then $k=\frac{1}{2}(b^2_{x}+b^2_{y})$. There are two geometrically
distinct types of solutions, namely cases $k=0$ and $k>0$ which correspond to vanishing and non--vanishing $\tau$.
Using the remaining coordinate freedom, these subclasses can be put into the canonical forms with
$a=1$, $b=0$ and $a=0$, $b=1$, respectively. This identifies two types of solutions:
\begin{itemize}
\item $k=0$: generalized pp--waves  \;\;\;\;\; $\tau=0$
\item $k>0$: generalized Kundt waves \;\;\;\;\; $\tau_{x}=-\frac{1}{x}$, $\tau_{y}=0$.
\end{itemize}
The subclass $k=0$ gives exactly the {\it pp}--waves (type N solutions) while
the case $k>0$ represents the Kundt waves for which the principal null vector ${k}$
is not covariantly constant.

When $\Lambda>0$, it is only possible for $k$ to be positive and there exists
just one canonical case. In other words, it is possible to use a coordinate
transformation to put either $a=1$, $b=0$ or $a=0$, $b=1$. Therefore the function
$\tau$ can always be transformed to either of the following two canonical forms
which are completely equivalent:
\begin{itemize}
\item $k>0$: generalized pp--waves or Kundt waves  case 1 or case 2.
\end{itemize}
These forms of $\tau$ reduce to the above two cases when $\Lambda=0$.
This family of solutions may be considered as a generalization of either
the {\it pp}--waves or the Kundt waves in the sense that they reduce to
either of these forms for the type N solutions in the appropriate limit
which depends also on the used coordinate system.

When $\Lambda<0$, there exists three distinct possibilities which are identified
by the sign $k$. If $k<0$, it is always possible to put $a=1$, $b=0$ to obtain
generalized {\it pp}--waves. Alternatively, if $k>0$, it is possible to put $a=0$, $b=1$
and hence to obtain generalized Kundt waves. Another interesting case arises here when $k=0$.
Due to the expression for $k$ \eqref{m03}, this occurs when $b=\sqrt{-\tfrac{\Lambda}{6}}a e^{i\theta}$
for an arbitrary function $\theta(u)$. Such solutions generalize the type N spacetimes
that have been described in detail by Siklos \cite{Sik:1985:Cam:} using a different
coordinate system. It can be concluded that for $\Lambda<0$ there exist three canonical
subfamilies of vacuum Kundt's solutions:
\begin{itemize}
\item $k<0$: generalized {\it pp}--waves  \;\;\;\;\;\;\;case 1,
\item $k>0$: generalized Kundt waves  \;\;\;\;\;case 2,
\item $k=0$: generalized  Siklos waves  \;\;\;\;\;\,case 3,
\end{itemize}
where by {\it case 3} we mean the $\tau$ in the form (we present here the complex form of $\tau$ which is more useful in calculations),
\begin{equation}\label{m0222}
\tau=-\sqrt{-\tfrac{\Lambda}{6}}e^{i\theta}\left(\frac{1+\sqrt{-\frac{\Lambda}{6}} e^{-i\theta}\zeta}{1+\sqrt{-\frac{\Lambda}{6}}e^{i\theta}\bar{\zeta}}\right),
\end{equation}
where it is clearly possible to remove the phase $e^{i\theta}$, so the canonical form of $\tau$ is
\begin{equation}\label{m0223}
\tau=-\sqrt{-\tfrac{\Lambda}{6}}\left(\frac{1+\sqrt{-\frac{\Lambda}{6}}\zeta}{1+\sqrt{-\frac{\Lambda}{6}}\bar{\zeta}}\right).
\end{equation}

Let us mention that since $a(u)$ and $b(u)$ in \eqref{m022} are arbitrary functions,
it is possible to construct composite spacetimes  in which these functions are non--zero
for different ranges of $u$.
When we study the cases for the other function $P$ \eqref{m011} we again obtain just those
equivalent three distinct canonical types mentioned above for $P$ \eqref{m01}, see \cite{GrifDochPod:2004:CLAQG:}.
The case investigated by Siklos is then equivalent to the case 3 and the other solutions on anti--de Sitter
spacetime cannot be related to each other.

\subsection{The transformation from real to complex coordinates}
It is more convenient for us to work in real coordinates, but
in general the complex coordinates are mainly used in literature about Kundt class.

The transformation between the complex coordinates $\zeta,\,\bar{\zeta}$ and real
coordinates $x,\,y$ is the following
\begin{equation}\begin{aligned}\label{m16}
\zeta=&\frac{1}{\sqrt{2}}(x+iy),\,\bar{\zeta}=\frac{1}{\sqrt{2}}(x-iy).
\end{aligned}\end{equation}

The functions $W$ and $a_{i}$ are connected as
\begin{equation}\begin{aligned}\label{m17}
W=&-a_{\zeta}=-\frac{1}{\sqrt{2}}(a_{x}-ia_{y}),\,\quad \overline{W}=&-a_{\bar{\zeta}}=-\frac{1}{\sqrt{2}}(a_{x}+ia_{y}),
\end{aligned}\end{equation}
and the functions $H$, $k$ and $\tau$ are then redefined as
\begin{equation}\begin{aligned}\label{m19}
H=-(\tau\bar{\tau}+\frac{\Lambda}{6})r^2+2gr+h,\quad k=&\frac{\Lambda}{6} a^2 +b\bar{b},\quad \tau=P(\log\frac{P}{Q})_{,\bar{\zeta}}
\end{aligned}\end{equation}
and the relation between real and complex expressions is,
\begin{equation}\begin{aligned}\label{m18}
\tau=&\frac{1}{\sqrt{2}}(\tau_{x}+i\tau_{y}),\,\bar{\tau}=\frac{1}{\sqrt{2}}(\tau_{x}-i\tau_{y}),\\
b=&\frac{1}{\sqrt{2}}(b_{x}+ib_{y}),\,\bar{b}=\frac{1}{\sqrt{2}}(b_{x}-ib_{y}).
\end{aligned}\end{equation}

The functions $P$ and $Q$ in complex coordinates read
\begin{equation}\begin{aligned}\label{m20}
P=&1+\frac{\Lambda}{6}\zeta\bar{\zeta},\quad Q=(1-\frac{\Lambda}{6}\zeta\bar{\zeta})a(u)+b\bar{\zeta}+\bar{b}\zeta.
\end{aligned}\end{equation}

\section{The gyratons on conformally flat spacetimes}
The solutions introduced in the first section are radiative spacetimes in which the rays are non--expanding, but
their wave surfaces have constant curvature proportional to the cosmological constant. The non--zero $\tau$
indicates that subsequent wave surfaces are locally rotated relative to each other.
As in \cite{GrifDochPod:2004:CLAQG:}, we will consider first backgrounds for which the functions $h$ and $a^{0}_{i}$ are taken
to be zero. In this limit the spacetime is conformally flat and is Minkowski, de Sitter and anti--de Sitter according to the
cosmological constant, i.e. $\Lambda=0$, $\Lambda>0$ and $\Lambda<0$.
The background metric is then given by \eqref{m0} with
\begin{equation}\label{LL}
H=-(\frac{1}{2}(\tau^2_{x}+\tau^2_{y})+\frac{\Lambda}{6})r^2,\,a_{i}=-\frac{2\tau_{i}}{P}r
\end{equation}
for different values of $\Lambda$ and differing expressions for $P$ and $\tau$.
In these background spacetimes it is possible to explicitly investigate the geometry
of the wave surfaces and the way in which they foliate the spacetime as it was done in \cite{GrifDochPod:2004:CLAQG:}.

In the following we will derive the solutions on these background spacetimes.
Before we will do so let us mention that the function $g$ is not arbitrary for
our solutions, it was shown also in \cite{GrifDochPod:2004:CLAQG:} that the
function $g$ has a structure
\begin{equation}\label{LL1}
g=\frac{P}{2}(\tau_{x} a^{0}_{x}+\tau_{y}a^{0}_{y})-\frac{\Lambda}{6} \sqrt{2}f_{x},
\end{equation}
where $f=\frac{1}{\sqrt{2}}(f_{x}+if_{y})$ and $f_{i}=f_{i}(u,x,y)$ is an arbitrary function satisfying the Laplace
equation.

\subsection{The solutions on the Minkowski background}
We first consider the case in which $\Lambda=0$.
The case 1 ($\tau=0$), represents the solution on Minkowski spacetime
presented in \cite{Fro-Fur:2005:PHYSR4:, Fro-Is-Zel:2005:PHYSR4:}.
The metric has compact form  ($P=1, H=2gr+h$)
\begin{equation}\label{a1}
\trgrad s^2=\trgrad x^2+\trgrad y^2 -2\trgrad u(\trgrad r + (2gr+h)\trgrad u-a^{0}_{i}\trgrad x^{i}).
\end{equation}
This solution is
generalization of the {\it pp}--waves which have plane wave surfaces and parallel rays.
The solution is well known and therefore we will pay attention to the other case 2,
for which $\tau$ has a form $\tau_{x}=-\frac{1}{x}$, $\tau_{y}=0$ and $k>0$.
The solution has a form
\begin{equation}\label{a2}
\trgrad s^2=\trgrad x^2+\trgrad y^2 -2\trgrad u(\trgrad r +(-\frac{r^2}{2x^2}+ 2gr+h)\trgrad u-\frac{2}{x}r\trgrad x-a^{0}_{i}\trgrad x^{i}).
\end{equation}
Hence  $\tau$ is non--zero, the wave surfaces $u$ rotate locally in the background Minkowski spacetime which
 was explicitly demonstrated in \cite{GrifDochPod:2004:CLAQG:}.
The solution itself is time--symmetric therefore the envelope of wave surfaces is a cylinder whose radius decreases to zero
at the speed of light and then increases. The complete family of wave surfaces has to be taken as the family of half--planes for which
$x\geq0$. The singularity at $x=0$ on the expanding cylinder can be interpreted as the caustic formed by the envelope of the family of wave
 surfaces, rotated one with respect to the other along the cylinder. No wave surfaces pass through points that are inside the expanding cylinder. The coordinates used in \eqref{a2} do not cover
this part of spacetime, however this is not necessary because in the general Kundt curved spacetime the expanding cylinder is curvature
envelope singularity through which the spacetime cannot be physically extended, \cite{GrifPod:2009:Cam:}.

\subsection{The solutions on de Sitter and anti--de Sitter backgrounds for the case 2}
Now we will consider the case 2 solutions in which $\Lambda\neq0$ and $\tau$ has the canonical form \eqref{m0223}.
In this case, the metric \eqref{m0} takes the form
\begin{equation}\begin{aligned}\label{a3}
\trgrad s^2=&-2\trgrad u\left(\trgrad r +(-\frac{1}{2}\frac{P^2}{x^2}r^2+ 2gr+h)\trgrad u-(2\frac{(1-\frac{\Lambda}{12}(x^2-y^2))}{Px}r+a^{0}_{x})\trgrad x\right.\\
&\left.-(-\frac{\frac{\Lambda}{3} y}{P}r+a^{0}_{y})\trgrad y\right)+\frac{1}{P^2}(\trgrad x^2+\trgrad y^2),
\end{aligned}\end{equation}
then it is convenient to put $r=\frac{Q^2}{P^2}v$ with $Q=\sqrt{2}x$, that the solutions on the backgrounds (de Sitter and anti--de Sitter)
can be expressed as
\begin{equation}\begin{aligned}\label{a4}
\trgrad s^2=&-\frac{4x^2}{P^2}\left(\trgrad u\trgrad v +(-v^2+ 2gv+\frac{P^2}{2x^2}h)\trgrad u^2\right)+2a^{0}_{i}\trgrad u\trgrad x^{i}+\frac{1}{P^2}(\trgrad x^2+\trgrad y^2),
\end{aligned}\end{equation}
where the function $g$ has the form \eqref{LL1}.
Again while the $\tau$ is non--zero the wave surfaces will rotate.

The expanding cylinder described above in Minkowski spacetime become an expanding torus in the closed de Sitter spacetime.
The wave surfaces are a family of spheres with constant area $4\pi a^2$. They are tangent to the expanding torus so that the
singularity can again be interpreted as a caustic formed from the envelope of wave surfaces. Since two spheres pass through each
point within the region covered by these coordinates it is appropriate to restrict the family of wave surfaces to the hemispheres
on which $x\geq0$ whose boundary is located on the expanding torus.

For the anti--de Sitter spacetime the wave surfaces are hyperboloidal. They are tangent to the singularity which is an expanding hyperboloid
that can be interpreted as an envelope of wave surfaces. Since it is only possible for one wave surface to pass through any point, it is
appropriate to take the wave surfaces as the family of semi--infinite hyperboloids on which $x\geq0$, which is obvious generalization
of the half--planes of the Minkowski case.
Finally, as for the case 2 in the Minkowski background the above wave surfaces in de Sitter and anti--de Sitter backgrounds are outside
the expanding torus or hyperboloid, respectively. For more details see \cite{GrifDochPod:2004:CLAQG:}.

\subsection{The solutions on de Sitter and anti--de Sitter backgrounds for the case 1}
In this section we will consider the case 1 solutions in which $\Lambda\neq0$ and $\tau$ has canonical form \eqref{m022}.
In this case, after we have used $r=(\frac{Q^2}{P^2})v$ and $Q=1-\frac{\Lambda}{12}(x^2+y^2)$, the metric
has the form
\begin{equation}\begin{aligned}\label{a5}
\trgrad s^2=&-2\frac{Q^2}{P^2}\left(\trgrad u\trgrad v +(-\frac{\Lambda}{6} v^2+ 2gv+\frac{P^2}{Q^2}h)\trgrad u^2\right)+2a^{0}_{i}\trgrad u\trgrad x^{i}+\frac{1}{P^2}(\trgrad x^2+\trgrad y^2),
\end{aligned}\end{equation}
again with the function $g$ in the form \eqref{LL1}.

The wave surfaces for $\Lambda>0$ are identical to those for the case 2 with $\Lambda>0$. This is consistent because
 the cases 1 and 2 are equivalent for  a positive cosmological constant.
 For $\Lambda<0$ the background universe is open and the analogues of the plane wave surfaces are hyperboloids
 which foliate the entire universe. For detailed analysis see \cite{GrifDochPod:2004:CLAQG:}.

\subsection{The solutions on anti--de Sitter background for the case 3}
Finally, let us consider the case 3 solutions in which $\Lambda$ is necessarily negative, $k=0$ and $\tau$ has
the canonical form \eqref{m0223} which we have left in complex form because it is more convenient in this case.
The metric then has a form in real coordinates, (the coefficient in front of $r^2$ vanishes in function $H$),
\begin{equation}\begin{aligned}\label{a6}
\trgrad s^2=&-2\frac{Q^2}{P^2}\left(\trgrad u\trgrad v +(2gv+\frac{P^2}{Q^2}h)\trgrad u^2\right)+2a^{0}_{i}\trgrad u\trgrad x^{i}+\frac{1}{P^2}(\trgrad x^2+\trgrad y^2),
\end{aligned}\end{equation}
where we used $r=(\frac{Q^2}{P^2})v$ and $Q=(1+\sqrt{-\frac{\Lambda}{6}}\zeta)(1+\sqrt{-\frac{\Lambda}{6}}\bar{\zeta})$,
and again with the function $g$ in the form \eqref{LL1}.
The waves surfaces are hyperboloidal and foliate the entire background  spacetime.
The type N cases of these solutions have been described in detail by Siklos \cite{Sik:1985:Cam:} and Podolsky \cite{Pod-rot:1998:CLAQG:}
in different coordinate systems. The gyraton on AdS presented in \cite{Fro-Zel:2005:PHYSR4:} are represenatives of the case 3
solutions but in different coordinate system.

\section{The derivation of the Einstein equations for the gyratons of type III}\label{equations}
In this section we will derive the Einstein equations. First, we will
present the Einstein tensor for the general metric \eqref{m0} with the metric functionsm7
$P$ \eqref{m01} and $H$ \eqref{m011} with general expressions for $\tau$ \eqref{m02}.

Then we will derive the Einstein equations with cosmological constant and the gyraton:
\begin{equation}
G_{\mu\nu}+\Lambda g_{\mu\nu}=\varkappa T^{\gyr}_{\mu\nu}\;.
\end{equation}
Here, $\Lambda$  and  $\varkappa=8\pi G$ are the cosmological and gravitational constants, respectively.

The gyraton is specified  by its stress-energy tensor defined as
\begin{equation}\label{mmm7}
\varkappa\, T^{\gyr}=j_{u}\,\trgrad u^2+2j_x\,\trgrad u\,\trgrad x+2j_y\,\trgrad u\,\trgrad y\;
\end{equation}
and it must satisfy the conservation law
\begin{equation}\label{gyrenergycons1}
  T^{\gyr}_{\;\,\mu\nu}{}^{\>;\nu}=0\;.
\end{equation}
The fields are characterized by functions ${P}$, ${H}$, ${a^{0}_i}$ which must be determined by the field equations, provided the gyraton sources ${j_u}$ and ${j_i}$ are prescribed. In fact, we know how the functions $P$ and $H$ look like but we will be able to derive them again
from source free components of Einstein equations.

\subsection{The Einstein tensor}
The non--zero components of the Einstein tensor for
the metric \eqref{m0} are:

\begin{equation}\label{EinsteinTT}
\begin{aligned}
G_{xy}&=-\frac{1}{2}\frac{1}{P}\left(P(\partial_{r}a_{x})(\partial_{r}a_{y})+P(\partial_{y}\partial_{r}a_{x})+P(\partial_{x}\partial_{r}a_{y})+2(\partial_{r}a_{x})(\partial_{y}P)+2(\partial_{r}a_{y})(\partial_{x}P)\right),\\
G_{ur}&= -\frac{1}{4}P^2 (\partial_{r}a_{x})^2-\frac{1}{4}P^2 (\partial_{r}a_{y})^2-\frac{1}{2}P^2 (\partial_{x}\partial_{r}a_{x})-\tfrac{1}{2}P^2 (\partial_{y}\partial_{r}a_{y})+\laplace\!\log P\;,\\
G_{xx}&=\frac{1}{4}\frac{1}{P^2}\left(P^2(\partial_{r}a_{x})^2-4P(\partial_{r}a_{x})(\partial_{x}P)+4P(\partial_{r}a_{y})(\partial_{y}P)
+4\partial^2_{r}H+3P^2(\partial_{r}a_{y})^2\right.\\
&\left.+4P^2(\partial_{y}\partial_{r}a_{y})\right)\;,\\
G_{yy}&=\frac{1}{4}\frac{1}{P^2}\left(P^2(\partial_{r}a_{y})^2+4P(\partial_{r}a_{x})(\partial_{x}P)-4P(\partial_{r}a_{y})(\partial_{y}P)
+4\partial^2_{r}H+3P^2(\partial_{r}a_{x})^2\right.\\
&\left.+4P^2(\partial_{x}\partial_{r}a_{x})\right)\;,\\
G_{ux}&=\frac{1}{2}b_{,y}-a_{x}\bigl(\laplace\!\log P-\partial^2_{r}H\bigr)+\partial_{r}H_{,x}+\frac{1}{2}\partial_{u}\partial_{r}a_{x}+\frac{1}{4}P^2 a_{x}\left[(\partial_{r}a_{x})^2+3(\partial_{r}a_{y})^2\right]\\
&+\frac{1}{2}P^2\left[a_{x}(\partial_{x}\partial_{r}a_{x})+a_{y}(\partial_{x}\partial_{r}a_{y})\right]+\frac{1}{2}P^2(\partial_{r}a_{y})(\partial_{x}a_{y})+a_{x}\left[P^2\partial_{y}\partial_{r}a_{y}+\frac{1}{2}\partial_{r}a_{y}(P^2)_{,y}\right]\\
&-a_{y}\left[P^2\partial_{y}\partial_{r}a_{x}+\frac{1}{2}\partial_{r}a_{x}(P^2)_{,y}\right]-\frac{1}{2}P^2(\partial_{r}a_{x})[a_{y}\partial_{r}a_{y}+\partial_{y}a_{y}]\;,\\
G_{uy}&=-\frac{1}{2}b_{,x}-a_{y}\bigl(\laplace\!\log P-\partial^2_{r}H\bigr)+\partial_{r}H_{,y}+\frac{1}{2}\partial_{u}\partial_{r}a_{y}+\frac{1}{4}P^2 a_{y}\left[(\partial_{r}a_{y})^2+3(\partial_{r}a_{x})^2\right]\\
&+\frac{1}{2}P^2\left[a_{x}(\partial_{y}\partial_{r}a_{x})+a_{y}(\partial_{y}\partial_{r}a_{y})\right]+\frac{1}{2}P^2(\partial_{r}a_{x})(\partial_{y}a_{x})+a_{y}\left[P^2\partial_{x}\partial_{r}a_{x}+\frac{1}{2}\partial_{r}a_{x}(P^2)_{,x}\right]\\
&-a_{x}\left[P^2\partial_{x}\partial_{r}a_{y}+\frac{1}{2}\partial_{r}a_{y}(P^2)_{,x}\right]-\frac{1}{2}P^2(\partial_{r}a_{y})[a_{x}\partial_{r}a_{x}+\partial_{x}a_{x}]\;,\\
G_{uu}&=\frac{1}{2}b^2+\laplace H +(\partial^2_{r}H)a^2+2a^i\partial_{r}H_{,i}+(\partial_r H)\,\trdiv a+\partial_{u}\trdiv a
+2H\laplace\!\log P\;,\\
&-2H(\partial_{r}a^{i})_{,i}+a^{i}\partial_{u}\partial_{r}a_{i}+P^2 b (a_{x}\partial_{r}a_{y}-a_{y}\partial_{r}a_{x})-\partial_{r}a^{i}H_{,i}\\
&+\frac{1}{2}P^2((\partial_{r}a_{x})a_{y}-(\partial_{r}a_{y})a_{x})^2-\frac{1}{2}P^2H\left[(\partial_{r}a_{x})^2+(\partial_{r}a_{y})^2\right],
\end{aligned}
\end{equation}
where
\begin{equation}\label{lapllogpcxx}
  \laplace\!\log P=P\bigl(P_{,xx}+P_{,yy}\bigr)-\bigl(P_{,x}^2+P_{,y}^2\bigr)\;
\end{equation}
and the scalar curvature is given by
\begin{equation}\label{Rss}
R=-2(\partial^2_{r}H)-\frac{3}{2}P^2\left[(\partial_{r}a_{x})^2+(\partial_{r}a_{y})^2\right]-2P^2 \partial_{x}\partial_{r}a_{x}
-2P^2\partial_{y}\partial_{r}a_{y}-2\laplace\log P.
\end{equation}

In the above components of the Einstein tensor and scalar curvature we have only used the fact that the functions $a_{i}$
are at most linear in $r$. Therefore expressions simplify a little bit, otherwise everything is in full generality.
Let us recall that the transversal space is identical with the transversal space for the ansatz for the gyratons on
direct product spacetimes and we use the geometry defined there, i.e. \cite{Kadlecova:2009:PHYSR4:}.

\subsection{The source free components of Einstein equations}
It is also important to show that the source--free components of the Einstein equations
are really vanishing, i.e. that the solutions we present here are solutions of these
Einstein equations.

First we start with the component $xy$ which vanishes only due to fact that
$Q_{,xy}=0$ and $P_{,xy}=0$. But the expression $G_{xy}=0$ is valuable from computational
point of view.
The component of the  Einstein equations $ur$ is zero due to the structure of $P$ and
the relation, \cite{GrifDochPod:2004:CLAQG:},
\begin{equation}
P^2\left[\left(\frac{\tau_{x}}{P}\right)_{,x}+\left(\frac{\tau_{y}}{P}\right)_{,y}\right]=\tau^2_{x}+\tau^2_{y}+\frac{2}{3}\Lambda.
\end{equation}
From components $xx$ and $yy$ we can derive the structure of $H$ and then we will be able to
determine explicit value of the scalar curvature.
When we sum up  the  $xx$ and $yy$ components of the Einstein equations we will get the
expression for the function $H$,
\begin{equation}
2(\partial^2_{r}H)=-2\Lambda-P^2(\partial_{r}a_{x})^2-P^2(\partial_{r}a_{y})^2-P^2\partial_{x}\partial_{r}a_{x}-P^2\partial_{y}\partial_{r}a_{y},
\end{equation}
which gives the function $H$ in the form
\begin{equation}
\partial^2_{r}H=-\frac{P^2}{Q^2}(a^2\frac{\Lambda}{3}+b^2_{x}+b^2_{y}),
\end{equation}
which is exactly the function $H$ \eqref{m011} when we use the relation
\begin{equation}
\frac{1}{2}(\tau^2_{x}+\tau^2_{y})+\frac{\Lambda}{6}=\frac{P^2}{Q^2}\left(a^2\frac{\Lambda}{6}+\frac{1}{2}(b^2_{x}+b^2_{y})\right).
\end{equation}
Then the scalar curvature $R$ \eqref{Rss} is exactly
\begin{equation}
R=4\Lambda.
\end{equation}

\subsection{The source Einstein equations}
Now, we will derive the source field equations.
The condition \eqref{gyrenergycons1} for the gyraton source gives
\begin{equation}\label{gyrenergycons22}
  -(\partial_r j_i)\,\trgrad x^i + \bigl(-\partial_r j_u+\trdiv j +a^i \partial_r j_i\bigr)\,\trgrad u =0\;,
\end{equation}
so that the source functions ${j_i}$ must be ${r}$-independent and ${j_u}$ has to have the structure
\begin{equation}\label{jdecomp3}
  j_u = r\,\trdiv j + \iota\;,
\end{equation}
the gyraton source \eqref{mmm7} is fully determined by three \mbox{${r}$-independent} functions ${\iota(u,x^j)}$ and ${j_i(u,x^j)}$.

Finally, the remaining nontrivial components of the Einstein equations are those involving the gyraton source \eqref{mmm7}. The $ui$-components give equations for the components with ${j_i}$ as,
\begin{equation}\begin{aligned}\label{jieqx}
 j_x= &\frac12\, f_{xy}{}^{:y}  + \partial_{r}H_{,x} -a_x\left(\frac{\Lambda}{3}-2\tau^2_{y}+P^2\left(\frac{\tau_{x}}{P}\right)_{,x}+2P^2\left(\frac{\tau_{y}}{P}\right)_{,y}+\left(\frac{\tau_{y}}{P}\right)(P^2)_{,y}\right)\\
&-a_{y}\left(2\tau_{x}\tau_{y}+P^2\left(\frac{\tau_{y}}{P}\right)_{,x}+2P^2\left(\frac{\tau_{x}}{P}\right)_{,y}+\left(\frac{\tau_{x}}{P}\right)(P^2)_{,y}\right)\\
&-P\left(\tau_{y}\partial_{x}a_{y}-\tau_{x}\partial_{y}a_{y}\right)-\frac{1}{P}\partial_{u}\tau_{x}
\end{aligned}\end{equation}
and
\begin{equation}\begin{aligned}\label{jieqy}
 j_y= &\frac12\, f_{yx}{}^{:x}  + \partial_{r}H_{,y}-a_y\left(\frac{\Lambda}{3}-2\tau^2_{x}+P^2\left(\frac{\tau_{y}}{P}\right)_{,y}+2P^2\left(\frac{\tau_{x}}{P}\right)_{,x}+\left(\frac{\tau_{x}}{P}\right)(P^2)_{,x}\right)\\
&-a_{x}\left(2\tau_{x}\tau_{y}+P^2\left(\frac{\tau_{x}}{P}\right)_{,y}+2P^2\left(\frac{\tau_{y}}{P}\right)_{,x}+\left(\frac{\tau_{y}}{P}\right)(P^2)_{,x}\right)\\
&-P\left(\tau_{x}\partial_{y}a_{y}-\tau_{y}\partial_{x}a_{x}\right)-\frac{1}{P}\partial_{u}\tau_{y},
\end{aligned}\end{equation}
where we have introduced the external derivative ${f_{ij}}$ of the 1-form ${a_i}$ as
\begin{equation}\label{fdeff}
  f_{ij} = a_{j,i}-a_{i,j} = (*\,\rot a)_{ij} \;.
\end{equation}

We denote the $f^{0}_{ij}=a^{0}_{j,i}-a^{0}_{i,j}$ as the 2--form $f^{0}_{ij}$  made from the pure gyratonic terms $a^{0}_{i}$.
We can show explicitly that $f_{ij}=f^{0}_{ij}$ because
\begin{equation}\label{fgyr}
    f_{xy}=2r\left[\left(\frac{\tau_{x}}{P}\right)_{,y}-\left(\frac{\tau_{y}}{P}\right)_{,x}\right]+f^{0}_{xy},
\end{equation}
where the expression in brackets is zero.

The expressions \eqref{jieqx} and \eqref{jieqy} are linear in $r$ with $r$--dependence hidden in $a_{i}$ and $H$.
The $r$--dependent part is trivial consequence of the field equations.
The $r$--independent part gives the equations for components of $a^{0}_{i}$
\begin{equation}\begin{aligned}\label{jieqx1}
 j_x= &\frac12\, f_{xy}{}^{:y}  + g_{,x} -a^{0}_x\left(\frac{\Lambda}{3}-2\tau^2_{y}+P^2\left(\frac{\tau_{x}}{P}\right)_{,x}+2P^2\left(\frac{\tau_{y}}{P}\right)_{,y}+\left(\frac{\tau_{y}}{P}\right)(P^2)_{,y}\right)\\
&-a^{0}_{y}\left(2\tau_{x}\tau_{y}+P^2\left(\frac{\tau_{y}}{P}\right)_{,x}+2P^2\left(\frac{\tau_{x}}{P}\right)_{,y}+\left(\frac{\tau_{x}}{P}\right)(P^2)_{,y}\right)\\
&-P\left(\tau_{y}\partial_{x}a^{0}_{y}-\tau_{x}\partial_{y}a^{0}_{y}\right)-\frac{1}{P}\partial_{u}\tau_{x}
\end{aligned}\end{equation}
and
\begin{equation}\begin{aligned}\label{jieqy1}
 j_y= &\frac12\, f_{yx}{}^{:x}  + g_{,x}-a^{0}_y\left(\frac{\Lambda}{3}-2\tau^2_{x}+P^2\left(\frac{\tau_{y}}{P}\right)_{,y}+2P^2\left(\frac{\tau_{x}}{P}\right)_{,x}+\left(\frac{\tau_{x}}{P}\right)(P^2)_{,x}\right)\\
&-a^{0}_{x}\left(2\tau_{x}\tau_{y}+P^2\left(\frac{\tau_{x}}{P}\right)_{,y}+2P^2\left(\frac{\tau_{y}}{P}\right)_{,x}+\left(\frac{\tau_{y}}{P}\right)(P^2)_{,x}\right)\\
&-P\left(\tau_{x}\partial_{y}a^{0}_{y}-\tau_{y}\partial_{x}a^{0}_{x}\right)-\frac{1}{P}\partial_{u}\tau_{y}.
\end{aligned}\end{equation}

Finally, the $uu$-component leads to the expression which is quadratic in~$r$.
But one can show that this quadratic term is trivial. The linear term gives,

\begin{equation}\label{jueqq}
\begin{split}
  \trdiv j =\,& 2\laplace g-(\tau^2_{x}+\tau^2_{y}+\frac{\Lambda}{3})(\trdiv a^{0}-\frac{4}{P}\tau^{i}a^{0}_{i})-\frac{4}{P}\tau^{i}g_{,i}
  -2(a^{0})^{i}(\tau^2_{x}+\tau^2_{y})_{,i}\\
  &+\frac{4}{P^2}\tau^{i}\partial_{u}\tau_{i}-2P^2\partial_{u}\left(\left(\frac{\tau_{x}}{P}\right)_{,x}+\left(\frac{\tau_{y}}{P}\right)_{,y}\right)+8g\left[\tau^2_{x}+\tau^2_{y}+\frac{5}{3}\Lambda\right],
\end{split}
\end{equation}
which is a consequence of \eqref{jdecomp3}, \eqref{jieqx1} and \eqref{jieqy1}.
The remaining $r$--independent part of the $uu$ component of the Einstein equations gives
the equation for $h$,
\begin{equation}\label{heq33}
\begin{split}
  \laplace h &+2h(\frac{2\Lambda}{3}+\tau^2_{x}+\tau^2_{y})+\frac{2}{P}\tau^{i}h_{,i}+(\tau^2_{x}+\tau^2_{y}+\frac{\Lambda}{6})(a^{0})^2=
      \iota \, - \frac12 b^2 - 4 (a^{0})^{i} g_{,i}\\
      &-(2\partial_u +g)\, \trdiv a+\frac{2}{P}(a^{0})^{i}\partial_{u}\tau_{i}-2(\tau_{x}a^{0}_{y}-\tau_{y}a^{0}_{x})^2.
\end{split}
\end{equation}

The equations apparently do not decouple and more investigation is needed to discuss
the solvability  and separability of those equations, i.e. discuss particular examples.

\section{Final remarks}
The main issue of our investigation was originally to look for the gyraton on de Sitter spacetime since
there was known a gyraton on anti--de Sitter spacetime \cite{Fro-Zel:2005:PHYSR4:} in the Siklos form.
The work presented in this chapter is mainly review of the theory of the Kundt class solutions of type III in real notation.
New results are only the gyratonic source field equations which are presented in completely general form. However further
discussion is needed.
In the future we want to investigate the geodesic motion of  the gyraton in the AdS/dS background spacetimes.

\newpage\cleardoublepage
\chapter{Higher dimensional gyratons on direct product spacetimes}
\section{Introduction}
Higher dimensional gravity is now very active research area. It is caused mainly by the fact that the gravity in $D>4$ behaves in many ways
differently and has some unexpected properties. The investigation of exact solutions of Einstein's equations in higher dimensions  might
help to understand to physical properties and to general features of the theory.

The Kundt family of solutions in higher-dimensions was recently studied in \cite{PodoZo:2009:CLAQG:} though several important subclasses of Kundt's family in higher dimensions have already been studied in detail. Namely, the well--known {\it pp}--waves which admit a covariantly constant null vector field were studied thoroughly in \cite{Coley:2008:CLAQG, CMPPPZ:2003:PHYSR4, CMPP:2004:CLAQG, CMPP2:2004:CLAQG, CFHP:2006:CLAQG}. Also the VSI and CSI spacetimes \cite{Prav-Prav:2002:CLAQG:, CMPP2:2004:CLAQG, CFHP:2006:CLAQG, Col-Her-Pel:2006:CLAQG:} for which all polynomial scalar invariants constructed from the Riemann tensor and its derivatives vanish and are constant, respectively, and of course the gyraton solutions which we mentioned in the introduction.

In this chapter we would like to investigate a subclass of \mbox{${D}$-dimensional} Kundt family of spacetimes in the presence of a special aligned electromagnetic field and gyratonic matter, as a subcases it would contain higher--dimensional Kundt waves and gyratons.
The main aim of our investigation was to find higher dimensional generalization of the gyratons on direct product spacetimes. They would
appear as a special subcase as we will see in the following text.
In the first section, we will introduce the notation, the general ansatz for the metric, the gauge freedoms and ansatz for the electromagnetic field and gyratonic matter. In the second section, we will formulate the  geometry on transversal space and calculate the Ricci tensor.
In the third section, we present the Maxwell  and Einstein equations in the presence of gyratonic matter and aligned electromagnetic field
which is the main result of our investigation.
This work is in progress and it is a result of collaboration with Pavel Krtou\v{s}, Andrei Zelnikov, Ji\v{r}\'{i} Podolsk\'{y}.

\section{The ansatz for the metric and matter}\label{sc:ansatz}

Kundt family consists of spacetimes which contain nonexpanding nontwisting shear-free null congruence ${k}$. Such spacetimes admit foliation by null hypersurfaces ${\Sigma_{\folph}}$ which are generated by the null congruence ${k}$. We additionally assume that there exists a foliation by timelike 2-planes ${T_{\folph}}$ which are \emph{preserved by the congruence ${k}$} and which are \emph{orthogonal to the transverse spaces} as will be specified below.

It is not surprising that causally behaving physical systems develop rather trivially along ${k}$ direction because of the character of the null congruence. Also, dynamics on different null hypersurfaces ${\Sigma_\folph}$ is highly independent. Therefore it is reasonable to expect that it is possible to reduce the field equations to ${d=D{-}2}$-dimensional space of orbits of ${k}$ independently for each hypersurface~${\Sigma_\folph}$. Such a space will be called \emph{transverse space} and our aim is to reduce all field equations to this space which would reflect the situation in other higher dimensional gyraton solutions.

\subsection{Notation and some geometrical properties of the transverse spaces}
First, we will introduce the notation which we will be used throughout of this higher dimensional chapter.

To distinguish the spacetime quantities from the transverse space quantities we use a label `${{}^\st}$' on the left of spacetime objects to indicate the dimension. Since we will work mainly on the transverse space, we skip a similar label for the transverse objects. We use Greek letters for spacetime indices and Latin letters for transverse indices. However, we try to escape to use indices as much as possible. For this reason `${\cdot}$' denotes the contraction, `${\trgrad}$' the spacetime gradient and external derivative, and `${\covd}$' the spacetime covariant derivative associated with the metric ${g}$. For example, ${k\cdot u=k^\mu\grad[\mu] u=k^\mu u_{,\mu}}$, ${(k\cdot\covd k)^\mu = k^\nu\covd_\nu k^\mu=k^\nu k^\mu{}_{;\nu}}$.

Also the sharp `${\rsix}$\,', and similarly the flat `${\lwix}$\,', indicates raising (or lowering) of the tensor indices using the spacetime metric ${g}$. Since we will use also transverse metric ${q}$ and raising and lowering of indices using these two metric is not, in general, identical, we always write down the spacetime operation explicitly.

We introduce the null coordinates $u$, $r$ and the coordinates on the transversal plane $x^{i}$ as it is done in other higher dimensional
gyraton solutions.

Let us mention briefly that to define the transverse space in a more technical language we should introduce the coordinates ${u}$ and ${r}$ more
precisely. The coordinate ${u}$ is adjusted to the null foliation ${\Sigma_\folph}$\footnote{%
The particular space is selected by changing the placeholder `${\folph}$' to a unique characterization of the space. E.g., ${F_x}$ is the space containing the spacetime point ${x}$.}. We call ${\Sigma_u}$ the hypersurface given by the value ${u}$ and ${\Sigma_x}$ the hypersurface containig the spacetime point ${x}$.

The intersections of hypersurfaces ${r=\text{const.}}$ and ${\Sigma_u}$ form a foliation of the spacetime by ${d}$-dimensional transverse spaces ${N_{u,r}}$.  Therefore, it is natural to identify all spaces ${N_{u,r}}$ with one \emph{typical transverse space~${N}$}.
To find such typical transverse space $N$ we have to identify points in transverse spaces ${N_{u,r}}$ with different values of ${u}$ and ${r}$. For given ${u}$ and different values of ${r}$ it is natural to identify points along orbits of the congruence ${k}$. For different values of ${u}$ one has to introduce a flow in ${u}$ direction which conserves the transverse foliation ${N_\folph}$ and also commutes with the flow along the null congruence ${k}$. It can be given by a vector field ${w}$ tangent to ${r=\text{const.}}$ which satisfies
    $w\cdot \grad r =0\comma w\cdot\grad u = 1\comma [k,w] = 0 \period$

The vector fields ${k}$ and ${w}$ thus span a 2-dimensional \emph{temporal surfaces} which form the foliation ${T_\folph}$. These surfaces intersect each of the transverse spaces exactly in one point and we can thus identify points of different transverse spaces using these temporal surfaces.

The simplest way how to distinguish the different transverse spaces is to choose the remaining ${d}$ coordinates ${x^i}$ to be constant along temporal planes, which may thus be denoted as ${T_{x^i}}$. With such a choice of the adjusted coordinates the vector fields ${k}$ and ${w}$ become coordinate fields
\begin{equation}\label{kwcoor}
    k=\cv{r} \comma w=\cv{u} \period
\end{equation}
Nevertheless, a particular choice of the transverse coordinates is not necessary. The key ingredients of the above geometrical construction is a choice of the transverse spaces ${N_\folph}$ using the coordinate ${r}$ and the identification of the different transverse spaces using the flow~${w}$.

Finally, we introduce \emph{temporal derivatives} along ${k}$ and~${w}$
\begin{equation}\label{tempders}
  \rder{X} = \lder{k} X\comma \uder{X}=\lder{w} X\commae
\end{equation}
which, when restricting to the typical transverse space ${N}$, turn to be just parametric derivatives with respect of ${r}$ and ${u}$, respectively.

In the following we will mostly work on the transversal spaces and we will use the coordinate fields \eqref{kwcoor}.
The more technical definition of the transversal space would be useful in defining the tensors on $2+d$ spacetime in
one of the next section. We will also refer to it throughout of the discussion of several gauges.

\subsection{The form of the metric}

It is known in four dimensions \cite{Step:2003:Cam:,GrifPod:2009:Cam:} and in higher dimensions \cite{KrtouPod:2010:} that  such  choice of transversal spaces and coordinates leads to the spacetime metric ${g}$ in the form\footnote{%
In tensorial expressions, we skip the tensor product, e.g., ${\grad u \grad u = \grad u \otimes \grad u}$, and `${\vee}$' denotes the symmetrical tensor product, for example, ${\grad u \vee a = \grad u\, a + a \grad u}$.
For convenience of the reader, the equivalent expressions for nontrivial metric components are:
${g_{uu} = -2H}$,
${g_{ur} = -1}$,
${g_{ui} = a_i}$,
${g_{ij} = q_{ij}}$.
The components ${g^{\mu\nu}}$ of the inverse metric are
${g^{rr} = 2H + a^2}$,
${g^{ur} = -1}$,
${g^{ui} = a^i}$,
${g^{ij} = q^{ij}}$.
}
\begin{equation}\label{metric}
  g = - 2 H \grad u \grad u - \grad u \vee \grad r + \grad u \vee a + q\commae
\end{equation}
and the inverse spacetime metric
\begin{equation}\label{invmetric}
  g^{\!-1} = (2 H+a^2)\, \cv{r}\, \cv{r} - \cv{r} \vee \cv{u} + \cv{r} \vee \vec{a} + q^{\!-1}\commae
\end{equation}
where ${H}$ is a scalar function, ${a}$ is a transverse 1-form, ${q}$ is the metric on the transverse space, and ${q^{-1}}$ is the inverse transverse metric. Here and in the following we employ `\,${\vec{}}$\;' to indicate raising tensor index of transverse 1-forms using the transverse metric ${q}$. All squares of 1-forms and vectors are transverse squares, i.e., performed using the metric~${q}$.

The spacetime metric ${g}$ is thus split into transverse objects ${H}$, ${a}$, and ${q}$ and we assume
\begin{equation}\label{assumptions}
  \rder q =0\comma
  \rder a =0\comma
\end{equation}
which is in fact result of our geometrical construction.
Thus, ${q}$ and ${a}$ are ${r}$-independent, however, both can be ${u}$-dependent. The metric function ${H}$ can depend on all coordinates.

\subsection{Various gauge freedoms}

The construction of the transverse spaces and splitting of the metric in the form \eqref{metric} is not unique. It contains three partially ambiguous choices---gauges. The choice of the coordinate ${u}$ ($u$-gauge), the choice of the coordinate ${r}$ ($r$-gauge), and the choice of the flow ${w}$ ($w$-gauge).

The foliation of null hypersurfaces ${\Sigma_\folph}$ defines the coordinate ${u}$ up to reparametrization
\begin{equation}\label{uresc}
  u \rightarrow \tilde{u}=f(u)\commae
\end{equation}
with one-to-one function ${f}$ of one variable. The reparametrization has to be accompanied by rescaling of ${k}$ and ${r}$, however, the transverse foliation ${N_\folph}$ and the temporal surfaces ${T_\folph}$ are unchanged. Different quantities transform as follows,
\begin{equation}\label{ugauge}
\begin{gathered}
  \tilde{u} = f\comma
  \tilde{r} = \frac1{f'}r\commae\\
  \tilde{k} = f' k\comma
  \tilde{w} = \frac1{f'}\Bigl(w+\frac{f''}{f'}rk\Bigr)\commae\\
  \tilde{H} = \frac1{f'{}^2}\Bigl(H+\frac{f''}{f'}r\Bigr)\comma
  \tilde{a} = \frac1{f'} a\comma
  \tilde{q} = q\period
\end{gathered}
\end{equation}
Clearly, ${u}$-gauge changes both derivatives along ${k}$ and ${w}$.

${r}$-gauge freedom is related to the fact that the affine parameter of the geodesic is defined up to a constant. The coordinate function ${r}$ is thus defined up to ${r}$-independent shift:
\begin{equation}\label{rshift}
  r \rightarrow \tilde{r}=r+\psi\comma \rder \psi = 0\period
\end{equation}
 If we additionally require that the temporal foliation ${T_\folph}$ remains unchanged we find that change of ${r}$ has to be accompanied by,
\begin{equation}\label{rgauge}
\begin{gathered}
  \tilde{u} = u\comma
  \tilde{r} = r+\psi\commae\\
  \tilde{k} = k\comma
  \tilde{w} = w - \uder\psi k\commae\\
  \tilde{H} = H-\uder\psi\comma
  \tilde{a} = a+\trgrad\psi\comma
  \tilde{q} = q\commae
\end{gathered}
\end{equation}
where ${\trgrad\psi}$ represents the gradient on the transverse space, i.e., only transverse components of the spacetime gradient ${\grad\psi}$.
We also see that ${r}$-gauge changes just the derivative along ${w}$.

Finally, ${w}$-gauge leaves the transverse spaces ${N_\folph}$ but changes the identification of them for different values of the coordinate ${u}$. When restricted to the typical transverse space ${N}$, it can be viewed as ${u}$-dependent (and ${r}$-independent) family of diffeomorphisms of ${N}$. Generator of this family of diffeomorphisms is exactly the vector field ${\vec\xi}$ by which the ${w}$-gauge modifies the flow ${w}$ \cite{KrtouPod:2010:},
\begin{equation}\label{wshift}
  w \rightarrow w+\vec\xi\comma \rder {\vec\xi} =0\period
\end{equation}
The corresponding changes of the metric quantities are\\
\begin{equation}\label{wgauge}
\begin{gathered}
  \tilde{u} = u\comma
  \tilde{r} = r\commae\\
  \tilde{k} = k\comma
  \tilde{w} = w + \vec\xi\commae\\
  \tilde{H} = H-a\cdot\xi-\frac12\xi^2\comma
  \tilde{{\vec{a}}} = \vec{a}+\vec\xi\comma
  \tilde{q}^{-1} = q^{\!-1}\period
\end{gathered}
\end{equation}
Here, ${\xi^2=\vec\xi\cdot q\cdot\vec\xi}$ is the transverse square of ${\vec\xi}$. As for ${r}$-gauge, only ${u}$ derivative is modified.

\subsection{The electromagnetic field}

We want to study gravitational field generated by null fluid and gyratonic matter in the presence of an aligned homogeneous electromagnetic field.

The alignment condition we impose is that the congruence ${k}$ is eigenvector of the Maxwell tensor ${F}$,
\begin{equation}\label{EMaligned}
  F\cdot k = E\; \lwix k\period
\end{equation}
As a consequence, the Maxwell 2-form ${F}$
has form
\begin{equation}\label{Maxwell}
  F = E\, \grad r\wedge \grad u+\grad u \wedge s + B \commae
\end{equation}
where ${s}$ is a transverse \mbox{1-form} and ${B}$ a transverse \mbox{2-form}. We identify the first term as an electric part of the field and ${B}$ as a (transverse) magnetic part, although the interpretation is not straightforward due to two-dimensional character of the temporal planes and ${d{=}D{-}2}\,$-dimensional character of the transverse spaces.

The stress-energy tensor has the structure\footnote{\label{TEMsplit}%
In components we have
${\kap T^\EM_{uu}=2 H \rho + \kap\epso (E a - s)^2}$,
${\kap T^\EM_{ur}=\rho}$,
${\kap T^\EM_{ui}=\tau a_i + \kap\epso(E\,a^j B_{ji} - E s_i - s^j B_{ji})}$, and
${\kap T^\EM_{ij}=\frac{\kap\epso}{2} E^2 q_{ij} + \kap\, T^{\mg}_{ij}}$.}
\begin{equation}\label{TEM}
\begin{split}
  \kap\, T^{\EM}&=
     \Bigl(2 H \rho + \kap\epso (E a - s)^2\Bigr)\grad u \grad u + \rho \grad u \vee\grad r\\
     &\quad+\grad u \vee \Bigl(\tau a + \kap\epso(E\,\vec{a}\cdot B - E s - \vec{s}\cdot B)\Bigr)\\
     &\quad+\frac{\kap\epso}{2} E^2 q + \kap\, T^{\mg}\commae
\end{split}\raisetag{14pt}
\end{equation}
where, e.g., ${\vec{s}\cdot B}$ is a transverse 1-form with components ${s^\nu B_{\nu\mu}}$. Here ${\kap}$ is Einstein's gravitational constant and ${\epso}$ permitivity of vacuum. The usual choices are gaussian one, ${\kap=8\pi}$, ${\epso=1/4\pi}$, or SI-like, ${\kap=1}$, ${\epso=1}$, respectively.
We also conveniently introduced scalar quantities ${\rho}$ and ${\tau}$ quadratic in ${E}$ and ${B}$,
\begin{equation}\label{rhotaudef}
\begin{gathered}
\rho = \frac{\kap\epso}2\bigl(E^2+B^2\bigr)\commae \quad\tau = \frac{\kap\epso}2\bigl(E^2-B^2\bigr)\commae\\
\end{gathered}
\end{equation}
where the square ${B^2}$ of the transverse magnetic 2-form includes the factor ${1/2}$
\begin{equation}\label{B2def}
    B^2=\frac12\, B_{\mu\kappa}B_{\nu\lambda}\,g^{\mu\nu}g^{\kappa\lambda}=\frac12\, B_{ik}B_{jl}\,q^{ij}q^{kl}\period
\end{equation}
Finally, the magnetic part ${T^{\mg}}$ of the transverse stress-energy tensor is constructed just from ${B}$,
\begin{equation}\label{Tmg}
  \frac1\epso\, T^{\mg}_{\mu\nu} = B_{\mu\kappa}B_{\nu\lambda}\,g^{\kappa\lambda} - \frac12\, B^2 q_{\mu\nu}\commae
\end{equation}

In a generic dimension, the stress-energy tensor is not tracefree. The trace is characterized by the quantity ${\tau}$,
\begin{equation}\label{Ttrace}
  \kap\, T^\EM_{\mu\nu}\,g^{\mu\nu}=(D-4)\,\tau\period
\end{equation}

\subsection{The gyratonic matter}

As a source of gravitational field, we admit a generic \emph{gyratonic matter} aligned with the congruence ${k}$. The gyratonic matter is a generalization of a null fluid allowing also inner spin. It is described phenomenologically by the stress-energy tensor
\begin{equation}\label{Tgyr}
  \kap\, T^\gyr = j_u\, \grad u\grad u + \grad u \vee j\commae
\end{equation}
with scalar energy density ${j_u}$ and the spinning part given by the transverse 1-form ${j}$. Clearly, for ${j=0}$ we obtain standard null fluid flying in the direction ${k}$.

We do not specify the field equation for this matter except that we assume local stress-energy conservation\footnote{%
In indices, ${g^{\mu\nu}\covd{}_\mu T^\gyr{}_{\mspace{-10mu}\nu\kappa}=0}$, cf.\ definition~\eqref{divdefix}.}
\begin{equation}\label{Tgyrcons}
  \div T^\gyr = 0\period
\end{equation}

\section{The $2+d$ splitting of spacetime}\label{sc:2+d}

We would like to formulate the field equations in terms of quantities on the transverse
space ${N}$. First step is a restriction of spacetime tensor quantities
to their transverse components.

\subsection{The transverse tensors}

Transverse tensors can be viewed in two closely related ways. They are quantities
from tangent space of the typical transverse space ${N}$ which can additionally depend on two parameters ${u}$ and ${r}$. In such a picture we use the Latin tensor indices.

Alternatively, they can be understood as spacetime tensors which are tangent to the foliation ${N_\folph}$. In this view, to grasp the notion of transverse tensors the space of all spacetime tensors have to be decomposed into a direct sum of tensors tangent to the transverse spaces ${N_\folph}$ and tensors tangent to the temporal surfaces ${T_\folph}$.
Then we can write down the projector ${p}$ to the transverse space
\begin{equation}\label{transproj}
  p = {}^\st\!\delta-k\,\grad r - w \, \grad u
\end{equation}
(${{}^\st\!\delta}$ being the identity spacetime tensor with components ${\delta^\mu_\nu}$).
It annihilates vectors ${k}$ and ${w}$ and 1-forms ${\grad u}$ and ${\grad r}$.
In adjusted coordinates ${\{u,r,x^i\}}$ this projection just cancels ${u}$ and ${r}$
components and leaves the transverse components untouched.

In the spacetime picture we use the Greek tensor indices even for transverse tensors.

Let us note that spaces of tensors tangent to the foliation ${N_\folph}$ depend, in general, on a choice of the gauge, in contrast with the tangent space of the transverse manifold ${N}$ which is independent of the gauge. However, the identification of these two pictures is gauge dependent. This dichotomy is reason why to keep both views of the transverse objects.

Let us stress that the transverse projection ${p}$ is not, in general, orthogonal. As a consequence, we have non-diagonal components of the metric ${a_\mu=g_{u\nu}p^{\nu}{}_{\mu}}$. However, our assumption guarantees that it can be made orthogonal just using gauge transformation. Indeed, ${a}$ is ${r}$-independent, cf.~\eqref{assumptions}, and therefore it can be eliminated using the ${w}$-gauge \eqref{wgauge}.

\subsection{The transverse derivatives}

Next, we study a relation between spacetime and transverse space derivatives. We already introduced temporal derivatives \eqref{tempders} along ${k}$ and ${w}$. In spacetime they correspond to Lie derivatives, in the transverse space they are just parametric derivatives with respect to ${r}$ and ${u}$. The spacetime gradient of a scalar function ${f}$ can thus be split into temporal and transverse parts
\begin{equation}\label{stgradsplit}
    \grad f = \rder f\, \grad r + \uder f\,\grad u+\trgrad f\period
\end{equation}
As a consequence, the transverse space gradient ${\trgrad f}$ is the ${p}$-projection of the spacetime gradient
\begin{equation}\label{stgradproj}
    \trgrad f = p\cdot \grad f 
    \period
\end{equation}

Moreover, under condition ${\rder q =0}$, the transverse space covariant derivative ${\trcovd A}$ of a transverse tensor ${A}$ is also given by ${p}$-projection of the spacetime derivative ${\covd A}$,
\begin{equation}\label{covdproj}
    \trcovd A = \covd^{}_\trpr A{\,}^{\trpr\dots}_{\trpr\dots}\commae
\end{equation}
as can be checked in adjusted coordinates inspecting Christoffel symbols \cite{PodoZo:2009:CLAQG:, KrtouPod:2010:}.

To be able to split a general spacetime covariant derivative of a spacetime tensor we write this tensor as a sum of its temporal and transverse parts. Employing Leibniz rule the spacetime covariant derivative leads to sum of terms with covariant derivatives of the temporal frame and transverse tensors. As an example, for a vector field~${v}$ we get
\begin{equation}\label{covdvexample}
\begin{split}
    \covd v &= \covd\bigl( v^r k + v^u w +v^\trpr \bigr)\\
            &= v^r \covd k + \grad v^r\, k + v^u \covd w + \grad v^u\, w + \covd \vec v\commae
\end{split}
\end{equation}
with\footnote{%
We introduced symbol ${\vec v}$ for the transverse projection ${v^\trpr}$ since in the term ${\covd\vec v}$ the transverse projection is performed before the covariant derivative. The expression ${\covd v^\trpr}$ is reserved for the transverse projection of the upper index of the covariant derivative ${\covd v}$, as, e.g., in equation \eqref{covdproj}.}
${\vec v = v^\trpr = p\cdot v}$.
Next we have to calculate all temporal and transverse projections of these terms.

For that, it is necessary to calculate derivatives of vectors ${k}$, ${w}$ and 1-forms ${\grad u}$, ${\grad r}$. General expressions can be found in \cite{KrtouPod:2010:}, employing the assumptions \eqref{assumptions} they give
\begin{align}
  \covd_\mu\! k^\nu &=
     \rder H\,\grad[\mu] u\; k^\nu\commae\label{covdk}\\
  \begin{split}
  \covd_\mu\! w^\nu &=
      \rder H\,\grad[\mu] r\; k^\nu -\rder H \grad[\mu] u\; w^\nu\\
      &\nquad+\Bigl( \bigl (2H+a^2\bigr) \rder H
              +\uder H + \vec a^\kappa \uder a_\kappa + \vec a^\kappa\trgrad[\kappa] H\Bigr)\,\grad[\mu]u\; k^\nu\\
      &\nquad+\Bigl( \trgrad[\mu] H - \frac12 \vec  a^\kappa \trgrad[\kappa] a_\mu + \frac12 \vec a^\kappa \uder q_{\kappa\mu}  \Bigr)\, k^\nu\\
      &\nquad+ \grad[\mu] u\,\bigl(\rder H a_\kappa + \uder a_\kappa +\trgrad[\kappa] H\bigr)\, q^{\kappa\nu}\\
      &\nquad+\Bigl(\frac12 \trgrad[\mu] a_{\kappa}+\uder q_{\mu\kappa}\Bigr)\, q^{\kappa\nu}\commae
  \end{split}\label{covdw}\raisetag{48pt}
\end{align}
and
\begin{align}
  \covd_\mu\!\grad[\nu] u &=
     \rder H\,\grad[\mu] u\,\grad[\nu] u\commae\label{covddu}\\
  \begin{split}
  \covd_\mu\!\grad[\nu] r &=
      -2\rder H\;\grad[(\mu] u\; \grad[\nu)] r\\
      &\nquad-\Bigl( \bigl (2H{+} a^2\bigr) \rder H
              +\uder H + \vec a^\kappa \uder a_\kappa + \vec a^\kappa\trgrad[\kappa] H\Bigr)\grad[\mu] u\,\grad[\nu] u\\
      &\nquad-\Bigl( 2\,\trgrad[(\mu] H - \vec a^\kappa \trgrad[\kappa] a_{(\mu} + \vec a^\kappa  \uder q_{\kappa(\mu} \Bigr) \grad[\nu)] u\\
      &\nquad-\frac12 \uder q_{\mu\nu} + \trcovd_{(\mu} a_{\nu)}\period
  \end{split}\label{covddr}\raisetag{48pt}
\end{align}
It is straightforward to read out different temporal and transverse projections of these covariant derivatives. All these projections are already expressed using only transverse quantities.

Finally, we have to deal with projections of spacetime covariant derivatives of the transverse tensors (i.e., of the terms as ${\covd \vec v}$ in the example above). The full transverse projections of such terms reduces to the transverse covariant derivatives according to \eqref{covdproj} above. Most of the temporal projections can be reduced to projections of the terms \eqref{covdk}--\eqref{covddr}, e.g.,
\begin{equation}\label{covdvexampleproj}
    \bigl(\covd \vec v\bigr)\cdot\grad u = - \bigl(\covd\grad u \bigr)\cdot \vec v\commae
\end{equation}
since ${\vec v\cdot\grad u =0}$. Only remaining terms are of type ${k\cdot\covd\vec v}$ and ${w\cdot\covd\vec v}$. They can be reduced to Lie derivative along ${k}$ and ${w}$, i.e., into temporal derivatives with respect to ${r}$ and ${u}$, respectively.
For tensors with more indices we would have a term with ${\covd w}$ for each upper index and with ${-\!\covd w}$ for each lower index.

The resulting splitting even of the simplest case of the covariant derivatives of just a vector or a 1-form leads to lengthy expressions which are not useful in full generality. We employ the described procedure in splitting the field equations into their transverse equivalent which, fortunately,  usually gives reasonable results thanks to a special structure of the field quantities and of the field equations.

\subsection{The Ricci tensor}

To express the Einstein equations on the transverse space we need to know the projections of the spacetime Ricci tensor ${\Ric}$. For general Kundt class, they have been calculated in components in \cite{PodoZo:2009:CLAQG:} and expressed in the covariant form in \cite{KrtouPod:2010:}. Assuming \eqref{assumptions}, the projections restricted on the transverse space ${N}$ have the form
\begin{align}
  \Ric_{rr}&=0\commae\notag\\
  \Ric_{ri} &= 0\commae\notag\\
  \Ric_{ru} &= {\rdder{H}}\commae\label{Riccisplit}\\
  \Ric_{uu} &=
    \trLB H + (\trgrad a)\bullet(\trgrad a)
    +2{\rdder H}\Bigl(H{+}\frac12 a^2\Bigr) \notag\\
    &\quad+\trdiv {\uder a} + \rder H \trdiv a + 2a\cdot \trgrad {\rder H}
    - {\uder q}\bullet {\uder q}
    - \rder H \theta_u - {\uder \theta}_u \commae\notag\\
  \Ric_{u\trpr} &= -\frac12 \trdiv\trgrad a + \trgrad \rder H
     +\frac12 \trdiv{\uder q} - \trgrad \theta_u \commae\notag\\
  \Ric_{\trpr\trpr} &= \trRic\commae\notag
\end{align}
where ${\trRic}$ is the Ricci tensor of the transverse metric ${q}$. ${\theta_u}$ is an expansion of the congruence ${w}$,
\begin{equation}\label{thetaudef}
    \theta_u = \frac12\, q^{ij}\, {\uder q}_{ij} = \trivol(\trvol)\,\uder{}\period
\end{equation}
Clearly, it characterizes the rate of ${u}$-change of the transverse volume element ${\trvol=(\Det q)^{\frac12}}$.
The transverse Laplace-Beltrami operator ${\trLB}$, the transverse divergence ${\trdiv}$, and the form product ${\bullet}$ will be discussed in detail in the following section.

The spacetime scalar curvature ${\scR}$ can be expressed in terms of the transverse scalar curvature ${\trscR}$ as
\begin{equation}\label{scRsplit}
    \scR = -2\rdder H+ \trscR\period
\end{equation}

\subsection{The geometry on transverse space}

To fix a notation and sign conventions we shortly review some definitions regarding the Hodge theory on transverse space.

The transverse space is ${d}$-dimensional Riemanian space with the metric ${q}$. We use this metric to lower and raise Latin indices. As we already mentioned, we use a tiny arrow above or below a symbol to emphasize a vector or 1-form character of the transverse object in the index-free notation.

Metric ${q}$ and a chosen orientation fixes the Levi-Civita tensor ${\eps}$ which allows to define the Hodge dual of an antisymmetric ${p}$-form:
\begin{equation}\label{hodge}
  (*\omega)_{a_{p+1}\dots a_d}=\frac1{p!}\,\omega^{a_1\dots a_p}\eps_{a_1\dots a_d}\period
\end{equation}
It satisfies
\begin{equation}\label{invhodge}
  * * \omega = (-1)^{p(d-p)}\omega\commae
\end{equation}
where we assumed the positive definiteness of ${q}$.

The inner product on antisymmetric ${p}$-forms is defined
\begin{equation}\label{bulletdef}
  \omega\bullet\sigma = \frac1{p!}\,\omega_{a_1\dots a_p} \sigma_{b_1\dots b_p}\, q^{a_1b_1}\dots q^{a_pb_p}\commae
\end{equation}
which satisfies ${\omega\wedge(*\sigma)=\sigma\wedge(*\omega)=(\omega\bullet\sigma)\,\eps}$.
We will use the definition \eqref{bulletdef} also for symmetric ${p}$-forms.

We define the transverse divergence of a general ${p}$-form
\begin{equation}\label{divdefix}
  (\trdiv \omega)_{a_1\dots a_{p-1}} = \trcovd{}^i \omega_{ia_1\dots a_{p-1}}\period
\end{equation}
For antisymmetric ${p}$-forms the divergence is, up to a sign, the standard co-derivative ${\delta}$:
\begin{equation}\label{divdelta}
  \trdiv\omega = -\delta\omega=-(-1)^{p}*^{\!-\!1}\!\trgrad\!*\omega\period
\end{equation}

We define Laplace--de~Rham operator on the antisymmetric forms as\footnote{%
In our convention ${\trlapl}$ is negative definite operator and it has the same sign as the Laplace--Beltrami operator, cf.\ eq.~\eqref{WBident}.}
\begin{equation}\label{LRdef}
  \trlapl=\trgrad \trdiv+\trdiv\trgrad\period
\end{equation}
which is related to the Laplace--Beltrami operator
\begin{equation}\label{LBdef}
  \trLB=q^{ij}\trcovd_i\trcovd_j
\end{equation}
through the Weitzenb\"ock--Bochner identity
\begin{equation}\label{WBident}
\begin{split}
  \trlapl\omega_{a_1\dots a_p}&= \trLB\omega_{a_1\dots a_p}
    -p\,\trRic{}_{n[a_1}\omega^n{}_{a_2\dots a_p]}\\
    &\qquad+{\textstyle\frac{p(p{-}1)}2}\; \trRiem_{mn[a_1a_2}\omega^{mn}{}_{a_3\dots a_p]}\period
\end{split}
\end{equation}

\section{The Einstein--Maxwell equations}\label{ssc:fieldeq}
In this section we derive gradually the divergence of
the gyratonic source, the Maxwell and Einstein equations and  we discuss the possible decoupling of our equations.

\subsection{The gyraton stress-energy conservation}

The divergence of the stress-energy tensor \eqref{Tgyr} can be written
in the form
\begin{equation}\label{divT}
  \kap \div T^\gyr = \bigl((-j_u+j\cdot\vec{a})\,\rder{} + \trdiv j\bigr) \grad u - (j)\,\rder{}\period
\end{equation}
The condition \eqref{Tgyrcons} can thus be solved by setting
\begin{equation}\label{iotadef}
  j_u = r\trdiv j + \iota,
\end{equation}
with
\begin{equation}\label{jiotacond}
  (\iota)\,\rder{}= 0 \comma (j)\,\rder{} =0\period
\end{equation}

\subsection{The Maxwell equations}

The transverse projections of components of the Maxwell equations have a form,
\begin{align}
 \grad F &=0 \comma\label{M1}\\
 \div F &= 0.\label{M2}
\end{align}

From the first Maxwell equation \eqref{M1} we get the following
restrictions on the electromagnetic field,
\begin{equation}\label{trMaxwell1}
\begin{gathered}
  \rder B =0\comma \trgrad B = 0\commae\\
  \rder s = - \trgrad E\comma  \trgrad s = \uder B \commae
\end{gathered}
\end{equation}
then the second Maxwell equation \eqref{M2} gives another
additional conditions
\begin{equation}\label{trMaxwell2}
\begin{gathered}
  \rder E =0\commae\\
  \bigl(E a +\vec{a}\cdot B -s\bigr)\,\rder{} = - \trdiv B \commae\\
  \trdiv\bigl(E a +\vec{a}\cdot B -s\bigr) = \uder E + \theta_u E.
\end{gathered}
\end{equation}
We observe that it is possible to solve the ${r}$ dependence of ${s}$ by setting
\begin{equation}\label{sigmadef}
  s= -r\,\trgrad E + \sigma\comma\rder\sigma=0\period
\end{equation}

The Maxwell equations are then equivalent to following expressions
\begin{subequations}\label{trMaxwell}
\begin{gather}
  \rder E =0 \comma \rder B =0\comma \rder\sigma = 0\commae\label{MEr}\\
  \uder B = \trgrad \sigma\comma\trgrad B = 0\commae\label{MEB}\\
  \trgrad E + \trdiv B = 0\commae\label{MEEB}\\
  \trdiv\bigl(E a +\vec{a}\cdot B -\sigma\bigr) = \uder E + \theta_u E\commae\label{MEEBs}
\end{gather}
\end{subequations}
where we used that ${\trdiv(\trdiv B)=0}$.

Let us note that as a consequence of the Maxwell equation we also get splitting of the 1-form ${\trgrad_i E\; q^{ij} B_{ja}}$ into its gradient and divergence part:
\begin{equation}\label{dEB}
\trgrad E \cdot q^{\!-1} \cdot B = \trdiv(E B)+\frac12\trgrad(E^2)\period
\end{equation}

\subsection{The Einstein equations}

Finally, we will derive the Einstein equations,
\begin{equation}\label{Einstein}
  \Ric -\frac12\scR+\Lambda g = \kap T^\tot,
\end{equation}
where  we assume the total stress-energy tensor in the form
\begin{equation}\label{Ttot}
  \kap T^\tot =  \rho \grad r\vee\grad u + j_u^\tot \grad u \grad u + \grad u \vee j^\tot + \kap T^\trspc\commae
\end{equation}
with the sources $j_{u}$, $j$ and $T^\trspc$ defined as
\begin{equation}\label{jtot}
\begin{gathered}
  j_u^\tot = 2H \rho + \kap\epso (Ea-s)^2 + r \trdiv j + \iota\commae\\
  j^\tot = -\rho a - \kap\epso (\vec{s} - E\vec{a})\cdot(E q + B)+ j\commae\\
  T^\trspc = \frac\epso2 E^2 q + T^\mg \period
\end{gathered}
\end{equation}
when the aligned electromagnetic field and gyratonic matter is present (${\rho}$ is given by \eqref{rhotaudef}).

Since the gyratonic stress-energy tensor is trace-free, we have
\begin{equation}\label{taudef}
  (D-4)\tau = \kap T^\tot{}^\mu_\mu \period
\end{equation}
Trace and trace-free part of the total transverse stress-energy tensor are
\begin{equation}\label{trace+tracefree}
\begin{gathered}
  \kap T^\trspc{}^i_i = 2\rho + (D-4)\, \tau\commae\\
  \frac1\epso T^\trspc_\TF{}_{ij} = B_{ik} B_{jl} \,q^{kl} - \frac{2}{d}\, B^2 q_{ij}\period
\end{gathered}
\end{equation}

Substituting \eqref{Riccisplit} and \eqref{Ttot} into \eqref{Einstein}, we easily check
that the components ${rr}$ and ${r\trpr}$ of the Einstein equations trivially vanishes. The ${ru}$ components gives
\begin{equation}\label{EEru}
  \frac12\trscR = \rho + \Lambda\period
\end{equation}
The trace of the Einstein equations implies
\begin{equation}\label{EEtrace}
  \frac12\trscR-\rdder H = -\frac{D-4}{D-2}\tau + \frac{D}{D-2}\Lambda\period
\end{equation}
Eliminating the transverse scalar curvature from the last two equations we obtain the equation for ${\rdder H}$,
\begin{equation}\label{hrdder}
    \rdder H =  \rho + \frac{D-4}{D-2}\,\tau -\frac2{D-2}\, \Lambda \period
\end{equation}
Taking into account the first two of the equations \eqref{trMaxwell} we find that ${H}$ can be written as
\begin{equation}\label{rdependofH}
    H = \frac12\Bigl(\rho + \frac{D-4}{D-2}\,\tau-\frac2{D-2}\, \Lambda\Bigr)\, r^2 + g\, r + h\commae
\end{equation}
where the functions\footnote{%
Here we use ${g}$ to denote the ${r}$-linear term of ${H}$---the same letter as for the spacetime metric ${g}$. Since the discussion of the field equations will be done on the transverse space a  threat of the confusion is minimal.}
${g}$ and ${h}$ are possibly ${u}$-dependent scalar functions on the transverse space.

The trace of the transverse part of the Einstein equations is linear combination of equations \eqref{EEru} and \eqref{EEtrace}. The trace-free part together with equation \eqref{EEru} gives the equation for the transverse metric,
\begin{equation}\label{EEq}
    \trRic = \frac{2}{D-2}(\rho+\Lambda)\,q + \kap T^\trspc_\TF\period
\end{equation}
As a consequence of a vanishing divergence of the Einstein tensor one obtains
\begin{equation}\label{divEB}
    \trdiv(EB) = \frac1d\,\trgrad\tau\period
\end{equation}

The ${u\trpr}$ component gives equations in the form,
\begin{equation}\label{EEuT}
\begin{split}
    &-\frac12 \trdiv f  + \rdder H a
    +\kap\epso (\vec{s} -E \vec{a}) \cdot (E q+ B) + \trgrad\rder H\\
    &\qquad\qquad
    =\trgrad \theta_u - \frac12 \trdiv \uder q + j\commae
\end{split}\raisetag{18pt}
\end{equation}
with 2-form ${f}$ being just a shorthand for ${\trgrad a}$
\begin{equation}\label{fdefff}
    f = \trgrad a\period
\end{equation}
This expression is linear in $r$ (with $r$-dependence hidden just in $\rder H$ and $\vec{s}$).
Using \eqref{divEB} it can be shown that $r$ term is a consequence of the already known field equations. The $r$ independent part gives the equation for $a$,
\begin{equation}\label{aeq}
\begin{split}
    &-\frac12 \trdiv f  + \rdder H a
    +\kap\epso (\vec{\sigma} -E \vec{a}) \cdot (E q+ B) + \trgrad g\\
    &\qquad\qquad
    =\trgrad \theta_u - \frac12 \trdiv \uder q + j\period
\end{split}
\end{equation}

Finally, the ${uu}$ component leads to an expression quadratic in $r$. Using \eqref{dEB}, \eqref{divEB}, and the fact that $E$ is harmonic (which is a consequence of \eqref{MEEB}) one can show that the quadratic term is trivial. The linear term gives,
\begin{equation}\label{laplgg}
\begin{split}
    &\trlapl g + 2 \vec{a}\cdot\trgrad\rdder H + (\trdiv a - \theta_u)\rdder H \\
    &\qquad\qquad+2\kap\epso (\vec{\sigma} -E \vec{a}) \cdot \trgrad E = \trdiv j\period
\end{split}
\end{equation}
It turns out that this equation is equivalent to divergence of \eqref{aeq}. Their difference leads to an expression equivalent to the trace of the $u$-derivative of \eqref{EEq}. To show that, it is useful to write down the geometrical relation
\begin{equation}\label{laplg}
  {\uder\trRic}_{ij}\;q^{ij} = -2\trlapl\theta_u + \frac12\trcovd^{\,i}\trcovd^{\,j}\uder{q}_{ij}\period
\end{equation}
which is a consequence of the fact that, thanks to relation ${\trcovd q =0}$, the transverse covariant derivative can be $u$-dependent.

The remaining $r$-independent part of  the ${uu}$ component of the Einstein equations gives the equation for $h$,
\begin{equation}\label{heqq}
\begin{split}
    &\trlapl h + 2 \vec{a}\cdot\trgrad g + \rdder H a^2 + (\trdiv a {-} \theta_u)\,g \\
    &\qquad - \kap\epso (\sigma {-}E a)^2 +\frac12 f^2 = \frac12 {\uder q}^2 + \trdiv a + \uder \theta_u + \iota\period
\end{split}\raisetag{8ex}
\end{equation}

\subsection{Decoupling the equations}\label{scc:decouple}

First, let us note that the equation for the transverse metric \eqref{EEq} and for the electromagnetic field \eqref{trMaxwell} are coupled, they cannot be solved one after another. It significantly complicates finding the solution. However, we can restrict the generality of the electromagnetic field in such a way that we obtain a solvable system describing development of a gyratonic matter accompanied by the gravitational wave in a non-dynamical electromagnetic field.

Namely, we will restrict to the cases, when the right-hand side of \eqref{EEq} is given just by tensors obtained in an algebraic way from the transverse metric ${q}$, possibly assuming some special topological or geometrical structure of the transverse space.

The initial motivation of this investigation was to find higher dimensional generalization of the gyratons on direct product
spacetimes where the electromagnetic field is characterized by two constants $E$ and $B$.
So we will discuss the situation when the electric field is taken to be constant, ${E=\text{const}}$. In the first case, we assume that the magnetic part is missing, ${B=0}$. In the second case we assume that the geometry of the transverse space is given by a product of two-dimensional spaces and the magnetic field is given by a linear combination of canonic 2-forms on these two-dimensional components.

In both these cases the right-hand-side of \eqref{EEq} depends only on a finite number constants characterizing the electromagnetic field and the preselected form of the transverse geometry. Choosing the electromagnetic constants we can find the transverse geometry and on this background solve other equations.

The restriction imposed on the electromagnetic field is not actually excessively strong. Taking gradient or divergence of the third of the equations \eqref{trMaxwell} we find that both ${E}$ and ${B}$ must by harmonic 0-form and 2-form, respectively,
\begin{equation}\label{harmEB}
    \trlapl E=0\comma \trlapl B =0 \period
\end{equation}
Imposing a finiteness of the fields in infinity of the transverse space or restricting to compact transverse space guarantees that ${E}$ is constant and ${B}$ nontrivial only for some topologically special spaces. If we additionally assume that both ${\rho}$ and ${\tau}$ are constants (which simplifies the structure of the function ${H}$, cf.\ eq.~\eqref{hrdder}), we obtain the condition ${B^2=\text{const}}$. Nontrivial harmonic \mbox{2-forms}~${B}$ with a constant square can exist only in very special spaces of which the direct-product spaces are significant representatives.

\section{Final remarks}
To conclude this chapter let us note that this work is new and still in progress \cite{Krtous:2010:}.
The remaining problems of the presented work are the discussion of the decoupling of the Einstein--Maxwell
equations as it is mentioned in the section \eqref{scc:decouple} and the discussion of these equations in different
gauges.
It would be interesting to study special gyratonic cases to understand more to the physical interpretation
of the higher dimensional gyraton spacetimes.

\newpage
\chapter{Conclusions and future prospects}
We have found and investigated gyraton solutions on various backgrounds.
Namely, we have found the gyratons on direct product spacetimes \cite{Kadlecova:2009:PHYSR4:}, the gyratons on
Melvin universe \cite{Kadlecova:2010:PHYSR4:} and its generalization with possible non--trivial cosmological
constant \cite{KadlKrt:2010:CLAQG:}. These solutions are of Petrov type II and their backgrounds are
generally of type D.

The solutions have similar but more complicated properties as the reviewed solutions in the introduction:
The Einstein--Maxwell equations reduce to a set of equations on the 2--dimensional
transverse space, in the Newman--Penrose formalism the new gyratonic terms $a_{i}$
generate the $\Psi_{3}$ and $\Phi_{12}$ components of the Weyl and Ricci scalars
(apart from the terms from the function $H$),
the property that the scalar polynomial invariants are the same for the full gyratonic
metric and for the background itself is also valid for the gyratons we have found.

We understand better the interpretation of gyratonic solutions. From the Einstein--Maxwell
equations we observed that the gyratonic terms $a_{i}$ are the consequence of the
gyratonic sources with internal rotation (spin). This internal rotation
was discussed in Minkowski spacetime \cite{Fro-Fur:2005:PHYSR4:, Fro-Is-Zel:2005:PHYSR4:}
 and either in anti--de Sitter \cite{Fro-Zel:2005:PHYSR4:}.
More interesting and new is character of the source $j_{u}$ which must be conserved.
The non--trivial divergence of $j$ describes an internal flow of the energy in the
gyraton beam which changes its internal energy $j_{u}$ with $r$. We call this effect as
some cooling which steadily decreases the energy density of the gyratonic beam.
This is not plausible physically and it leads to unnatural causal behaviour of the source.

We have also studied the type III Kundt spacetimes in four dimensions to find
gyraton solutions on de Sitter spacetime. We have investigated the Einstein
equations for these gyratonic spacetimes. This work is still in progress \cite{KadlKr:2010:}.

Another direction mentioned in the thesis was generalization of gyratons on direct product
spacetimes to higher dimensions where we had to deal with more complicated structure of the transversal
space.

\vspace{1em}
The work in progress and our future perspectives involve:
(i) include the electromagnetic field to the Kundt type III spacetimes,
(ii) discussion of the character of dS and AdS gyratons,
(iii) extension of the gyraton solution to the Robinson--Trautman expanding class of spacetimes,
 (iv) looking for more general gyratonic solutions of type II and III in the Kundt class of spacetimes,
  (v) find the higher dimensional generalization to known solutions in four dimensions,
   (vi) study the geodesics motion in the newly found gyraton solutions which is completely an open problem.



\cleardoublepage
\newpage
\bibliographystyle{unsrt}
\bibliography{mynames}
\end{document}